\documentclass[11pt]{book}
\usepackage{a4,exscale,epsf}
\usepackage{epsfig}
\usepackage{latexsym}  %
\usepackage{amssymb} 
\usepackage{deluxetable} %
\usepackage{chap}%
\usepackage{changebar}
\usepackage{makeidx}
\makeindex
\usepackage{fancyhdr}         %
\usepackage{natbib}
\newfont{\sfsl}{cmssqi8 scaled 1200}
\newfont{\sfslp}{cmssqi8 scaled 1250}
\newfont{\sfsls}{cmssqi8 scaled 900}
\newfont{\sfsln}{cmssqi8 scaled 1350}
\newfont{\sfslz}{cmssqi8 scaled 1500}
\newfont{\sfslzz}{cmssqi8 scaled 1150}
\newfont{\sfslms}{cmssqi8 scaled 1000}	%
\newfont{\sfsla}{cmssqi8 scaled 3000}
\newfont{\sfslb}{cmssqi8 scaled 2200}
\newcommand{\gcs}{{\sfsln HIFLUGCS}} %
\newcommand{\gcss}{{\sfslms HIFLUGCS}} %
\newcommand{\ash}{{\,\mbox{arcsinh}\,}}

\newcommand{\ashn}{{\,\mbox{arcsinh}}} %

\newcommand{\ro}{{\rm ROSAT}}
\newcommand{\ra}{{\rm RASS}}
\newcommand{\ps}{{\rm PSPC}}
\newcommand{\as}{{\rm ASCA}}
\newcommand{\xs}{{\rm XSPEC}}
\newcommand{\xmm}{{\rm XMM-Newton}}

\newcommand{\cha}{{\rm Chandra}}
\newcommand{\icg}{{\rm ICG}}

\newcommand{\met}{g^{\mu\nu}}
\newcommand{\rt}{R_{\mu\nu\sigma}^\tau}
\newcommand{\rta}{R_{\mu\nu\alpha}^\alpha}
\newcommand{\rtu}{R_{\mu\nu}}
\newcommand{\rto}{R^{\mu\nu}}
\newcommand{\eto}{G^{\mu\nu}}

\newcommand{\eit}{T^{\mu\nu}}
\newcommand{\of}{\Omega_f}
\newcommand{\om}{\Omega_{\rm m}}
\newcommand{\omn}{\Omega_{\rm m,0}}
\newcommand{\ok}{\Omega_{\rm k}}
\newcommand{\ol}{\Omega_{\Lambda}}
\newcommand{\oln}{\Omega_{\Lambda,0}}
\newcommand{\ob}{\Omega_{\rm b}}
\newcommand{\oc}{\Omega_{\rm Cluster}}
\newcommand{\obc}{\Omega_{\rm b, Cluster}}
\newcommand{\roc}{\rho_{\rm c}}
\newcommand{\rom}{{\rho_{\rm m}}}
\newcommand{\drm}{{p_{\rm m}}}
\newcommand{\dc}{\delta_{\rm c}}
\newcommand{\dcv}{\delta_{\rm c}^{\rm v}}

\newcommand{\sx}{S_{\rm X}}
\newcommand{\fx}{f_{\rm X}}
\newcommand{\fxl}{f_{\rm X,lim}}
\newcommand{\flim}{f_{\rm lim}}
\newcommand{\cx}{C_{\rm X}}
\newcommand{\lx}{L_{\rm X}}

\newcommand{\lnu}{L_{\nu}}
\newcommand{\epsnu}{\epsilon_{\nu}}
\newcommand{\lamnu}{\Lambda_{\nu}}
\newcommand{\lbol}{L_{\rm Bol}}
\newcommand{\fbol}{f_{\rm Bol}}
\newcommand{\ncl}{N_{\rm Cl}}
\newcommand{\rx}{r_{\rm X}}
\newcommand{\rc}{r_{\rm c}}
\newcommand{\rch}{R_{\rm ch}}
\newcommand{\rab}{r_{\rm A}}
\newcommand{\mab}{M_{\rm A}}
\newcommand{\mpr}{m_{\rm p}}
\newcommand{\rogal}{\rho_{\rm gal}}
\newcommand{\rog}{\rho_{\rm gas}}
\newcommand{\ngas}{n_{\rm gas}}
\newcommand{\rot}{\rho_{\rm tot}}
\newcommand{\fg}{f_{\rm gas}}
\newcommand{\rov}{\rho_{\rm vir}}

\newcommand{\nh}{n_{\rm H}}
\newcommand{\nhcol}{n_{\rm h}}
\newcommand{\nhe}{n_{\rm He}}
\newcommand{\nel}{n_{\rm e}}
\newcommand{\emi}{E_{\rm m}}
\newcommand{\pg}{P_{\rm gas}}
\newcommand{\tg}{T_{\rm gas}}
\newcommand{\tx}{T_{\rm X}}
\newcommand{\mt}{M_{\rm tot}}
\newcommand{\mti}{M_{\rm tot,min}}
\newcommand{\mtz}{M_{200}}
\newcommand{\mtzi}{M_{200,i}}
\newcommand{\mtf}{M_{500}}
\newcommand{\mg}{M_{\rm gas}}
\newcommand{\mgz}{M_{{\rm gas}, 200}}
\newcommand{\mgf}{M_{{\rm gas}, 500}}
\newcommand{\fgz}{f_{{\rm gas}, 200}}
\newcommand{\fgf}{f_{{\rm gas}, 500}}
\newcommand{\vmax}{V_{\rm max}}
\newcommand{\vmaxi}{V_{{\rm max},i}}
\newcommand{\dlm}{D_{\rm L, max}}
\newcommand{\dl}{D_{\rm L}}
\newcommand{\dm}{D_{\rm M}}
\newcommand{\da}{D_{\rm A}}
\newcommand{\zm}{z_{\rm max}}
\newcommand{\bii}{b_{\rm II}}
\newcommand{\lii}{l_{\rm II}}
\newcommand{\csh}{C_{\rm s}}
\newcommand{\sll}{\sigma_{\log\lx}}
\newcommand{\slm}{\sigma_{\log\mt}}

\newcommand{\fobsl}{f^{\rm OF}_{{\rm lim}\,[0.1-2.4\,\rm keV]}}
\newcommand{\lobs}{L^{\rm OF}_{[0.1-2.4\,\rm keV]}}

\newcommand{\lsou}{L^{\rm SF}_{[0.1-2.4\,\rm keV]}}
\newcommand{\md}{{\rm MD}}
\newcommand{\ngx}{{N_{\rm gx}}}
\newcommand{\lbj}{{L_{Bj}}}

\newcommand{\mpc}{h_{50}^{-1}\,{\rm Mpc}}
\newcommand{\kpc}{h_{50}^{-1}\,{\rm kpc}}
\newcommand{\kev}{{\rm keV}}
\newcommand{\eh}{0.5-2.0\,{\rm keV}}
\newcommand{\eb}{0.1-2.4\,{\rm keV}}
\newcommand{\ek}{0.64-2.36\,{\rm keV}}
\newcommand{\ebol}{0.01-40\,{\rm keV}}
\newcommand{\cts}{{\rm cts\,s^{-1}}}
\newcommand{\esc}{\times10^{-11}\,{\rm ergs\,s^{-1}\,cm^{-2}}}
\newcommand{\escl}{\times10^{-12}\,{\rm ergs\,s^{-1}\,cm^{-2}}}

\newcommand{\esct}{10^{-11}\,{\rm ergs\,s^{-1}\,cm^{-2}}}
\newcommand{\escc}{\times10^{-10}\,{\rm ergs\,s^{-1}\,cm^{-2}}}
\newcommand{\esl}{h_{50}^{-2}\, 10^{40}\,{\rm ergs\,s^{-1}}}
\newcommand{\esll}{h_{50}^{-2}\, 10^{44}\,{\rm ergs\,s^{-1}}}
\newcommand{\esls}{h_{50}^{-2}\, 10^{43}\,{\rm ergs\,s^{-1}}}
\newcommand{\es}{{\rm ergs\,s^{-1}}}
\newcommand{\erg}{{\rm ergs}}
\newcommand{\msug}{h_{50}^{-5/2}\,M_{\odot}}
\newcommand{\msu}{h_{50}^{-1}\,M_{\odot}}
\newcommand{\mlsu}{h_{50}\,M_{\odot}/L_{\odot}}

\newcommand{\kb}{k_{\rm B}}
\begin{document}
\title{ \vspace{-7cm} \centerline{ \sf LUDWIG--MAXIMILIANS--UNIVERSIT\"AT
M\"UNCHEN }
\vspace{-0mm} \centerline{ \epsffile{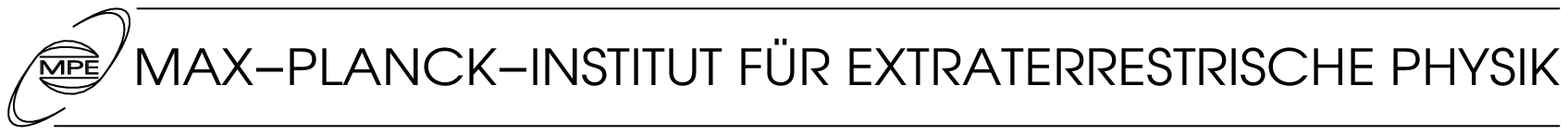}}
\hspace{1mm}  \\
\hspace{1mm}  \\
\hspace{1mm}  \\
\hspace{1mm}  \\
\hspace{1mm}  \\
\hspace{1mm}  \\
\hspace{1mm}  \\
\hspace{1mm}  \\
\hspace{1mm}  \\
\Huge Cosmological Implications and Physical Properties of an
X-Ray Flux-Limited Sample of Galaxy Clusters}

\author{\Large Dissertation der Fakult\"at f\"ur Physik\\
\Large der\\ \Large Ludwig-Maximilians-Universit\"at M\"unchen
\\ \\ \\ \\
\Large Vorgelegt von Thomas H. Reiprich aus Neuwied\\}
\date{M\"unchen, den 27.\ Juli 2001}
\maketitle
\clearpage
\phantom{x}
\clearpage
{\large
\noindent
\vspace{16cm}\phantom{x}\\
1. Gutachter: Prof.\ G. E. Morfill

\vspace{0.5cm}

\noindent
2. Gutachter: Prof.\ R. Bender

\vspace{1cm}

\noindent
Tag der m\"undlichen Pr\"ufung: 10. Dezember 2001
}
\clearpage
\phantom{x}
\clearpage
{\large
\noindent
\vspace{4cm}\phantom{x}\\
\center \emph{To my angel}
\vspace{13cm}
\phantom{x}\\

\vspace{1cm}

\noindent
}
\setcounter{chapter}{-1} %
\chapter{Zusammenfassung}
\label{zusam}

Ein Hauptziel dieser Arbeit ist die Bestimmung der mittleren
Materiedichte im Universum. Die Materiedichte ist ein wesentlicher
kosmologischer Parameter, der die Zukunft des Universums als Ganzem
mitbestimmt. Es wurde dazu
eine r\"ontgen\-selektierte und r\"ontgen\-flu\ss \-be\-grenzte Stichprobe 
der 63 r\"ontgenhellsten Galaxienhaufen am Himmel (ohne das galaktische
Band, genannt \gcs ) zusammengestellt, basierend auf der \ro\
Himmelsdurchmusterung. Die Flu\ss grenze betr\"agt $2\,\esc$ im Energieband
$\eb$. Anhand mehrerer Tests wurde gezeigt,
da\ss\ eine hohe Vollst\"andigkeit erreicht wurde. Diese Stichprobe kann,
aufgrund der hoch angesetzten Flu\ss grenze, f\"ur eine Vielzahl von
Anwendungen benutzt werden, die eine statistische Galaxienhaufenstichprobe
ben\"otigen, ohne Korrekturen an das effektive Durchmusterungsvolumen
anbringen zu m\"ussen.

Zur Bestimmung von
Fl\"ussen und physikalischen Haufenparametern wurden haupts\"achlich
tief belichtete pointierte Beobachtungen benutzt. Es wurde gezeigt, da\ss\
zwischen der R\"ontgen\-leucht\-kraft und der gravitativen Masse
eine enge Korrelation besteht, wobei \gcs\ und eine
erweitere Stichprobe von 106 Galaxienhaufen benutzt wurde.
Die Relation und die Streuung wurden quantifiziert mit Hilfe verschiedener
Anpassungsmethoden. Ein Vergleich mit einfachen
und erweiterten theoretischen und numerischen Vorhersagen
zeigt insgesamt \"Ubereinstimmung. In gro\ss en
R\"ontgenhaufendurchmusterungen oder Simulationen dunkler Materie kann diese
Relation direkt f\"ur Konvertierungen zwischen der R\"ontgen\-leucht\-kraft
und der gravitativen Masse angewendet werden.

Daten des Galaxienhaufens Abell 1835, aufgenommen w\"ahrend der
`Performance Verification' Phase des k\"urzlich gestarteten
R\"ontgensatellitenobservatoriums \xmm\ wurden ausgewertet, um die in
dieser Arbeit benutzte Annahme, da\ss\ das Intrahaufengas in den
\"au\ss eren Gebieten des Haufens isotherm ist, zu testen.
Es wurde gefunden, da\ss\ das gemessene
\"au\ss ere Temperaturprofil konsistent mit einem isothermen Profil ist.
In den inneren Regionen wurde ein klarer Abfall der Gastemperatur um einen
Faktor zwei gefunden.

Physikalische Eigenschaften der Galaxienhaufenstichprobe wurden untersucht,
indem Relationen zwischen verschiedenen Haufenparametern analysiert wurden.
Die Gesamteigenschaften sind gut verstanden, aber im Detail ergaben sich
Abweichungen von einfachen Erwartungen. Es wurde gefunden, da\ss\ der
Anteil der Gasmasse an der Gesamtmasse nicht als Funktion der Temperatur des
Intrahaufengases variiert. F\"ur Galaxiengruppen ($\kb\tx\lesssim 2$\,keV)
wurde jedoch ein steiler Abfall dieses Anteils gefunden. Keine klare Tendenz
f\"ur eine Variation des Oberfl\"achenhelligkeitsprofils, d.h.\ $\beta$,
als Funktion der Temperatur wurde beobachtet. Es wurde gefunden, da\ss\
die
Relation zwischen der R\"ontgenleuchtkraft und der Temperatur
steiler als von einfachen
selbst\"ahnlichen Modellen erwartet verl\"auft, wie be\-reits in
fr\"uheren Arbeiten festgestellt. Allerdings wurden keine klaren
Abweichungen von der Form eines Potenzgesetzes bis zu einer
gemessenen Gastemperatur $\kb\tx=0.7$\,keV gefunden. Die hier
gefundene Relation zwischen der Gesamtmasse und der Temperatur ist steiler
als von selbst\"ahnlichen Modellen erwartet und die Normierung ist niedriger
im Vergleich zu hydrodynamischen Simulationen, in \"Ubereinstimmung mit
fr\"uheren Resultaten. Vorgeschlagene Szenarien, darunter Heiz- und
K\"uhlprozesse, zur Beschreibung dieser Abweichungen und Schwierigkeiten
bei dem Beobachtungsproze\ss\ wurden dargestellt. Es scheint, da\ss\ eine
\"Uberlagerung verschiedener Effekte, m\"oglicherweise inklusive
einer Ver\"anderung der mittleren Entstehungsrotverschiebung als Funktion
der Galaxienhaufenmasse, ben\"otigt ist, um die hier vorgelegten
Beobachtungen zu beschreiben.

Unter Benutzung von \gcs\ wurde
die gravitative Massenfunktion in dem Massenintervall
$3.5\times 10^{13}< \mtz < 5.2\times 10^{15}\,\msu$ bestimmt.
Vergleich mit Press--Schechter Massenfunktionen f\"uhrte dazu, da\ss\
die mittlere Materiedichte im Universum und die Amplitude der
Dichtefluktuationen eng eingegrenzt werden konnten.
Das gro\ss e \"uberdeckte Massenintervall erlaubte eine individuelle
Eingrenzung der Parameter.
Im einzelnen wurde gefunden, da\ss\ 
$\om=0.12^{+0.06}_{-0.04}$ und $\sigma_8=0.96^{+0.15}_{-0.12}$ (90\,\%
Konfidenzintervall statistische Unsicherheit).
Dieses Resultat ist konsistent mit zwei weiteren unterschiedlichen
Absch\"atzungen f\"ur $\om$ in dieser Arbeit.
Der mittlere Anteil des Intrahaufengases an der Gesamtmasse von
Galaxienhaufen, bestimmt mit
Hilfe einer erweiterten Stichprobe von 106 Haufen, kombiniert mit Vorhersagen
der Theorie der Elemententstehung deutet an, da\ss\ $\om\lesssim 0.34$.
Das Masse zu Licht Verh\"altnis in den Haufen multipliziert mit 
der mittleren Leuchtkraftdichte impliziert $\om\approx 0.15$.
Eine Anzahl von Tests auf systematische Unsicherheiten wurde duchgef\"uhrt,
darunter ein Vergleich der Press--Schechter Massenfunktion
mit den neuesten Resultaten von gro\ss en Vielteilchenrechnungen. Diese
Tests ergaben Abweichungen kleiner als die statistischen Unsicherheiten.
Zum Vergleich wurden die Werte der besten Anpassung von $\om$ f\"ur gegebenes
$\sigma_8$ bestimmt, was zu der Relation $\sigma_8=0.43\,\om^{-0.38}$
f\"uhrte.

Die Massenfunktion wurde integriert, um den Anteil an der gesamten
gravitativen Masse im Universum zu bestimmen,
der in Galaxienhaufen enthalten ist.
Normiert auf die kriti\-sche Dichte ergab sich
$\oc = 0.012^{+0.003}_{-0.004}$ f\"ur Galaxienhaufenmassen gr\"o\ss er als 
$6.4^{+0.7}_{-0.6}\times 10^{13}\,\msu$. Dies impliziert mit dem hier
bestimmten Wert f\"ur $\om$, da\ss\ sich ca.\ 90\,\% der Gesamtmasse des
Universums au\ss erhalb von virialisierten Haufenregionen befindet.
Auf \"ahnliche Weise wurde gefunden, da\ss\ der Anteil des Intrahaufengases
an der Gesamtmasse des Universums mit
$\obc = 0.0015^{+0.0002}_{-0.0001}\,h_{50}^{-1.5}$
f\"ur Gasmassen gr\"o\ss er als
$6.9^{+1.4}_{-1.5}\times 10^{12}\,\msug$ sehr klein ist.

\tableofcontents
\chapter{Introduction}
\label{intro}
\pagenumbering{arabic}

Observational cosmology is currently a very active field of astronomical
research. The aim is the determination of the past, present,
and future status of the universe. {\bf Optical} 
observations of the magnitude--redshift
relation of distant 
supernovae (SNe) indicate an \emph{accelerating} universe,
i.e.\ $2\,\ol>\om$ \citep[e.g.,][]{pag99},
where $\om$ is the normalized mean matter density of the universe
and $\ol$ the normalized cosmological constant. 
In the {\bf microwave} regime, measurements of the fluctuations in the Cosmic
Microwave Background (CMB) from recent satellite and ballon borne
experiments indicate a \emph{flat} geometry, i.e.\ $\om+\ol = 1$
\citep[e.g.,][]{dab00}. 
From the {\bf X-ray} side, clusters of galaxies, as the most massive
collapsed objects
known in the universe, are the ideal and most commonly
used cosmological probes and indicate a \emph{low density} universe,
i.e.\ $\om<1$, almost independent of $\ol$. 

The SNe and CMB measurements clearly are very important cosmological
tests, nevertheless both
suffer from inherent problems. The SNe are used as standard candles but the
detailed physical processes that take place
during a SN explosion are not well understood even at redshift $z\approx 0$.
Furthermore additional dimming of the apparent brightness caused by
gas and dust in the host galaxy of a
distant SN is difficult to quantify. The presence of outliers in the
SN distribution is worrying and a large number of objects is needed to
clearly identify such events (early results based on a smaller number
of objects actually seemed to indicate $\ol \approx 0$,
\citealt{pgg97}).
CMB measurements yield information on the
state of the universe at $z\sim1000$, whereas galaxy clusters yield
information from relatively low redshifts ($0 < z \lesssim 1$).
It is thus important to take advantage of an independent method using
galaxy clusters as performed in this work.

Galaxy clusters can be used in a variety of ways to constrain cosmological
parameters.
For instance one may determine the typical mass to light ratio in clusters
and multiply it by the measured total luminosity
density to determine the mean mass density
\citep[e.g.,][and Sect.~\ref{relat_d}]{bld95}. The underlying assumption is
that the mass to light ratio in clusters is a good approximation to the 
mass to light ratio of the universe.
Another method uses
the amount of mass contained in the intracluster
gas
as compared to the total gravitational cluster mass
to set an upper limit on $\om$ by comparison to predictions for the baryon
density from the
theory of nucleosynthesis \citep[e.g.,][and Sect.~\ref{funct_d}]{wne93}.
This approach assumes that the gas fraction in clusters resembles the
baryon fraction in the universe. Furthermore 
within the framework of hierarchical structure formation -- small
objects, e.g.\ galaxies, form first and assemble to larger structures,
e.g.\ galaxy clusters, afterwards under the influence of gravity -- the merger
rate depends also on $\om$ \citep[e.g.,][]{lc93}. A comparison of observed
cluster substructure frequencies with predictions of specific models therefore
allows in principal to put constraints on important parameters
\citep[e.g.,][]{sbr00}. Moreover within this framework
analytical prescriptions have been developped, which allow statistical
predictions of the cosmic mass distribution for given cosmological models
\citep[e.g.,][]{ps74,bce91,b91,lc93,ks96,sba00}. These predictions
have been tested 
against a number of $N$-body simulations and in general
good agreement has been found over wide mass ranges
\citep[e.g.,][]{efw88,lc94,gbq99,jfw01}\footnote{A more detailed
discussion is presented in Sect.~\ref{back:tmf}.}. In this work
the mean density of the universe is estimated utilizing three different
methods. Most weight is given to the method, where
the local
cluster mass distribution is determined and compared to predictions,
because the mass distribution is
the most fundamental predicted quantity from cosmological
models and simulations of structure formation. 

The observed galaxy
cluster mass function, i.e.\ the cluster number density as a function
of mass, is particularly sensitive to the fundamental cosmological parameters
$\om$ and amplitude of the density fluctuations (e.g., \citealt{ha91,bc92}).
Previous local galaxy cluster mass functions have been derived by
\citet{bc93}, \citet{bgg93}, \citet{gbg98}, and
\citet[for galaxy groups]{gg00}.
\citet{bc93} used the galaxy richness (measure of the cluster galaxy content,
Sect.~\ref{back:galax}) to relate to cluster
masses for optical observations and an X-ray temperature--mass relation
to convert the temperature function given by \citet{ha91} to a mass
function. \citet{bgg93}, \citet{gbg98}, and \citet{gg00} used velocity
dispersions for optically selected samples to determine the mass function.
Here a different strategy is used for the determination of the mass function.
A statistical cluster sample is constructed taking advantage of the
availability of an all-sky X-ray imaging survey. Furthermore the
large number of archival cluster observations is exploited allowing detailed
gravitational mass determinations through X-ray imaging and X-ray
spectroscopy for the clusters included in the sample.

The overall physical processes determining the main properties of
clusters
and their appearance in X-rays
are well understood.
Intracluster gas (\icg ) is trapped and heated to $10^7$--$10^8\,\rm K$ in the
cluster gravitational potential. Thermal bremsstrahlung emission from this
\icg\ makes clusters luminous X-ray sources. Only bright quasars
exceed the typical cluster luminosities of $\sim 10^{45}\,\es$.

The X-ray luminosity is well
correlated with cluster mass (as will be shown in this work) as opposed to
the measured galaxy richness, which has often been employed as selection
criterium for optically selected cluster samples.
Therefore X-ray selection effectively selects galaxy clusters by their mass.
This property is vital for the construction of the mass function.

The \ro\ All-Sky Survey (\ra ), being the only all-sky survey carried out by
an X-ray satellite with imaging capabilities to date
\citep[e.g.,][]{t93}, has
yielded a wealth of newly discovered X-ray sources \citep[e.g.,][]{vab99}.
A variety of galaxy cluster catalogs, together covering the whole sky,
have been homogeneously built from the \ra\ (see references in
Chap.~\ref{sample}). These catalogs are utilized in this work for the
construction of an X-ray selected and X-ray flux-limited sample of the
brightest galaxy clusters in the sky
(excluding a strip of $\pm 20$\,deg from the galactic plane, in order
to ensure a high completeness). To perform a detailed characterization
of the thereby selected galaxy clusters,
including the determination of the gravitational cluster mass,
high quality (deep exposure)
pointed observations of the \ps\ detector onboard the \ro\ observatory
are then analyzed and intracluster gas temperatures,
mainly determined with the \as\ satellite because of its superior
spectral resolution and larger sensitive energy range compared to \ro ,
are compiled from the literature.

Within the framework of hierarchical structure formation the properties
of galaxy clusters are also
expected to follow certain scaling relations and simulations have shown
that scaled dark matter profiles look similar \citep[e.g.,][]{nfw95}.
Despite the well established overall understanding of galaxy clusters,
in detail 
deviations from this picture have been found from observationally
accessible quantities, e.g., by the relation between X-ray luminosity
and intracluster gas temperature \citep[e.g.,][]{dsj93}, and the
gas properties in the center of groups of galaxies
\citep[e.g.,][]{pcn99}. A variety of models has been suggested
to explain these deviations (Sect.~\ref{fg_d}). Tests of detailed
predictions of these models unfortunately are still compromised by
observational difficulties. For instance the observed gas
mass fraction has been found in the recent literature
to either stay constant, decrease, or increase as a function of
cluster temperature (Sect.~\ref{fg_d}).
The homogeneously
selected and analyzed cluster sample presented here, comprising more
than 100 galaxy groups
and clusters, is therefore used to determine physical quantities like
the X-ray luminosity, intracluster gas density distribution,
temperature, and mass, as well as the gravitational mass over a wide
temperature range from 0.7 to 13\,keV. The relations between these
quantities are analyzed and compared to predicted relations.

The structure of this work is as follows. In Chap.~\ref{back} the 
different components of galaxy clusters are introduced with emphasis
on the gas and gravitational mass determination. Furthermore some relevant
cosmological background is given, including the calculation of model
mass functions. Last not least the observing instruments are briefly
described with
most of the weight assigned to \ro\ and the \ra , according to their
importance for this work. The sample construction is described in
Chap.~\ref{sample}. The details of the data reduction and analysis, and
the determination of relevant quantities are given in Chap.~\ref{anal}.
The gas temperature structure is very important for the X-ray
mass determination of clusters. Therefore due to its importance for
the present investigation the temperature profile for an example cluster
is determined using brandnew data from the X-ray satellite mission \xmm ,
which are in many respects superior to \ro\ and \as\ data (Chap.~\ref{a1835}).
Before the mass function is determined in Chap.~\ref{resu} -- 
being the first galaxy cluster mass function constructed from
an X-ray selected and X-ray flux-limited sample based on the \ra\ --  
the physical properties of the
cluster sample are examined. Especially the correlation between X-ray
luminosity and gravitational mass is of major importance here. In
Chap.~\ref{disc} tests of the sample completeness are performed and the
cluster masses determined here are compared to independent determinations.
The relations found between physical cluster properties are discussed.
The mass function is compared to
previous determinations and the cosmological implications of this mass
function are presented, including a fit to model mass functions.
Tight constraints on $\om$ are derived.
Previous work indicated that the mass fraction contained in galaxy
clusters may comprise already a fairly large fraction of the total mass
in the universe \citep[e.g.,][]{fhp98}. The well determined mass function
given in this work is therefore used to determine the mass fraction in
bound objects above a minimum mass to test these results.

\chapter{Theoretical Background}
\label{back}
\section{Galaxy Clusters}
\label{back:clus}

Clusters of galaxies are believed to consist of four main
components. As indicated by the name galaxy clusters have been
discovered as conglomerates of \emph{galaxies}. The space between
these galaxies is not empty but contains huge amounts of
\emph{intracluster gas} (\icg ). The largest portion of the total
gravitating mass in clusters, however, exists in the form of
\emph{dark matter}.
A possible forth component is a population of highly
\emph{relativistic electrons}, i.e.\ electrons having velocities close
to the speed of light. Some characteristics of these components are
briefly summarized in this Section \citep[for a review
see, e.g.,][]{s86}. Since this work mainly deals with the
intracluster gas and its implications for the dark matter content,
these two components are awarded more attention.

\subsection{Cluster Galaxies}
\label{back:galax}

How many galaxies make a cluster?
An assembly of
more than 4--5 galaxies is called a galaxy group \citep[e.g.,][]{h82},
$\sim 100$ galaxies make a cluster, and $\sim 1\,000$ galaxies a rich
cluster.
These rough numbers exclude `dwarf' galaxies, which are difficult to
count due to their faintness, except in the most nearby clusters.
\citet{a58}
introduced the richness as a measure for clusters. The richness is
determined by the number of galaxies above background fulfilling
certain criteria. The two main criteria are that only galaxies
be counted that a) are not more than two magnitudes fainter than the third
brightest member galaxy, and b) have a projected distance from the
center not larger than the Abell radius\footnote{$h_{50}$ is defined in
Sect.~\ref{back:practi}. 1\,pc $= 3.085678\times 10^{18}$\,cm.}
$\rab\equiv 3\,\mpc$.

The galaxy population in clusters differs from the field
population, i.e.\ galaxies not contained in clusters, especially in
the following properties.
\begin{itemize}
\item Morphology. The relative number of elliptical (E) and lens
shaped (S0) galaxies in clusters is larger and the relative number of
spiral galaxies is smaller than in the field \citep[e.g.,][]{d84,o92}.
\item Color. Spirals and irregular galaxies in clusters are redder on
average than the same types in the field \citep[e.g.,][]{o92}.
\item Gas content. Especially spirals close to the cluster center contain
less amounts of neutral hydrogen than spirals in the field
\citep[e.g.,][]{cgb90}. 
\item cD galaxies. These giant elliptical galaxies are
found in the center of most groups and clusters. The most striking
property of these cD galaxies is a very extended halo of low
surface brightness \citep[e.g.,][]{mms64}.
\end{itemize}

\subsection{Intracluster Gas}
\label{back:icg}

The intracluster gas is the most massive visible component of galaxy
clusters. Its mass exceeds the (gravitating) mass contained in the
cluster galaxies by a factor of $\sim$ 2--5. The temperature,
$\tg$, of the \icg\
is in the range $1\lesssim\kb\tg\lesssim 15\,\rm keV$
(here $1\,{\rm keV}$ corresponds to $1.16045\times 10^{7}\,\rm K$). The central
gas number density, $\ngas (0)$, is in the range $10^{-3}$--$10^{-1}\rm
\,particles\,cm^{-3}$. The collisionally ionized plasma is optically thin
and emits thermal radiation in X-rays. For $\tg \gtrsim 2\,\rm keV$ the main
component is bremsstrahlung (free-free transitions), for lower
temperatures recombination (free-bound transitions) and line emission
(bound-bound transitions) become more important. The emissivity
depends on the density$^2$.
A parameterized radial gas density distribution
can be determined analytically from the observed surface brightness
distribution.
Numerical deprojections using onion shell models are
also applied \citep[e.g.,][]{fhc81}. The procedure of the analytic
deprojection is outlined in Sect.~\ref{back:rog}. The observational
determination of $\tg$ is described in Sect.~\ref{back:tg}.

\subsubsection{Gas Density}
\label{back:rog}

Assuming King's (\citeyear{k62}) approximation to an isothermal sphere
for the galaxy density distribution, $\rogal$, leads to an analytical
representation of 
the radial gas density distribution
\citep[the `standard $\beta$ model', e.g.,][]{cf76,sb77,gft78,jf84,s86},
\begin{equation}
\rog(r)=\rog(0)\left(1+\frac{r^{2}}{\rc^{2}}\right)^{-\frac{3}{2}\beta}\,,
\label{back:bm1}
\end{equation}
by using $\rog \propto \rogal^\beta$, as implied by assuming the gas to
be ideal, isothermal, and in hydrostatic equilibrium, and the galaxies to
have an isotropic velocity dispersion,
where $\beta$ denotes the ratio of the specific kinetic energies of the
galaxies and the gas. 
The shape of the gas density distribution is therefore determined by the
core radius, $\rc$, and the shape parameter, $\beta$.
The asumptions leading to the $\beta$ model may be violated in detail. The
justification for its wide spread usage comes from the fact that the surface
brightness profile derived from it (see below) represents the measured
profile well in the relevant radial ranges.
The gas mass,
\begin{equation}
\mg(<r)=4\pi\int_0^r \rog(r) r^2 dr\,,
\label{back:gm1}
\end{equation}
may for illustrative purposes be approximated for large radii
and small $\beta$ values by
\begin{equation}
\mg(<r)\approx \frac{4\pi\rog(0)\rc^3}{-3\beta +3}\left (
\frac{r}{\rc}\right )^{-3\beta
+3} \quad : \quad \frac{r}{\rc}\gg 1 \wedge \beta < 1\,.
\label{back:gm2}
\end{equation}
The main constituents of the \icg\ are Hydrogen
and Helium, where a good approximation for the number densities is
$\nhe=\nh/10$.
Due to the high temperature the gas can be considered
completely ionized and the mean molecular weight
including the electrons
\begin{equation}
\mu\approx\frac{1+2\sum\limits_{Z>1}w_{\rm Z}Z}{2+\sum\limits_{Z>1}w_{\rm Z}(Z+1)}\,,
\label{back:mu}
\end{equation}
where $Z$ is the atomic number  and $w_{\rm Z}$ the relative weight
(e.g., here $w_2=0.1$ and $w_{\rm Z}=0$ for $Z>2$). Therefore one has
$\mu\approx0.61$ and
\begin{equation}
\rog\approx1.17\,\nel\,\mpr\,.
\label{back:ne}
\end{equation}
Because of this proportionality between electron number density, $\nel$, and
gas density it follows from (\ref{back:bm1})
\begin{equation}
\nel (r)=\nel (0)\left(1+\frac{r^{2}}{\rc^{2}}\right)^{-\frac{3}{2}\beta}\,.
\label{back:bm2}
\end{equation}
Before the connection between the observable surface brightness and
the gas density is made a few more important quantities are
introduced. The luminosity, $\lnu$, i.e.\ the energy radiated per unit
time at the frequency $\nu$ is given by
\begin{equation}
\lnu=\int_{V} \epsilon_{\nu}\,dV\,,
\label{back:le}
\end{equation}
where the emissivity
\begin{equation}
\epsnu =\nel\,\nh\,\lamnu (\tg,A)\,.
\label{back:em}
\end{equation}
The emission coefficient, $\lamnu$, mainly depends on the gas
temperature and metallicity, $A$. However, it varies only
weakly in the energy range where \ro\ is sensitive
\citep[e.g.,][]{b95}, for the relevant
cluster gas temperature range (2--10\,keV).
The emission measure is defined as
\begin{equation}
\emi\equiv \int_{V}\nel^2\,dV\,.
\label{back:es}
\end{equation}
For the X-ray surface brightness, i.e.\ the number of photons detected
in a defined energy range per unit time and per unit solid angle, one has
\begin{equation}
\sx\propto\int_{-\infty}^{\infty}\nel^{2}\,dl\,,
\label{back:sx1}
\end{equation}
where the integration is along the line of sight ($l=0$ at the cluster
center). With~(\ref{back:bm2}) it follows
\begin{equation}
\sx\propto\int_{-\infty}^{\infty}\left(1+\frac{r^{2}}{\rc^{2}}\right)^{-3\beta}dl\,.
\label{back:sx2}
\end{equation}
This integral can be reduced to a form solved in, e.g.,
\citet[Integral No.\ 39]{bs80} and one finds
\begin{equation}
\sx (R)=\sx (0)\left(1+\frac{R^{2}}{\rc^{2}}\right)^{-3\beta+\frac{1}{2}}\,,
\label{back:sx}
\end{equation}
where $R$ denotes the projected distance from the cluster center.
$\sx (0)$ depends on $\nel (0)$, $\rc$, $\beta$, $\lamnu (\tg ,A)$, and
redshift, $z$. Equation~(\ref{back:sx}) is used as a fitting formula to fit
the observed surface brightness profile. With the obtained fit
parameter values for $\sx (0)$, $\rc$, and $\beta$ the gas density
profile can be determined with~(\ref{back:bm1}), where $\rog(0)$ is
obtained from (\ref{back:ne}). 
The important step for the determination of the gas density
distribution from (\ref{back:sx}) is the emission mechanism
(\ref{back:em}), which is well understood. The $\beta$ model has been applied
successfully already for many years, but also other models have been
used, e.g., gas density distributions \citep[e.g.,][]{mss98} based on the Navarro-Frenk-White
profile (\citealt{nfw96}, \citeyear{nfw97}), which is a
fitting formula that represents well the cluster dark matter distribution found in
$N$-body simulations for varying cosmological models (but see
Sect.~\ref{back:dm}). 

Some clusters exhibit a central excess emission not well approximated
by (\ref{back:sx}).
To get a more accurate
decription of the gas density profile in such cases, a double
$\beta$ model has
been used by different authors \citep[e.g.,][]{ief96,mme99} to fit
the data, where the surface brightness takes
the form $\sx=\sx{_{,1}}+\sx{_{,2}}$. The motivation is to have one
component accounting for the central excess emission and the other
component accounting for the overall cluster emission.
It follows from the proportionality
(\ref{back:sx1}) that the gas density can then be determined from
$\nel=[\nel{_{,1}}^2+\nel{_{,2}}^2]^{1/2}$. It has been shown, however,
that the gas mass determination is not biased by the presence of central
excess emission for instance by \citet{r98}, who compared gas masses
determined using single and double $\beta$ models.

It is worth noting that a new method to determine the gas mass in
clusters is becoming more and more important \citep[e.g.,][]{cjg96}, which uses the
distortion of the CMB photon spectrum caused by inverse Compton
scattering on the hot \icg , the Sunyaev--Zeldovich effect
\citep{zs69,sz70}.

\subsubsection{Gas Temperature}
\label{back:tg}

When clusters of galaxies had been discovered as strong X-ray emitters
more than 30 years ago \citep[for references of the first
detections and interpretations see, e.g.,][]{s86} several possible emission
mechanisms were discussed. The detection of line emission due to highly
ionized iron in the X-ray spectra \citep[e.g.,][]{mcd76,ssb77},
however, made clear that the major contribution is thermal
emission. The main mechanism to heat the intracluster gas to the
high temperatures observed is expected to be shocks. These shocks
are caused by the gravitational assembling of the cluster from subunits.

The electron temperature can be determined by fitting model
spectra (folded with the instrument response) to the observed
X-ray spectra (Chap.~\ref{a1835}). Assuming electrons and ions to be
in thermal equilibrium this
X-ray temperature corresponds to the gas temperature. Within
$2\,\mpc$ \citet{fl97} have shown that this assumption should be
satisfied. Since X-ray temperatures are seldom available for radii larger than
$2\,\mpc$ they should generally be good indicators of the gas temperatures.
Several spectral codes for hot, optically thin plasmas have
been published \citep[e.g.,][]{rs77,mkl95,sbl01}. 

The general dependence of the gas
temperature on the distance from the cluster center has been discussed
controversely
recently utilizing data from various satellites
\citep[e.g.,][]{f97,mfs98,ibe99,w00,ib00}. 
Including the latest findings from \xmm\ (M. Arnaud,
private communication; Chap.~\ref{a1835}) the
gas seems to be isothermal out to at least half the virial radius. In
the very central part, where processes related to cooling flows
\citep[e.g.,][and references therein]{f94} or cD galaxies \citep[e.g.,][and references therein]{m00,mef01} may become important, a
temperature drop is often found.
\subsection{Dark Matter}
\label{back:dm}

The sum of the mass of all visible galaxies does by far not provide enough
gravitational attraction to hold these galaxies in a cluster
\citep[e.g.,][]{z33}. Now, after the detection of the large amounts of
gas present in clusters, does the gas mass suffice to retain the
galaxies and the gas? The answer is
no. Assuming the laws of gravitation to be the same at the distance and at
the scale of clusters
still about
3/4 of the mass is `missing'.

Several candidates for this `dark' matter
have been and are being discussed. While, for instance, observations
of the large scale clustering of galaxies rule out neutrinos
(candidates for Hot Dark Matter, HDM) as forming
the only component of the dark matter \citep[e.g.,][]{wfd83}, the recent strong evidence 
that neutrinos with finite rest mass do exist \citep[e.g.,][]{sk98}
leaves the possibility that at least part of the missing mass is provided
by neutrinos. One of the frequently cited possible Cold Dark Matter
(CDM) particles is the axion \citep[e.g.,][]{ow93}; also the heavier neutralino and
gravitino are often discussed \citep[e.g.,][]{ow97}.

Clusters of galaxies form a natural
laboratory -- obviously quite a bit larger than  any experiment that
could be built on Earth -- filled abundantly with dark
matter particles and
may therefore be utilized to actually place constraints on the nature
of dark matter candidates. 
Recently, e.g., \citet{ss00} suggested that elastic collisions of
weakly self interacting particles may provide an explanation for the
discrepancy between simulated CDM halos and observations of galaxies
and clusters of galaxies. The discrepancy arises when radial dark
matter profiles from simulations of collisionless dark matter
particles \citep[e.g.,][]{nfw96} are compared to dark matter profiles indicated by
rotation curves of dwarf galaxies (e.g., \citealt{bu95}; but see
\citealt{kkb98})  and by radial gas density profiles
of clusters \citep[e.g.,][]{mss98}\footnote{Note, however, that
\citet{ysw00} have shown that simulations, placed in a
cosmological context, do not allow a simple model of
dark matter particles with a finite cross section for elastic
collisions to account for the discrepancy in dwarf galaxies
\emph{and} clusters \emph{simultaneously}.}.

A more empirical dark matter density profile suggested by
\citet{bu95} better reproduces the data on dwarf galaxies and also on
the gas density distribution in clusters \citep[e.g.,][]{wx00}.

This work mainly concentrates on the observational determination of the
amount of gravitational mass.
Therefore in this paragraph a widely used method for this determination,
which has also been used
here, is decribed. The basic assumption is that the \icg\
is in hydrostatic equilibrium, i.e.\
\begin{equation}
\frac{d\pg(r)}{dr}=\frac{-\rog(r) G\mt (<r)}{r^{2}}\,,
\label{back:hy}
\end{equation}
where $\pg$ represents the gas pressure, $G$ the gravitational
constant, and $\mt$ the cluster's gravitational mass.
With the ideal gas equation,
\begin{equation}
\pg=\frac{\kb}{\mu \mpr}\rog \tg\,,
\label{back:id}
\end{equation}
this leads to
\begin{equation}
\mt(<r)=-\frac{\kb\tg(r)r^{2}}{\mu \mpr G}\left(\frac{1}{\rog
(r)}\frac{d\rog (r)}{dr}+\frac{1}{\tg (r)}\frac{d\tg (r)}{dr}\right)\,.
\label{back:ma1}
\end{equation}
Inserting (\ref{back:bm1}) and, based on the recent findings of \xmm\
(Sect.~\ref{back:tg}), assuming the cluster gas to be isothermal yields
\begin{equation}
\mt (<r)=\frac{3\kb\tg r^{3}\beta}{\mu \mpr G}\left(\frac{1}{r^2+\rc
^2}\right)\,.
\label{back:ma2}
\end{equation}
This equation is used throughout for the determination of the
gravitational mass.
The influence of a possible non isothermality of
the cluster gas on the results is discussed in Chap.~\ref{disc}.
The assumption of hydrostatic equilibrium can be
motivated by considering that the sound speed in the \icg\ $\csh\approx
1000\,{\rm km\,s^{-1}}\approx 1\,{\rm Mpc\,Gyr^{-1}}$.
The time a sound wave needs to cross the cluster
therefore is small compared to the cooling time and the time scales for
suggested heating mechanisms \citep[e.g.,][]{s86}. More quantitatively
$N$-body/hydrodynamic simulations have shown that as long as extreme
merger situations are excluded, where the assumption of hydrostatic
equilibrium probably breaks down, the mass estimates based on the
hydrostatic equation give unbiased, accurate results with an uncertainty of
14--29\,\% \citep[e.g.,][]{s96,emn96}.
The influence of bulk flows of the gas within the cluster and
magnetic fields has been neglected.
Cluster wide magnetic fields in a range of reasonable strengths
about 1\,$\mu$G have been shown in magneto-hydrodynamic simulations to
provide only a minor non thermal pressure support \citep[e.g.,][]{ds00},
which is negligible for an X-ray mass determination based on
(\ref{back:ma1}).

Two other basic independent methods exist currently to estimate
cluster gravitational masses. The oldest one utilizes the velocity
dispersion of the cluster galaxies, and the youngest takes advantage
of the alteration of images of background galaxies due to the
gravitational field excerted by the cluster.	The latter method can
be subdivided into weak and strong gravitational lensing, the latter
utilizing multiple and/or strongly distorted images of one or a few
background galaxies and the
former using a statistical approach on weak distortions of 
many background galaxies. Comparisons between
the three mass determinations on low to medium redshift clusters seem
to generally find good agreement between the velocity dispersion, X-ray,
and weak lensing methods, whereas the strong lensing methods -- probing
the very center of clusters -- yield factors of 2--4 higher masses
\citep[e.g.,][]{a98,wcf98}. The good agreement for the cluster masses
at large radii is encouraging, since for the current work the overall cluster
mass is important. In Sect.~\ref{mest_d} masses for clusters determined here are
compared to independent optical and X-ray estimates. The influence of
weak systematic differences on the estimation of cosmological parameters is
tested in Sect.~\ref{func_pred}.

\subsection{Relativistic Electrons}
\label{back:rele}

Extended diffuse synchrotron emission has been detected in radio
images of galaxy clusters indicating the presence of highly
relativistic electrons \citep[e.g.,][see, e.g., \citealt{gfg01} for a
recent comparison of radio and X-ray cluster properties]{w70} and
large scale magnetic fields \citep[see also][]{ckb01}.
Another indication is the detection of emission in excess of the
bremsstrahlung and line emission expected for the intracluster gas
temperature and metallicity on the soft ($\lesssim 0.5\,\rm
keV$) and hard ($\gtrsim 10\,\rm keV$) side, if inverse Compton
scattering of CMB photons is the
emission mechanism \citep[e.g.,][]{sa99}. However, the significance
and especially the abundance of the soft excess is still debated vigorously
\citep[e.g.,][]{lbm99,bbk99a,bbk99b}. There seems to be less confusion
about the hard excess being present in a few clusters
\citep[e.g.,][]{f99}.
Cluster mergers have been discussed as being
responsible for the production of relativistic electrons, since part
of the released energy may go into particle acceleration
\citep[e.g.,][]{r99}. Observationally some indications for a
correlation between the
presence of diffuse radio emission and substructure in clusters have
been found \citep[e.g.,][]{sbr00}.
The total energy contained in relativistic particles, however, is likely to
be small compared to the total thermal energy content
($\sim 5\times 10^{44}\,\es \times \sim 10^{10}\,{\rm yr}\approx 2\times
10^{62}\,\erg$) of a typical cluster \citep[e.g.,][]{s01}. Therefore the
pressure supplied by these particles is negligible for the mass
determination.

\section{Cosmology}
\label{back:cosmo}

One of the main goals of the present work is to constrain important
cosmological parameters.
Sects.~\ref{back:basic} and \ref{back:practi}
give a brief overview of the relevant theoretical background.
See, for
instance, \citet{m58}, \citet{w72}, \citet{s88}, \citet{cpt92}, \citet{r95},
\citet{p99}, and \citet{gs00} for more extensive information and details. In
Sect.~\ref{back:MF} some observational and theoretical tools are given
to turn knowledge about cluster gravitational masses into tests of 
cosmological models.

\subsection{Some Basics}
\label{back:basic}
In an attempt to motivate the formulae used throughout this work
the Einstein field equations, the Robertson--Walker Metric, the
Friedmann--Lema\^{\i}tre equation, and a few other important relations are
introduced.

The law of conservation of energy and momentum in special relativity (SR)
takes the form
\begin{equation}
\frac{\partial}{\partial x^\mu}\eit\equiv\partial_\mu \eit=0\, ,
\label{srt}
\end{equation}
where $\mu,\nu=\{0,1,2,3\}$ and, e.g., $x^0=ct$, $x^1=x$,
$x^2=y$, $x^3=z$ are contravariant coordinates, where the vacuum speed of
light $c\equiv 299792.458\,\rm km\,s^{-1}$. $\eit$ is the
energy-momentum tensor, describing the distribution and displacement
of energy and momentum. In general relativity (GR) SR can be
approximated locally, e.g., a free falling elevator of small
dimensions (system of inertia). If (\ref{srt}) is to be written in GR
for a general coordinate system then the partial derivatives have to
be replaced by absolute (covariant) derivatives
\begin{equation}
\nabla_\mu \eit=0\, .
\label{art}
\end{equation}
Note that (\ref{art}) is an invariant relation but in general cannot be interpreted as conservation
of energy and momentum anymore since the energy and momentum induced by
gravitation are not included in $\eit$. Now the hypothesis that the
distribution and displacement of (non gravitational) energy and
momentum is related to the geometry of space-time is introduced. Let
the geometry be described by the Riemann curvature tensor, $\rt$, and
the metric tensor, $\met$, then the simplest way to express this
hypothesis and fulfil (\ref{art}) is
\begin{equation}
\eit\propto\rto-\frac{1}{2}R\met\equiv\eto\, ,
\label{fg1}
\end{equation}
where the Ricci tensor $\rtu=\rta$ and the curvature scalar
$R=\met\rtu$ and $\eto$ is the Einstein tensor.
The simplest extension of (\ref{fg1}) to fulfil (\ref{art}) is obtained by adding a constant
\begin{equation}
\eit\propto\eto+\Lambda\met\, ,
\label{fg2}
\end{equation}
where $\Lambda$ is the so called cosmological constant.
The constant of proportionality is obtained by considering the limit
of weak gravitational fields where Einstein's theory must go over to
Newton's. \emph{Einstein's field equations} then read
\begin{equation}
-\frac{8\pi G}{c^4}\eit=\eto+\Lambda\met\, .
\label{fg3}
\end{equation}
A $\Lambda>0$ counteracts
gravity and therefore theoretically allows a static universe, which
was the main reason why Einstein introduced it into the field
equations, before it was discovered that the universe actually is
expanding.

The next building block needed is the metric. Assuming that the
universe is isotropic for \emph{every} observer comoving with matter,
the line element of the metric is given by
\begin{equation}
ds^2=c^2dt^2-R(t)^2\left(\frac{dr^2}{1-kr^2}+r^2d\theta^2+r^2\sin^2\theta\,d\phi^2\right)\, ,
\label{rw}
\end{equation}
where $R(t)$ is the scale factor at cosmic time $t$ (not to be
confused with the curvature scalar), $r,\theta,\phi$ are unitless comoving
coordinates, and the curvature index $k=\{0,\pm 1\}$. The metric
defined by (\ref{rw}) is often referred to as \emph{Robertson--Walker
metric}.

Inserting the metric tensor given by (\ref{rw}) in Einstein's field
equations (\ref{fg3}) yields the
Friedmann--Lema\^{\i}tre equation. This equation can, however,
illustratively also be found within Newton's theory by considering a
particle on a spherical shell with radius $R$ encompassing a mass $M(<R)$
\begin{equation}
\frac{1}{2}\dot{R}^2-G\frac{M}{R}=\rm const.\, ,
\label{fla2}
\end{equation}
with
\begin{equation}
M=4/3\pi\rom R^3\,,
\label{eq:mr}
\end{equation}
where $\rom$ is the mean matter density within $R$,
\begin{equation}
\dot{R}^2-\frac{8\pi G}{3}\rom R^2=\rm const.
\label{fla3}
\end{equation}
follows.
Including $\Lambda$ und the constant one finds the
\emph{Friedmann--Lema\^{\i}tre  equation}
\begin{equation}
\frac{\dot{R^2}}{R^2}=\frac{8\pi G}{3}\rom+\frac{\Lambda}{3}-\frac{kc^2}{R^2}\, .
\label{fla4}
\end{equation}
In the same picture another important relation is (illustratively) obtained by considering
\begin{equation}
\ddot{R}=-G\frac{M}{R^2}\, ,
\label{flb1}
\end{equation}
with (\ref{eq:mr})
\begin{equation}
\ddot{R}=-\frac{4\pi G}{3}\rom R
\label{flb2}
\end{equation}
follows. Including the pressure, $\drm$, and $\Lambda$ one finds
\begin{equation}
\ddot{R}=-\frac{4\pi G}{3}(\rom + \frac{3\drm}{c^2}) R + \frac{\Lambda}{3}R\, .
\label{flb3}
\end{equation}
Note that (\ref{flb3}) in general cannot be obtained directly by
differentiating (\ref{fla4}) because $\rom$ may depend on $t$.
Instead utilizing (\ref{fla4}) and (\ref{flb3}), and setting
$\drm=0$ it is found that
$\rom R^3=\rm const.$,
and
\begin{equation}
\om +\ol +\ok =1\, ,
\label{oms}
\end{equation}
where the normalized cosmic matter density $\om (t)\equiv 8\pi G\rom (t)
/(3H(t)^2)$, the normalized cosmological constant $\ol (t) \equiv
\Lambda/(3H(t)^2)$, and the normalized curvature index
$\ok (t)\equiv -kc^2/(H(t)^2R(t)^2)$. The Hubble parameter
$H(t)\equiv \dot{R}(t)/R(t)$.
For a flat universe ($k=0$) one has $\om +\ol =1$.
For the deceleration parameter $q_0\equiv
-\ddot{R}(t_0)/(R(t_0)H_0^2)$, where $H_0\equiv H(t_0)$, one finds in
a similar way
\begin{equation}
q_0=\frac{\omn}{2}-\oln\,,
\label{q0}
\end{equation}
where $\omn\equiv\om(t_0)$ and $\oln\equiv\ol(t_0)$.
For $q_0>0$ ($q_0<0$) the expansion of the universe is decelerating
(accelerating).
The
critical cosmic matter density is defined as
\begin{equation}
\roc (t)\equiv \frac{\rom (t)}{\om (t)} \equiv \frac{3\,H(t)^2}{8\,\pi\,G}\,.
\label{roc}
\end{equation}
Without subscripting from now on present day
values, i.e.\ $t=t_0\equiv$ today, will be assumed for $\om$, $\ol$,
and $\ok$. 

\subsection{Practical Formulae}
\label{back:practi}

For the physical distance, $d$, of two objects on the
hypersphere of constant cosmic time, $t$, follows from (\ref{rw})
\begin{equation}
d=R(t)\int_0^r\frac{dr'}{\sqrt{1-kr'^2}}=
\left\{\begin{array}{l@{\quad:\quad}l}
R(t)\arcsin r & k=+1\,, \\ R(t)\,r & k=0\,, \\ R(t)\ash r &
k=-1\,.\end{array} \right .
\label{eq:gr:d1}
\end{equation}
For $k=0$ one therefore finds that the physical volume of a sphere
with radius $d$ is given by
\begin{equation}
V(r)=\frac{4}{3}\,\pi R^3(t)\,r^3\quad : \quad \ok=0\,.
\label{eq:gr:v1}
\end{equation}
The redshift, $z$, compares a photon's wavelength measured in the observer rest
frame, $\lambda_0$, with the wavelength emitted in the
source rest frame, $\lambda_1$, by
\begin{equation}
z\equiv\frac{\lambda_0-\lambda_1}{\lambda_1}\,.
\label{eq:gr:z1}
\end{equation}
The redshift is caused by a change in the scale factor between the
time of emission, $t_1$, and absorption in the detector,
$t_0$. Therefore one has
\begin{equation}
z=\frac{R(t_0)}{R(t_1)}-1\,.
\label{eq:gr:z2}
\end{equation}
The luminosity distance is defined as
\begin{equation}
\dl\equiv\sqrt\frac{\lbol}{4\pi \fbol}\,,
\label{eq:gr:dl1}
\end{equation}
where $\fbol$ is the observed bolometric energy flux and $\lbol$ is
the energy per unit time emitted by the source, the bolometric luminosity.
Furthermore with $R(t_0)\equiv R_0$
\begin{equation}
\dl=R_0\,r\,(1+z)\,.
\label{eq:gr:dl2}
\end{equation}
The angular diameter distance and the proper motion distance are related to $\dl$ by
\begin{equation}
\da=\dl\,(1+z)^{-2}\ \quad {\rm and}\quad \  \dm=\dl\,(1+z)^{-1}\,,
\label{eq:gr:da}
\end{equation}
respectively.
The dependence of the Hubble parameter on redshift is given by
\begin{equation}
H(z)^2=H^2_0\,E(z)^2\,,\quad {\rm where}\quad E(z)=[\om (1+z)^3+\ok
(1+z)^2+\ol]^{1/2}\,.
\label{h0e}
\end{equation}
Setting $\Lambda=0$ in (\ref{fla4}) the important Mattig equation can be
derived for non vanishing $\rom$
\begin{equation}
R_0r= \frac{c}{H_0}
\frac{zq_0+(q_0-1)(\sqrt{2q_0z+1}-1)}{q_0^2\,(1+z)}\quad : \quad
\ol=0\wedge \om> 0\,.
\label{eq:gr:mt1}
\end{equation}
For $\ol =0$ and $\om =1$ (\ref{q0}) yields $q_0=1/2$ and therefore
\begin{equation}
R_0r= \frac{2c}{H_0}\left(1-\frac{1}{\sqrt{z+1}}\right)\quad : \quad
\ol=0\wedge \om =1\,.
\label{eq:gr:mt2}
\end{equation}
Using (\ref{eq:gr:dl2}) one finds
\begin{equation}
z= \frac{H_0}{2c}\dl-\frac{1}{2}+ \sqrt{\frac{H_0}{2c}\dl+\frac{1}{4}}\quad : \quad
\ol=0\wedge \om =1\,.
\label{eq:gr:z3}
\end{equation}

The more general (valid also for $\ol\neq 0$) relation between
distance measure and redshift is given by
\begin{equation}
\dm=\frac{c}{H_0}
\left\{\begin{array}{l@{\quad:\quad}l}
\vert \ok\vert^{-1/2}\sin [\vert \ok\vert^{1/2} F(z;\om,\ol)] & k=+1\,, \\
F(z;\om,\ol) & k=0\,, \\
\ok^{-1/2}\sinh [\ok^{1/2} F(z;\om,\ol)] & k=-1\,,\end{array} \right .
\label{back:dz}
\end{equation}
where 
\begin{equation}
F(z;\om,\ol)\equiv \int_0^z [(1+z')^2(1+\om z')-z'(2+z')\ol]^{-1/2}dz'\,.
\label{back:f}
\end{equation}
And comoving volumes for $k\neq 0$ can be calculated by
\begin{equation}
V(\dm)=\frac{4\pi c^3}{2H_0^3\ok}
\left\{\begin{array}{l@{\quad :\quad }l}
G_{\dm}(1+\ok G_{\dm}^2)^{1/2}-\vert \ok\vert^{-1/2}\arcsin (G_{\dm}\vert \ok\vert^{1/2}) & k=+1\,, \\
G_{\dm}(1+\ok G_{\dm}^2)^{1/2}-\ok^{-1/2}\ashn (G_{\dm}\ok^{1/2}) & k=-1\,,\end{array} \right .
\label{back:vdm}
\end{equation}
where 
\begin{equation}
G_{\dm}\equiv \frac{H_0\,\dm}{c}\,.
\label{back:g}
\end{equation}
Remember that for $k=0$ one simply has $V(\dm)=4/3\pi \dm^3$ from
(\ref{eq:gr:v1}). 

Throughout $H_0=50\,h_{50}\,\rm km\,s^{-1}\,Mpc^{-1}$, $h_{50}=1$, 
$\om=1$, $\ol=0$, and $\drm =0$ is assumed if not stated
otherwise.
Note that the determination of physical cluster parameters has a
negligible dependence on $\om$ and $\ol$ for the small low redshift range
under consideration here, as will be shown later. Therefore it is justified to
determine the parameters for this specific model but to
discuss the results also in the context of other models.
\subsection{Mass Function}
\label{back:MF}

A major motivation of this study is to determine the mean density of the universe,
which is one of the key quantities that determines the fate of the universe.
In this work this aim is achieved by determining an unprecedentedly accurate
observational galaxy cluster mass function and comparing it to predicted
mass functions.
In this Section the basics for an observational determination of the
mass function are outlined and it is shown how
mass functions are predicted from cosmological models.

\subsubsection{Observational Determination}
\label{back:omf}

The commonly used definition of the galaxy cluster mass function is analogous to the
definition of the luminosity function (e.g., \citealt{s76}):
the mass function, $\phi(M)$, denotes the number of
clusters, $N$, per unit comoving volume, $dV$, per unit mass in the interval
$[M,M+dM]$, i.e.\ 
\begin{equation}
\phi(M)\equiv N(M)/(dV\,dM)\equiv dn(M)/dM\,.
\label{eq:gr:mf}
\end{equation}
Assuming constant density the
classical $\vmax$ estimator
\citep[e.g.,][]{s68,f76,bst88} can be used for the estimation of
luminosity functions, i.e.\
\begin{equation}
\hat\phi(L)=\frac{1}{\Delta L}\sum^{N}_{i=1}\,\frac{1}{\vmaxi}\,.
\label{eq:gr:lf}
\end{equation}
$\vmax$ is the
maximum comoving volume within which a cluster with given luminosity for a
given survey flux limit and sky coverage could have been detected.
Combining (\ref{eq:gr:v1}) and (\ref{eq:gr:dl2}), and replacing
$4\,\pi$ by the actual solid angle covered by the survey (this `sky
coverage' may in general be a function of flux), $\omega$,  one
finds that the surveyed volume enclosed by a cluster
sitting at redshift $z$ is given by
\begin{equation}
V(z)= \frac{\omega}{3}\,\left(\frac{\dl}{1+z}\right)^3\,,
\label{eq:gr:v2}
\end{equation}
where $\dl$ is calculated by combining (\ref{eq:gr:dl2}) and
(\ref{eq:gr:mt2}). The maximum surveyed volume that a cluster with luminosity
$L$ could possibly enclose is given by
\begin{equation}
\vmax (L)= \frac{\omega}{3}\,\left(\frac{\dlm}{1+\zm}\right)^3\,,
\label{eq:gr:v3}
\end{equation}
where $\dlm$ is determined by (\ref{eq:gr:dl1}), replacing the measured
flux with the flux limit, $\flim$. $\zm\equiv z (\dlm)$ is
given by (\ref{eq:gr:z3}). Note, however, that in
general only a finite energy band is available for flux
measurements. Due to the different redshifts of detector ($z=0$) and
source the emitted spectrum for a given source rest frame (SF) energy range
differs from the measured spectrum in the observer rest frame energy
range (OF) and a correction factor has to be applied leading (for the
\ro\ band) to
\begin{equation}
\dlm= \sqrt\frac{\lsou}{4\,\pi\,K(\tg ,\zm)\,\fobsl}\,,
\label{eq:gr:dlm}
\end{equation}\\
where $K(\tg ,z)\equiv \lsou/\lobs$ for a cluster at redshift $z$.
Unless noted otherwise in this work fluxes are quoted in the
OF energy range and luminosities in the SF energy range.

\subsubsection{Theoretical Determination}
\label{back:tmf}

In this Section the currently most important formalism for the prediction
of the abundance of massive objects is introduced.
It is one of the basic ingredients to describe the growth of structure
in the universe. This formalism is
used in Sect.~\ref{func_pred} to obtain constraints on cosmological
parameters.

The following prescription for the mass function (Eq.~\ref{eq:ps})
is based on the
work of \citet{ps74}. Various assumptions of this `PS' mass function
have been relaxed by a number of works. 
See, e.g., \citet{sba00} for a compilation.
The underlying idea of the \emph{extended} PS formalism \citep{bce91} is to
filter the
initial mass density field on successively smaller scales. If a filtered region
around the comoving location $\vec{r}$ exceeds a given density threshold it
is assumed to end up as a collapsed object of the enclosed mass. Note
that in this way the counting of objects contained in larger objects
is avoided. This `cloud in cloud' problem was present in the original PS
recipe and required the ad hoc introduction of a factor of 2.
Under the assumption of a random initial density fluctuation field
this picture is analogous to a random walk with an absorbing barrier,
where the filter scale is interpreted as the time variable
(in this picture time increases from large to small scales, i.e.\ from small
to large fluctuations) and the
density contrast, $\delta(\vec{r})\equiv\rho(\vec{r})/\bar\rho-1$, filtered
on the corresponding filter scale, where
$\bar\rho$ is the mean comoving density,
is interpreted as the spatial coordinate. 
The mass fraction in collapsed objects above a minimum mass
is associated with the fraction of volume
elements which have a density contrast above the threshold.

Three of the basic assumptions of this approach are the following.
First, the \emph{initial} density fluctuations are Gaussian and are
uniquely described by their power spectrum, $P(\vec{k})$, where $\vec{k}$
is the comoving wave number.
Second, in order to get analytic results a top hat filter in Fourier space
(sharp $k$ space filter) is used.
A third assumption, entering through the density contrast threshold, is
that all objects form by a spherical collapse.

As yet there seems to be no
compelling observational evidence against the assumption
of Gaussianity (e.g., \citealt{wbb01}).
The second condition, which requires 
unattractive oscillating filters in configuration
space, has been relaxed recently by \citet{sba00}, who derived
mass functions for more realistic filter functions.
These new mass functions, however, are rather complicated to apply and
the standard mass function has been used in this work to allow direct
comparison to previous results.
Note that these two assumptions imply that each point of each
trajectory of a random walk has no memory of the past if followed from
small to large scales\footnote{Note
that the use of, e.g., a Gaussian instead of the sharp
$k$ space filter
does introduce some correlations between different scales
-- with the drawback of either having only numerical solutions
\citep{bce91}
or a non negligible increase of complexity \citep{sba00}.}
(in this picture the time coordinate is
reversed). This means
a galaxy sized object at an early time before most clusters formed,
say $z=3$, has no
information if it will end up in a cluster or in a void at $z=0$
\citep[e.g.,][]{w96}.
However, there is evidence that the formation of objects
depends on their environment. For instance $N$-body simulations indicate that objects
often align with large scale filaments \citep[e.g.,][]{w96} and
observations show differences in
the galaxy populations in the field and in clusters (Sect.~\ref{back:galax}).
If this property of the extended PS
formalism were realistic then possibly the influence of the environment
would have to be only effective between $z=3$ and $z=0$.
A model constructed completely free of this assumption might allow to
better understand the discrepancies found if individual volumes
(actually mass particles) are followed up in time in simulations and
the final masses are compared to the predictions
\citep[e.g.,][]{w96}, i.e.\ a comparison on the halo by halo basis, and 
possibly also the violation of the PS prediction that objects grow
monotonically.
As to the assumption of spherical collapse, simulations based on the
hierarchical scenario rather indicate
that objects grow by accretion of matter along filaments and/or merging
\citep[e.g.,][]{cwj99,gkk01}.
Many cluster systems have also been observed which seem to be in a state
of merging \citep[e.g.,][]{msv99}.
This and the violated predictions mentioned above have led
to the suggestion of replacing the spherical collapse model with ellipsoidal
collapse \citep[e.g.,][]{smt01}.

For the purpose of
applying the analytical mass function as a fitting formula to
observational mass functions, these discrepancies are of minor
importance as long as good agreement to simulated mass functions is
shown, i.e.\ if there is good statistical agreement for simulated
and predicted distributions.
And really, good agreement has been found for many years
\citep[e.g.,][]{efw88,wef93,lc94,mjw96}.
Just recently large
simulations covering very large mass ranges have convincingly
shown slight deviations at the high and low mass end
\citep[e.g.,][]{gbq99,jfw01}. However, the importance of these
deviations for the
current investigation is shown to be small in Sect.~\ref{func_pred}.
In summary the justification for the usage of the standard formalism to
be outlined below for the present work comes from the sufficient agreement
with $N$-body simulations.

The PS formalism to predict cluster mass functions for given cosmological
models may be summarized as follows (e.g., \citealt{brt99}).
To allow easier comparison with the theoretical literature on this
subject in this Section $h_{100}=h_{50}/2$ is used.
The mass function is then given by
\begin{eqnarray}
\frac{dn(M)}{dM}=\sqrt{\frac{2}{\pi}}\,\frac{\bar\rho_0}{M}\,\frac{\dc(z)}{\sigma(M)^2}
\,\left\vert\frac{d\sigma(M)}{dM}\right\vert
\exp\left(-\frac{\dc(z)^2}{2\,\sigma(M)^2}\right)\,.
\label{eq:ps}
\end{eqnarray}
Here $M$ represents the object (cluster) virial mass and
\begin{equation}
\bar\rho_0 =
2.7755\times 10^{11}\,\om\,h_{100}^2\,M_\odot\,\rm
Mpc^{-3}
\label{eq:mmd}
\end{equation}
is the present mean matter density. The linear
overdensity (density contrast threshold) computed at present $\dc(z)=\dcv(z)\,D(0)\,D(z)^{-1}$,
where the linear overdensity at the time of virialization, $\dcv(z)$,
is computed using the spherical collapse model
summarized in \citet{ks96},
\begin{equation}
\dcv(z)=\frac{3\,(12\,\pi)^{2/3}}{20}\approx 1.686\quad :\quad \om = 1
\label{eq:ks1}
\end{equation}
and
\begin{equation}
\dcv(z)\approx \frac{3\,(12\,\pi)^{2/3}}{20}\{1+0.0123\log[\of(z)]\}
\quad :\quad \om<1 \wedge \ok=0\,,
\label{eq:ks2}
\end{equation}
where
\begin{equation}
\of(z)=\frac{\om\,(1+z)^3}{\om\,(1+z)^3+(1-\om-\ol)\,(1+z)^2+\ol}\,,
\label{eq:ks3}
\end{equation}
where $z$ is the redshift of cluster formation. In this work the recent
formation approximation is adopted and the observed cluster redshift is
assumed to be the formation redshift. With the assumptions (\ref{eq:ks2})
the second term in the denominator always vanishes.
The linear growth factor
\begin{equation}
D(z)=2.5\,\om\,E(z)\int_z^\infty (1+z')\,E(z')^{-3}\,dz'\,.
\label{eq:lgf}
\end{equation}
The initial linear variance of the cosmic mass density fluctuations,
\begin{equation}
\sigma(R)^2=\frac{1}{2\,\pi^2}\int_0^\infty k^2\,P(k)\,
\vert W(k\,R)\vert^2\,dk\,,
\label{eq:ilv}
\end{equation}
where spherical symmetry has been assumed and the power spectrum is
taken as
\begin{equation}
P(k)\propto k^n\,T(k)^2\,,
\label{eq:pk}
\end{equation}
may be expressed as
\begin{equation}
\sigma(M)^2=\sigma_8^2\frac{\int_0^\infty
k^{2+n}\,T(k)^2\,\vert W(k\,R(M))\vert^2\,dk}{\int_0^\infty
k^{2+n}\,T(k)^2\,\vert W(k\,8\,h_{100}^{-1}\,{\rm Mpc})\vert^2\,dk}\,,
\label{eq:sigma}
\end{equation}
where $\sigma_8$ represents the amplitude of density fluctuations within a
radius of
$8\,h_{100}^{-1}\,{\rm Mpc}$. Recent measurements of the CMB
anisotropies indicate that the 
primordial power spectral index,
$n$, has a value close to 1
\citep[e.g.,][]{bab00,jab01,phl01,wtz01,dab01} and is therefore set to 1
throughout. For the transfer function the
fitting formula for CDM power spectra provided by \citet{bbk86} is used
\begin{eqnarray}
T(k)\equiv T(q(k))=\ln(1+2.34q)/(2.34q)\nonumber\\
\times\,[1+3.89q+(16.1q)^2+(5.46q)^3+(6.71q)^4]^{-1/4}
\label{eq:transf}
\end{eqnarray}
for
\begin{equation}
q(k)\equiv k/(\Gamma\,h_{100}\,\rm Mpc^{-1})\,,
\label{eq:qk}
\end{equation}
where the shape parameter is given by (modified to account for a small
normalized baryon density $\ob > 0$, \citealt{s95})
\begin{equation}
\Gamma = \om\,h_{100}\,\left(\frac{2.7\,\rm
K}{T_0}\right)^{2}\,\exp\,\left(-\ob-\sqrt{\frac{h_{100}}{0.5}}\,\frac{\ob}{\om}\right)\,.
\label{eq:gama}
\end{equation}
The temperature of the CMB
$T_0=2.726\,\rm K$ \citep{mcc94} and $\ob\,h_{100}^{2}=0.0193$
\citep{bt98}, for the latter equation and (\ref{eq:gama}) $h_{100}=0.71$
\citep{mhf00} will
be used in the comparison to observed mass functions. 
The comoving filter radius
\begin{equation}
R(M)=[3M/(4\pi\bar\rho_0)]^{1/3}
\label{eq:filterrad}
\end{equation}
for
the top hat filter function in configuration space, which is given in $k$
space by
\begin{equation}
W(x)=3\,(\sin x - x\cos x )/x^3\quad : \quad x\equiv kR\,,
\label{eq:filterfunc}
\end{equation}
which is
adopted in this analysis, because the cluster masses will be
determined with a top hat filter, too (Sect.~\ref{back:dm}).
It is customary to use this
filter despite the fact that the extended PS formalism, which
predicts the correct normalization, has been derived using
the sharp $k$ space filter. Again the justification comes from the
comparison to $N$-body simulations.

The two important parameters here, which will be constrained by a comparison
to the observed mass function later, are the mean density of the universe,
$\bar\rho_0$, i.e.\ $\om$ when normalized by the critical density, and
the amplitude of density fluctuations within a radius of 
$8\,h_{100}^{-1}\,{\rm Mpc}$, $\sigma_8$. 

\section{Instruments}
\label{back:inst}

This work is based on observations performed by the X-ray satellites
\ro\ \citep[e.g.,][]{t93}, \as\ \citep[e.g.,][]{tih94}, and \xmm\
\citep[e.g.,][]{jla01}. Since \ro\ is the most important instrument
for this study, it is decribed briefly below. Some details on \xmm\
are given in Chap.~\ref{a1835}.

A Delta rocket put \ro\ into an orbit 580\,km above the Earth on June
1st 1990. After the calibration measurements the first (and up to now
last) All-Sky Survey (\ra
, e.g., \citealt{vab99}) 
with an imaging X-ray telescope was performed with an average exposure
of about 500\,s. From 1991 till 
1999 many much deeper exposures were taken of special fields (pointed
observations) upon request by guest observers. 

\begin{figure}[htb]
\begin{center}
\psfig{file=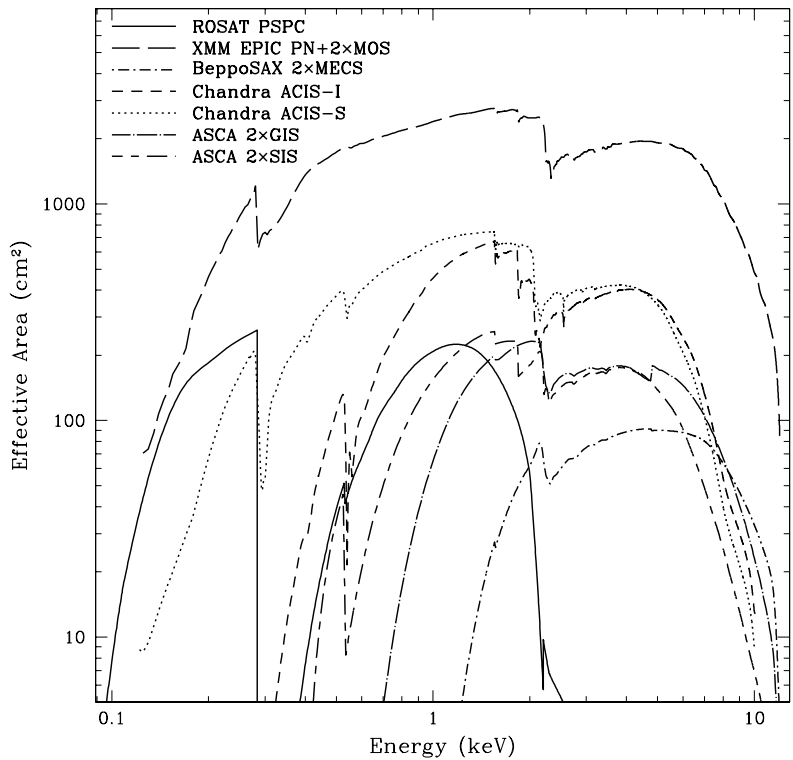,width=11cm,angle=0,clip=}
\end{center}
\caption{Comparison of the effective area of various instruments
(Figure provided by V. Burwitz). Notice the log scale.}\label{fig:effar}
\end{figure}

The main component of \ro\ is the Wolter I telescope, which maps the
X-ray photons onto one out of three focal plane detectors. It consists
of four gold coated nested mirrors. The front part of the mirrors is
slightly parabolically and the back part slightly hyperbolically
shaped. Either one out of
two proportional counters (Position Sensitive Proportional Counter,
\ps ) or a channel plate detector (High Resolution Imager, HRI) can be
placed in the focal plane. Additionally \ro\ carries a UV Camera (Wide
Field Camera, WFC). The instrument parameters energy
resolution, position resolution and field of view are
given in Tab.~\ref{tab:inst}. The effective area as a function of
energy is compared to other missions in Fig.~\ref{fig:effar}. The high
(particle induced) background rejection efficiency ($>$ 99\,\%) and the large field of
view of the \ps\ allow to trace the source emission out to large
apparent radii, an important feature for the study of extended objects
like galaxy clusters.
\begin{deluxetable}{lccc}
\tabletypesize{\footnotesize}
\tablewidth{0pt}
\tablecaption{Instrument parameters\label{tab:inst}} 
\tablehead{Satellite & Energy resolution& Position resolution &
Field of view (diameter)\\ & FWHM [eV] & FWHM [arcsec]  & [arcmin]}
\startdata
\ro\ \ps\  &    410\,@\,1.0\,keV & 25\,@\,1.0\,keV & 120 \\ 
\ro\ HRI  & \nodata & 5 & 40\\ 
\ra\  &    410\,@\,1.0\,keV & 40\,@\,1.0\,keV & unlimited \\ 
\as\ GIS  & 210\,@\,1.5\,keV & 180\tablenotemark{a} & 50 \\
\xmm\  EPIC-pn & 110\,@\,1.5\,keV & 7\,@\,1.5\,keV & 30 \\
\enddata
\tablenotetext{a}{Half power diameter.}
\end{deluxetable}

The average positional resolution of the
\ra\ is lower compared to the on-axis resolution of the pointed observations
performed with the \ps\ (Tab.~\ref{tab:inst}). This is due to a
decreasing resolution with increasing off-axis angle and the fact that \ra\
photons have been collected at various different detector positions. Despite
this lower resolution the nearby clusters which are relevant for this work
still clearly appear as extended
sources in the \ra\ which discriminates them from other X-ray sources
like stars and AGN.

\begin{figure}[htb]
\begin{center}
\psfig{file=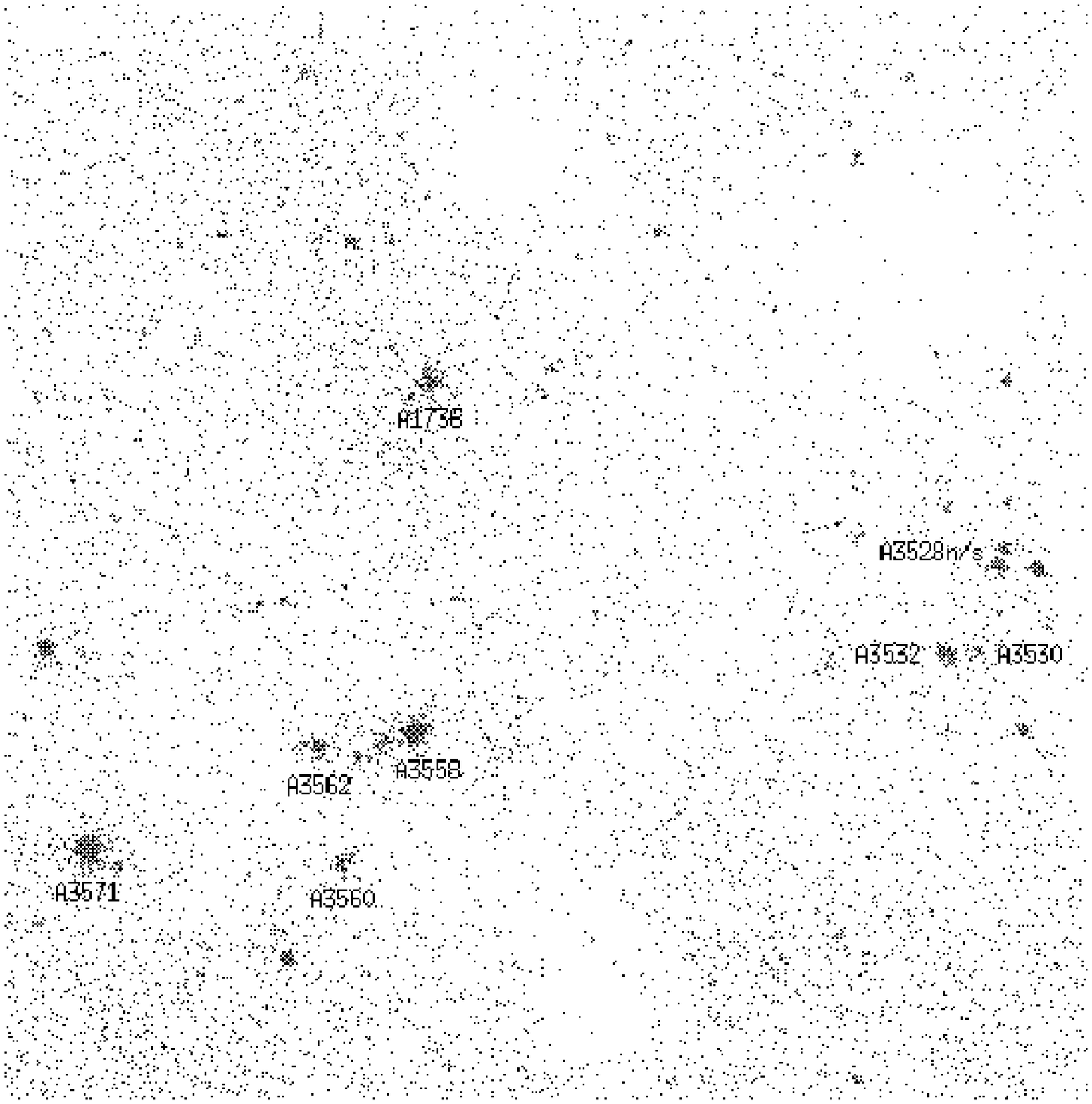,width=14cm,angle=0,clip=}
\end{center}
\caption{Raw \ra\ image of the Shapley supercluster with names of the main
individual galaxy clusters overlaid. The box size is 13.8\,deg. The apparent
large scale variations of the background are caused by differences in the
exposure time (find a better version of this image at www.reiprich.net).}
\label{fig:shapley}
\end{figure}

Figure~\ref{fig:effar} also shows that the major X-ray missions
are sensitive exactly in the energy range where clusters of galaxies have
their emission maximum. Because of this and their high luminosity
the nearby galaxy
clusters clearly stick out of the background.
This can be appreciated in Fig.~\ref{fig:shapley}, where the region of the
Shapley supercluster of galaxies is shown. Most of the dominant sources
in this high cluster density region are galaxy clusters. On average it is
estimated that about 10\,\% of all detected sources in the \ra\ are galaxy
clusters.

\chapter{Sample}
\label{sample}

The mass function measures
the cluster number density as a function of mass. Therefore any
cluster fulfilling the selection criteria and
not included in the sample distorts the result systematically.
It is then obvious that for the construction of the mass function
it is vital to use a
homogeneously selected and highly complete sample of objects,
and additionally the selection must be closely related to cluster
mass. In this work the \ra , where one single instrument
has surveyed the whole sky, has been chosen as the basis for the
sample construction. Using the X-ray emission from the hot
intracluster medium for cluster selection minimizes projection
effects and the good correlation between X-ray luminosity and
gravitational mass convincingly demonstrates that X-ray cluster
surveys have the important property of being mass selective
(Sect.~\ref{relat}). 

Another aim of this work is the characterization of physical
cluster properties. For such a statistical investigation it
is important not to introduce any bias caused by selection
effects. It is therefore desirable to select clusters by objective
criteria, as performed here. 

Several cluster catalogs have already been constructed
from the \ra\ with high completeness down to low flux limits (see refs
below). These have been utilized for the selection of candidates. Low
thresholds have been set for the initial selection in order
not to miss any cluster due
to measurement uncertainties. These candidates
have been homogeneously reanalyzed, using higher quality \ro\ \ps\
pointed observations whenever possible (Chap.~\ref{anal}). A flux
limit well above the limit for candidate selection has then been
applied to define the new flux-limited sample of the brightest
clusters in the sky. 

In detail the candidates
emerged from the following input catalogs.
Table~\ref{tbl:cand} lists the selection criteria and the number of
clusters selected from each of the catalogs, that are contained in the
final sample.  
\begin{itemize}
\item [1)] The REFLEX (\ro -ESO Flux-Limited X-ray) galaxy cluster survey
\citep{bsg01} covers the
southern hemisphere (declination $\delta \leq$
+2.5\,deg; galactic latitude $\vert \bii \vert \geq 20.0$\,deg) with a
flux limit  $\fxl(\eb) = 3.0\,\escl$.
\item [2)] The NORAS (Northern \ro\ All-Sky) galaxy cluster survey \citep{bvh00}
contains clusters showing extended emission in the RASS in the
northern hemisphere  ($\delta \geq$
0.0\,deg; $\vert \bii \vert \geq 20.0$\,deg) with
count rates $\cx(\eb)\geq 0.06\,\cts$. 
\item [3)] NORAS II (J. Retzlaff et al., in prep.)\ is the continuation of the
NORAS survey project. It includes point like sources and aims for a
flux limit $\fxl(\eb) = 2.0\,\escl$.
\item [4)] The BCS (\ro\ Brightest Cluster Sample) \citep{eeb98} covers the northern
hemisphere ($\delta \geq$
0.0\,deg; galactic latitude $\vert \bii \vert \geq 20.0$\,deg) with
$\fxl(\eb) = 4.4\,\escl$ and redshifts $z\leq 0.3$.
\item [5)] The \ra\ 1 Bright Sample of Clusters of Galaxies \citep{gbg99} covers the
south galactic cap region in the southern hemisphere ($\delta <$
+2.5\,deg; $\bii < -20.0$\,deg) with an effective flux limit $\fxl(\eh)$
between $\sim$\,3 and $4\,\escl$. 
\item [6)] XBACs (X-ray Brightest Abell-type Clusters of galaxies) \citep{evb96} is an
all-sky sample of \citet{a58}/ACO \citep{aco89} clusters limited to high
galactic latitudes $\vert \bii \vert \geq 20.0$\,deg with nominal ACO
redshifts $z\leq 0.2$ and X-ray fluxes $\fx(\eb) > 5.0\,\escl$.
\item [7)] An all-sky list of Abell/ACO/ACO-supplementary clusters
(H. B\"ohringer 1999, private communication) with count rates
$\cx(\eh)\ge  0.6\,\cts$.
\item [8)] Early type galaxies with measured \ra\ count rates from a magnitude
limited sample of \citet{bdb99} have been been checked in order not to
miss any X-ray faint groups. 
\item [9)] All clusters from the sample of \citet{lef89} and \citet{esf90},
where clusters had been compiled from various X-ray missions, have
been checked.
\end{itemize}

\begin{deluxetable}{lcccccccc}
\tabletypesize{\scriptsize}
\tablecaption{Selection of candidates\label{tbl:cand}}
\tablewidth{0pt}
\tablehead{
\colhead{Sample}    &  \multicolumn{3}{c}{$\cx\ [\cts]$} &   \colhead{}   &
\multicolumn{2}{c}{$\fx\ [\esct]$} & \colhead{$\ncl$} & \colhead{Ref.}\\
\cline{2-4} \cline{6-7} \\
\colhead{} & \colhead{$\eb$}   &
\colhead{$\eh$}   & \colhead{$\ek$} & 
\colhead{} &\colhead{$\eb$} &\colhead{$\eh$} &
\colhead{} & \colhead{}
}
\startdata
REFLEX       & \nodata & 0.9 & \nodata &  & 1.7 & \nodata  & 33  &1\\
NORAS        & \nodata & 0.7 & \nodata &  & 1.7 & \nodata  & 25  &2\\
NORAS II     & \nodata & 0.7 & \nodata &  & 1.7 & \nodata  & 4   &3\\
BCS          & 1.0 & \nodata & \nodata &  & 1.7 & \nodata  & 1   &4\\
\ra\ 1       & \nodata & \nodata & \nodata &  & \nodata & 1.0 & 0 &5\\
XBACs        & 1.0 & \nodata & \nodata &  & 1.7 &  \nodata  & 0  &6\\
Abell/ACO    & \nodata & 0.7 & \nodata &  & \nodata & \nodata & 0 &7\\
early type   & \nodata & \nodata & 0.7 &  & \nodata & \nodata & 0 &8\\
previous sat\tablenotemark{a} & \nodata & \nodata & \nodata &  & \nodata & \nodata  & 0 &9\\
 \enddata%

\tablenotetext{a}{All clusters from this catalog have been flagged as candidates.}

\tablecomments{Only one of the criteria, count rate or flux, has to be
met for a cluster to be selected as candidate. The catalogs are
listed in search sequence, therefore $\ncl$ gives the number of
candidates additionally selected from the current catalog and
contained in the final flux-limited sample. So in the case of NORAS a
cluster is selected as candidate if it fulfils $\cx(\eh)\geq 0.7\,\cts$
or $\fx(\eb)\geq 1.7\esc$ and has not already been selected from
REFLEX. This candidate is counted under $\ncl$ if it meets the
selection criteria for \gcss .}

\tablerefs{
(1) \citealt{bsg01}; (2) \citealt{bvh00}; (3) J. Retzlaff et al.,
in prep.; (4) \citealt{eeb98}; (5) \citealt{gbg99}; (6)
\citealt{evb96}; (7) H. B\"ohringer 1999, private
communication; (8) \citealt{bdb99}; (9) \citealt{lef89,esf90}.}%

\end{deluxetable}

The main criterion
for candidate selection, a flux threshold $1.7\esc$, has been chosen
to allow for measurement uncertainties. E.g., for REFLEX clusters with
$1.5\esc \leq \fx \leq 2.5\esc$ the mean statistical flux error is less than
8\,\%.
With an additional mean systematic error of 6\,\%, caused by
underestimation of fluxes due to the comparatively low \ra\ exposure
times\footnote{This has been measured by comparing the count rates
determined using pointed observations of clusters in this work to
count rates for the same clusters determined in REFLEX and NORAS. If
count rates are compared also for fainter clusters, not relevant for
the present work, the mean systematic
error increases to about 9\,\% \citep{bvh00}.},
the flux threshold of $1.7\esc$ for candidate selection then ensures
that no clusters are
missed for a final flux limit $\fxl = 2.0\esc$.

Almost none of the fluxes given in the input catalogs have been
calculated using a measured X-ray temperature, but mostly using
gas temperatures estimated from an $\lx$--$\tx$ relation.
In order to be independent of this additional
uncertainty clusters have also be
selected as candidates if they exceed a count rate threshold which
corresponds to $\fx(\eb)=2.0\esc$ for a typical cluster temperature,
$\tg=4\,\rm keV$, and redshift, $z=0.05$, and for an exceptionally
high neutral hydrogen column
density, e.g., in the NORAS case $\nhcol=1.6\,\times
10^{21}\rm cm^{-2}$.

Most of the samples mentioned above excluded the area on the sky close to the
galactic plane as well as the area of the Magellanic Clouds. In order to
construct a highly complete sample from the candidate list the
following selection criteria that successful clusters must fulfil have
been applied:
\begin{itemize}
\item [1)] \emph{redetermined} flux $\fx(\eb)\ge 2.0\esc$,
\item [2)] galactic latitude $\vert \bii\vert \ge 20.0$\,deg,
\item [3)] projected position outside the excluded 324\,deg$^2$
area of the Magellanic Clouds (see Tab.~\ref{tab:mc}),
\item [4)] projected position outside the excluded 98\,deg$^2$
region of the Virgo galaxy cluster (see Tab.~\ref{tab:mc})\footnote{The
large scale X-ray
background of the irregular and very extended X-ray emission of the
Virgo cluster makes the undiscriminating detection/selection of
clusters in this area diffcult.
Candidates excluded due to this criterium are Virgo, M86, and M49.}.
 \end{itemize}

\begin{deluxetable}{lccc}
\tabletypesize{\footnotesize}
\tablewidth{0pt}
\tablecaption{Regions of the sky not sampled in \gcs \label{tab:mc}} 
\tablehead{ {\rm Region}& {\rm RA Range }  & {\rm DEC Range} & {\rm
Area (sr)}} 
\startdata
{\rm LMC 1}  & 58  $\to 103^o$  & $-63 \to -77^o$   & 0.0655 \\                     
{\rm LMC 2}  & 81  $\to 89^o $ & $-58 \to -63^o$   & 0.0060 \\                      
{\rm LMC 3}  & 103 $\to 108^o$  & $-68 \to -74^o$   & 0.0030 \\                     
{\rm SMC 1}  &358.5 $\to 20^o$  & $-67.5 \to -77^o$ & 0.0189 \\                     
{\rm SMC 2}  &356.5 $\to 358.5^o$ & $-73 \to -77^o$   & 0.0006 \\                   
{\rm SMC 3}  & 20  $\to 30^o$   & $-67.5 \to -72^o$ & 0.0047 \\                     
{\rm Virgo}  & 182.7  $\to 192.7^o$   & $7.4 \to 17.4^o$ &  0.0297\\                     
{\rm Milky Way}\tablenotemark{a}  & 0  $\to 360^o$ ($\lii$) & $-20 \to 20^o$ ($\bii$)  & 4.2980 \\                     
\enddata
\tablenotetext{a}{Galactic coordinates.}
\tablecomments{Excised areas for the Magellanic Clouds are the same as
in \citet{bsg01}, because REFLEX forms the basic input catalog in the
southern hemisphere.}
\end{deluxetable}

These selection criteria are fulfilled by 63 candidates. The
advantages of the redetermined fluxes over the fluxes from the input
catalogs are summarized in Chap.~\ref{anal}.
In Tab.~\ref{tbl:cand} one notes that 98\,\% of all clusters in
\gcs\ have been flagged as candidates in REFLEX, NORAS, or in the
candidate list for NORAS II; these surveys are not only all based on
the \ra\ but all use the same algorithm for the
count rate determination, further substantiating the homogeneous
candidate selection for \gcs .

The fraction of available \ro\ \ps\ pointed observations for clusters
included in \gcs\ equals 86\,\%. The actually used fraction is
slightly reduced to 75\,\% because some clusters appear extended
beyond the \ps\ field of view and therefore \ra\ data had to be
used. The fraction of clusters with published \as\ temperatures equals 87\,\%.
If a lower flux limit had been chosen the fraction of available
\ps\ pointed observations and published \as\ temperatures
would have been decreased thereby increasing the uncertainties in the derived
cluster parameters.
Furthermore this value for the flux limit ensures that no 
corrections, due to low exposure in the \ra\ or high galactic hydrogen column
density, need to be applied for the effective area covered. This can
be seen by the
effective sky coverage in the REFLEX survey area: for a flux limit
$\fxl(\eb) = 2.0\,\esc$ and a minimum of 30 source counts the sky
coverage amounts to 99\,\%.
The clear advantage is that the \gcs\ catalog can be used in a
straightforward manner in statistical analyses, because the effective
area is the same for all clusters and simply equals the covered solid
angle on the sky.

\begin{figure}[thbp]
\centering\begin{tabular}{l}
\hspace{-1.3cm}
\psfig{file=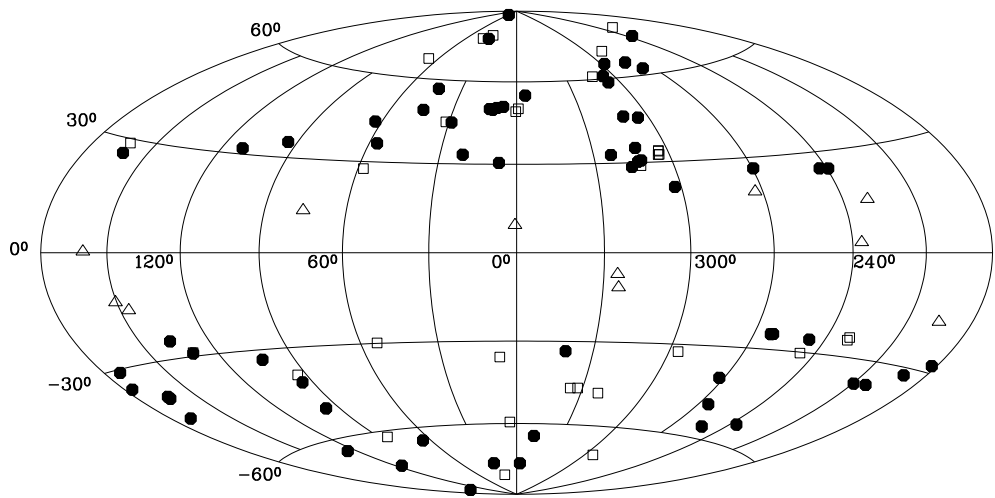,width=11cm,angle=270,clip=}
\end{tabular}
\caption{Aitoff projection of the 63 \gcs\ galaxy clusters in galactic coordinates
(filled circles). Additionally shown are 11 clusters above the flux limit
but with $\vert \bii\vert < 20.0\,\rm deg$ (open triangles) and 32
clusters with fluxes below the flux limit (open squares).}\label{aito}
\end{figure}

The distribution of clusters included in \gcs\ projected onto the
sky is shown in Fig.~\ref{aito}.
The sky coverage for the cluster sample equals 26\,721.8\,deg$^2$
(8.13994\,sr),  about two thirds of the sky.
The cluster names, coordinates and redshifts
are listed in Tab.~\ref{tbl:data1}.
Further properties of the cluster sample and completeness tests are
discussed in Sect.~\ref{disc}. 

For later analyses which do not necessarily require a complete sample,
e.g., correlations between physical parameters, 43 clusters (not
included in \gcs ) from the candidate list have been combined with
\gcs\ to form an `extended sample' of 106 clusters. This sample is not
a purely flux-limited sample with clearly defined selection criteria.
Nevertheless this extended sample is not dominated by subjective
selection and therefore one may take advantage of the
increased statistics. The difference between the results of relations
using \gcs\ and the extended sample is discussed in Sect.~\ref{relat}.

\chapter{Reduction and Analysis}
\label{anal}

This Chapter describes the derivation of the basic quantities in this work,
e.g., count rates, fluxes,
luminosities, and mass estimates for the galaxy clusters.
These and other relevant cluster parameters are tabulated along
with their uncertainties.

\section{Flux Determination}\label{fluxd}

Measuring the count rate of galaxy clusters is an important step in constructing a
flux-limited cluster sample. The count rate determination performed here is
based on the growth curve analysis method \citep{bvh00}, with
modifications adapted to the higher photon statistics available here. The main
features of the method
as well as the modifications are outlined below.

The instrument used is the \ro\ \ps\ \citep{pbh87}, with a low internal
background ideally suited for this study which needs good signal to
noise of the outer, low surface brightness regions of the
clusters. Mainly pointed observations from the public archive at
MPE have been 
used. If the cluster is extended beyond the \ps\ field of view
making a proper background 
determination difficult or if there is no pointed \ps\ observation available,
\ra\ data have been used.
The \ro\ hard energy band (channels $52-201\approx \eh$) has been used
for all count
rate measurements because of the higher background in the soft band.
The reduction of the soft X-ray background is about a factor of four in the
hard band, while still 60--100\,\% of the cluster emission is detected
\citep{bvh00}. Therefore the signal to noise ratio is multiplied by a factor
0.92--1.25 if the hard band is used.

Two X-ray cluster centers are
determined by finding the two-dimensional `center of mass' of the photon
distribution iteratively for an aperture radius of 3 and 7.5 arcmin around the
starting position. The small aperture yields the center representing the
position of the cluster's peak emission and therefore probably indicates the
position where the cluster's potential well is deepest. This center is used for
the regional selection, e.g.\ $\vert \bii\vert \ge 20.0$. The more
globally defined center with 
the larger aperture is used for the subsequent analysis tasks since for the
mass determination it is most important to have a good estimate of the slope of
the surface brightness profile in the outer parts of the cluster.

\begin{figure}[thbp]
\centering\begin{tabular}{l}
\psfig{file=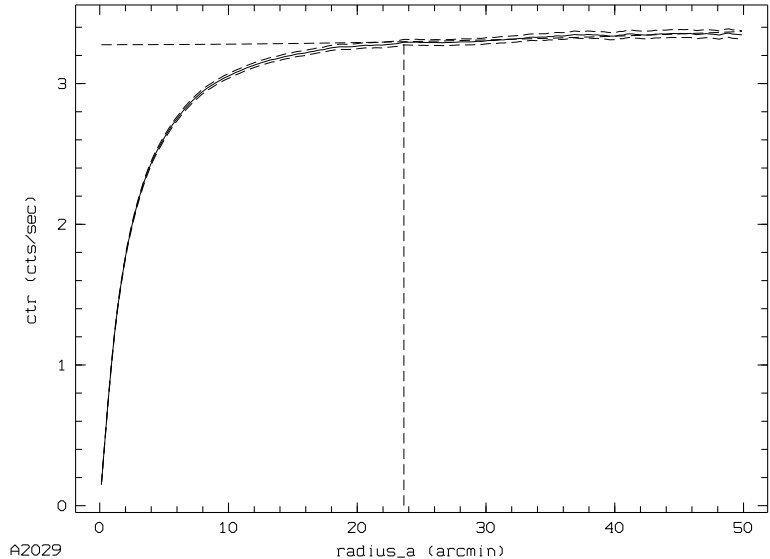,width=8cm,angle=270,clip=}
\end{tabular}
\caption{Cumulative source count rate as a function of radius (solid line) for the cluster
A2029 (pointed observation). The vertical dashed line indicates the
outer significance radius, $\rx$. The dashed lines just above and
below the source count rate indicate the 1-$\sigma$ Poissonian
errors.}\label{a2029}
\end{figure}

The background surface brightness is determined in a
ring outside the cluster emission. To minimize the influence of
discrete sources the ring is subdivided into twelve parts of equal area and a
sigma clipping is performed. To determine the count rate the area around the
global center is divided into concentric rings. For pointed observations 200
rings with a width of 15 arcsec each are used. Due to the lower photon statistics
a width of 30 arcsec is used for \ra\ data and the number of rings
depends on the field size extracted ($100 - 300$ rings for field sizes of
$2\times 2\,\rm deg^2 - 8\times 8\,\rm deg^2$). Each photon is divided by the
vignetting and deadtime corrected exposure time of the skypixel where it has
been
detected and these ratios are summed up in each ring yielding the ring count
rate. From this value the background count rate for the respective ring area is
subtracted yielding a source ring count rate. These individual source ring count
rates are integrated with increasing radius yielding the (cumulative)
source count rate for a
given radius (Fig.~\ref{a2029}). Obvious contaminating point sources
have been excluded manually.
The cut-out regions have then been assigned the average surface
brightness of the ring.
If a cluster has been found to be clearly made up of
two major components, for instance A3395n/s,
these components have been treated separately.
This procedure ensures that double clusters are not treated as a
single entity for which spherical symmetry is assumed.
For the same reason strong
subtructure has been excluded in the same manner as contaminating
point sources. In this work the aim is to characterize all cluster
properties consistently and homogeneously. Therefore if strong substructure
is identified then it is excluded for the
flux/luminosity \emph{and} mass determination.

\begin{figure}[thbp]
\centering\begin{tabular}{l}
\psfig{file=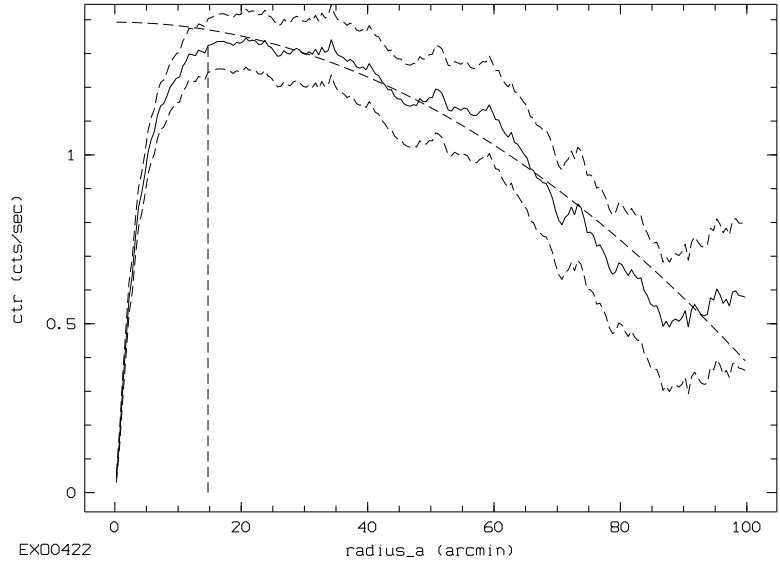,width=8cm,angle=270,clip=}
\end{tabular}
\caption{Cumulative source count rate as a function of radius for the cluster
EXO0422, shown as an extreme example (\ra\ data). The parabolic dashed
line indicates the best fit parabola for count rates larger than
$\rx$.}\label{exo}
\end{figure}

\begin{figure}[thbp]
\centering\begin{tabular}{l}
\psfig{file=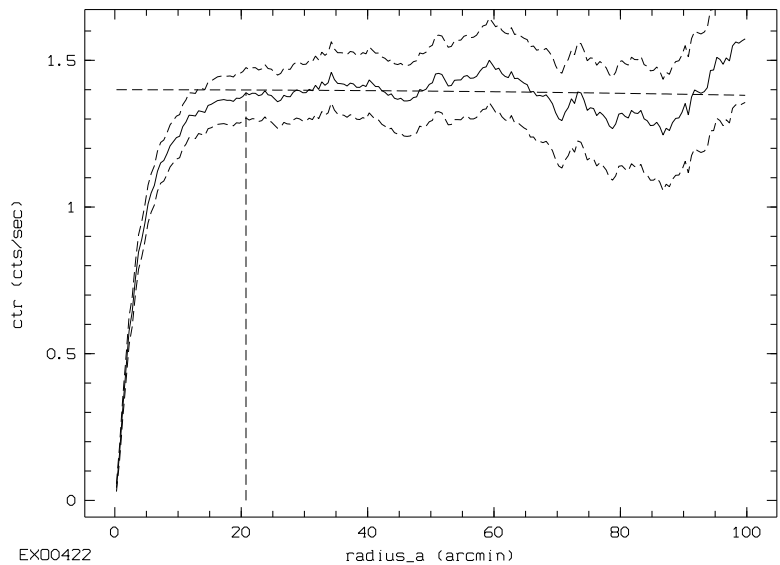,width=8cm,angle=270,clip=}
\end{tabular}
\caption{Corrected cumulative source count rate as a function of radius for the cluster
EXO0422. The count rate correction is less than 5\,\%.}\label{exoc}
\end{figure}

An outer 
significance radius of the cluster, $\rx$, is determined at the position from
where on the Poissonian
1-$\sigma$ error rises faster than the source count rate. Usually the source count
rate settles into a nearly horizontal line for radii larger than
$\rx$. In some cases, however, the source count rate
seems to increase or
decrease roughly quadratically for radii larger than $\rx$ indicating a possibly
under- or overestimated background (Fig.~\ref{exo}). Therefore
a parabola of the form 
$y=mx^2+b$ has been fitted to the source count rate for radii larger
than $\rx$ and the measured background has been corrected.
An example for a corrected source count rate profile is shown in Fig.~\ref{exoc}.

\begin{figure}[thbp]
\centering\begin{tabular}{l}
\psfig{file=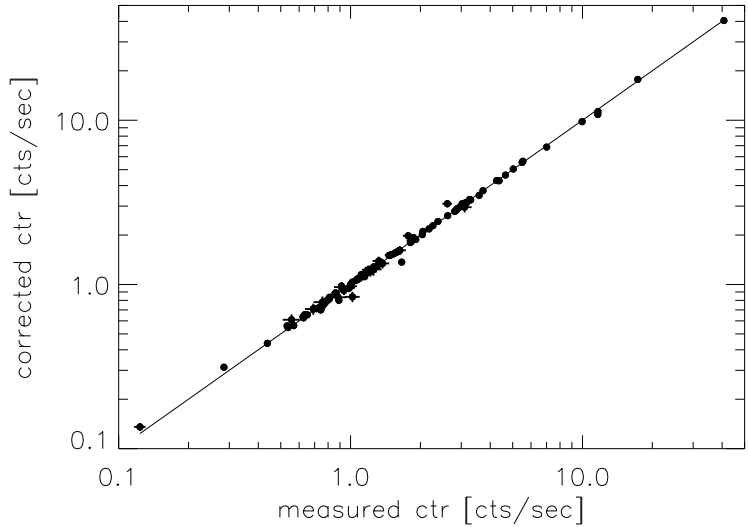,width=11cm,angle=0,clip=}
\end{tabular}
\caption{Comparison of measured and corrected source count rates for
the extended sample of 106 galaxy clusters. The
solid line indicates equality.}\label{counts}
\end{figure}

Figure~\ref{counts} shows for the extended sample (106 clusters) that
the difference between measured and 
corrected source count rate is generally very small. Nevertheless an
inspection of each count rate profile has been performed, to decide
whether the measured or corrected count rate is adopted as the final
count rate, to avoid artificial corrections due to large scale
variations of the background (especially in the large \ra\
fields). The count rates are given in Tab.~\ref{tbl:data1}. 

The conversion factor for the count rate to flux conversion depends on the
hydrogen column density, $\nhcol$, on the cluster gas temperature, $\tg$, on the
cluster gas metallicity, $A$, on the cluster redshift, $z$, and on the respective
detector responses for the two different \ps s used. The $\nhcol$ value is taken as
the value inferred from 21\,cm radio measurements for our galaxy at the
projected cluster position (\citealt{dl90};
included in the EXSAS 
software package, \citealt{zbb98}; photoelectric absorption cross
sections are taken from \citealt{mm83}).
Gas temperatures have been estimated by compiling X-ray temperatures, $\tx$, 
from the literature, giving preference to temperatures measured
with the \as\ satellite. 
For clusters where no \as\ measured temperature has been
available,
$\tx$ measured with previous X-ray satellites have been used.
The X-ray temperatures and corresponding references are given in
Tab.~\ref{tbl:data2}. 
For two
clusters included in \gcs\ no measured temperature has been found in the
literature and the $\lx (<2\,\mpc)-\tx$ relation of \citet{m98} has
been used. The relation 
for non cooling flow corrected luminosities and cooling flow
corrected/emission 
weighted temperatures has been chosen, because the luminosities
determined here have not been corrected for cooling flows and cooling
flow corrected temperatures should be a better estimate of the cluster
virial temperature. For $h_{50}=1$ this relation reads
\begin{equation}
\tx=\left[{\frac{\lx(<2\,\rm Mpc)}{6.28\times 10^{44}\,\es}}\right]^{1/2.09}\times 6\rm \,keV\,.
\label{eq:te:lt}
\end{equation}
Since the conversion from count
rate to flux depends only weakly on $\tg$
in the \ro\ energy band for the relevant temperature
range a cluster temperature $\kb \tg = 4\,\kev$ has been assumed in a first step
to determine $\lx (<2\,\mpc)$ for the clusters where no gas
temperature has been found in the literature. With this luminosity the
gas temperature has been estimated. 
The metallicity is set to 0.35 times the solar value for all
clusters \citep[e.g.,][see also Chap.~\ref{a1835}]{arb92}.
The redshifts have been compiled from
the literature and are given in
Tab.~\ref{tbl:data1} together with the corresponding references. With these quantities
and the count rates given in Tab.~\ref{tbl:data1} fluxes in the observer
rest frame energy range $\eb$ have
been calculated applying a modern version of a Raymond--Smith spectral code
\citep{rs77}.
The results are listed in Tab.~\ref{tbl:data1}. The flux calculation has also been checked
using \xs\ \citep{a96} by folding the model spectrum created with the
parameters given above with the detector response and adjusting the
normalization to reproduce the observed count rate.
It is found that for 90\,\% of the clusters the deviation between the two
results for the flux measurement is less than 1\,\%.
Luminosities in the source rest frame energy range $\eb$
have then been calculated within \xs\ by adjusting the normalization to
reproduce the initial flux measurements. 

The improvements of the flux determination performed here compared to
the input catalogs in general are now summarized.
\begin{itemize}
\item [1)] Due to the use of a high fraction of pointed observations the
photon statistics is on average much better, e.g, for the 33 clusters
contained in REFLEX and \gcs\ one finds a mean of 841 and 19580 source photons,
respectively. Consequently the cluster emission has been traced out to
larger radii for \gcs . 
\item [2)] The higher photon statistics has allowed a proper exclusion of
contaminating point sources (stars, active galactic nuclei (AGN), etc.)
and substructure, and the
separation of double clusters.
\item [3)] An iterative background correction has been performed.
\item [4)] A measured X-ray temperature is used for the flux determination in
most cases.
\end{itemize}

Simulations have shown that even for the
\gcs\ clusters with the lowest number of photons the determined flux
shows no significant trend with redshift in the relevant redshift range
\citep{irb01}.

The parameters of the table columns of Tab.~\ref{tbl:data1} are
described as follows. Column (1) lists the cluster name. Names have
been truncated to at most eight characters. Columns (2)
and (3) give the equatorial coordinates of the cluster center used for the
regional selection for the epoch J2000 in decimal degrees. Column (4)
gives the heliocentric cluster redshift. Column (5) lists the column density
of neutral galactic hydrogen in units of $10^{20}\,\rm
atoms\,cm^{-2}$. Column (6) gives the count rate in the
channel range 52--201 which corresponds to about (the energy
resolution of the \ps\ is limited) the energy range
$\eh$ in units of $\cts$. Column (7) lists the relative
($1-\sigma$ Poissonian) error of the count rate,
the flux, and the luminosity in
percent. Column (8) gives the significance radius in $\mpc$. Column (9)
lists the flux in the energy range $\eb$ in units of $\esct$. Column
(10) gives the luminosity in the energy range $\eb$ in units of $\esll$.
 Column (11) gives the bolometric luminosity (energy range $\ebol$) in
units of $\esll$. Column (12) indicates whether a \ra\ (R) or a
pointed (P) \ro\ \ps\ observation has been used. Column (13) lists the
code for the redshift reference decoded at the end of the table.
\begin{deluxetable}{lrrrrrrrrrrcc}
\tablecolumns{13} 
\tabletypesize{\footnotesize}
\tablecaption{Cluster properties \label{tbl:data1}}
\tablewidth{0pt}
\tablehead{
\colhead{Cluster}	& \colhead{R.A.}	& \colhead{Dec.}	&
\colhead{$z$}	& \colhead{$\nhcol$}	& \colhead{$\cx$}	&
\colhead{$\Delta$}	& \colhead{$\rx$}	& \colhead{$\fx$}	&
\colhead{$\lx$}	& \colhead{$\lbol$}	& \colhead{Obs}	&
\colhead{Ref}	\\
\colhead{(1)}	& \colhead{(2)}	& \colhead{(3)}	&
\colhead{(4)}	& \colhead{(5)}	& \colhead{(6)}	&
\colhead{(7)}	& \colhead{(8)}	& \colhead{(9)}	&
\colhead{(10)}	& \colhead{(11)}	& \colhead{(12)}	&
\colhead{(13)}
}
\startdata
A0085	&  10.4632 & $ -9.3054$ & 0.0556 &  3.58 &  3.488 &  0.6 & 2.13 &   7.429 &  9.789 &  24.448 & P & 2 \\ 
A0119	&  14.0649 & $ -1.2489$ & 0.0440 &  3.10 &  1.931 &  0.9 & 2.68 &   4.054 &  3.354 &   7.475 & P & 1 \\ 
A0133	&  15.6736 & $-21.8806$ & 0.0569 &  1.60 &  1.058 &  0.8 & 1.52 &   2.121 &  2.944 &   5.389 & P & 3 \\ 
NGC507	&  20.9106 & $ 33.2553$ & 0.0165 &  5.25 &  1.093 &  1.3 & 0.88 &   2.112 &  0.247 &   0.326 & P & 5 \\ 
A0262	&  28.1953 & $ 36.1528$ & 0.0161 &  5.52 &  4.366 &  3.8 & 1.48 &   9.348 &  1.040 &   1.533 & R & 6 \\ 
A0400	&  44.4152 & $  6.0170$ & 0.0240 &  9.38 &  1.146 &  1.1 & 1.85 &   2.778 &  0.686 &   1.033 & P & 8 \\ 
A0399	&  44.4684 & $ 13.0462$ & 0.0715 & 10.58 &  1.306 &  5.4 & 3.18 &   3.249 &  7.070 &  17.803 & R & 6 \\ 
A0401	&  44.7384 & $ 13.5796$ & 0.0748 & 10.19 &  2.104 &  1.1 & 3.81 &   5.281 & 12.553 &  34.073 & P & 6 \\ 
A3112	&  49.4912 & $-44.2367$ & 0.0750 &  2.53 &  1.502 &  1.1 & 2.18 &   3.103 &  7.456 &  16.128 & P & 1 \\ 
FORNAX	&  54.6686 & $-35.3103$ & 0.0046 &  1.45 &  5.324 &  5.6 & 0.53 &   9.020 &  0.082 &   0.107 & P+R & 4 \\ 
2A0335	&  54.6690 & $  9.9713$ & 0.0349 & 18.64 &  3.028 &  0.8 & 1.54 &   9.162 &  4.789 &   7.918 & P & 10 \\ 
IIIZw54	&  55.3225 & $ 15.4076$ & 0.0311 & 16.68 &  0.708 &  7.7 & 1.27 &   2.001 &  0.831 &   1.226 & R & 11 \\ 
A3158	&  55.7282 & $-53.6301$ & 0.0590 &  1.06 &  1.909 &  1.5 & 1.94 &   3.794 &  5.638 &  12.779 & P & 1 \\ 
A0478	&  63.3554 & $ 10.4661$ & 0.0900 & 15.27 &  1.827 &  0.6 & 3.12 &   5.151 & 17.690 &  49.335 & P & 6 \\ 
NGC1550	&  64.9066 & $  2.4151$ & 0.0123 & 11.59 &  1.979 &  5.4 & 0.71 &   4.632 &  0.302 &   0.407 & R & 13 \\ 
EXO0422	&  66.4637 & $ -8.5581$ & 0.0390 &  6.40 &  1.390 &  6.2 & 1.32 &   3.085 &  2.015 &   3.283 & R & 10 \\ 
A3266	&  67.8410 & $-61.4403$ & 0.0594 &  1.48 &  2.879 &  0.7 & 2.99 &   5.807 &  8.718 &  23.663 & P & 4 \\ 
A0496	&  68.4091 & $-13.2605$ & 0.0328 &  5.68 &  3.724 &  0.7 & 1.78 &   8.326 &  3.837 &   7.306 & P & 8 \\ 
A3376	&  90.4835 & $-39.9741$ & 0.0455 &  5.01 &  1.115 &  1.4 & 2.86 &   2.450 &  2.174 &   4.077 & P & 4 \\ 
A3391	&  96.5925 & $-53.6938$ & 0.0531 &  5.42 &  0.999 &  1.9 & 1.98 &   2.225 &  2.681 &   5.857 & P & 4 \\ 
A3395s	&  96.6920 & $-54.5453$ & 0.0498 &  8.49 &  0.836 &  3.8 & 1.45 &   2.009 &  2.131 &   4.471 & P & 4 \\ 
A0576	& 110.3571 & $ 55.7639$ & 0.0381 &  5.69 &  1.374 &  6.8 & 2.32 &   3.010 &  1.872 &   3.518 & R & 6 \\ 
A0754	& 137.3338 & $ -9.6797$ & 0.0528 &  4.59 &  1.537 &  1.6 & 1.91 &   3.366 &  3.990 &  11.967 & P & 6 \\ 
HYDRA-A	& 139.5239 & $-12.0942$ & 0.0538 &  4.86 &  2.179 &  0.6 & 1.66 &   4.776 &  5.930 &  11.520 & P & 13 \\ 
A1060	& 159.1784 & $-27.5212$ & 0.0114 &  4.92 &  4.653 &  3.3 & 0.95 &   9.951 &  0.554 &   0.945 & R & 6 \\ 
A1367	& 176.1903 & $ 19.7030$ & 0.0216 &  2.55 &  2.947 &  0.8 & 1.55 &   6.051 &  1.206 &   2.140 & P & 8 \\ 
MKW4	& 181.1124 & $  1.8962$ & 0.0200 &  1.86 &  1.173 &  1.7 & 1.23 &   2.268 &  0.390 &   0.543 & P & 10 \\ 
ZwCl1215	& 184.4220 & $  3.6604$ & 0.0750 &  1.64 &  1.081 &  1.3 & 2.55 &   2.183 &  5.240 &  11.656 & P & 19 \\ 
NGC4636	& 190.7084 & $  2.6880$ & 0.0037 &  1.75 &  3.102 &  7.2 & 0.39 &   4.085 &  0.023 &   0.027 & R & 13 \\ 
A3526	& 192.1995 & $-41.3087$ & 0.0103 &  8.25 & 11.655 &  2.2 & 1.64 &  27.189 &  1.241 &   2.238 & R & 15 \\ 
A1644	& 194.2900 & $-17.4029$ & 0.0474 &  5.33 &  1.853 &  5.1 & 1.85 &   4.030 &  3.876 &   7.882 & R & 8 \\ 
A1650	& 194.6712 & $ -1.7572$ & 0.0845 &  1.54 &  1.218 &  6.6 & 3.17 &   2.405 &  7.308 &  17.955 & R & 6 \\ 
A1651	& 194.8419 & $ -4.1947$ & 0.0860 &  1.71 &  1.254 &  1.2 & 2.03 &   2.539 &  8.000 &  18.692 & P & 22 \\ 
COMA	& 194.9468 & $ 27.9388$ & 0.0232 &  0.89 & 17.721 &  1.4 & 4.04 &  34.438 &  7.917 &  22.048 & R & 8 \\ 
NGC5044	& 198.8530 & $-16.3879$ & 0.0090 &  4.91 &  3.163 &  0.5 & 0.56 &   5.514 &  0.193 &   0.246 & P & 24 \\ 
A1736	& 201.7238 & $-27.1765$ & 0.0461 &  5.36 &  1.631 &  6.3 & 2.47 &   3.537 &  3.223 &   5.682 & R & 25 \\ 
A3558	& 201.9921 & $-31.5017$ & 0.0480 &  3.63 &  3.158 &  0.5 & 2.11 &   6.720 &  6.615 &  14.600 & P & 1 \\ 
A3562	& 203.3984 & $-31.6678$ & 0.0499 &  3.91 &  1.367 &  0.9 & 2.01 &   2.928 &  3.117 &   6.647 & P & 4 \\ 
A3571	& 206.8692 & $-32.8553$ & 0.0397 &  3.93 &  5.626 &  0.7 & 2.35 &  12.089 &  8.132 &  20.310 & P & 21 \\ 
A1795	& 207.2201 & $ 26.5944$ & 0.0616 &  1.20 &  3.132 &  0.3 & 2.14 &   6.270 & 10.124 &  27.106 & P & 6 \\ 
A3581	& 211.8852 & $-27.0153$ & 0.0214 &  4.26 &  1.603 &  3.2 & 0.64 &   3.337 &  0.657 &   0.926 & P & 28 \\ 
MKW8	& 220.1596 & $  3.4717$ & 0.0270 &  2.60 &  1.255 &  8.4 & 1.90 &   2.525 &  0.789 &   1.355 & R & 29 \\ 
A2029	& 227.7331 & $  5.7450$ & 0.0767 &  3.07 &  3.294 &  0.6 & 2.78 &   6.938 & 17.313 &  50.583 & P & 6 \\ 
A2052	& 229.1846 & $  7.0211$ & 0.0348 &  2.90 &  2.279 &  1.0 & 1.14 &   4.713 &  2.449 &   4.061 & P & 6 \\ 
MKW3S	& 230.4643 & $  7.7059$ & 0.0450 &  3.15 &  1.578 &  1.0 & 1.39 &   3.299 &  2.865 &   5.180 & P & 10 \\ 
A2065	& 230.6096 & $ 27.7120$ & 0.0721 &  2.84 &  1.227 &  6.1 & 3.09 &   2.505 &  5.560 &  12.271 & R & 6 \\ 
A2063	& 230.7734 & $  8.6112$ & 0.0354 &  2.92 &  2.038 &  1.3 & 2.13 &   4.232 &  2.272 &   4.099 & P & 8 \\ 
A2142	& 239.5824 & $ 27.2336$ & 0.0899 &  4.05 &  2.888 &  0.9 & 3.09 &   6.241 & 21.345 &  64.760 & P & 6 \\ 
A2147	& 240.5628 & $ 15.9586$ & 0.0351 &  3.29 &  2.623 &  3.2 & 1.87 &   5.522 &  2.919 &   6.067 & P & 8 \\ 
A2163	& 243.9433 & $ -6.1436$ & 0.2010 & 12.27 &  0.773 &  1.5 & 3.15 &   2.039 & 34.128 & 123.200 & P & 31 \\ 
A2199	& 247.1586 & $ 39.5477$ & 0.0302 &  0.84 &  5.535 &  1.8 & 2.37 &  10.642 &  4.165 &   7.904 & R & 8 \\ 
A2204	& 248.1962 & $  5.5733$ & 0.1523 &  5.94 &  1.211 &  1.6 & 3.29 &   2.750 & 26.938 &  68.989 & P & 6 \\ 
A2244	& 255.6749 & $ 34.0578$ & 0.0970 &  2.07 &  1.034 &  2.1 & 2.64 &   2.122 &  8.468 &  21.498 & P & 6 \\ 
A2256	& 255.9884 & $ 78.6481$ & 0.0601 &  4.02 &  2.811 &  1.4 & 3.09 &   6.054 &  9.322 &  22.713 & P & 6 \\ 
A2255	& 258.1916 & $ 64.0640$ & 0.0800 &  2.51 &  0.976 &  1.2 & 3.22 &   2.022 &  5.506 &  13.718 & P & 6 \\ 
A3667	& 303.1362 & $-56.8419$ & 0.0560 &  4.59 &  3.293 &  0.7 & 2.81 &   7.201 &  9.624 &  24.233 & P & 1 \\ 
S1101	& 348.4941 & $-42.7268$ & 0.0580 &  1.85 &  1.237 &  0.9 & 1.64 &   2.485 &  3.597 &   5.939 & P & 35 \\ 
A2589	& 350.9868 & $ 16.7753$ & 0.0416 &  4.39 &  1.200 &  1.3 & 1.46 &   2.591 &  1.924 &   3.479 & P & 37 \\ 
A2597	& 351.3318 & $-12.1246$ & 0.0852 &  2.50 &  1.074 &  1.2 & 1.43 &   2.213 &  6.882 &  13.526 & P & 6 \\ 
A2634	& 354.6201 & $ 27.0269$ & 0.0312 &  5.17 &  1.096 &  1.6 & 1.79 &   2.415 &  1.008 &   1.822 & P & 6 \\ 
A2657	& 356.2334 & $  9.1952$ & 0.0404 &  5.27 &  1.148 &  0.9 & 1.52 &   2.535 &  1.771 &   3.202 & P & 8 \\ 
A4038	& 356.9322 & $-28.1415$ & 0.0283 &  1.55 &  2.854 &  1.3 & 1.35 &   5.694 &  1.956 &   3.295 & P & 4 \\ 
A4059	& 359.2541 & $-34.7591$ & 0.0460 &  1.10 &  1.599 &  1.3 & 1.72 &   3.170 &  2.872 &   5.645 & P & 36 \\
\tableline
\sidehead{\hspace{4.21cm}\vspace{1mm} Clusters from the extended sample not included in \gcss .} 
\tableline \\[-3.15mm]
A2734	&   2.8389 & $-28.8539$ & 0.0620 &  1.84 &  0.710 &  2.5 & 1.74 &   1.434 &  2.365 &   4.357 & P & 1 \\ 
A2877	&  17.4796 & $-45.9225$ & 0.0241 &  2.10 &  0.801 &  1.2 & 1.06 &   1.626 &  0.405 &   0.714 & P & 4 \\ 
NGC499	&  20.7971 & $ 33.4587$ & 0.0147 &  5.25 &  0.313 &  2.5 & 0.30 &   0.479 &  0.045 &   0.051 & P & 5 \\ 
AWM7	&  43.6229 & $ 41.5781$ & 0.0172 &  9.21 &  7.007 &  2.0 & 1.58 &  16.751 &  2.133 &   3.882 & R & 7 \\ 
PERSEUS	&  49.9455 & $ 41.5150$ & 0.0183 & 15.69 & 40.723 &  0.8 & 3.30 & 113.731 & 16.286 &  40.310 & R & 9 \\ 
S405	&  58.0078 & $-82.2315$ & 0.0613 &  7.65 &  0.781 &  8.2 & 2.14 &   1.800 &  2.899 &   5.574 & R & 12 \\ 
3C129	&  72.5602 & $ 45.0256$ & 0.0223 & 67.89 &  1.512 &  5.6 & 1.61 &  10.566 &  2.242 &   4.996 & R & 10 \\ 
A0539	&  79.1560 & $  6.4421$ & 0.0288 & 12.06 &  1.221 &  1.3 & 1.37 &   3.182 &  1.135 &   1.935 & P & 14 \\ 
S540	&  85.0265 & $-40.8431$ & 0.0358 &  3.53 &  0.788 &  5.0 & 0.84 &   1.611 &  0.887 &   1.353 & R & 4 \\ 
A0548w	&  86.3785 & $-25.9340$ & 0.0424 &  1.79 &  0.136 &  5.4 & 0.73 &   0.234 &  0.183 &   0.240 & P & 15 \\ 
A0548e	&  87.1596 & $-25.4692$ & 0.0410 &  1.88 &  0.771 &  1.8 & 2.12 &   1.551 &  1.117 &   1.870 & P & 15 \\ 
A3395n	&  96.9005 & $-54.4447$ & 0.0498 &  5.42 &  0.699 &  3.9 & 1.37 &   1.555 &  1.650 &   3.461 & P & 4 \\ 
UGC03957	& 115.2481 & $ 55.4319$ & 0.0340 &  4.59 &  0.936 &  6.0 & 0.94 &   1.975 &  0.980 &   1.531 & R & 16 \\ 
PKS0745	& 116.8837 & $-19.2955$ & 0.1028 & 43.49 &  1.268 &  1.0 & 2.44 &   6.155 & 27.565 &  70.604 & P & 17 \\ 
A0644	& 124.3553 & $ -7.5159$ & 0.0704 &  5.14 &  1.799 &  1.0 & 4.02 &   3.994 &  8.414 &  22.684 & P & 6 \\ 
S636	& 157.5151 & $-35.3093$ & 0.0116 &  6.42 &  3.102 &  4.9 & 1.18 &   5.869 &  0.341 &   0.446 & R & 18 \\ 
A1413	& 178.8271 & $ 23.4051$ & 0.1427 &  1.62 &  0.636 &  1.6 & 2.39 &   1.289 & 11.090 &  28.655 & P & 6 \\ 
M49	& 187.4437 & $  7.9956$ & 0.0044 &  1.59 &  1.259 &  1.0 & 0.27 &   1.851 &  0.015 &   0.019 & P & 15 \\ 
A3528n	& 193.5906 & $-29.0130$ & 0.0540 &  6.10 &  0.560 &  2.3 & 1.51 &   1.263 &  1.581 &   2.752 & P & 1 \\ 
A3528s	& 193.6708 & $-29.2254$ & 0.0551 &  6.10 &  0.756 &  1.6 & 1.35 &   1.703 &  2.224 &   3.746 & P & 20 \\ 
A3530	& 193.9211 & $-30.3451$ & 0.0544 &  6.00 &  0.438 &  2.8 & 1.55 &   0.987 &  1.252 &   2.317 & P & 21 \\ 
A3532	& 194.3375 & $-30.3698$ & 0.0539 &  5.96 &  0.797 &  1.8 & 1.64 &   1.797 &  2.235 &   4.483 & P & 21 \\ 
A1689	& 197.8726 & $ -1.3408$ & 0.1840 &  1.80 &  0.712 &  1.1 & 2.36 &   1.454 & 20.605 &  60.707 & P & 23 \\ 
A3560	& 203.1119 & $-33.1355$ & 0.0495 &  3.92 &  0.714 &  2.5 & 2.00 &   1.519 &  1.601 &   2.701 & P & 26 \\ 
A1775	& 205.4582 & $ 26.3820$ & 0.0757 &  1.00 &  0.654 &  1.8 & 2.02 &   1.290 &  3.175 &   5.735 & P & 27 \\ 
A1800	& 207.3408 & $ 28.1038$ & 0.0748 &  1.18 &  0.610 &  7.9 & 1.98 &   1.183 &  2.840 &   5.337 & R & 28 \\ 
A1914	& 216.5035 & $ 37.8268$ & 0.1712 &  0.97 &  0.729 &  1.4 & 2.35 &   1.454 & 17.813 &  56.533 & P & 6 \\ 
NGC5813	& 225.2994 & $  1.6981$ & 0.0064 &  4.19 &  0.976 &  6.7 & 0.17 &   1.447 &  0.025 &   0.029 & R & 13 \\ 
NGC5846	& 226.6253 & $  1.6089$ & 0.0061 &  4.25 &  0.569 &  2.3 & 0.21 &   0.851 &  0.014 &   0.016 & P & 13 \\ 
A2151w	& 241.1465 & $ 17.7252$ & 0.0369 &  3.36 &  0.754 &  1.9 & 1.46 &   1.568 &  0.917 &   1.397 & P & 8 \\ 
A3627	& 243.5546 & $-60.8430$ & 0.0163 & 20.83 &  9.962 &  3.0 & 2.20 &  31.084 &  3.524 &   8.179 & R & 30 \\ 
TRIANGUL	& 249.5758 & $-64.3557$ & 0.0510 & 12.29 &  4.294 &  0.7 & 2.54 &  11.308 & 12.508 &  37.739 & P & 32 \\ 
OPHIUCHU	& 258.1115 & $-23.3634$ & 0.0280 & 20.14 & 11.642 &  2.0 & 2.29 &  35.749 & 11.953 &  37.391 & R & 33 \\ 
ZwCl1742	& 266.0623 & $ 32.9893$ & 0.0757 &  3.56 &  0.889 &  4.4 & 1.83 &   1.850 &  4.529 &   9.727 & R & 34 \\ 
A2319	& 290.2980 & $ 43.9484$ & 0.0564 &  8.77 &  5.029 &  1.0 & 3.57 &  12.202 & 16.508 &  47.286 & P & 6 \\ 
A3695	& 308.6991 & $-35.8135$ & 0.0890 &  3.56 &  0.836 &  9.2 & 2.58 &   1.739 &  5.882 &  12.715 & R & 1 \\ 
IIZw108	& 318.4752 & $  2.5564$ & 0.0494 &  6.63 &  0.841 &  7.3 & 2.20 &   1.884 &  1.969 &   3.445 & R & 5 \\ 
A3822	& 328.5438 & $-57.8668$ & 0.0760 &  2.12 &  0.964 &  7.3 & 3.18 &   1.926 &  4.758 &   9.877 & R & 1 \\ 
A3827	& 330.4869 & $-59.9641$ & 0.0980 &  2.84 &  0.953 &  5.8 & 1.78 &   1.955 &  7.963 &  20.188 & R & 1 \\ 
A3888	& 338.6255 & $-37.7343$ & 0.1510 &  1.20 &  0.546 &  2.4 & 1.52 &   1.096 & 10.512 &  30.183 & P & 23 \\ 
A3921	& 342.5019 & $-64.4286$ & 0.0936 &  2.80 &  0.626 &  1.7 & 2.43 &   1.308 &  4.882 &  11.023 & P & 12 \\ 
HCG94	& 349.3041 & $ 18.7060$ & 0.0417 &  4.55 &  0.820 &  1.0 & 2.09 &   1.775 &  1.324 &   2.319 & P & 36 \\ 
RXJ2344	& 356.0723 & $ -4.3776$ & 0.0786 &  3.54 &  0.653 &  1.4 & 1.61 &   1.385 &  3.661 &   7.465 & P & 12 \\ 
\enddata

\tablecomments{
The columns are described in detail at the end of Section~\ref{fluxd}.
}

\tablerefs{
(1) \citealt{1996A&A...310....8K}.
(2) \citealt{mkd96}. %
(3) Median of 9 galaxy redshifts compiled from \citealt{1989spce.book.....L,1991AJ....101..783M,1996ApJS..107..201L,1998ApJ...502..134W}.
(4) \citealt{aco89}.
(5) \citealt{1999ApJS..121..287H}.
(6) \citealt{sr87}.
(7) \citealt{1994AJ....107..427D}.
(8) \citealt{1993AJ....106.1273Z}.
(9) \citealt{1992A&AS...95..129P}.
(10) \citealt{1992NEDR....1....1N}.
(11) \citealt{bvh00}.
(12) \citealt{gbg99}.
(13) \citealt{1991RC3...C......0D}.
(14) \citealt{1990ApJS...74....1Z}.
(15) \citealt{1996MNRAS.279..349d}.
(16) \citealt{1988PASP..100.1423M}.
(17) \citealt{1995ApJ...454...44Y}.
(18) \citealt{1995A&A...297...56G}.
(19) \citealt{eeb98}.
(20) Median of 8 galaxy redshifts compiled from \citealt{1991RC3...C......0D,1995AJ....110..463Q,1998A&AS..129..399K}.
(21) \citealt{1990AJ.....99.1709V}.
(22) \citealt{1992MNRAS.259...67A}.
(23) \citealt{1990ApJS...72..715T}.
(24) \citealt{1998AJ....116....1D}.
(25) \citealt{1988AJ.....95..985D}.
(26) \citealt{1987RMxAA..14...72M}.
(27) Median of 13 galaxy redshifts compiled from \citealt{1983AJ.....88.1285K,1990ApJS...74....1Z,1992NEDR....1....1N,1995A&AS..110...19D,1995AJ....110...32O}.
(28) \citealt{1992ApJ...384..404P}.
(29) \citealt{1994AJ....108..361A}.
(30) \citealt{1996Natur.379..519K}.
(31) \citealt{1995A&A...293..337E}.
(32) \citealt{es91a}.
(33) \citealt{lef89}.
(34) \citealt{1976ApJ...206..364U}.
(35) \citealt{1991ApJS...76..813S}.
(36) \citealt{1992ApJ...399..353H}.
(37) \citealt{1991AJ....102.1581B}.
}

\end{deluxetable}

\section{Mass Determination}\label{massd}

The gravitational mass has been determined as outlined in
Sect.~\ref{back:dm} utilizing the surface brightness profile
(Sect.~\ref{back:rog}) and the
gas temperature (Sects.~\ref{back:tg}, \ref{fluxd}). Since the general
cluster temperature structure for radii larger than about half the
virial radius has not been studied well observationally until now,
isothermality is assumed.
The influence of the isothermal assumption on the results is discussed in
Chap.~\ref{disc}.

Having determined the integrated mass as a function of radius using (\ref{back:ma2}), a
physically meaningful fiducial radius for the mass measurement has to
be defined.
The radii commonly used are either the Abell
radius, $r_{200}$, or $r_{500}$. The Abell radius is fixed at
$\rab\equiv 3\,\mpc$.
The radius $r_{200}$ ($r_{500}$) is the radius within which the mean
gravitational mass density is equal to a multiple of the critical density
of the universe,
$\langle \rot \rangle = 200\, (500)\,\roc$.
It has been shown that a correction for redshift
is not necessary for the nearby clusters included in \gcs\
\citep{frb00} and the zero redshift value for $\roc$ (see Eqs.~\ref{roc}, \ref{h0e})
is used for all calculations,
i.e.\ $\roc=4.6975\times 10^{-30}\rm g\,cm^{-3}$, unless noted otherwise.
In Sect.~\ref{func_pred} it is shown that the results are not affected if
instead the cluster redshift is incorporated and
a strong evolution model is assumed.

In order to treat clusters of different size in a homogeneous way 
the cluster mass is determined at a characteristic density but also the
mass determined formally
at a fixed radius is given for comparison. Spherical collapse models predict a
virial density
$\langle \rov \rangle \approx 178\,\roc$ for $\om=1$
\citep[e.g.,][]{ks96}, so a pragmatic 
approximation to the virial mass is to use
$r_{200}$ as the outer boundary. Simulations performed by
\citet{emn96} have shown, however, that isothermal X-ray mass
measurements may be biased 
towards high masses for $ r > r_{500}$. Furthermore for
most of the clusters in \gcs\ (86\,\%) up to $r_{500}$ no extrapolation outside
the significantly 
detected cluster emission is necessary, i.e.\ $r_{500} < \rx$, whereas
the fraction is lower
for $r_{200}$ (25\,\%) and $\rab$ (17\,\%).
In summary the most accurate results are expected for $\mt (<r_{500}) \equiv
M_{500}$, but for a comparison to predicted mass functions $M_{200}$ is
the more appropriate value (see Sect.~\ref{back:tmf}).
Results for all determined masses and their corresponding radii are
given in Tab.~\ref{tbl:data2}.
\begin{figure}[thbp]
\centering\begin{tabular}{l}
\psfig{file=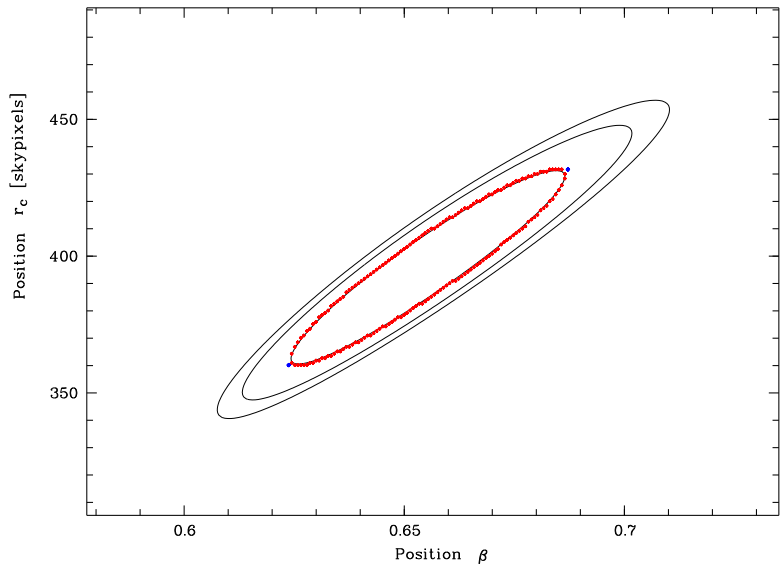,width=8cm,angle=270,clip=}
\end{tabular}
\caption{Confidence contours of the fit parameters $\rc$ and $\beta$ for the
cluster A3532 as a typical example. Shown are the 68, 90, and 95\,\%
confidence levels (for two interesting parameters). The points on the
68\,\% error ellipse mark the value
pairs used for the calculation of the uncertainty of the mass
determination.}\label{a3532rcb}
\end{figure}

A major source of uncertainty comes from the temperature
measurements. However, this (statistical) error is less than 5\,\% for
one third of the clusters, therefore also other sources of error have
to be taken into account, in particular one cannot neglect the
uncertainties of the fit parameter values when assessing the statistical errors
of the mass measurements.
Therefore mass errors have been calculated by varying the
fit parameter values, $\beta$ and
$\rc$, along their 68\,\% confidence level error
ellipse (Fig.~\ref{a3532rcb}) and using the upper and lower bound of the quoted temperature
ranges. The statistical mass error range has then been defined 
between the maximum and minimum mass.
Note that a simple error propagation applied to (\ref{back:ma2}) would
underestimate the uncertainty on $\mtz$ and $\mtf$, since $r_{200}$
and $r_{500}$ also
depend on $\tg$, $\beta$, and (weakly) $\rc$.
The individual mass errors have been used in
subsequent calculations, unless noted otherwise. In log space
errors in general have been
transformed as 
$\Delta\log(x)=\log(e)\,(x^+-x^-)/(2\,x)$, where $x^+$ and 
$x^-$ denote the upper and lower boundary of the quantity's error range,
respectively.
A mean statistical error in the gravitational mass estimate of 23\,\%
for clusters 
included in \gcs\ and a mean error of 27\,\% for the extended
sample has been found.

The parameters of the table columns of Tab.~\ref{tbl:data2} are
described as follows. Column (1) lists the cluster name. Column (2)
gives the $\beta$ parameter value and the corresponding 68\,\% c.l.\
statistical uncertainty for two interesting parameters. Column (3)
gives the core radius in $\kpc$ and the corresponding
uncertainty. Column (4) lists the X-ray temperature along with its
error. Columns (5) and (7) give $\mtf$ and $\mtz$ and their 
uncertainties in units of $10^{14}\,\msu$. Columns (6) and (8) list
$r_{500}$ and $r_{200}$ and their uncertainties in
$\mpc$.
Column (9) gives $\mab\equiv \mt(<\rab)$ in units of $10^{14}\,\msu$. 
Column (10) lists the
code for the temperature reference decoded at the end of the table.
Temperatures for codes 1--7 have been determined with
\as , code 8 with \ro , code 9 with EXOSAT, code 10 with Einstein, and
code 11 with a \ro --\as\ $\lx$--$\tx$ relation. Temperatures for code
11 are enclosed in parentheses and the corresponding errors have been
calculated using the scatter in the $\lx$--$\tx$ relation.

In the following the parameters of the columns of Tab.~\ref{tbl:data3}
are described.
Column (1) lists the cluster name.
Column (2) gives the central gas density in units of $10^{-25}\,\rm g\,cm^{-3}$.
In column (3) and (4) the gas masses are given in units of
$10^{13}\,\msug$. The errors include the
uncertainty of $r_{500}$ and $r_{200}$.
The respective gas mass fractions ($\fg\equiv\mg\mt^{-1}$) are listed in
columns (5) and (6). The errors
are determined from the (symmetrized) uncertainty of $\mt$.
\begin{deluxetable}{lccccccccc}
\tablecolumns{10} 
\tabletypesize{\footnotesize}
\tablecaption{Cluster properties \label{tbl:data2}}
\tablewidth{0pt}
\tablehead{
\colhead{Cluster}	& 
\colhead{$\beta$}	& \colhead{$\rc$}		& \colhead{$\tx$}		&
 \colhead{$\mtf$}	& \colhead{$r_{500}$}	&
\colhead{$\mtz$}	& \colhead{$r_{200}$}	& \colhead{$\mab$}		&  
\colhead{Ref}	\\
\colhead{(1)}	& \colhead{(2)}	& \colhead{(3)}	&
\colhead{(4)}	& \colhead{(5)}	& \colhead{(6)}	&
\colhead{(7)}	& \colhead{(8)}	& \colhead{(9)}	&
\colhead{(10)}
}
\startdata
A0085	& $0.532^{+0.004}_{-0.004}$ & $  83^{+  3}_{-  3}$ & $ 6.90^{+0.40}_{-0.40}$ & $ 6.84^{+ 0.66}_{- 0.66}$ & $1.68^{+0.05}_{-0.06}$ & $10.80^{+ 1.12}_{- 1.04}$ & $2.66^{+0.09}_{-0.09}$ & 12.21 & 1 \\[0.9mm] 
A0119	& $0.675^{+0.026}_{-0.023}$ & $ 501^{+ 28}_{- 26}$ & $ 5.60^{+0.30}_{-0.30}$ & $ 6.23^{+ 0.92}_{- 0.76}$ & $1.63^{+0.08}_{-0.07}$ & $10.76^{+ 1.50}_{- 1.39}$ & $2.66^{+0.11}_{-0.13}$ & 12.24 & 1 \\[0.9mm] 
A0133	& $0.530^{+0.004}_{-0.004}$ & $  45^{+  2}_{-  2}$ & $ 3.80^{+2.00}_{-0.90}$ & $ 2.78^{+ 2.51}_{- 0.95}$ & $1.24^{+0.30}_{-0.16}$ & $ 4.41^{+ 4.00}_{- 1.52}$ & $1.97^{+0.47}_{-0.27}$ & 6.71 & 9 \\[0.9mm] 
NGC507	& $0.444^{+0.005}_{-0.005}$ & $  19^{+  1}_{-  1}$ & $ 1.26^{+0.07}_{-0.07}$ & $ 0.41^{+ 0.04}_{- 0.04}$ & $0.66^{+0.02}_{-0.02}$ & $ 0.64^{+ 0.07}_{- 0.06}$ & $1.04^{+0.04}_{-0.04}$ & 1.86 & 2 \\[0.9mm] 
A0262	& $0.443^{+0.018}_{-0.017}$ & $  42^{+ 12}_{- 10}$ & $ 2.15^{+0.06}_{-0.06}$ & $ 0.90^{+ 0.10}_{- 0.09}$ & $0.86^{+0.03}_{-0.03}$ & $ 1.42^{+ 0.15}_{- 0.13}$ & $1.35^{+0.05}_{-0.04}$ & 3.17 & 2 \\[0.9mm] 
A0400	& $0.534^{+0.014}_{-0.013}$ & $ 154^{+  9}_{-  9}$ & $ 2.31^{+0.14}_{-0.14}$ & $ 1.28^{+ 0.17}_{- 0.15}$ & $0.96^{+0.04}_{-0.04}$ & $ 2.07^{+ 0.30}_{- 0.25}$ & $1.53^{+0.08}_{-0.06}$ & 4.10 & 2 \\[0.9mm] 
A0399	& $0.713^{+0.137}_{-0.095}$ & $ 450^{+132}_{-100}$ & $ 7.00^{+0.40}_{-0.40}$ & $10.00^{+ 3.73}_{- 2.48}$ & $1.91^{+0.21}_{-0.18}$ & $16.64^{+ 6.61}_{- 4.32}$ & $3.07^{+0.36}_{-0.30}$ & 16.24 & 1 \\[0.9mm] 
A0401	& $0.613^{+0.010}_{-0.010}$ & $ 246^{+ 11}_{- 10}$ & $ 8.00^{+0.40}_{-0.40}$ & $10.27^{+ 1.08}_{- 0.93}$ & $1.92^{+0.07}_{-0.05}$ & $16.59^{+ 1.62}_{- 1.62}$ & $3.07^{+0.09}_{-0.10}$ & 16.21 & 1 \\[0.9mm] 
A3112	& $0.576^{+0.006}_{-0.006}$ & $  61^{+  3}_{-  3}$ & $ 5.30^{+0.70}_{-1.00}$ & $ 5.17^{+ 1.17}_{- 1.45}$ & $1.53^{+0.11}_{-0.16}$ & $ 8.22^{+ 1.79}_{- 2.31}$ & $2.43^{+0.16}_{-0.25}$ & 10.16 & 1 \\[0.9mm] 
FORNAX	& $0.804^{+0.098}_{-0.084}$ & $ 174^{+ 17}_{- 15}$ & $ 1.20^{+0.04}_{-0.04}$ & $ 0.87^{+ 0.22}_{- 0.16}$ & $0.84^{+0.07}_{-0.06}$ & $ 1.42^{+ 0.36}_{- 0.27}$ & $1.35^{+0.11}_{-0.09}$ & 3.20 & 2 \\[0.9mm] 
2A0335	& $0.575^{+0.004}_{-0.003}$ & $  33^{+  1}_{-  1}$ & $ 3.01^{+0.07}_{-0.07}$ & $ 2.21^{+ 0.10}_{- 0.09}$ & $1.15^{+0.02}_{-0.02}$ & $ 3.51^{+ 0.16}_{- 0.15}$ & $1.83^{+0.03}_{-0.03}$ & 5.76 & 2 \\[0.9mm] 
IIIZw54	& $0.887^{+0.320}_{-0.151}$ & $ 289^{+124}_{- 73}$ & ($ 2.16^{+0.35}_{-0.30}$) & $ 2.36^{+ 2.22}_{- 0.90}$ & $1.18^{+0.29}_{-0.17}$ & $ 3.93^{+ 3.83}_{- 1.54}$ & $1.89^{+0.48}_{-0.29}$ & 6.32 & 11 \\[0.9mm] 
A3158	& $0.661^{+0.025}_{-0.022}$ & $ 269^{+ 20}_{- 19}$ & $ 5.77^{+0.10}_{-0.05}$ & $ 7.00^{+ 0.52}_{- 0.42}$ & $1.69^{+0.04}_{-0.03}$ & $11.29^{+ 0.95}_{- 0.68}$ & $2.69^{+0.08}_{-0.06}$ & 12.61 & 3 \\[0.9mm] 
A0478	& $0.613^{+0.004}_{-0.004}$ & $  98^{+  2}_{-  2}$ & $ 8.40^{+0.80}_{-1.40}$ & $11.32^{+ 1.78}_{- 2.81}$ & $1.99^{+0.10}_{-0.18}$ & $17.89^{+ 2.95}_{- 4.35}$ & $3.13^{+0.18}_{-0.27}$ & 17.12 & 1 \\[0.9mm] 
NGC1550	& $0.554^{+0.049}_{-0.037}$ & $  45^{+ 15}_{- 11}$ & $ 1.43^{+0.04}_{-0.03}$ & $ 0.69^{+ 0.12}_{- 0.09}$ & $0.78^{+0.04}_{-0.04}$ & $ 1.09^{+ 0.20}_{- 0.14}$ & $1.23^{+0.07}_{-0.06}$ & 2.64 & 5 \\[0.9mm] 
EXO0422	& $0.722^{+0.104}_{-0.071}$ & $ 142^{+ 40}_{- 30}$ & $ 2.90^{+0.90}_{-0.60}$ & $ 2.89^{+ 2.39}_{- 1.14}$ & $1.26^{+0.28}_{-0.19}$ & $ 4.63^{+ 3.84}_{- 1.82}$ & $2.00^{+0.44}_{-0.31}$ & 6.96 & 9 \\[0.9mm] 
A3266	& $0.796^{+0.020}_{-0.019}$ & $ 564^{+ 21}_{- 20}$ & $ 8.00^{+0.50}_{-0.50}$ & $14.17^{+ 1.94}_{- 1.84}$ & $2.14^{+0.09}_{-0.10}$ & $23.76^{+ 3.23}_{- 2.91}$ & $3.45^{+0.15}_{-0.14}$ & 20.47 & 1 \\[0.9mm] 
A0496	& $0.484^{+0.003}_{-0.003}$ & $  30^{+  1}_{-  1}$ & $ 4.13^{+0.08}_{-0.08}$ & $ 2.76^{+ 0.11}_{- 0.11}$ & $1.24^{+0.02}_{-0.02}$ & $ 4.35^{+ 0.18}_{- 0.17}$ & $1.96^{+0.03}_{-0.03}$ & 6.66 & 2 \\[0.9mm] 
A3376	& $1.054^{+0.101}_{-0.083}$ & $ 755^{+ 69}_{- 60}$ & $ 4.00^{+0.40}_{-0.40}$ & $ 6.32^{+ 2.11}_{- 1.59}$ & $1.64^{+0.17}_{-0.15}$ & $11.96^{+ 3.82}_{- 2.98}$ & $2.75^{+0.26}_{-0.25}$ & 13.20 & 1 \\[0.9mm] 
A3391	& $0.579^{+0.026}_{-0.024}$ & $ 234^{+ 24}_{- 22}$ & $ 5.40^{+0.60}_{-0.60}$ & $ 5.18^{+ 1.31}_{- 1.08}$ & $1.53^{+0.12}_{-0.11}$ & $ 8.41^{+ 2.13}_{- 1.81}$ & $2.44^{+0.19}_{-0.19}$ & 10.35 & 1 \\[0.9mm] 
A3395s	& $0.964^{+0.275}_{-0.167}$ & $ 604^{+173}_{-118}$ & $ 5.00^{+0.30}_{-0.30}$ & $ 8.82^{+ 4.79}_{- 2.61}$ & $1.83^{+0.29}_{-0.20}$ & $15.34^{+ 8.79}_{- 4.74}$ & $2.99^{+0.49}_{-0.35}$ & 15.42 & 1 \\[0.9mm] 
A0576	& $0.825^{+0.432}_{-0.185}$ & $ 394^{+221}_{-125}$ & $ 4.02^{+0.07}_{-0.07}$ & $ 5.36^{+ 4.42}_{- 1.66}$ & $1.55^{+0.34}_{-0.18}$ & $ 8.96^{+ 8.01}_{- 2.91}$ & $2.50^{+0.60}_{-0.31}$ & 10.86 & 3 \\[0.9mm] 
A0754	& $0.698^{+0.027}_{-0.024}$ & $ 239^{+ 17}_{- 16}$ & $ 9.50^{+0.70}_{-0.40}$ & $16.37^{+ 2.91}_{- 1.84}$ & $2.25^{+0.13}_{-0.09}$ & $26.19^{+ 4.45}_{- 2.95}$ & $3.57^{+0.18}_{-0.15}$ & 21.94 & 1 \\[0.9mm] 
HYDRA-A	& $0.573^{+0.003}_{-0.003}$ & $  50^{+  1}_{-  1}$ & $ 4.30^{+0.40}_{-0.40}$ & $ 3.76^{+ 0.58}_{- 0.55}$ & $1.38^{+0.07}_{-0.07}$ & $ 5.94^{+ 0.91}_{- 0.84}$ & $2.17^{+0.11}_{-0.10}$ & 8.21 & 1 \\[0.9mm] 
A1060	& $0.607^{+0.040}_{-0.034}$ & $  94^{+ 15}_{- 13}$ & $ 3.24^{+0.06}_{-0.06}$ & $ 2.66^{+ 0.34}_{- 0.28}$ & $1.23^{+0.05}_{-0.04}$ & $ 4.24^{+ 0.55}_{- 0.47}$ & $1.95^{+0.08}_{-0.08}$ & 6.54 & 2 \\[0.9mm] 
A1367	& $0.695^{+0.035}_{-0.032}$ & $ 383^{+ 24}_{- 22}$ & $ 3.55^{+0.08}_{-0.08}$ & $ 3.34^{+ 0.36}_{- 0.32}$ & $1.32^{+0.05}_{-0.04}$ & $ 5.69^{+ 0.63}_{- 0.56}$ & $2.14^{+0.08}_{-0.07}$ & 8.08 & 2 \\[0.9mm] 
MKW4	& $0.440^{+0.004}_{-0.005}$ & $  11^{+  1}_{-  1}$ & $ 1.71^{+0.09}_{-0.09}$ & $ 0.64^{+ 0.06}_{- 0.06}$ & $0.76^{+0.02}_{-0.03}$ & $ 1.00^{+ 0.10}_{- 0.09}$ & $1.20^{+0.04}_{-0.03}$ & 2.51 & 2 \\[0.9mm] 
ZwCl1215	& $0.819^{+0.038}_{-0.034}$ & $ 431^{+ 28}_{- 25}$ & ($ 5.58^{+0.89}_{-0.78}$) & $ 8.79^{+ 3.00}_{- 2.29}$ & $1.83^{+0.19}_{-0.18}$ & $14.52^{+ 4.92}_{- 3.67}$ & $2.93^{+0.30}_{-0.27}$ & 14.91 & 11 \\[0.9mm] 
NGC4636	& $0.491^{+0.032}_{-0.027}$ & $   6^{+  3}_{-  2}$ & $ 0.76^{+0.01}_{-0.01}$ & $ 0.22^{+ 0.03}_{- 0.02}$ & $0.53^{+0.02}_{-0.02}$ & $ 0.35^{+ 0.04}_{- 0.04}$ & $0.85^{+0.03}_{-0.03}$ & 1.24 & 4 \\[0.9mm] 
A3526	& $0.495^{+0.011}_{-0.010}$ & $  37^{+  5}_{-  4}$ & $ 3.68^{+0.06}_{-0.06}$ & $ 2.39^{+ 0.15}_{- 0.13}$ & $1.18^{+0.02}_{-0.02}$ & $ 3.78^{+ 0.23}_{- 0.18}$ & $1.87^{+0.04}_{-0.03}$ & 6.07 & 2 \\[0.9mm] 
A1644	& $0.579^{+0.111}_{-0.074}$ & $ 300^{+128}_{- 92}$ & $ 4.70^{+0.90}_{-0.70}$ & $ 4.10^{+ 2.64}_{- 1.41}$ & $1.42^{+0.26}_{-0.18}$ & $ 6.73^{+ 4.54}_{- 2.38}$ & $2.27^{+0.43}_{-0.31}$ & 8.98 & 10 \\[0.9mm] 
A1650	& $0.704^{+0.131}_{-0.081}$ & $ 281^{+104}_{- 71}$ & $ 6.70^{+0.80}_{-0.80}$ & $ 9.62^{+ 4.91}_{- 2.92}$ & $1.88^{+0.28}_{-0.21}$ & $15.60^{+ 8.08}_{- 4.85}$ & $3.01^{+0.45}_{-0.35}$ & 15.56 & 1 \\[0.9mm] 
A1651	& $0.643^{+0.014}_{-0.013}$ & $ 181^{+ 10}_{- 10}$ & $ 6.10^{+0.40}_{-0.40}$ & $ 7.45^{+ 1.00}_{- 0.95}$ & $1.73^{+0.07}_{-0.08}$ & $11.91^{+ 1.60}_{- 1.52}$ & $2.75^{+0.12}_{-0.13}$ & 13.01 & 1 \\[0.9mm] 
COMA	& $0.654^{+0.019}_{-0.021}$ & $ 344^{+ 22}_{- 21}$ & $ 8.38^{+0.34}_{-0.34}$ & $11.99^{+ 1.28}_{- 1.29}$ & $2.03^{+0.07}_{-0.08}$ & $19.38^{+ 2.08}_{- 1.97}$ & $3.22^{+0.11}_{-0.11}$ & 18.01 & 2 \\[0.9mm] 
NGC5044	& $0.524^{+0.002}_{-0.003}$ & $  11^{+  1}_{-  1}$ & $ 1.07^{+0.01}_{-0.01}$ & $ 0.41^{+ 0.01}_{- 0.01}$ & $0.66^{+0.01}_{-0.01}$ & $ 0.65^{+ 0.01}_{- 0.01}$ & $1.04^{+0.01}_{-0.01}$ & 1.87 & 2 \\[0.9mm] 
A1736	& $0.542^{+0.147}_{-0.092}$ & $ 374^{+178}_{-130}$ & $ 3.50^{+0.40}_{-0.40}$ & $ 2.19^{+ 1.23}_{- 0.74}$ & $1.15^{+0.18}_{-0.15}$ & $ 3.78^{+ 2.41}_{- 1.34}$ & $1.87^{+0.34}_{-0.25}$ & 6.22 & 1 \\[0.9mm] 
A3558	& $0.580^{+0.006}_{-0.005}$ & $ 224^{+  5}_{-  5}$ & $ 5.50^{+0.40}_{-0.40}$ & $ 5.37^{+ 0.70}_{- 0.64}$ & $1.55^{+0.07}_{-0.06}$ & $ 8.64^{+ 1.12}_{- 1.03}$ & $2.46^{+0.10}_{-0.10}$ & 10.56 & 1 \\[0.9mm] 
A3562	& $0.472^{+0.006}_{-0.006}$ & $  99^{+  5}_{-  5}$ & $ 5.16^{+0.16}_{-0.16}$ & $ 3.68^{+ 0.24}_{- 0.23}$ & $1.37^{+0.03}_{-0.03}$ & $ 5.83^{+ 0.38}_{- 0.36}$ & $2.16^{+0.05}_{-0.04}$ & 8.10 & 3 \\[0.9mm] 
A3571	& $0.613^{+0.010}_{-0.010}$ & $ 181^{+  7}_{-  7}$ & $ 6.90^{+0.20}_{-0.20}$ & $ 8.33^{+ 0.56}_{- 0.53}$ & $1.79^{+0.04}_{-0.04}$ & $13.31^{+ 0.90}_{- 0.85}$ & $2.85^{+0.06}_{-0.06}$ & 14.04 & 1 \\[0.9mm] 
A1795	& $0.596^{+0.003}_{-0.002}$ & $  78^{+  1}_{-  1}$ & $ 7.80^{+1.00}_{-1.00}$ & $ 9.75^{+ 2.01}_{- 1.90}$ & $1.89^{+0.12}_{-0.14}$ & $15.39^{+ 3.17}_{- 2.92}$ & $2.99^{+0.19}_{-0.20}$ & 15.46 & 1 \\[0.9mm] 
A3581	& $0.543^{+0.024}_{-0.022}$ & $  35^{+  5}_{-  4}$ & $ 1.83^{+0.04}_{-0.04}$ & $ 0.96^{+ 0.09}_{- 0.09}$ & $0.87^{+0.02}_{-0.03}$ & $ 1.52^{+ 0.16}_{- 0.13}$ & $1.38^{+0.05}_{-0.04}$ & 3.30 & 5 \\[0.9mm] 
MKW8	& $0.511^{+0.098}_{-0.059}$ & $ 107^{+ 70}_{- 42}$ & $ 3.29^{+0.23}_{-0.22}$ & $ 2.10^{+ 0.86}_{- 0.52}$ & $1.14^{+0.13}_{-0.10}$ & $ 3.33^{+ 1.45}_{- 0.83}$ & $1.79^{+0.24}_{-0.17}$ & 5.60 & 5 \\[0.9mm] 
A2029	& $0.582^{+0.004}_{-0.004}$ & $  83^{+  2}_{-  2}$ & $ 9.10^{+1.00}_{-1.00}$ & $11.82^{+ 2.14}_{- 1.99}$ & $2.01^{+0.11}_{-0.12}$ & $18.79^{+ 3.40}_{- 3.17}$ & $3.20^{+0.18}_{-0.19}$ & 17.62 & 1 \\[0.9mm] 
A2052	& $0.526^{+0.005}_{-0.005}$ & $  37^{+  2}_{-  2}$ & $ 3.03^{+0.04}_{-0.04}$ & $ 1.95^{+ 0.07}_{- 0.06}$ & $1.10^{+0.02}_{-0.01}$ & $ 3.10^{+ 0.09}_{- 0.11}$ & $1.75^{+0.01}_{-0.02}$ & 5.30 & 3 \\[0.9mm] 
MKW3S	& $0.581^{+0.008}_{-0.007}$ & $  66^{+  3}_{-  3}$ & $ 3.70^{+0.20}_{-0.20}$ & $ 3.06^{+ 0.32}_{- 0.30}$ & $1.29^{+0.05}_{-0.04}$ & $ 4.84^{+ 0.51}_{- 0.47}$ & $2.03^{+0.07}_{-0.07}$ & 7.16 & 1 \\[0.9mm] 
A2065	& $1.162^{+0.734}_{-0.282}$ & $ 690^{+360}_{-186}$ & $ 5.50^{+0.40}_{-0.40}$ & $13.44^{+16.12}_{- 5.17}$ & $2.10^{+0.63}_{-0.31}$ & $23.37^{+29.87}_{- 9.42}$ & $3.43^{+1.09}_{-0.54}$ & 20.21 & 1 \\[0.9mm] 
A2063	& $0.561^{+0.011}_{-0.011}$ & $ 110^{+  7}_{-  6}$ & $ 3.68^{+0.11}_{-0.11}$ & $ 2.84^{+ 0.23}_{- 0.19}$ & $1.25^{+0.04}_{-0.03}$ & $ 4.54^{+ 0.36}_{- 0.31}$ & $1.99^{+0.06}_{-0.04}$ & 6.86 & 2 \\[0.9mm] 
A2142	& $0.591^{+0.006}_{-0.006}$ & $ 154^{+  6}_{-  6}$ & $ 9.70^{+1.50}_{-1.10}$ & $13.29^{+ 3.45}_{- 2.41}$ & $2.10^{+0.17}_{-0.14}$ & $21.04^{+ 5.46}_{- 3.69}$ & $3.31^{+0.26}_{-0.20}$ & 19.05 & 1 \\[0.9mm] 
A2147	& $0.444^{+0.071}_{-0.046}$ & $ 238^{+103}_{- 65}$ & $ 4.91^{+0.28}_{-0.28}$ & $ 2.99^{+ 0.92}_{- 0.63}$ & $1.28^{+0.12}_{-0.10}$ & $ 4.84^{+ 1.64}_{- 1.03}$ & $2.03^{+0.21}_{-0.15}$ & 7.21 & 2 \\[0.9mm] 
A2163	& $0.796^{+0.030}_{-0.028}$ & $ 519^{+ 31}_{- 29}$ & $13.29^{+0.64}_{-0.64}$ & $31.85^{+ 4.24}_{- 3.74}$ & $2.81^{+0.12}_{-0.11}$ & $51.99^{+ 6.96}_{- 6.13}$ & $4.49^{+0.19}_{-0.18}$ & 34.18 & 3 \\[0.9mm] 
A2199	& $0.655^{+0.019}_{-0.021}$ & $ 139^{+ 10}_{- 10}$ & $ 4.10^{+0.08}_{-0.08}$ & $ 4.21^{+ 0.33}_{- 0.29}$ & $1.43^{+0.04}_{-0.03}$ & $ 6.73^{+ 0.52}_{- 0.51}$ & $2.27^{+0.06}_{-0.06}$ & 8.92 & 2 \\[0.9mm] 
A2204	& $0.597^{+0.008}_{-0.007}$ & $  67^{+  3}_{-  3}$ & $ 7.21^{+0.25}_{-0.25}$ & $ 8.67^{+ 0.67}_{- 0.57}$ & $1.82^{+0.05}_{-0.04}$ & $13.79^{+ 0.96}_{- 1.00}$ & $2.89^{+0.06}_{-0.08}$ & 14.34 & 3 \\[0.9mm] 
A2244	& $0.607^{+0.016}_{-0.015}$ & $ 126^{+ 11}_{- 10}$ & $ 7.10^{+5.00}_{-2.20}$ & $ 8.65^{+11.47}_{- 3.89}$ & $1.82^{+0.59}_{-0.33}$ & $13.78^{+18.02}_{- 6.20}$ & $2.89^{+0.92}_{-0.52}$ & 14.33 & 10 \\[0.9mm] 
A2256	& $0.914^{+0.054}_{-0.047}$ & $ 587^{+ 40}_{- 37}$ & $ 6.60^{+0.40}_{-0.40}$ & $12.83^{+ 2.38}_{- 2.00}$ & $2.07^{+0.12}_{-0.11}$ & $21.81^{+ 4.07}_{- 3.54}$ & $3.36^{+0.19}_{-0.20}$ & 19.34 & 1 \\[0.9mm] 
A2255	& $0.797^{+0.033}_{-0.030}$ & $ 593^{+ 35}_{- 32}$ & $ 6.87^{+0.20}_{-0.20}$ & $10.90^{+ 1.15}_{- 0.95}$ & $1.96^{+0.07}_{-0.05}$ & $18.65^{+ 2.01}_{- 1.67}$ & $3.18^{+0.11}_{-0.09}$ & 17.54 & 3 \\[0.9mm] 
A3667	& $0.541^{+0.008}_{-0.008}$ & $ 279^{+ 10}_{- 10}$ & $ 7.00^{+0.60}_{-0.60}$ & $ 6.88^{+ 1.08}_{- 1.02}$ & $1.68^{+0.08}_{-0.09}$ & $11.19^{+ 1.76}_{- 1.65}$ & $2.69^{+0.13}_{-0.14}$ & 12.50 & 1 \\[0.9mm] 
S1101	& $0.639^{+0.006}_{-0.007}$ & $  56^{+  2}_{-  2}$ & $ 3.00^{+1.20}_{-0.70}$ & $ 2.58^{+ 1.76}_{- 0.88}$ & $1.22^{+0.23}_{-0.16}$ & $ 4.08^{+ 2.78}_{- 1.39}$ & $1.92^{+0.36}_{-0.25}$ & 6.38 & 9 \\[0.9mm] 
A2589	& $0.596^{+0.013}_{-0.012}$ & $ 118^{+  8}_{-  7}$ & $ 3.70^{+2.20}_{-1.10}$ & $ 3.14^{+ 3.44}_{- 1.35}$ & $1.29^{+0.37}_{-0.22}$ & $ 5.01^{+ 5.41}_{- 2.15}$ & $2.06^{+0.56}_{-0.35}$ & 7.33 & 9 \\[0.9mm] 
A2597	& $0.633^{+0.008}_{-0.008}$ & $  58^{+  2}_{-  2}$ & $ 4.40^{+0.40}_{-0.70}$ & $ 4.52^{+ 0.72}_{- 1.11}$ & $1.47^{+0.07}_{-0.14}$ & $ 7.14^{+ 1.14}_{- 1.72}$ & $2.31^{+0.11}_{-0.20}$ & 9.27 & 1 \\[0.9mm] 
A2634	& $0.640^{+0.051}_{-0.043}$ & $ 364^{+ 44}_{- 39}$ & $ 3.70^{+0.28}_{-0.28}$ & $ 3.15^{+ 0.78}_{- 0.60}$ & $1.29^{+0.10}_{-0.09}$ & $ 5.35^{+ 1.34}_{- 1.04}$ & $2.10^{+0.17}_{-0.14}$ & 7.77 & 2 \\[0.9mm] 
A2657	& $0.556^{+0.008}_{-0.007}$ & $ 119^{+  5}_{-  5}$ & $ 3.70^{+0.30}_{-0.30}$ & $ 2.83^{+ 0.43}_{- 0.39}$ & $1.25^{+0.06}_{-0.06}$ & $ 4.52^{+ 0.68}_{- 0.62}$ & $1.99^{+0.10}_{-0.09}$ & 6.84 & 1 \\[0.9mm] 
A4038	& $0.541^{+0.009}_{-0.008}$ & $  59^{+  4}_{-  4}$ & $ 3.15^{+0.03}_{-0.03}$ & $ 2.16^{+ 0.09}_{- 0.08}$ & $1.14^{+0.02}_{-0.02}$ & $ 3.41^{+ 0.14}_{- 0.11}$ & $1.80^{+0.03}_{-0.01}$ & 5.67 & 3 \\[0.9mm] 
A4059	& $0.582^{+0.010}_{-0.010}$ & $  90^{+  5}_{-  5}$ & $ 4.40^{+0.30}_{-0.30}$ & $ 3.95^{+ 0.52}_{- 0.48}$ & $1.40^{+0.06}_{-0.06}$ & $ 6.30^{+ 0.83}_{- 0.76}$ & $2.22^{+0.09}_{-0.09}$ & 8.52 & 1 \\[0.9mm]
\tableline
\sidehead{\hspace{3.94cm}\vspace{1mm} Clusters from the extended sample not included in \gcss .} 
\tableline \\[-2.6mm]
A2734	& $0.624^{+0.034}_{-0.029}$ & $ 212^{+ 26}_{- 23}$ & ($ 3.85^{+0.62}_{-0.54}$) & $ 3.49^{+ 1.25}_{- 0.89}$ & $1.34^{+0.15}_{-0.12}$ & $ 5.67^{+ 1.98}_{- 1.48}$ & $2.14^{+0.22}_{-0.21}$ & 7.97 & 11 \\[0.9mm] 
A2877	& $0.566^{+0.029}_{-0.025}$ & $ 190^{+ 19}_{- 17}$ & $ 3.50^{+2.20}_{-1.10}$ & $ 2.61^{+ 3.32}_{- 1.24}$ & $1.22^{+0.39}_{-0.23}$ & $ 4.24^{+ 5.28}_{- 2.00}$ & $1.95^{+0.60}_{-0.38}$ & 6.57 & 10 \\[0.9mm] 
NGC499	& $0.722^{+0.034}_{-0.030}$ & $  24^{+  2}_{-  2}$ & $ 0.72^{+0.03}_{-0.02}$ & $ 0.36^{+ 0.05}_{- 0.04}$ & $0.63^{+0.03}_{-0.02}$ & $ 0.58^{+ 0.08}_{- 0.06}$ & $1.00^{+0.04}_{-0.04}$ & 1.73 & 4 \\[0.9mm] 
AWM7	& $0.671^{+0.027}_{-0.025}$ & $ 173^{+ 18}_{- 15}$ & $ 3.75^{+0.09}_{-0.09}$ & $ 3.79^{+ 0.38}_{- 0.32}$ & $1.38^{+0.05}_{-0.04}$ & $ 6.08^{+ 0.62}_{- 0.52}$ & $2.19^{+0.08}_{-0.06}$ & 8.35 & 2 \\[0.9mm] 
PERSEUS	& $0.540^{+0.006}_{-0.004}$ & $  64^{+  2}_{-  2}$ & $ 6.79^{+0.12}_{-0.12}$ & $ 6.84^{+ 0.29}_{- 0.26}$ & $1.68^{+0.02}_{-0.02}$ & $10.80^{+ 0.46}_{- 0.41}$ & $2.66^{+0.04}_{-0.04}$ & 12.20 & 2 \\[0.9mm] 
S405	& $0.664^{+0.263}_{-0.133}$ & $ 459^{+262}_{-159}$ & ($ 4.21^{+0.67}_{-0.59}$) & $ 3.91^{+ 3.56}_{- 1.57}$ & $1.40^{+0.33}_{-0.22}$ & $ 6.75^{+ 6.80}_{- 2.81}$ & $2.27^{+0.60}_{-0.37}$ & 9.09 & 11 \\[0.9mm] 
3C129	& $0.601^{+0.260}_{-0.131}$ & $ 318^{+178}_{-107}$ & $ 5.60^{+0.70}_{-0.60}$ & $ 5.68^{+ 5.58}_{- 2.29}$ & $1.58^{+0.40}_{-0.25}$ & $ 9.30^{+ 9.51}_{- 3.85}$ & $2.53^{+0.67}_{-0.42}$ & 11.08 & 9 \\[0.9mm] 
A0539	& $0.561^{+0.020}_{-0.018}$ & $ 148^{+ 13}_{- 12}$ & $ 3.24^{+0.09}_{-0.09}$ & $ 2.33^{+ 0.21}_{- 0.19}$ & $1.18^{+0.03}_{-0.03}$ & $ 3.74^{+ 0.35}_{- 0.34}$ & $1.87^{+0.05}_{-0.06}$ & 6.04 & 2 \\[0.9mm] 
S540	& $0.641^{+0.073}_{-0.051}$ & $ 130^{+ 38}_{- 29}$ & ($ 2.40^{+0.38}_{-0.34}$) & $ 1.83^{+ 0.83}_{- 0.54}$ & $1.08^{+0.14}_{-0.12}$ & $ 2.93^{+ 1.34}_{- 0.87}$ & $1.72^{+0.23}_{-0.19}$ & 5.13 & 11 \\[0.9mm] 
A0548w	& $0.666^{+0.194}_{-0.111}$ & $ 198^{+ 90}_{- 62}$ & ($ 1.20^{+0.19}_{-0.17}$) & $ 0.63^{+ 0.48}_{- 0.23}$ & $0.76^{+0.16}_{-0.11}$ & $ 1.06^{+ 0.84}_{- 0.41}$ & $1.23^{+0.26}_{-0.19}$ & 2.64 & 11 \\[0.9mm] 
A0548e	& $0.480^{+0.013}_{-0.013}$ & $ 118^{+ 12}_{- 11}$ & $ 3.10^{+0.10}_{-0.10}$ & $ 1.74^{+ 0.15}_{- 0.15}$ & $1.07^{+0.03}_{-0.04}$ & $ 2.77^{+ 0.27}_{- 0.23}$ & $1.68^{+0.06}_{-0.05}$ & 4.95 & 3 \\[0.9mm] 
A3395n	& $0.981^{+0.619}_{-0.244}$ & $ 672^{+383}_{-203}$ & $ 5.00^{+0.30}_{-0.30}$ & $ 8.70^{+ 9.53}_{- 3.19}$ & $1.82^{+0.51}_{-0.26}$ & $15.47^{+18.82}_{- 6.07}$ & $2.99^{+0.92}_{-0.46}$ & 15.55 & 1 \\[0.9mm] 
UGC03957	& $0.740^{+0.133}_{-0.086}$ & $ 142^{+ 45}_{- 33}$ & ($ 2.58^{+0.41}_{-0.36}$) & $ 2.51^{+ 1.50}_{- 0.83}$ & $1.20^{+0.21}_{-0.15}$ & $ 4.02^{+ 2.41}_{- 1.33}$ & $1.91^{+0.33}_{-0.23}$ & 6.35 & 11 \\[0.9mm] 
PKS0745	& $0.608^{+0.006}_{-0.006}$ & $  71^{+  2}_{-  2}$ & $ 7.21^{+0.11}_{-0.11}$ & $ 8.88^{+ 0.35}_{- 0.28}$ & $1.83^{+0.03}_{-0.01}$ & $14.12^{+ 0.56}_{- 0.53}$ & $2.91^{+0.04}_{-0.04}$ & 14.58 & 3 \\[0.9mm] 
A0644	& $0.700^{+0.011}_{-0.011}$ & $ 203^{+  7}_{-  7}$ & $ 7.90^{+0.80}_{-0.80}$ & $12.50^{+ 2.29}_{- 2.11}$ & $2.06^{+0.12}_{-0.12}$ & $19.83^{+ 3.79}_{- 3.23}$ & $3.24^{+0.21}_{-0.17}$ & 18.33 & 1 \\[0.9mm] 
S636	& $0.752^{+0.217}_{-0.123}$ & $ 344^{+130}_{- 86}$ & ($ 1.18^{+0.19}_{-0.17}$) & $ 0.61^{+ 0.44}_{- 0.22}$ & $0.75^{+0.15}_{-0.10}$ & $ 1.16^{+ 0.90}_{- 0.44}$ & $1.26^{+0.27}_{-0.18}$ & 2.93 & 11 \\[0.9mm] 
A1413	& $0.660^{+0.017}_{-0.015}$ & $ 179^{+ 12}_{- 11}$ & $ 7.32^{+0.26}_{-0.24}$ & $10.20^{+ 0.93}_{- 0.82}$ & $1.92^{+0.05}_{-0.05}$ & $16.29^{+ 1.49}_{- 1.31}$ & $3.05^{+0.09}_{-0.08}$ & 16.03 & 3 \\[0.9mm] 
M49	& $0.592^{+0.007}_{-0.007}$ & $  11^{+  1}_{-  1}$ & $ 0.95^{+0.02}_{-0.01}$ & $ 0.41^{+ 0.02}_{- 0.01}$ & $0.66^{+0.01}_{-0.01}$ & $ 0.65^{+ 0.04}_{- 0.02}$ & $1.04^{+0.02}_{-0.01}$ & 1.87 & 4 \\[0.9mm] 
A3528n	& $0.621^{+0.034}_{-0.030}$ & $ 178^{+ 17}_{- 16}$ & $ 3.40^{+1.66}_{-0.64}$ & $ 2.89^{+ 2.84}_{- 0.94}$ & $1.26^{+0.32}_{-0.15}$ & $ 4.65^{+ 4.54}_{- 1.48}$ & $2.00^{+0.51}_{-0.23}$ & 7.00 & 8 \\[0.9mm] 
A3528s	& $0.463^{+0.013}_{-0.012}$ & $ 101^{+  9}_{-  8}$ & $ 3.15^{+0.89}_{-0.59}$ & $ 1.69^{+ 0.87}_{- 0.50}$ & $1.05^{+0.16}_{-0.12}$ & $ 2.70^{+ 1.39}_{- 0.80}$ & $1.67^{+0.25}_{-0.19}$ & 4.86 & 8 \\[0.9mm] 
A3530	& $0.773^{+0.114}_{-0.085}$ & $ 421^{+ 75}_{- 61}$ & $ 3.89^{+0.27}_{-0.25}$ & $ 4.52^{+ 1.52}_{- 1.05}$ & $1.47^{+0.15}_{-0.13}$ & $ 7.64^{+ 2.72}_{- 1.80}$ & $2.36^{+0.26}_{-0.20}$ & 9.82 & 7 \\[0.9mm] 
A3532	& $0.653^{+0.034}_{-0.029}$ & $ 282^{+ 27}_{- 24}$ & $ 4.58^{+0.19}_{-0.17}$ & $ 4.77^{+ 0.70}_{- 0.52}$ & $1.49^{+0.07}_{-0.05}$ & $ 7.79^{+ 1.16}_{- 0.91}$ & $2.38^{+0.12}_{-0.10}$ & 9.88 & 7 \\[0.9mm] 
A1689	& $0.690^{+0.011}_{-0.011}$ & $ 163^{+  7}_{-  6}$ & $ 9.23^{+0.28}_{-0.28}$ & $15.49^{+ 1.18}_{- 1.00}$ & $2.20^{+0.06}_{-0.05}$ & $24.68^{+ 1.70}_{- 1.76}$ & $3.50^{+0.07}_{-0.10}$ & 21.13 & 3 \\[0.9mm] 
A3560	& $0.566^{+0.033}_{-0.029}$ & $ 256^{+ 30}_{- 27}$ & ($ 3.16^{+0.51}_{-0.44}$) & $ 2.16^{+ 0.79}_{- 0.56}$ & $1.14^{+0.12}_{-0.11}$ & $ 3.59^{+ 1.30}_{- 0.95}$ & $1.84^{+0.20}_{-0.18}$ & 5.92 & 11 \\[0.9mm] 
A1775	& $0.673^{+0.026}_{-0.023}$ & $ 260^{+ 19}_{- 18}$ & $ 3.69^{+0.20}_{-0.11}$ & $ 3.61^{+ 0.50}_{- 0.34}$ & $1.36^{+0.06}_{-0.05}$ & $ 5.91^{+ 0.83}_{- 0.56}$ & $2.17^{+0.09}_{-0.07}$ & 8.21 & 3 \\[0.9mm] 
A1800	& $0.766^{+0.308}_{-0.139}$ & $ 392^{+223}_{-132}$ & ($ 4.02^{+0.64}_{-0.56}$) & $ 4.75^{+ 4.64}_{- 1.85}$ & $1.49^{+0.38}_{-0.23}$ & $ 7.97^{+ 8.31}_{- 3.17}$ & $2.39^{+0.65}_{-0.37}$ & 10.08 & 11 \\[0.9mm] 
A1914	& $0.751^{+0.018}_{-0.017}$ & $ 231^{+ 11}_{- 10}$ & $10.53^{+0.51}_{-0.50}$ & $21.43^{+ 2.39}_{- 2.16}$ & $2.46^{+0.09}_{-0.08}$ & $33.99^{+ 4.06}_{- 3.43}$ & $3.88^{+0.16}_{-0.13}$ & 26.20 & 3 \\[0.9mm] 
NGC5813	& $0.766^{+0.179}_{-0.103}$ & $  25^{+  9}_{-  6}$ & ($ 0.52^{+0.08}_{-0.07}$) & $ 0.24^{+ 0.17}_{- 0.08}$ & $0.55^{+0.11}_{-0.07}$ & $ 0.38^{+ 0.27}_{- 0.13}$ & $0.87^{+0.16}_{-0.12}$ & 1.32 & 11 \\[0.9mm] 
NGC5846	& $0.599^{+0.016}_{-0.015}$ & $   7^{+  1}_{-  1}$ & $ 0.82^{+0.01}_{-0.01}$ & $ 0.33^{+ 0.02}_{- 0.02}$ & $0.61^{+0.01}_{-0.01}$ & $ 0.53^{+ 0.03}_{- 0.03}$ & $0.97^{+0.02}_{-0.01}$ & 1.63 & 4 \\[0.9mm] 
A2151w	& $0.564^{+0.014}_{-0.013}$ & $  68^{+  5}_{-  5}$ & $ 2.40^{+0.06}_{-0.06}$ & $ 1.52^{+ 0.12}_{- 0.10}$ & $1.02^{+0.03}_{-0.02}$ & $ 2.42^{+ 0.18}_{- 0.18}$ & $1.61^{+0.03}_{-0.04}$ & 4.51 & 3 \\[0.9mm] 
A3627	& $0.555^{+0.056}_{-0.044}$ & $ 299^{+ 56}_{- 49}$ & $ 6.02^{+0.08}_{-0.08}$ & $ 5.63^{+ 0.95}_{- 0.68}$ & $1.57^{+0.09}_{-0.06}$ & $ 9.20^{+ 1.61}_{- 1.16}$ & $2.51^{+0.14}_{-0.10}$ & 11.03 & 3 \\[0.9mm] 
TRIANGUL	& $0.610^{+0.010}_{-0.010}$ & $ 279^{+ 11}_{- 10}$ & $ 9.60^{+0.60}_{-0.60}$ & $13.42^{+ 1.70}_{- 1.55}$ & $2.10^{+0.09}_{-0.09}$ & $21.54^{+ 2.73}_{- 2.36}$ & $3.34^{+0.14}_{-0.11}$ & 19.35 & 1 \\[0.9mm] 
OPHIUCHU	& $0.747^{+0.035}_{-0.032}$ & $ 279^{+ 23}_{- 22}$ & $10.26^{+0.32}_{-0.32}$ & $20.25^{+ 2.51}_{- 2.10}$ & $2.41^{+0.10}_{-0.08}$ & $32.43^{+ 4.05}_{- 3.38}$ & $3.83^{+0.16}_{-0.13}$ & 25.32 & 2 \\[0.9mm] 
ZwCl1742	& $0.717^{+0.073}_{-0.053}$ & $ 232^{+ 46}_{- 38}$ & ($ 5.23^{+0.84}_{-0.73}$) & $ 6.88^{+ 3.06}_{- 1.96}$ & $1.68^{+0.22}_{-0.18}$ & $11.05^{+ 4.93}_{- 3.16}$ & $2.67^{+0.35}_{-0.28}$ & 12.42 & 11 \\[0.9mm] 
A2319	& $0.591^{+0.013}_{-0.012}$ & $ 285^{+ 15}_{- 14}$ & $ 8.80^{+0.50}_{-0.50}$ & $11.16^{+ 1.39}_{- 1.20}$ & $1.97^{+0.08}_{-0.07}$ & $18.07^{+ 2.12}_{- 2.06}$ & $3.16^{+0.11}_{-0.13}$ & 17.17 & 1 \\[0.9mm] 
A3695	& $0.642^{+0.259}_{-0.117}$ & $ 399^{+254}_{-149}$ & ($ 5.29^{+0.85}_{-0.74}$) & $ 5.57^{+ 5.29}_{- 2.16}$ & $1.57^{+0.39}_{-0.24}$ & $ 9.32^{+ 9.56}_{- 3.74}$ & $2.53^{+0.67}_{-0.40}$ & 11.12 & 11 \\[0.9mm] 
IIZw108	& $0.662^{+0.167}_{-0.097}$ & $ 365^{+159}_{-105}$ & ($ 3.44^{+0.55}_{-0.48}$) & $ 2.96^{+ 2.00}_{- 1.02}$ & $1.27^{+0.24}_{-0.16}$ & $ 5.04^{+ 3.60}_{- 1.80}$ & $2.06^{+0.40}_{-0.28}$ & 7.47 & 11 \\[0.9mm] 
A3822	& $0.639^{+0.150}_{-0.093}$ & $ 351^{+160}_{-111}$ & ($ 4.90^{+0.78}_{-0.69}$) & $ 4.97^{+ 3.30}_{- 1.75}$ & $1.51^{+0.28}_{-0.20}$ & $ 8.26^{+ 5.64}_{- 3.02}$ & $2.43^{+0.46}_{-0.34}$ & 10.29 & 11 \\[0.9mm] 
A3827	& $0.989^{+0.410}_{-0.192}$ & $ 593^{+248}_{-149}$ & ($ 7.08^{+1.13}_{-0.99}$) & $16.35^{+17.02}_{- 6.76}$ & $2.25^{+0.60}_{-0.37}$ & $27.44^{+29.53}_{-11.46}$ & $3.62^{+0.99}_{-0.60}$ & 22.44 & 11 \\[0.9mm] 
A3888	& $0.928^{+0.084}_{-0.066}$ & $ 401^{+ 46}_{- 40}$ & ($ 8.84^{+1.41}_{-1.24}$) & $22.00^{+ 9.28}_{- 6.28}$ & $2.48^{+0.31}_{-0.26}$ & $35.74^{+15.07}_{-10.38}$ & $3.96^{+0.49}_{-0.44}$ & 26.85 & 11 \\[0.9mm] 
A3921	& $0.762^{+0.036}_{-0.030}$ & $ 328^{+ 26}_{- 23}$ & $ 5.73^{+0.24}_{-0.23}$ & $ 8.46^{+ 1.13}_{- 0.96}$ & $1.80^{+0.08}_{-0.07}$ & $13.80^{+ 1.87}_{- 1.59}$ & $2.89^{+0.12}_{-0.12}$ & 14.37 & 3 \\[0.9mm] 
HCG94	& $0.514^{+0.007}_{-0.006}$ & $  86^{+  4}_{-  4}$ & $ 3.45^{+0.30}_{-0.30}$ & $ 2.28^{+ 0.36}_{- 0.34}$ & $1.17^{+0.06}_{-0.06}$ & $ 3.62^{+ 0.56}_{- 0.51}$ & $1.84^{+0.09}_{-0.09}$ & 5.90 & 6 \\[0.9mm] 
RXJ2344	& $0.807^{+0.033}_{-0.030}$ & $ 301^{+ 20}_{- 18}$ & ($ 4.73^{+0.76}_{-0.66}$) & $ 6.91^{+ 2.30}_{- 1.69}$ & $1.68^{+0.17}_{-0.14}$ & $11.27^{+ 3.74}_{- 2.80}$ & $2.69^{+0.27}_{-0.25}$ & 12.58 & 11 \\[0.9mm] 
\enddata

\tablecomments{
The columns are described in detail at the end of Section~\ref{massd}.
}

\tablerefs{
(1) \citealt{mfs98}.
(2) \citealt{fmt98}.
(3) \citealt{w00}.
(4) \citealt{m97}.
(5) \citealt{irb01}.
(6) \citealt{frb00}.
(7) This work.
(8) \citealt{s96a}.
(9) \citealt{es91a}.
(10) \citealt{dsj93}.
(11) Estimated from the $\lx$--$\tx$ relation given by \citealt{m98}.
}

\end{deluxetable}

\begin{deluxetable}{lcrrcc}
\tablecolumns{6} 
\tabletypesize{\footnotesize}
\tablecaption{Cluster properties \label{tbl:data3}}
\tablewidth{0pt}
\tablehead{
\colhead{Cluster}	& 
\colhead{$\rog(0)$}	& \colhead{$\mgf$}	& \colhead{$\mgz$}		&
 \colhead{$\fgf$}	& \colhead{$\fgz$}	\\
\colhead{(1)}	& \colhead{(2)}	& \colhead{(3)}	&
\colhead{(4)}	& \colhead{(5)}	& \colhead{(6)}
}
\startdata
A0085	& 0.337 & $ 16.834^{+  1.076}_{ -1.177}$ & $ 32.372^{+  2.041}_{ -2.222}$ & $0.246\pm 0.024$ & $0.300\pm 0.030$ \\[0.9mm] 
A0119	& 0.026 & $ 11.633^{+  0.984}_{ -0.888}$ & $ 22.878^{+  1.373}_{ -1.533}$ & $0.187\pm 0.025$ & $0.213\pm 0.029$ \\[0.9mm] 
A0133	& 0.421 & $  5.377^{+  2.126}_{ -1.089}$ & $ 10.414^{+  4.055}_{ -2.080}$ & $0.194\pm 0.121$ & $0.236\pm 0.148$ \\[0.9mm] 
NGC507	& 0.226 & $  0.648^{+  0.055}_{ -0.057}$ & $  1.390^{+  0.118}_{ -0.119}$ & $0.159\pm 0.016$ & $0.216\pm 0.022$ \\[0.9mm] 
A0262	& 0.158 & $  1.936^{+  0.370}_{ -0.322}$ & $  4.180^{+  0.823}_{ -0.662}$ & $0.215\pm 0.022$ & $0.293\pm 0.029$ \\[0.9mm] 
A0400	& 0.039 & $  2.141^{+  0.189}_{ -0.177}$ & $  4.345^{+  0.355}_{ -0.302}$ & $0.167\pm 0.021$ & $0.210\pm 0.028$ \\[0.9mm] 
A0399	& 0.042 & $ 18.017^{+  3.889}_{ -3.344}$ & $ 32.014^{+  5.680}_{ -4.984}$ & $0.180\pm 0.056$ & $0.192\pm 0.063$ \\[0.9mm] 
A0401	& 0.111 & $ 24.426^{+  1.716}_{ -1.428}$ & $ 44.912^{+  2.628}_{ -2.828}$ & $0.238\pm 0.023$ & $0.271\pm 0.026$ \\[0.9mm] 
A3112	& 0.544 & $ 10.515^{+  1.313}_{ -1.613}$ & $ 19.206^{+  2.234}_{ -3.049}$ & $0.203\pm 0.052$ & $0.234\pm 0.058$ \\[0.9mm] 
FORNAX	& 0.018 & $  0.397^{+  0.031}_{ -0.032}$ & $  0.629^{+  0.097}_{ -0.091}$ & $0.046\pm 0.010$ & $0.044\pm 0.010$ \\[0.9mm] 
2A0335	& 1.066 & $  5.091^{+  0.182}_{ -0.161}$ & $  9.265^{+  0.267}_{ -0.341}$ & $0.230\pm 0.010$ & $0.264\pm 0.011$ \\[0.9mm] 
IIIZw54	& 0.039 & $  2.596^{+  0.840}_{ -0.624}$ & $  3.955^{+  1.199}_{ -0.958}$ & $0.110\pm 0.073$ & $0.101\pm 0.069$ \\[0.9mm] 
A3158	& 0.076 & $ 13.600^{+  0.795}_{ -0.753}$ & $ 23.884^{+  1.264}_{ -1.056}$ & $0.194\pm 0.013$ & $0.211\pm 0.015$ \\[0.9mm] 
A0478	& 0.502 & $ 23.507^{+  1.858}_{ -2.850}$ & $ 40.781^{+  3.147}_{ -4.781}$ & $0.208\pm 0.042$ & $0.228\pm 0.047$ \\[0.9mm] 
NGC1550	& 0.145 & $  0.806^{+  0.183}_{ -0.147}$ & $  1.510^{+  0.323}_{ -0.249}$ & $0.117\pm 0.018$ & $0.139\pm 0.022$ \\[0.9mm] 
EXO0422	& 0.128 & $  3.894^{+  1.428}_{ -0.997}$ & $  6.220^{+  2.216}_{ -1.535}$ & $0.135\pm 0.082$ & $0.134\pm 0.082$ \\[0.9mm] 
A3266	& 0.045 & $ 26.348^{+  1.654}_{ -1.780}$ & $ 44.048^{+  2.291}_{ -2.260}$ & $0.186\pm 0.025$ & $0.185\pm 0.024$ \\[0.9mm] 
A0496	& 0.651 & $  6.775^{+  0.244}_{ -0.305}$ & $ 13.770^{+  0.639}_{ -0.592}$ & $0.246\pm 0.010$ & $0.317\pm 0.013$ \\[0.9mm] 
A3376	& 0.020 & $  8.873^{+  1.059}_{ -1.111}$ & $ 14.548^{+  1.075}_{ -1.109}$ & $0.140\pm 0.041$ & $0.122\pm 0.035$ \\[0.9mm] 
A3391	& 0.050 & $  8.577^{+  1.306}_{ -1.158}$ & $ 16.622^{+  2.134}_{ -2.104}$ & $0.165\pm 0.038$ & $0.198\pm 0.046$ \\[0.9mm] 
A3395s	& 0.025 & $  9.609^{+  1.588}_{ -1.438}$ & $ 14.857^{+  2.202}_{ -2.118}$ & $0.109\pm 0.046$ & $0.097\pm 0.043$ \\[0.9mm] 
A0576	& 0.031 & $  5.981^{+  1.377}_{ -1.154}$ & $  9.674^{+  1.916}_{ -1.962}$ & $0.112\pm 0.063$ & $0.108\pm 0.066$ \\[0.9mm] 
A0754	& 0.086 & $ 14.934^{+  1.188}_{ -0.960}$ & $ 24.392^{+  2.019}_{ -1.664}$ & $0.091\pm 0.013$ & $0.093\pm 0.013$ \\[0.9mm] 
HYDRA-A	& 0.634 & $  7.874^{+  0.633}_{ -0.653}$ & $ 14.281^{+  1.190}_{ -1.085}$ & $0.209\pm 0.031$ & $0.240\pm 0.035$ \\[0.9mm] 
A1060	& 0.092 & $  2.270^{+  0.257}_{ -0.225}$ & $  4.054^{+  0.440}_{ -0.405}$ & $0.085\pm 0.010$ & $0.096\pm 0.011$ \\[0.9mm] 
A1367	& 0.025 & $  5.197^{+  0.294}_{ -0.272}$ & $  9.817^{+  0.375}_{ -0.367}$ & $0.156\pm 0.016$ & $0.173\pm 0.018$ \\[0.9mm] 
MKW4	& 0.570 & $  1.033^{+  0.097}_{ -0.087}$ & $  2.225^{+  0.209}_{ -0.184}$ & $0.163\pm 0.015$ & $0.222\pm 0.021$ \\[0.9mm] 
ZwCl1215	& 0.052 & $ 14.585^{+  1.795}_{ -1.775}$ & $ 23.257^{+  2.560}_{ -2.232}$ & $0.166\pm 0.050$ & $0.160\pm 0.047$ \\[0.9mm] 
NGC4636	& 0.328 & $  0.086^{+  0.025}_{ -0.017}$ & $  0.175^{+  0.050}_{ -0.037}$ & $0.039\pm 0.004$ & $0.050\pm 0.006$ \\[0.9mm] 
A3526	& 0.286 & $  3.382^{+  0.350}_{ -0.332}$ & $  6.783^{+  0.683}_{ -0.575}$ & $0.141\pm 0.008$ & $0.180\pm 0.010$ \\[0.9mm] 
A1644	& 0.043 & $  9.463^{+  4.242}_{ -2.921}$ & $ 18.907^{+  7.091}_{ -5.272}$ & $0.231\pm 0.114$ & $0.281\pm 0.145$ \\[0.9mm] 
A1650	& 0.084 & $ 15.770^{+  4.817}_{ -3.728}$ & $ 26.523^{+  6.834}_{ -5.777}$ & $0.164\pm 0.067$ & $0.170\pm 0.070$ \\[0.9mm] 
A1651	& 0.153 & $ 15.147^{+  1.156}_{ -1.209}$ & $ 26.313^{+  1.877}_{ -2.036}$ & $0.203\pm 0.027$ & $0.221\pm 0.029$ \\[0.9mm] 
COMA	& 0.061 & $ 21.482^{+  1.482}_{ -1.640}$ & $ 38.281^{+  2.613}_{ -2.359}$ & $0.179\pm 0.019$ & $0.198\pm 0.021$ \\[0.9mm] 
NGC5044	& 0.672 & $  0.421^{+  0.007}_{ -0.010}$ & $  0.809^{+  0.021}_{ -0.019}$ & $0.103\pm 0.002$ & $0.125\pm 0.003$ \\[0.9mm] 
A1736	& 0.025 & $  5.825^{+  2.436}_{ -1.805}$ & $ 13.133^{+  4.808}_{ -3.715}$ & $0.266\pm 0.119$ & $0.347\pm 0.172$ \\[0.9mm] 
A3558	& 0.085 & $ 13.832^{+  0.988}_{ -0.957}$ & $ 26.421^{+  1.840}_{ -1.720}$ & $0.258\pm 0.032$ & $0.306\pm 0.038$ \\[0.9mm] 
A3562	& 0.110 & $  7.655^{+  0.456}_{ -0.448}$ & $ 16.014^{+  0.964}_{ -0.860}$ & $0.208\pm 0.013$ & $0.275\pm 0.018$ \\[0.9mm] 
A3571	& 0.144 & $ 17.277^{+  0.775}_{ -0.681}$ & $ 31.022^{+  1.244}_{ -1.203}$ & $0.208\pm 0.014$ & $0.233\pm 0.015$ \\[0.9mm] 
A1795	& 0.499 & $ 16.766^{+  1.514}_{ -1.566}$ & $ 29.603^{+  2.690}_{ -2.732}$ & $0.172\pm 0.035$ & $0.192\pm 0.038$ \\[0.9mm] 
A3581	& 0.314 & $  1.473^{+  0.155}_{ -0.145}$ & $  2.785^{+  0.334}_{ -0.284}$ & $0.153\pm 0.015$ & $0.184\pm 0.018$ \\[0.9mm] 
MKW8	& 0.051 & $  2.371^{+  1.019}_{ -0.721}$ & $  4.761^{+  1.806}_{ -1.309}$ & $0.113\pm 0.037$ & $0.143\pm 0.049$ \\[0.9mm] 
A2029	& 0.564 & $ 25.113^{+  2.332}_{ -2.322}$ & $ 45.532^{+  3.748}_{ -4.166}$ & $0.212\pm 0.037$ & $0.242\pm 0.042$ \\[0.9mm] 
A2052	& 0.521 & $  4.176^{+  0.201}_{ -0.166}$ & $  8.132^{+  0.334}_{ -0.373}$ & $0.214\pm 0.007$ & $0.262\pm 0.009$ \\[0.9mm] 
MKW3S	& 0.314 & $  5.337^{+  0.348}_{ -0.362}$ & $  9.641^{+  0.639}_{ -0.589}$ & $0.174\pm 0.018$ & $0.199\pm 0.020$ \\[0.9mm] 
A2065	& 0.039 & $ 15.090^{+  2.965}_{ -2.682}$ & $ 20.342^{+  4.423}_{ -4.768}$ & $0.112\pm 0.089$ & $0.087\pm 0.073$ \\[0.9mm] 
A2063	& 0.122 & $  5.223^{+  0.334}_{ -0.279}$ & $  9.932^{+  0.551}_{ -0.509}$ & $0.184\pm 0.013$ & $0.219\pm 0.016$ \\[0.9mm] 
A2142	& 0.268 & $ 33.630^{+  4.346}_{ -3.659}$ & $ 60.598^{+  7.866}_{ -6.013}$ & $0.253\pm 0.056$ & $0.288\pm 0.063$ \\[0.9mm] 
A2147	& 0.034 & $  7.417^{+  2.132}_{ -1.537}$ & $ 16.939^{+  4.283}_{ -3.092}$ & $0.248\pm 0.065$ & $0.350\pm 0.096$ \\[0.9mm] 
A2163	& 0.102 & $ 69.134^{+  4.558}_{ -4.557}$ & $108.565^{+  6.794}_{ -6.813}$ & $0.217\pm 0.027$ & $0.209\pm 0.026$ \\[0.9mm] 
A2199	& 0.162 & $  7.507^{+  0.512}_{ -0.432}$ & $ 12.806^{+  0.796}_{ -0.749}$ & $0.178\pm 0.013$ & $0.190\pm 0.015$ \\[0.9mm] 
A2204	& 0.987 & $ 23.639^{+  1.472}_{ -1.287}$ & $ 41.924^{+  2.411}_{ -2.575}$ & $0.273\pm 0.020$ & $0.304\pm 0.022$ \\[0.9mm] 
A2244	& 0.234 & $ 15.681^{+  7.540}_{ -4.036}$ & $ 27.889^{+ 12.847}_{ -7.019}$ & $0.181\pm 0.161$ & $0.202\pm 0.178$ \\[0.9mm] 
A2256	& 0.051 & $ 23.288^{+  1.651}_{ -1.628}$ & $ 35.923^{+  2.141}_{ -2.166}$ & $0.182\pm 0.031$ & $0.165\pm 0.029$ \\[0.9mm] 
A2255	& 0.032 & $ 18.433^{+  1.182}_{ -1.008}$ & $ 31.898^{+  1.599}_{ -1.335}$ & $0.169\pm 0.016$ & $0.171\pm 0.017$ \\[0.9mm] 
A3667	& 0.066 & $ 19.756^{+  1.909}_{ -1.844}$ & $ 40.301^{+  3.284}_{ -3.679}$ & $0.287\pm 0.044$ & $0.360\pm 0.055$ \\[0.9mm] 
S1101	& 0.554 & $  4.366^{+  1.040}_{ -0.727}$ & $  7.315^{+  1.732}_{ -1.163}$ & $0.169\pm 0.086$ & $0.179\pm 0.091$ \\[0.9mm] 
A2589	& 0.120 & $  4.967^{+  2.071}_{ -1.211}$ & $  9.060^{+  3.596}_{ -2.128}$ & $0.158\pm 0.121$ & $0.181\pm 0.136$ \\[0.9mm] 
A2597	& 0.711 & $  7.714^{+  0.649}_{ -0.980}$ & $ 12.988^{+  1.118}_{ -1.635}$ & $0.171\pm 0.035$ & $0.182\pm 0.036$ \\[0.9mm] 
A2634	& 0.021 & $  4.509^{+  0.647}_{ -0.560}$ & $  8.958^{+  0.933}_{ -0.898}$ & $0.143\pm 0.031$ & $0.167\pm 0.037$ \\[0.9mm] 
A2657	& 0.098 & $  4.896^{+  0.445}_{ -0.432}$ & $  9.398^{+  0.789}_{ -0.807}$ & $0.173\pm 0.025$ & $0.208\pm 0.030$ \\[0.9mm] 
A4038	& 0.259 & $  4.064^{+  0.189}_{ -0.179}$ & $  7.727^{+  0.396}_{ -0.323}$ & $0.188\pm 0.007$ & $0.227\pm 0.008$ \\[0.9mm] 
A4059	& 0.202 & $  6.362^{+  0.590}_{ -0.512}$ & $ 11.650^{+  0.975}_{ -0.968}$ & $0.161\pm 0.020$ & $0.185\pm 0.023$ \\[0.9mm]
\tableline
\sidehead{\hspace{1.74cm}\vspace{1mm} Clusters from the extended sample not included in \gcss .} 
\tableline \\[-2.4mm]
A2734	& 0.063 & $  6.351^{+  1.259}_{ -1.038}$ & $ 11.721^{+  1.980}_{ -1.786}$ & $0.182\pm 0.056$ & $0.207\pm 0.063$ \\[0.9mm] 
A2877	& 0.028 & $  2.574^{+  1.305}_{ -0.748}$ & $  5.072^{+  2.313}_{ -1.373}$ & $0.099\pm 0.086$ & $0.120\pm 0.103$ \\[0.9mm] 
NGC499	& 0.204 & $  0.086^{+  0.010}_{ -0.009}$ & $  0.131^{+  0.019}_{ -0.018}$ & $0.024\pm 0.003$ & $0.023\pm 0.003$ \\[0.9mm] 
AWM7	& 0.088 & $  5.379^{+  0.461}_{ -0.424}$ & $  9.163^{+  0.714}_{ -0.676}$ & $0.142\pm 0.013$ & $0.151\pm 0.014$ \\[0.9mm] 
PERSEUS	& 0.632 & $ 19.605^{+  0.646}_{ -0.941}$ & $ 37.157^{+  1.388}_{ -1.417}$ & $0.287\pm 0.012$ & $0.344\pm 0.014$ \\[0.9mm] 
S405	& 0.024 & $  7.638^{+  3.933}_{ -2.581}$ & $ 15.243^{+  6.282}_{ -4.520}$ & $0.195\pm 0.128$ & $0.226\pm 0.161$ \\[0.9mm] 
3C129	& 0.034 & $  9.078^{+  2.882}_{ -2.280}$ & $ 17.611^{+  4.391}_{ -3.627}$ & $0.160\pm 0.111$ & $0.189\pm 0.136$ \\[0.9mm] 
A0539	& 0.061 & $  3.767^{+  0.266}_{ -0.269}$ & $  7.286^{+  0.435}_{ -0.456}$ & $0.161\pm 0.014$ & $0.195\pm 0.018$ \\[0.9mm] 
S540	& 0.078 & $  2.440^{+  0.777}_{ -0.599}$ & $  4.283^{+  1.226}_{ -0.975}$ & $0.134\pm 0.050$ & $0.146\pm 0.055$ \\[0.9mm] 
A0548w	& 0.019 & $  0.697^{+  0.287}_{ -0.207}$ & $  1.333^{+  0.402}_{ -0.348}$ & $0.112\pm 0.064$ & $0.126\pm 0.075$ \\[0.9mm] 
A0548e	& 0.054 & $  3.090^{+  0.259}_{ -0.271}$ & $  6.484^{+  0.583}_{ -0.475}$ & $0.177\pm 0.016$ & $0.234\pm 0.021$ \\[0.9mm] 
A3395n	& 0.020 & $  8.853^{+  2.287}_{ -1.918}$ & $ 14.014^{+  2.251}_{ -2.045}$ & $0.102\pm 0.075$ & $0.091\pm 0.073$ \\[0.9mm] 
UGC03957	& 0.093 & $  2.497^{+  0.668}_{ -0.524}$ & $  3.927^{+  1.092}_{ -0.864}$ & $0.099\pm 0.046$ & $0.098\pm 0.046$ \\[0.9mm] 
PKS0745	& 0.970 & $ 24.105^{+  1.020}_{ -0.791}$ & $ 42.225^{+  1.487}_{ -1.487}$ & $0.271\pm 0.010$ & $0.299\pm 0.012$ \\[0.9mm] 
A0644	& 0.152 & $ 17.316^{+  1.322}_{ -1.413}$ & $ 27.890^{+  2.228}_{ -2.054}$ & $0.139\pm 0.024$ & $0.141\pm 0.025$ \\[0.9mm] 
S636	& 0.014 & $  0.968^{+  0.414}_{ -0.275}$ & $  2.023^{+  0.645}_{ -0.504}$ & $0.158\pm 0.084$ & $0.175\pm 0.101$ \\[0.9mm] 
A1413	& 0.193 & $ 19.436^{+  1.381}_{ -1.297}$ & $ 32.883^{+  2.197}_{ -2.132}$ & $0.191\pm 0.016$ & $0.202\pm 0.017$ \\[0.9mm] 
M49	& 0.259 & $  0.076^{+  0.004}_{ -0.003}$ & $  0.134^{+  0.008}_{ -0.006}$ & $0.019\pm 0.001$ & $0.021\pm 0.001$ \\[0.9mm] 
A3528n	& 0.066 & $  4.681^{+  1.716}_{ -0.820}$ & $  8.515^{+  3.026}_{ -1.359}$ & $0.162\pm 0.106$ & $0.183\pm 0.119$ \\[0.9mm] 
A3528s	& 0.094 & $  4.604^{+  1.379}_{ -0.910}$ & $  9.928^{+  2.823}_{ -1.916}$ & $0.273\pm 0.111$ & $0.368\pm 0.149$ \\[0.9mm] 
A3530	& 0.025 & $  5.677^{+  0.794}_{ -0.755}$ & $  9.851^{+  1.139}_{ -0.980}$ & $0.126\pm 0.036$ & $0.129\pm 0.038$ \\[0.9mm] 
A3532	& 0.045 & $  7.523^{+  0.704}_{ -0.577}$ & $ 13.693^{+  1.042}_{ -0.898}$ & $0.158\pm 0.020$ & $0.176\pm 0.023$ \\[0.9mm] 
A1689	& 0.334 & $ 28.143^{+  1.461}_{ -1.266}$ & $ 45.453^{+  2.130}_{ -2.183}$ & $0.182\pm 0.013$ & $0.184\pm 0.013$ \\[0.9mm] 
A3560	& 0.033 & $  4.346^{+  0.949}_{ -0.783}$ & $  8.958^{+  1.644}_{ -1.489}$ & $0.201\pm 0.062$ & $0.250\pm 0.078$ \\[0.9mm] 
A1775	& 0.061 & $  7.440^{+  0.659}_{ -0.521}$ & $ 13.270^{+  1.012}_{ -0.827}$ & $0.206\pm 0.024$ & $0.225\pm 0.026$ \\[0.9mm] 
A1800	& 0.036 & $  7.573^{+  3.471}_{ -2.443}$ & $ 13.020^{+  4.657}_{ -3.619}$ & $0.159\pm 0.109$ & $0.163\pm 0.118$ \\[0.9mm] 
A1914	& 0.218 & $ 30.019^{+  1.679}_{ -1.733}$ & $ 45.586^{+  2.963}_{ -2.574}$ & $0.140\pm 0.015$ & $0.134\pm 0.015$ \\[0.9mm] 
NGC5813	& 0.175 & $  0.052^{+  0.021}_{ -0.018}$ & $  0.076^{+  0.040}_{ -0.034}$ & $0.022\pm 0.011$ & $0.020\pm 0.010$ \\[0.9mm] 
NGC5846	& 0.472 & $  0.052^{+  0.004}_{ -0.004}$ & $  0.090^{+  0.009}_{ -0.008}$ & $0.015\pm 0.001$ & $0.017\pm 0.001$ \\[0.9mm] 
A2151w	& 0.159 & $  2.300^{+  0.153}_{ -0.133}$ & $  4.313^{+  0.254}_{ -0.257}$ & $0.151\pm 0.011$ & $0.178\pm 0.013$ \\[0.9mm] 
A3627	& 0.037 & $ 10.468^{+  1.316}_{ -1.088}$ & $ 21.250^{+  2.145}_{ -1.937}$ & $0.186\pm 0.027$ & $0.231\pm 0.035$ \\[0.9mm] 
TRIANGUL	& 0.097 & $ 29.978^{+  2.191}_{ -1.874}$ & $ 54.917^{+  3.637}_{ -3.232}$ & $0.223\pm 0.027$ & $0.255\pm 0.030$ \\[0.9mm] 
OPHIUCHU	& 0.128 & $ 25.759^{+  1.966}_{ -1.907}$ & $ 40.138^{+  3.336}_{ -3.094}$ & $0.127\pm 0.014$ & $0.124\pm 0.014$ \\[0.9mm] 
ZwCl1742	& 0.096 & $ 10.444^{+  2.350}_{ -2.016}$ & $ 17.049^{+  3.655}_{ -3.021}$ & $0.152\pm 0.055$ & $0.154\pm 0.056$ \\[0.9mm] 
A2319	& 0.098 & $ 31.209^{+  2.536}_{ -2.171}$ & $ 59.361^{+  3.837}_{ -4.071}$ & $0.280\pm 0.032$ & $0.328\pm 0.038$ \\[0.9mm] 
A3695	& 0.040 & $ 12.950^{+  6.632}_{ -4.499}$ & $ 25.006^{+  9.973}_{ -7.707}$ & $0.233\pm 0.156$ & $0.268\pm 0.192$ \\[0.9mm] 
IIZw108	& 0.027 & $  5.307^{+  2.342}_{ -1.561}$ & $ 10.347^{+  3.808}_{ -2.789}$ & $0.179\pm 0.092$ & $0.205\pm 0.110$ \\[0.9mm] 
A3822	& 0.040 & $ 10.313^{+  4.526}_{ -3.179}$ & $ 19.723^{+  7.175}_{ -5.558}$ & $0.208\pm 0.105$ & $0.239\pm 0.125$ \\[0.9mm] 
A3827	& 0.051 & $ 21.863^{+  5.124}_{ -4.719}$ & $ 31.169^{+  7.551}_{ -6.849}$ & $0.134\pm 0.097$ & $0.114\pm 0.085$ \\[0.9mm] 
A3888	& 0.101 & $ 23.011^{+  2.863}_{ -2.531}$ & $ 31.556^{+  4.412}_{ -3.993}$ & $0.105\pm 0.037$ & $0.088\pm 0.031$ \\[0.9mm] 
A3921	& 0.067 & $ 13.019^{+  0.942}_{ -0.896}$ & $ 21.075^{+  1.268}_{ -1.252}$ & $0.154\pm 0.019$ & $0.153\pm 0.019$ \\[0.9mm] 
HCG94	& 0.105 & $  3.641^{+  0.350}_{ -0.367}$ & $  7.222^{+  0.717}_{ -0.657}$ & $0.159\pm 0.024$ & $0.200\pm 0.030$ \\[0.9mm] 
RXJ2344	& 0.072 & $  9.458^{+  1.164}_{ -1.048}$ & $ 14.619^{+  1.584}_{ -1.460}$ & $0.137\pm 0.040$ & $0.130\pm 0.038$ \\[0.9mm] 
\enddata

\tablecomments{
The columns are described in detail at the end of Section~\ref{massd}.
}

\end{deluxetable}
\chapter{Temperature Structure of Abell 1835}
\label{a1835}
\section{Motivation}

As mentioned in Sects.~\ref{back:tg} and \ref{back:dm} knowledge about
the temperature structure of the intracluster gas is
vital for the determination of the gravitational cluster mass using the
hydrostatic equation. Since for the aim of this work the mass is the
most fundamental cluster
property here I take advantage of the availability of data from the
recently launched satellite observatory \xmm\ \citep[e.g.,][]{jla01} on
the galaxy cluster Abell 1835 to study the temperature structure.

\xmm\ is a cornerstone mission of the European Space Agency's (ESA) Horizon
2000 program. The most important advantage of this largest scientific
satellite that has been launched by ESA to date, compared to previous
missions is the huge effective area (Fig.~\ref{fig:effar}). Compared to \ro\
the sensitive energy range and the spectral resolution has also been
significantly improved and compared to \as\ the spatial resolution is
much better (Tab.~\ref{tab:inst}). Compared to the recently launched satellite
\cha\ \citep[e.g.,][]{wos96}, whose main virtue is the high spatial resolution
of $\lesssim 1$\,arcsec FWHM, \xmm\ has the -- especially
for extended objects crucial -- advantages of a larger field of view, and a
larger effective area.

Data
from previous satellites have been used
extensively in order to answer the question whether or not clusters
show a trend of decreasing gas temperatures with increasing radius in
the outer cluster parts in general \citep[e.g.,][]{mfs98,ibe99,w00,ib00}. The
findings are contradictory. A significant temperature decrease would
lead to an overestimate of total cluster masses if the gas is assumed to
be isothermal (Sect.~\ref{mest_x}). With the availability of cluster data obtained by
\xmm\ it should now be possible in principal to determine the
temperature profile
accurately out to large physical radii. The large effective area, the good
simultaneous spatial 
and spectral resolution combined with its field of view make
\xmm\ the ideal instrument to resolve the previous
discrepancies.

Abell 1835 is an especially interesting galaxy cluster. First of all
because of its medium redshift, $z=0.252$. Only a few detailed temperature
profiles for clusters around this redshift have been published up to
now and this distance also allows to trace the cluster emission far out
in a single observation.
Secondly A1835 is believed to host one of the most massive
cooling flows with 
mass accretion rates $\dot M\sim 2\,000\,M_{\odot}\,\rm
yr^{-1}$ \citep{afe96}. Following one of the major cluster specific
\xmm\ discoveries
these mass accretion rates have been significantly reduced
recently due to high resolution spectroscopy evidence obtained with
the Reflection Grating Spectrometer (RGS) onboard \xmm\ \citep{ppk00}. 
These facts make A1835 one of the crucial clusters to test the
predictions of cooling flow and alternative models.
Furthermore strong gravitational lensing has been observed \citep[e.g.,][]{a98}
and the Sunyaev-Zeldovich effect has been measured
for this cluster \citep[e.g.,][]{maa00}. The latter measurement in combination
with X-ray data has been used to constrain $H_0$, with the largest
uncertainty being the uncertainty in the gas temperature available
previously \citep{maa00}. With a flux $\fx(\eb)=1.2\esc$ A1835 is not included
in \gcs\ or in the extended sample but
may still be called a fairly bright cluster.

In this Chapter the temperature profile of A1835 is determined from the central regions
out to large radii utilizing data from the EPIC-pn camera \citep{sbd01} onboard
\xmm .

\section{Data Reduction and Analysis}

A1835 was observed during the performance verification
phase. The data are public and have been retrieved from
the \xmm\ archive at MPE.
The basic observation parameters are listed in Tab.~\ref{tab:a1835}.

\begin{figure}
\hspace{2.0cm}
\psfig{figure=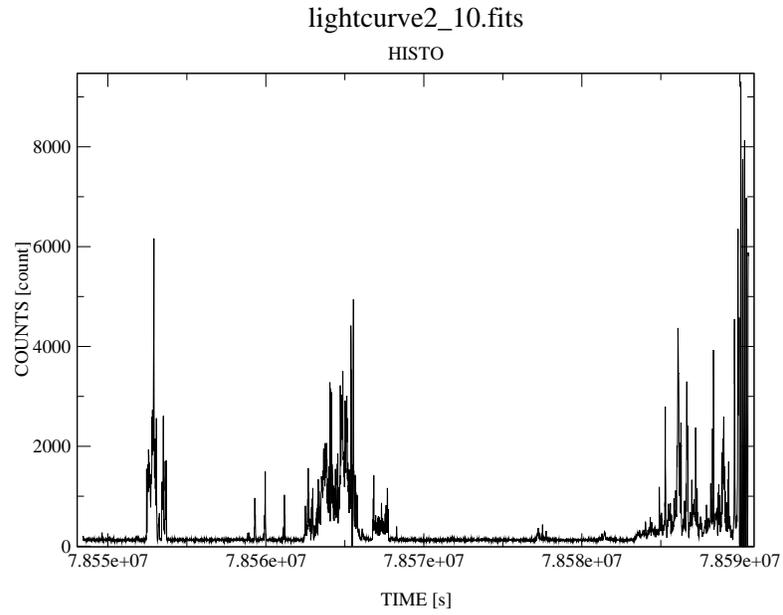,width=9cm,angle=270}
\caption{Lightcurve binned in 10\,s intervals.
\label{lc}}
\end{figure}

\begin{figure}
\hspace{2.0cm}
\psfig{figure=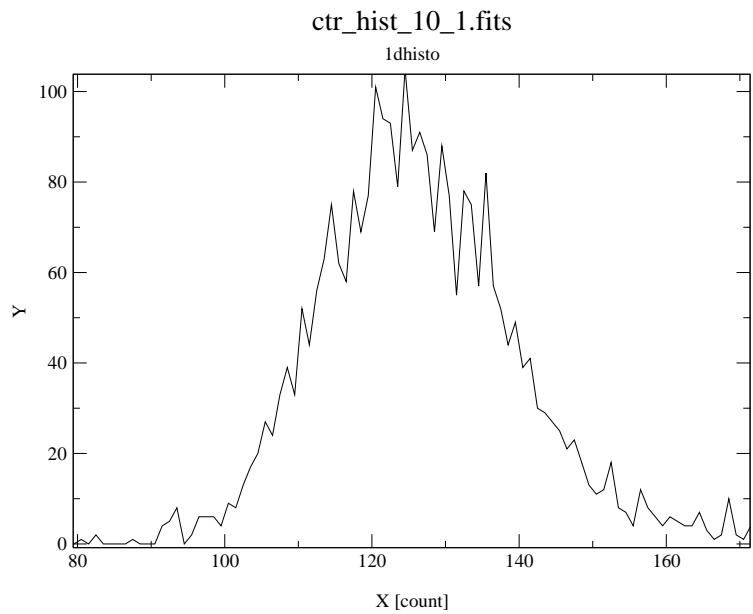,width=9cm,angle=270}
\caption{Central part of the count rate [counts per 10\,s] distribution.
\label{ch}}
\end{figure}

\begin{figure}
\begin{center}
\psfig{figure=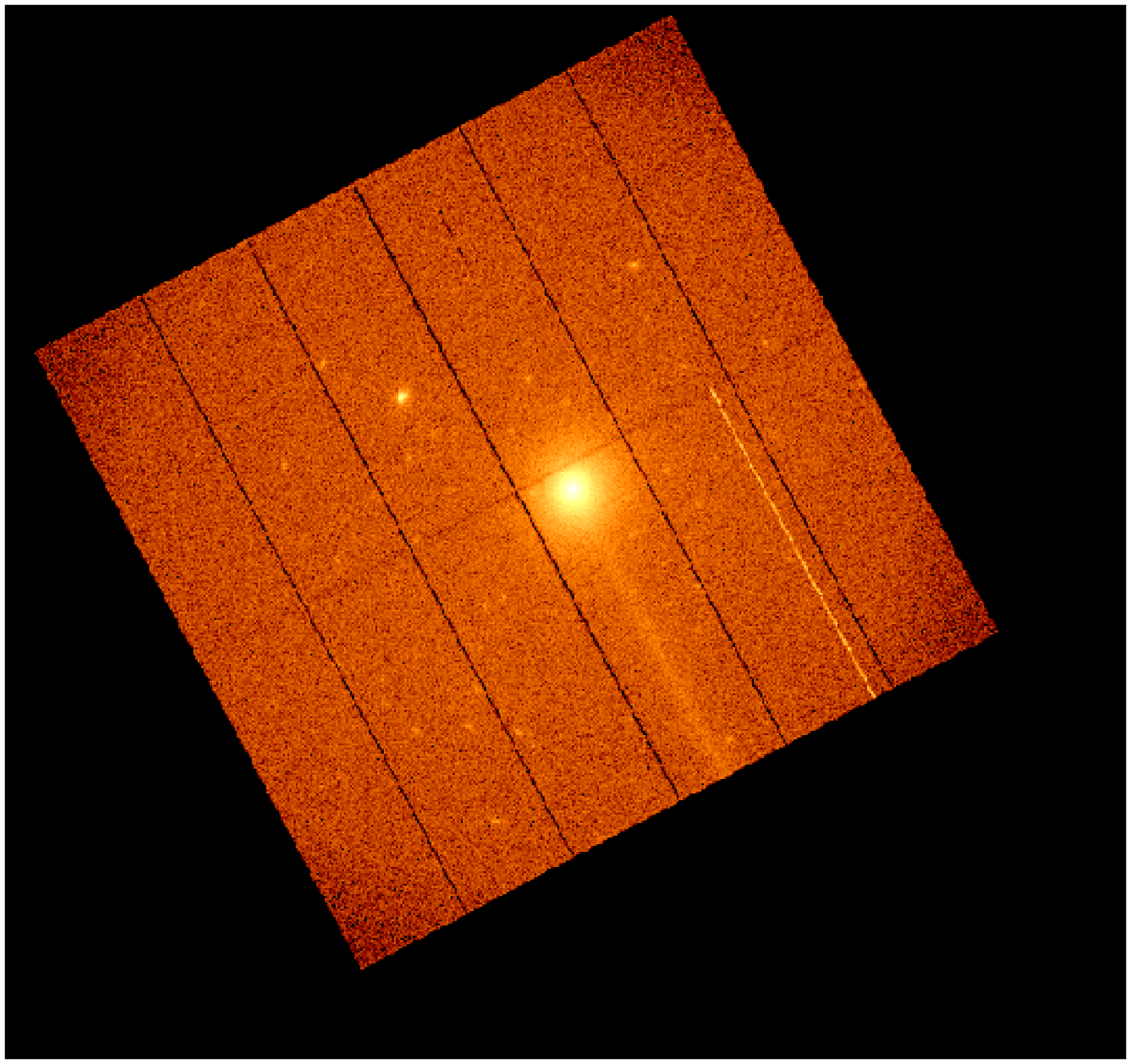,width=7.6cm}
\psfig{figure=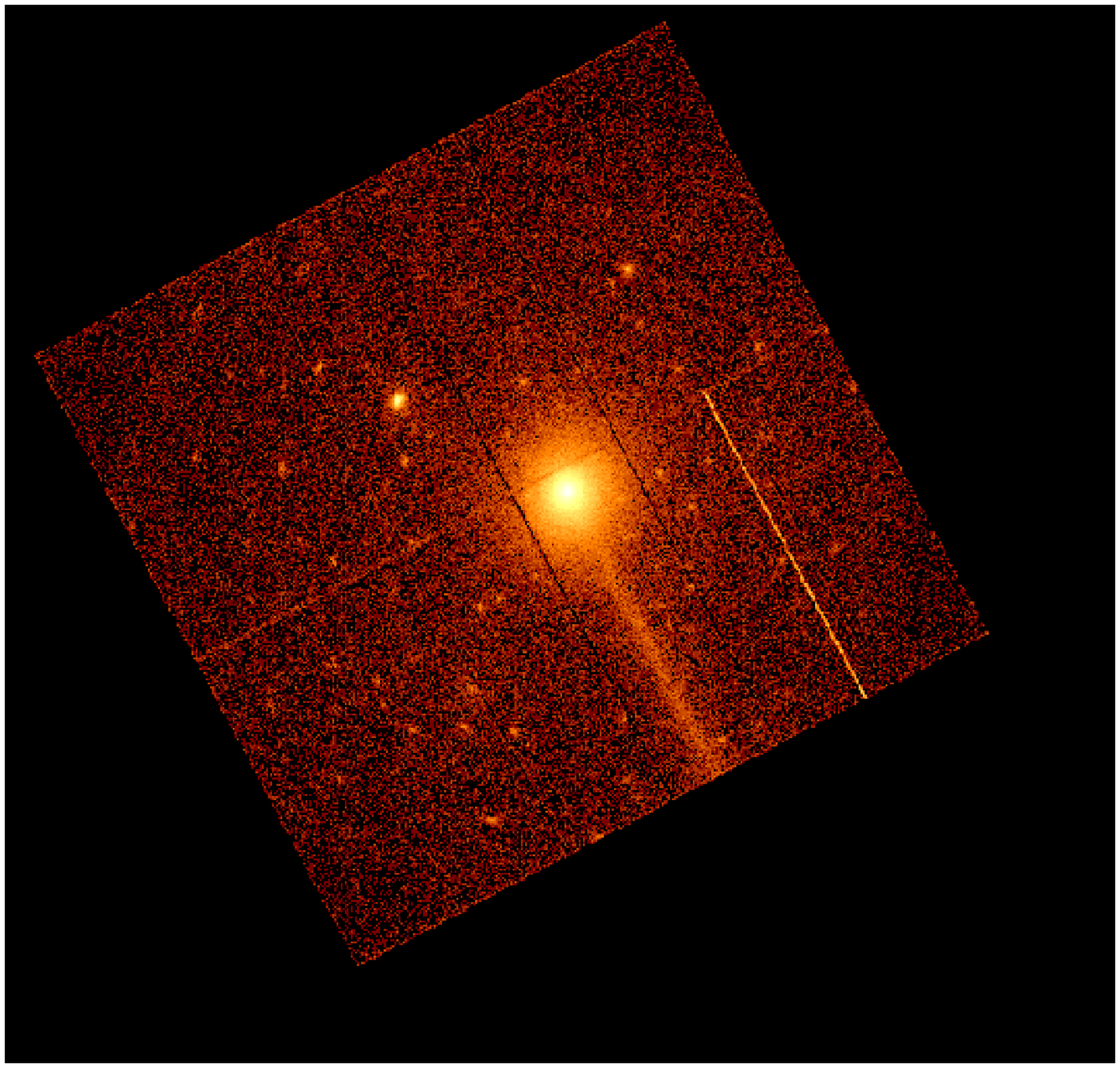,width=7.6cm}
\end{center}
\caption{EPIC-pn images of A1835 in sky coordinates in the energy band
0.35--10\,keV. The
field of view is almost
30\,arcmin. Left: Full 42\,ks observation. Right: 25\,ks
Good Time Intervals after application of the count rate thresholds.
\label{ia}}
\end{figure}

Data taken with the Chandra and \xmm\ observatories both suffer
from times of very high background composed of `soft protons'.
In Fig.~\ref{lc} the lightcurve of the EPIC-pn observation of A1835 is
shown\footnote{The events file has been created with SAS version
xmmsas\_20001220\_1901.}. 
To obtain good statistics all events with energies above
0.3\,keV have been 
selected. Most of the time the count rate, $\cx$, is stable at
about 125 counts
per 10\,s. But occasionally huge flares are visible, reaching
peak count rates more than 50 times higher than normal. Additionally smaller
flares and times when the count rate drops to zero (counting
mode times) appear. To increase the signal to noise ratio
count rate thresholds ($100\leq \cx \leq 150$ counts
per 10\,s) for selection of Good Time Intervals (GTI) have been
set after inspection of the count rate histogram (Fig.~\ref{ch}). 
The improvement is obvious in Fig.~\ref{ia}. Dozens of
sources pop out of the background in the screened image. Since the
four corners of the EPIC-pn camera are shielded against X-ray photons,
the image on the right hand side in Fig.~\ref{ia} also shows that most of the remaining
background is Particle Induced Background (PIB;
background induced directly or through instrumental lines by 
high energy particles).

\begin{deluxetable}{lcccccccc}
\tabletypesize{\footnotesize}
\tablecaption{Observation parameters\label{tab:a1835}}
\tablewidth{0pt}
\tablehead{
\colhead{ID} & \colhead{Date} &
\colhead{R.A.} & \colhead{Dec.} & \colhead{Duration} &
\colhead{Exposure} &
\colhead{GTI} & \colhead{Detector} &
\colhead{Filter} 
}
\startdata
0098010101 & 2000-06-28 & 210.2629 & 2.8507 & 56\,ks & 42\,ks & 25\,ks &
EPIC-pn FF & Thin \\ 
\enddata
\end{deluxetable}
For the spectral analysis
all obvious point sources have been excluded as well as hot
pixels and columns. Furthermore a correction has been applied to account
for events counted during read out (out of time events, most obviously
visible as the `jet' extending from the cluster center parallel to the hot
column, Fig.~\ref{ia}). The correction is done by creating an events file within the
XMM SAS (Science Analysis System), where
events have been redistributed randomly along columns keeping all
other info, e.g., pattern and time\footnote{The time information in this events
file is accurate up to one read out cycle.}. The CTI (Charge Transfer
Inefficiency) correction is done
after the redistribution. Then a spectrum (or image) created from this
events file is multiplied by a mode dependent factor ($=0.06$ for Full
Frame mode, FF) and subtracted from the original spectrum (or image). Neglecting the
influence of out of time events may lead to
wrong temperature estimates since these events are assigned an
incorrect CTI correction.
Moreover,
assume the cluster center has a lower temperature than the outer parts
(as will actually be found later), then a spectrum taken from a ring of the
outer low surface brightness region is contaminated by photons having
a lower energy on average and the temperature estimate may be biased
low.

The subtraction of the (remaining) background is performed using
spectra created in the same way as outlined above. Various methods
may be applied for background subtraction. The major drawbacks of two
simple methods for the observation here are described as follows. One
possibility is to use a `blank sky' observation like the Lockman Hole
observation and create a spectrum from the same detector region as the
source spectrum, scaling by the source spectrum's exposure time. The
disadvantage of this method is that it ignores variation in the Cosmic
X-ray Background (CXB). The CXB is mainly composed of A) isotropic
extragalactic emission
(dominant for photon energies $E\gtrsim 2$\,keV) and B) anisotropic emission
from galactic diffuse
sources and the `hot bubble' around the solar system (important for
$E\lesssim 2$\,keV). Galactic photoelectric absorption (important for
$E\lesssim 1$\,keV) causes an additional anisotropy.
The CXB therefore can be considered as constant in time but varying
with pointing direction (at soft energies) \citep[e.g.,][]{i97}. A further
drawback is that a
possible temporal variation of the normalization and/or spectrum of the
PIB is
ignored. Last not least if this method is used one must ensure that
the flare filtering criteria are equivalent for science and background
observation. This can be achieved, e.g., by assuming a constant PIB
and selecting an energy band
containing exclusively PIB and soft proton events (possibly
$E>12$\,keV) with the drawback of a
significant loss in statistics. 
Another simple method for the background subtraction is to use a ring
outside the cluster emission in
the science observation, taking advantage of the medium redshift of
A1835 and \xmm 's field of view. Here the drawbacks are that the
structure on
the detector of the PIB is ignored. Using this
method also the question arises whether or not one should apply a
vignetting correction for the background spectrum. If a correction is
applied then the actually not vignetted PIB is
overestimated. If no correction is applied then the vignetted CXB is
underestimated. 

More complicated methods are possible circumventing some of the above
problems \citep[e.g.,][]{mn01,paa01}. Here
the background from not vignetting corrected data in a ring
where the contribution from cluster emission is negligible
(7--11\,arcmin from cluster center)
has been taken because the PIB dominates the background. The main structural
feature of the PIB is a fluorescent copper line at 8\,keV, which is
almost absent at the detector center but increases towards the outer
parts of the EPIC-pn detector. Therefore events with $7.9 \le
E \le 8.1$\,keV have been excluded from the fit.
With this method the CXB and PIB are corrected for in one step, since
the CXB is not expected to change drastically for this small angular
distance and variations in the PIB spectrum with time have no effect
since exactly equivalent time intervals are used. Note, however,
that the absolute contribution of the CXB is underestimated due to the
vignetting increase with increasing radius.
In the outer parts the fitted value of the column
density of neutral hydrogen, $\nhcol$, drops below the galactic value, which
indicates that leaving $\nhcol$ as a free parameter 
compensates for the underestimated CXB. However, the decrease of the
$\nhcol$ value is certainly not an ideal correction for this
effect, since the spectral changes invoked by both effects probably
do not cancel each other exactly. On the other hand
using a vignetting corrected background for this method, where
photons have been corrected individually, has resulted in
worse fits and underestimated temperatures in the outer parts because
of the wrongly corrected PIB.  This is due to the dominance of the PIB at high
energies and to the steeper vignetting increase with radius for higher
energies.

Only singles, i.e.\ photons having deposited their energy in a single
pixel on the CCD, have been selected for the spectral analysis. Most
events are single events, but the just recently made available response
matrix for double
events will allow to increase the statistics at least for higher energies
and test the effects of a possible
dependence of the single to double fraction on detector position soon.
A 5-$\sigma$ binning for the spectra has been used to ensure a
significant signal in each bin. Choosing a  3-$\sigma$ binning changes
the fit results only well within the statistical errors. The energy
range used has nominally been set to 0.3--10.0\,keV. In the
outer low
surface brightness parts the number of significant high energy bins
decreases. The fit parameter values have therefore been
determined partly within smaller energy ranges. If two or more
different temperatures are present for a specific radial bin then the
availability of only a weak signal at high energies may bias the
estimated temperature in the outermost radial bin low. The
presence of a lower signal to noise ratio at higher energies is
independent of the binning method.
The spectra have been modeled and fitted within \xs\ using
a single temperature MeKaL model \citep{mkl95}
including photoelectric absorption
(wabs*mekal). The temperature, metal abundance, absorption column
density, and normalization have been left as
free parameters.
Leaving the redshift as a free parameter yields values consistent with
the optically determined redshift and with encouragingly small
uncertainties.
The 2001-05-07 release of EPIC-pn response matrices has
been used.
The spectra have not been deprojected, i.e.\ in the innermost radial
bins also contributions from physically larger radii are
included. However, the very steep radial surface brightness profile of
A1835 ensures that these contaminating contributions are small.
Changing the background normalization by $\pm 10$\,\% has a
negligible effect on the fit parameter values. Inclusion of arf files
that correct the source spectra for telescope vignetting also affects
the parameter values only insignificantly.

\section{Results and Discussion}\label{a1835_res}

\begin{figure}
\begin{center}
\psfig{figure=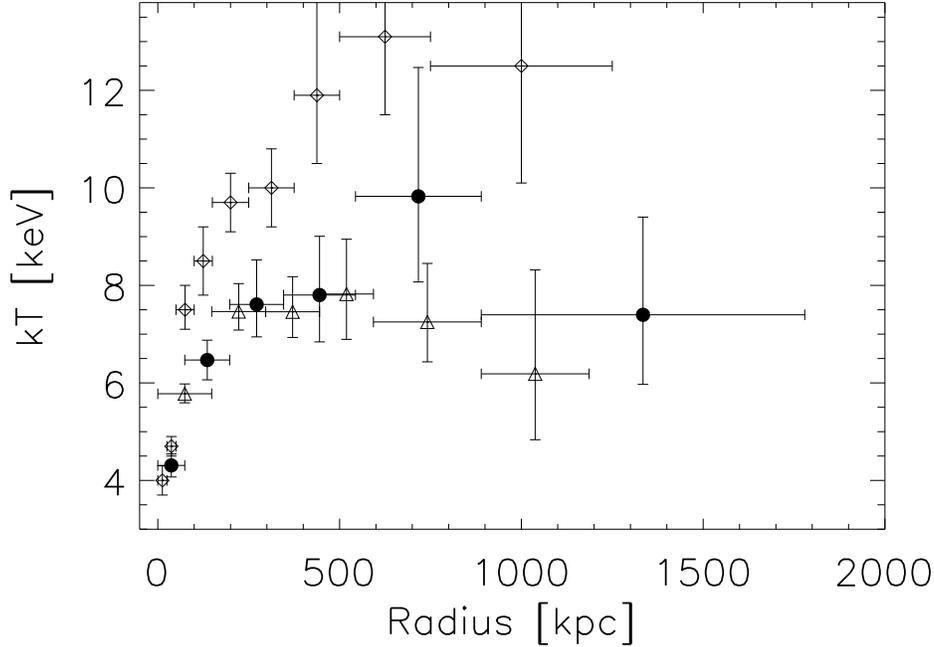,width=14cm,angle=0}
\end{center}
\caption{Temperature profiles as obtained with different
detectors. Filled circles: EPIC-pn, this work; open triangles:
EPIC-MOS2 \citep{mn01}; open
diamonds: Chandra \citep{saf01}. Vertical error bars
indicate the 90\,\% confidence level
statistical uncertainty for one interesting parameter for the first
two works and the 1-$\sigma$ statistical uncertainty for the last
work. Horizontal bars indicate the radial bin size.
\label{TR}}
\end{figure}
The best fit cluster temperatures for six radial bins are shown in
Fig.~\ref{TR} as filled 
circles. The intracluster gas in the center ($r<200$\,kpc) is
significantly cooler than in the outer parts, which may be caused
either by a cooling flow \citep[e.g.,][]{f94} or by the Interstellar Medium
(ISM) of the central brightest cluster galaxy \citep[e.g.,][]{mef01}.
Within the statistical
uncertainty the temperature profile for $r>200$\,kpc can be considered
isothermal, in agreement with the assumption made for the mass determination
throughout this work. Although at present a temperature decrease in the
very outer part obviously cannot be excluded. The main limiting factor for a
precise determination of the outer temperature is the
relatively high PIB.

Also shown in Fig.~\ref{TR} is the temperature profile determined
using data from the EPIC-MOS2 camera onboard \xmm\ by \citet{mn01}
(open triangles). There is perfect
agreement between the results from the two EPIC cameras out to $\sim
600$\,kpc. For larger radii the temperature estimates are well
consistent. 

The open diamonds in Fig.~\ref{TR} additionally indicate the
temperature estimates obtained for A1835 with the Chandra satellite by
\citet{saf01}. The agreement in the very
center ($r<100$\,kpc) between the Chandra and EPIC-pn results is very
good. For larger radii Chandra gives systematically higher
temperatures, but note that 1-$\sigma$ errors are shown
for Chandra.

Using a two temperature model does not result in a significant
improvement in any of the EPIC-pn fits, indicating the absence of
evidence for gas with $\kb \tx$ below $\sim 2$\,keV. This is in
agreement with the results obtained with the RGS \citep{ppk00}.

\begin{figure}
\begin{center}
\psfig{figure=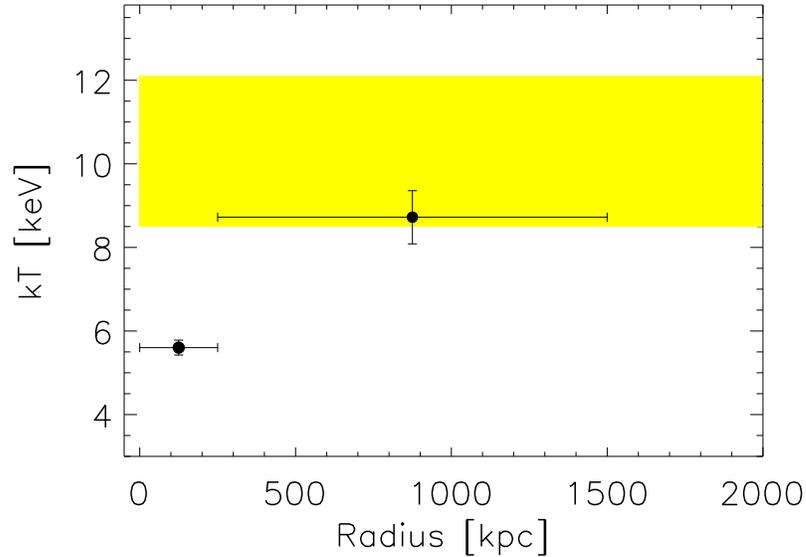,width=12cm,angle=0}
\end{center}
\caption{Overall gas temperature as estimated with EPIC-pn (upper
filled circle; this work)
and \as\ (shaded area; \citealt{afe96}).
\label{OT}}
\end{figure}
To obtain a good estimate of the overall cluster temperature outside 
the central region, where the temperature is assumed to be constant
-- consistent with observations (Fig.~\ref{TR}), a spectrum has been extracted
within $250\le
r\le1\,500$\,kpc. The best fit temperature for this region
$\kb \tx=8.7\pm0.6$\,keV (90\,\% c.l.). In Fig.~\ref{OT} this result is
compared to the previous estimate from \as\ (shaded area;
\citealt{afe96}).  The \as\ data suggest a higher temperature
($9.8^{+2.3}_{-1.3}$\,keV)
but the results are consistent. Note that the errors have been
decreased significantly with EPIC-pn.
\begin{figure}
\begin{center}
\psfig{figure=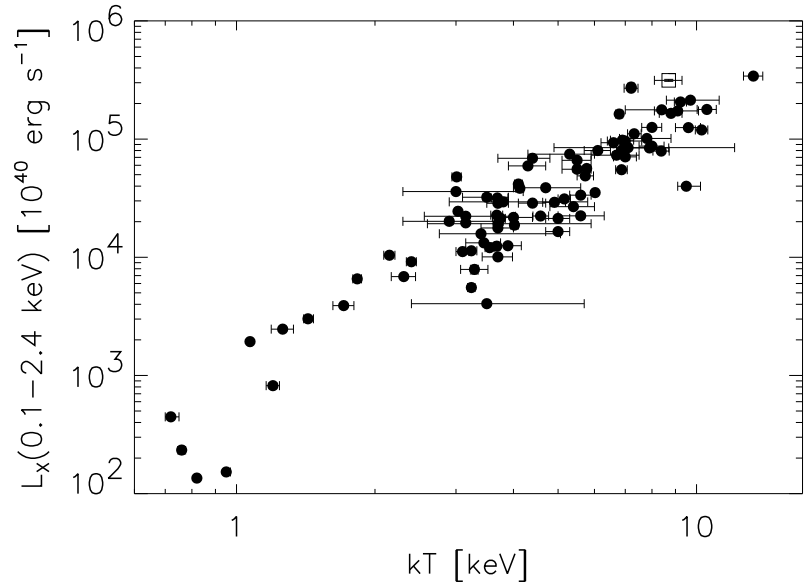,width=12cm,angle=0}
\end{center}
\caption{Luminosity and temperature of A1835 (open square) as compared to the
$\lx$--$\tx$ relation of  88 clusters
compiled during construction of \gcs .
\label{LT}}
\end{figure}
\begin{figure}
\begin{center}
\psfig{figure=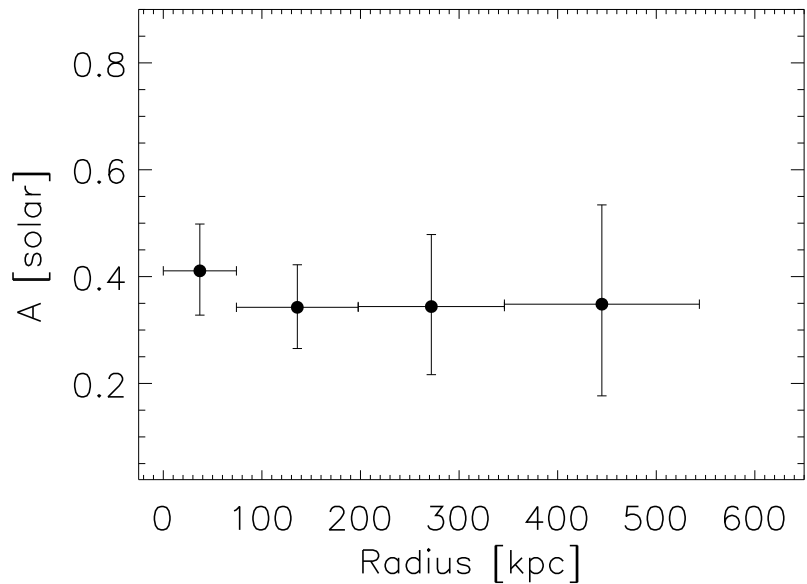,width=12.5cm,angle=0}
\end{center}
\caption{Metal abundance profile as determined with the EPIC-pn camera.
\label{AP}}
\end{figure}

Comparing this overall cluster temperature and A1835's X-ray luminosity as
measured with a \ro\ \ps\ pointed observation ($\rx = 2.85$\,Mpc) with
the $\lx$--$\tx$
relation of the part of the clusters in extended sample (Sect.~\ref{sample}) that have measured
temperatures one finds that within the
scatter it falls nicely onto this relation (Fig.~\ref{LT}). Moreover the exceptionally
strong central cool emission in A1835 may boost the total X-ray
luminosity, which would explain the slightly higher luminosity than
expected from the relation.

The metal abundance profile relative to solar abundances \citep{ag89} is
found to be flat within the statistical errors (Fig.~\ref{AP}). The
mean abundance is about 0.35 solar, in agreement with the value adopted
in the flux determination (Sect.~\ref{fluxd}).

\section{Conclusions}

The intracluster gas in the central region is significantly cooler
than in the outer part of A1835.
The innermost circle with radius 74\,kpc yields a temperature estimate
$\tx=4.3\pm 0.2$\,keV.
The gas in the outer part is consistent with being isothermal, in agreement
with the assumption made for the mass determination
throughout this work.
No indications for gas temperatures below $\sim 2$\,keV have been found.
The metal abundance profile is consistent with being flat with
$A\approx 0.35$.
The cluster temperature within $250 \le r\le
1\,500$\,kpc has been determined as $\kb \tx=8.7\pm0.6$\,keV.
This temperature is almost exactly equal to the temperature of the
innermost circle multiplied by a factor of two. However, it is
possible that the temperature decreases further a bit for even smaller radii.
The significantly reduced error of the overall cluster temperature
will allow to put tighter constraints on $H_0$ in combination with the SZ
effect.
Assuming the cluster temperature to stay isothermal beyond
1\,500\,kpc an estimate of the total gravitational  mass has been
obtained utilizing the gas density profile determined
from a \ro\ \ps\ pointed observation. 
Under the assumption of hydrostatic equilibrium it
is found using (\ref{back:ma2}) that
$\mt = 1.3\pm0.2\times 10^{15}\,\msu$ within
$r_{500} = 2.10$\,Mpc.

\chapter{Results}
\label{resu}

In this Chapter it is shown that a tight correlation exists between the
gravitational cluster mass and the X-ray luminosity. This ensures that
\gcs\ is essentially selected by cluster mass. Also relations between
other physical
properties of the clusters are shown. In the last part of
this Chapter the cluster mass function is constructed, including the
proper treatment of the scatter in the $\lx$--$\mt$ relation.

\section{Mass--Luminosity Relation}\label{relat}

\begin{figure}[thbp]
\centering\begin{tabular}{l}
\psfig{file=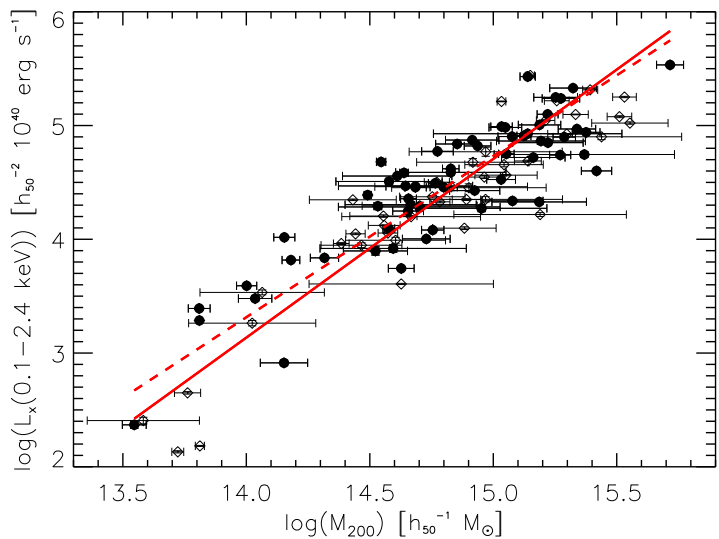,width=11cm,angle=0,clip=}
\end{tabular}
\caption{Gravitational mass--X-ray luminosity relation (solid line) for the
extended sample of 106 galaxy clusters. The dashed line gives the best
fit relation for the 63 clusters included in \gcs\ (filled circles only). The bisector fit
results are shown.
One-$\sigma$
statistical error bars are plotted for both axes, however, only the mass errors
are larger than the symbol sizes.}\label{mtlx}
\end{figure}

Since
one of the main aims of this work is the construction of a mass
function from a flux-limited sample
it is important to
test for a correlation between X-ray luminosity and gravitational mass. In
Fig.~\ref{mtlx} $\lx$, given in the \ro\ energy band, is plotted as a
function of $\mtz$, showing clearly the existence of a tight (linear
Pearson correlation coefficient $=$ 0.92) correlation, as expected.

To quantify the mass--luminosity relation, a linear regression
fit in log--log space has 
been performed. The method used allows for intrinsic scatter and errors in both
variables \citep{ab96}.
Tables~\ref{tab:lm1}--\ref{tab:lm5} give the results for different fit methods, where
minimization has been performed in vertical, horizontal, and
orthogonal direction, and the bisector result is given, which bisects the best
fit results of vertical and horizontal minimization.
The fits have been performed using the form
\begin{equation}
\log \left(\frac{\lx (\eb )}{\esl
}\right)=A+\alpha\log\left(\frac{\mtz}{\msu}\right)\,.
\label{eq:lmform}
\end{equation}
It is found, as noted in general by previous authors (e.g., \citealt{ifa90}), that the
chosen fitting method has a significant influence on the best fit
parameter values\footnote{This also implies that for a proper comparison of relations which have been
quantified by many different authors, e.g., the $\lx$--$\tx$ relation,
one and the same fitting statistic ought to be
used
(e.g., \citealt{wxf99}).}.
For different applications different fitting methods may be required.
In this work the appropriate relation for the application under
consideration is always indicated.

The difference between the fit results for 106 (Tab.~\ref{tab:lm1})
and 63 (Tab.~\ref{tab:lm2})
clusters may indicate a scale dependence of the $\lx$--$\mt$ relation,
since the difference is slightly larger than the uncertainty evaluated
with the bootstrap method. The small number of low luminosity clusters
in \gcs\ compared to the extended sample may be responsible for the
less steep relation obtained using the \gcs\ clusters only.
Note that only two out of the six clusters with $\lx<\esls$ are included
in \gcs .
To reliably detect any deviations from the power law shape of the
$\lx$--$\mt$ relation, however, even more clusters with $\mt <
10^{14}\,\msu$ (and possibly $\mt >
3\times 10^{15}\,\msu$) would need to be sampled. In Sect.~\ref{relat_d}
this will be further discussed.
As will be seen later, in
the procedure used here for the comparison
of observed and predicted mass functions the precise
shape of the $\lx$--$\mt$ relation is not important.

\begin{deluxetable}{lcccc}
\tabletypesize{\footnotesize}
\tablecaption{Fit parameter values\label{tab:lm1}}
\tablewidth{0pt}
\tablehead{
\colhead{Fit}    &  {$\alpha$} &   \colhead{$\Delta\alpha$}   &
\colhead{$A$} & \colhead{$\Delta A$}
}
\startdata
BCES($L\mid M$) & 1.496 & 0.089 & -17.741 & 1.320 \\ 
bootstrap & 1.462 & 0.089 & -17.238 & 1.327 \\ 
BCES($M\mid L$) & 1.652 & 0.085 & -20.055 & 1.261 \\ 
bootstrap & 1.672 & 0.086 & -20.357 & 1.278 \\ 
BCES-Bisector & 1.571 & 0.083 & -18.857 & 1.237  \\ 
bootstrap & 1.562 & 0.083 & -18.717 & 1.239 \\ 
BCES-Orthogonal & 1.606 & 0.086 & -19.375 & 1.283 \\ 
bootstrap & 1.609 & 0.088 & -19.419 & 1.308 \\ 
 \enddata
\tablecomments{Best fit parameter values and standard deviations for
the extended sample (106
clusters) for a fit of the form (\ref{eq:lmform}).
The rows denoted `bootstrap' give the results obtained for 10\,000
bootstrap resamplings.}
\end{deluxetable}
\begin{deluxetable}{lcccc}
\tabletypesize{\footnotesize}
\tablecaption{Fit parameter values\label{tab:lm2}}
\tablewidth{0pt}
\tablehead{
\colhead{Fit}    &  {$\alpha$} &   \colhead{$\Delta\alpha$}   &
\colhead{$A$} & \colhead{$\Delta A$}
}
\startdata
BCES($L\mid M$) & 1.310 & 0.103 & -14.935 & 1.526 \\
bootstrap & 1.256 & 0.103 & -14.146 & 1.531 \\
BCES($M\mid L$) & 1.538 & 0.105 & -18.320 & 1.568 \\
bootstrap & 1.584 & 0.113 & -18.995 & 1.681 \\
BCES-Bisector & 1.418 & 0.097 & -16.536 & 1.434 \\
bootstrap & 1.407 & 0.096 & -16.368 & 1.427 \\
BCES-Orthogonal & 1.460 & 0.105 & -17.157 & 1.559 \\
bootstrap & 1.468 & 0.110 & -17.274 & 1.633 \\
\enddata
\tablecomments{Same as Tab.~\ref{tab:lm1} but for the purely
flux-limited sample (63 clusters).}
\end{deluxetable}
\begin{deluxetable}{lcccc}
\tabletypesize{\footnotesize}
\tablecaption{Fit parameter values\label{tab:lm3}}
\tablewidth{0pt}
\tablehead{
\colhead{Fit}    &  {$\alpha$} &   \colhead{$\Delta\alpha$}   &
\colhead{$A$} & \colhead{$\Delta A$}
}
\startdata
BCES($L\mid M$) & 1.756 & 0.091 & -21.304 & 1.350 \\
bootstrap & 1.719 & 0.090 & -20.746 & 1.338 \\
BCES($M\mid L$) & 1.860 & 0.084 & -22.836 & 1.246 \\
bootstrap & 1.881 & 0.084 & -23.144 & 1.250 \\
BCES-Bisector & 1.807 & 0.084 & -22.053 & 1.251 \\
bootstrap & 1.797 & 0.084 & -21.899 & 1.241 \\
BCES-Orthogonal & 1.835 & 0.085 & -22.473 & 1.260 \\
bootstrap & 1.841 & 0.085 & -22.563 & 1.270 \\
\enddata
\tablecomments{Same as Tab.~\ref{tab:lm1} but for $\lbol$.}
\end{deluxetable}
\begin{deluxetable}{lcccc}
\tabletypesize{\footnotesize}
\tablecaption{Fit parameter values\label{tab:lm4}}
\tablewidth{0pt}
\tablehead{
\colhead{Fit}    &  {$\alpha$} &   \colhead{$\Delta\alpha$}   &
\colhead{$A$} & \colhead{$\Delta A$}
}
\startdata
BCES($L\mid M$) & 1.504 & 0.089 & -17.545 & 1.298 \\
bootstrap & 1.469 & 0.089 & -17.042 & 1.300 \\
BCES($M\mid L$) & 1.652 & 0.086 & -19.708 & 1.254 \\
bootstrap & 1.671 & 0.086 & -19.992 & 1.260 \\
BCES-Bisector & 1.575 & 0.084 & -18.590 & 1.228 \\
bootstrap & 1.565 & 0.083 & -18.445 & 1.224 \\
BCES-Orthogonal & 1.609 & 0.087 & -19.075 & 1.274 \\
bootstrap & 1.611 & 0.088 & -19.109 & 1.290 \\
\enddata
\tablecomments{Same as Tab.~\ref{tab:lm1} but for  $\mtf$.}
\end{deluxetable}
\begin{deluxetable}{lcccc}
\tabletypesize{\footnotesize}
\tablecaption{Fit parameter values\label{tab:lm5}}
\tablewidth{0pt}
\tablehead{
\colhead{Fit}    &  {$\alpha$} &   \colhead{$\Delta\alpha$}   &
\colhead{$A$} & \colhead{$\Delta A$}
}
\startdata
BCES($L\mid M$) & 2.001 & 0.124 & -25.484 & 1.858 \\
bootstrap & 1.933 & 0.126 & -24.474 & 1.897 \\
BCES($M\mid L$) & 2.488 & 0.127 & -32.761 & 1.902 \\
bootstrap & 2.519 & 0.130 & -33.223 & 1.940 \\
BCES-Bisector & 2.223 & 0.118 & -28.793 & 1.773 \\
bootstrap & 2.193 & 0.119 & -28.357 & 1.785 \\
BCES-Orthogonal & 2.404 & 0.129 & -31.509 & 1.938 \\
bootstrap & 2.415 & 0.133 & -31.667 & 1.984 \\
\enddata
\tablecomments{Same as Tab.~\ref{tab:lm1} but for  $\mab$ for
comparison (mass errors have not been taken into account here).}
\end{deluxetable}
\begin{deluxetable}{lccc}
\tabletypesize{\footnotesize}
\tablecaption{Measured scatter\label{tab:scat}}
\tablewidth{0pt}
\tablehead{
\colhead{Scatter}    & \colhead{($L\mid M$)}  &   \colhead{($M\mid L$)}   &
\colhead{Bisector}
}
\startdata
\sidehead{\hspace{0.2cm}\vspace{1mm} 106 clusters included in the extended sample.} 
\tableline \\[-2.6mm]
$\slm$ & 0.21 & 0.21 & 0.21 \\
$\sll$ & 0.31 & 0.34 & 0.32 \\
$\sigma_{\log L/M}$ & 0.17 & 0.18 & 0.17 \\[0.9mm]
\tableline
\sidehead{\hspace{0.9cm}\vspace{1.0mm} 63 clusters included in \gcss .} 
\tableline \\[-2.6mm]
$\slm$ & 0.22 & 0.21 & 0.21 \\
$\sll$ & 0.29 & 0.32 & 0.30 \\
$\sigma_{\log L/M}$ & 0.18 & 0.18 & 0.17 \\
\enddata
\tablecomments{Scatter measured for different relations.}
\end{deluxetable}

The best fit relation for $\lbol$ (Tab.~\ref{tab:lm3}) will be discussed
in Sect.~\ref{relat_d}. One notes that this relation is steeper than the one
for $\lx(\eb)$. This is
caused by the fact that the bolometric emission coefficient depends stronger
on the gas temperature. As expected the relation for $\mtf$
(Tab.~\ref{tab:lm4}) has a
very similar slope and a higher normalization than the one for $\mtz$. It is
given here for a possible comparison with subsequent works as well as the
relation with the formally determined masses within an Abell radius
(Tab.~\ref{tab:lm5}), which is steeper because low luminosity clusters
are assigned much higher masses.

When constructing the mass function, the overall (measurement plus
intrinsic) scatter in the 
$\lx$--$\mt$ relation may become important (Sect.~\ref{funct}). After verifying that
the scatter is approximately Gaussian in log space the scatter has been
measured as given in Tab.~\ref{tab:scat}.
The scatter in $\lx(\eb)$, $\mtz$, and orthogonal to the best fit line is
given by $\sll$, $\slm$, and $\sigma_{\log L/M}$, respectively. The
measured scatter
does not vary strongly depending on the chosen fitting method. The slight
differences are expected qualitatively. For instance $\sll$ is smaller
for ($L\mid M$) compared to ($M\mid L$) because the best fit is determined
by minimizing the residuals in $\log\lx$ for the former.

\section{Intracluster Gas Fraction}\label{fgas}

An important parameter for the understanding of cluster physics is
the amount of intracluster gas relative to the total mass,
since for instance heating and cooling processes are likely to affect
this gas fraction. The mean gas mass fraction within $r_{200}$ for the
extended sample of 106 clusters has been found as
$\langle\fg\rangle=0.19\pm 0.08\,h_{50}^{-3/2}$.
This parameter is also important for cosmology as outlined in
Sect.~\ref{func_pred}.
In the next two Sections it is tested whether there are systematic
variations of the gas fraction.

\subsection{Variation of Gas Fraction with Mass}\label{fgas_m}

In the following relations between various quantities for the 106
galaxy clusters will be shown.
Since for 18 clusters the gas temperature has been estimated
using a luminosity--temperature relation, these clusters are
omitted here and in all other Sections when the temperature is
one of the quantities of interest.
For a better illustration of possible deviations from pure power
laws in some plots non parametric regression fits using a smoothed spline
function \citep[e.g.,][Sect.~5.6]{si96} are shown. These fits have been
calculated with the standard statistics software package S-Plus using
5 degrees of freedom and assigning each point the same weight.
\begin{figure}[thp]
\centering\begin{tabular}{l}
\psfig{file=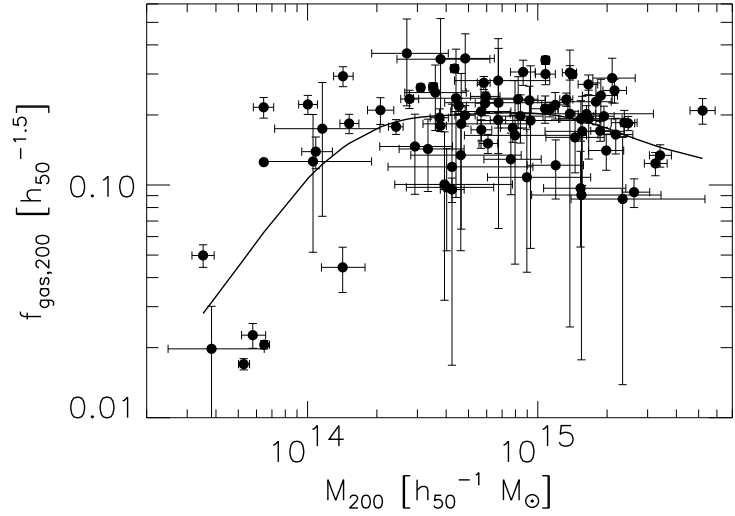,width=11cm,angle=0,clip=}
\end{tabular}
\caption{Gas mass fraction as a function of gravitational
mass. The solid line denotes the result of a smoothed
spline fit, indicating a break around
$2\times 10^{14}\,\msu$.}\label{fg_mt}
\end{figure}
\begin{figure}[thp]
\centering\begin{tabular}{l}
\psfig{file=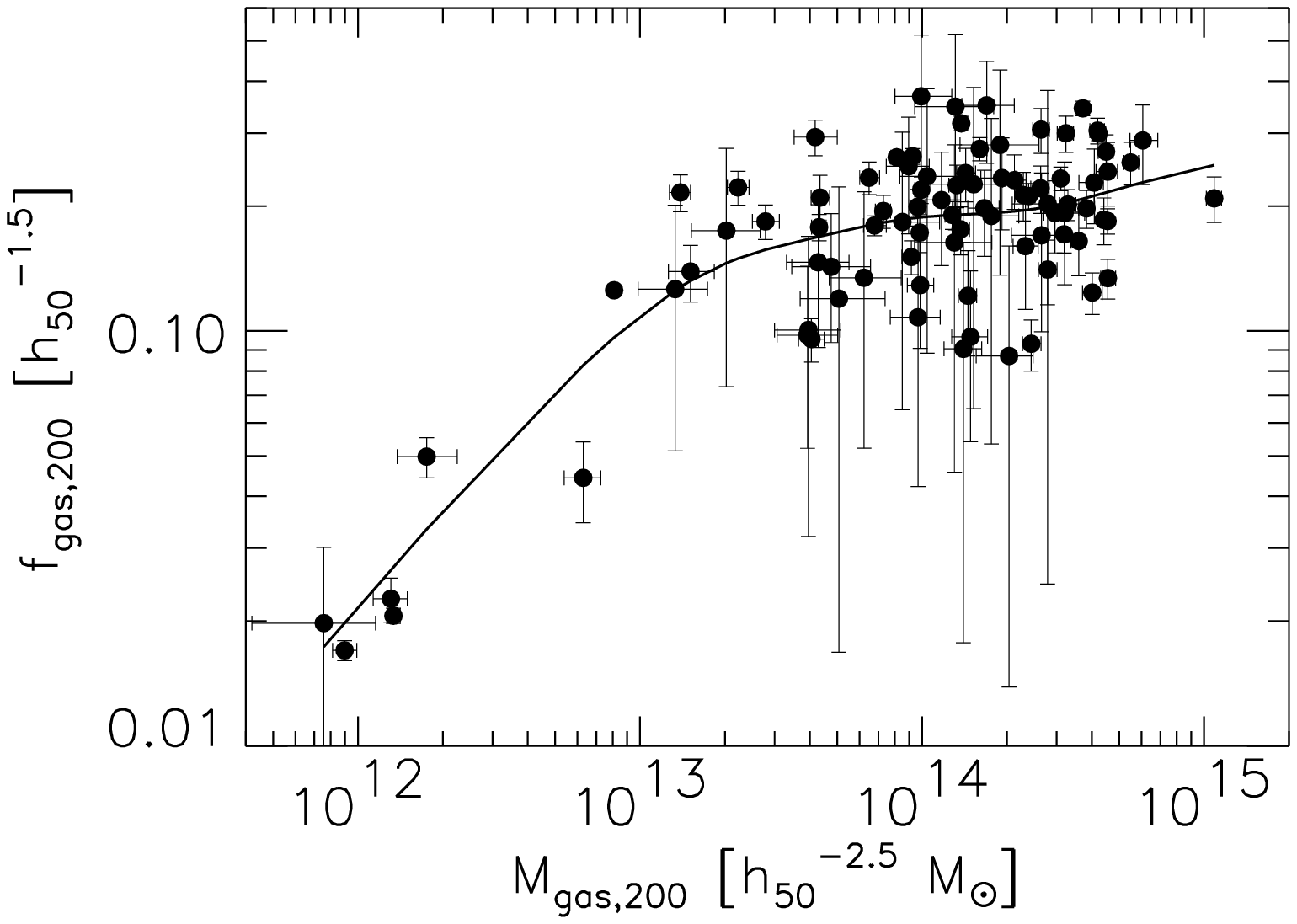,width=11cm,angle=0,clip=}
\end{tabular}
\caption{Gas mass fraction as a function of gas
mass. The solid line denotes the result of a smoothed
spline fit, indicating a break around
$2\times 10^{13}\,\msu$.}\label{fg_mg}
\end{figure}
\begin{figure}[thp]
\centering\begin{tabular}{l}
\psfig{file=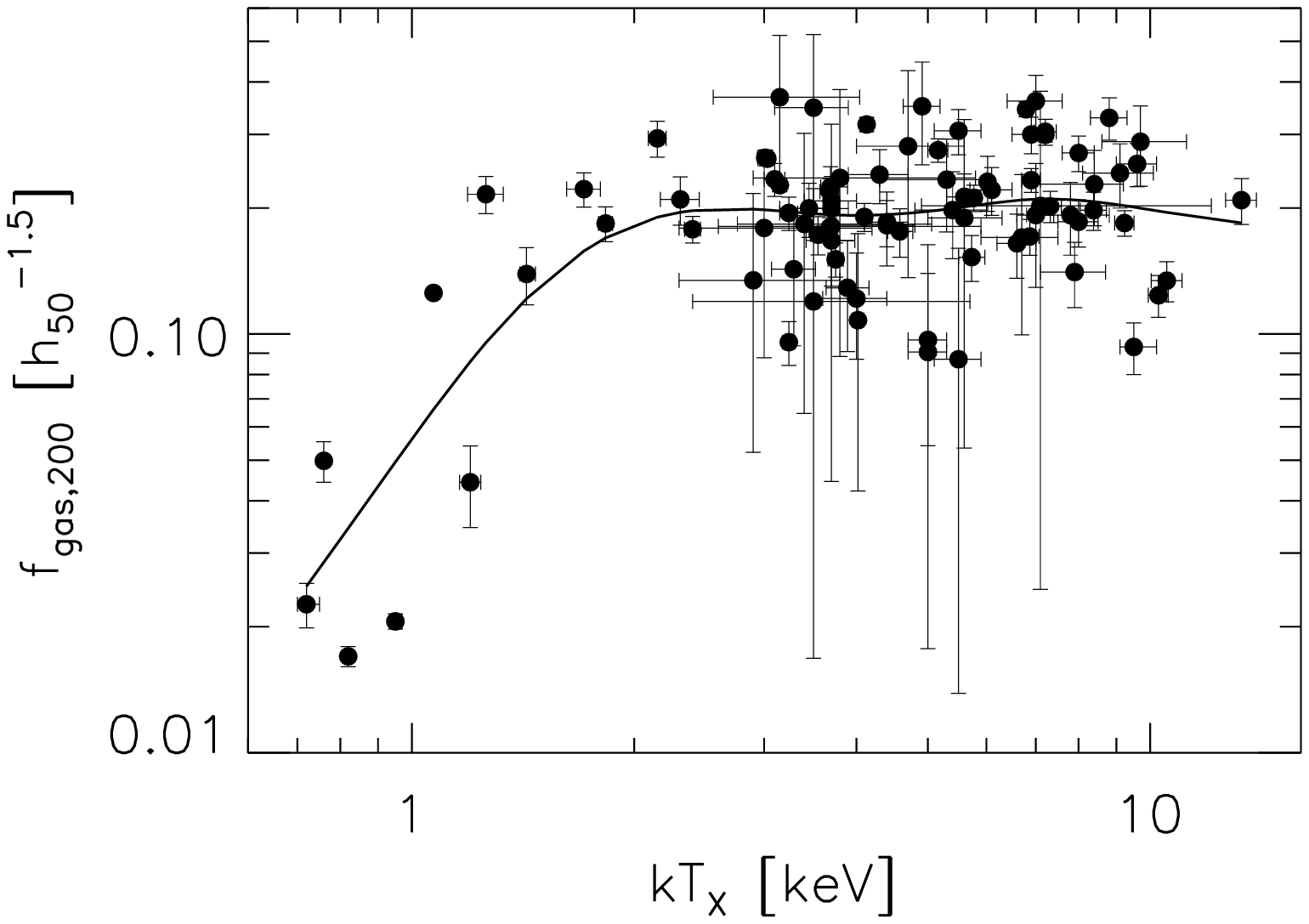,width=11cm,angle=0,clip=}
\end{tabular}
\caption{Gas mass fraction as a function of gas temperature.
The solid line denotes the result of a smoothed
spline fit, indicating a break around
2\,keV.}\label{fg_tem}
\end{figure}

In Fig.~\ref{fg_mt} the gas mass fraction is shown as a function of the
gravitational mass. Note that
the main uncertainty of $\mgz$ comes from the uncertainty of $r_{200}$. Therefore
the errors of $\mgz$ and $\mtz$ are strongly correlated and combining
these errors via standard error propagation is not a useful
option for assessing the uncertainty of $\fg$. Since for a given
radius the error of the gas mass
estimation is much smaller it is neglected here. Rather the error of
$\fg$ is determined from the error of $\mtz$ (including the
uncertainty of $r_{200}$).

It is clear from Fig.~\ref{fg_mt} that $\fg\,h_{50}^{1.5}\approx$ 10--30\,\%
for most of
the clusters. No clear trend with the gravitational mass is apparent,
as long as $\mt\gtrsim 2\times 10^{14}\,\msu$. For smaller masses, however,
a drop of the gas fraction is indicated, even though caused by only 5--6
clusters. A similar behavior is noted if $\fg$ is plotted as a function
of the gas mass (Fig.~\ref{fg_mg}) and of the gas temperature
(Fig.~\ref{fg_tem}). Possible reasons for this sudden drop of $\fg$ towards
smaller systems are discussed in Sect.~\ref{fg_d}. 

\subsection{Variation of Gas Fraction with Radius}\label{fgas_r}

To characterize the extent of the gas relative to the dark matter
the ratio $\fgz/\fgf$ is introduced. This ratio compares the gas fraction
at two characteristic radii, i.e.\ at two radii of constant overdensity.
Obviously, since the overdensity decreases with increasing radius,
$\fgz/\fgf>1$ indicates that the gas is more extended than the dark
matter.

\begin{figure}[thp]
\centering\begin{tabular}{l}
\psfig{file=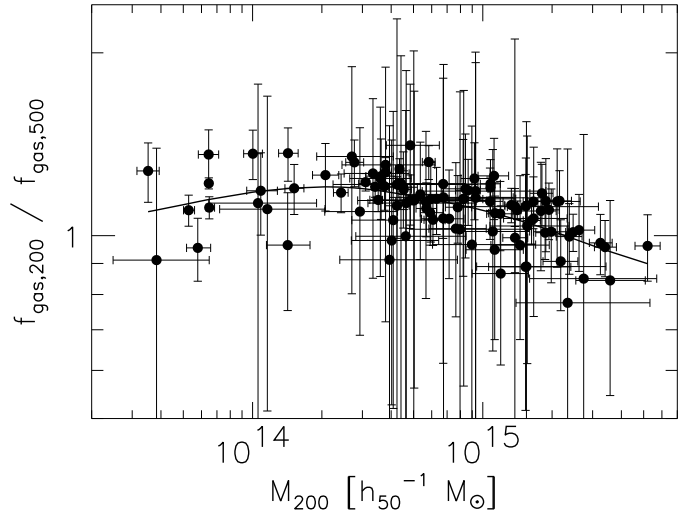,width=11cm,angle=0,clip=}
\end{tabular}
\caption{Gas fraction within $r_{200}$ divided by the gas fraction
within $r_{500}$ as a function of $\mtz$. For most of the clusters the gas
fraction increases with radius but for some of the high mass clusters there
are indications for a decrease. The result of a smoothed spline fit is shown as
solid line.}\label{fg2_fg5_m}
\end{figure}
In Fig.~\ref{fg2_fg5_m} this ratio is plotted as a function of gravitational
mass. Note that
the errors of $\fgz$ and $\fgf$ are strongly correlated.
Only the error in $\fgf$ is therefore taken into account for the
calculation of the uncertainty of $\fgz / \fgf$. The statistical errors
still seem larger than the actual scatter of the points. This may be
caused by the fact that the uncertainty in $r_{500}$ is not neglected.

One notes that for most of the clusters the gas fraction is larger at
larger radii. For some clusters, however, the dark matter seems to be more
extended than the gas. Note that taking into account the error bars no
cluster has $\fgz / \fgf$ \emph{significantly} lower than 1.
Interestingly, however, there appears to be a systematic trend
for the larger systems that the gas extent becomes smaller with increasing
mass relative to the dark matter extent.
This apparent trend will be discussed further in Sect.~\ref{fg_d}.

\section{Relations between Shape Parameters, Temperature,
Luminosity, and Mass}\label{tlm}

In this Section various relations between the parameters
describing the surface brightness (and therefore gas density) profile,
$\beta$ and $\rc$, the gas temperature, X-ray luminosity, gas and gravitating
mass will be presented. The implications of the trends seen here will be
discussed in Sect.~\ref{tlm_d}.

\begin{figure}[thp]
\centering\begin{tabular}{l}
\psfig{file=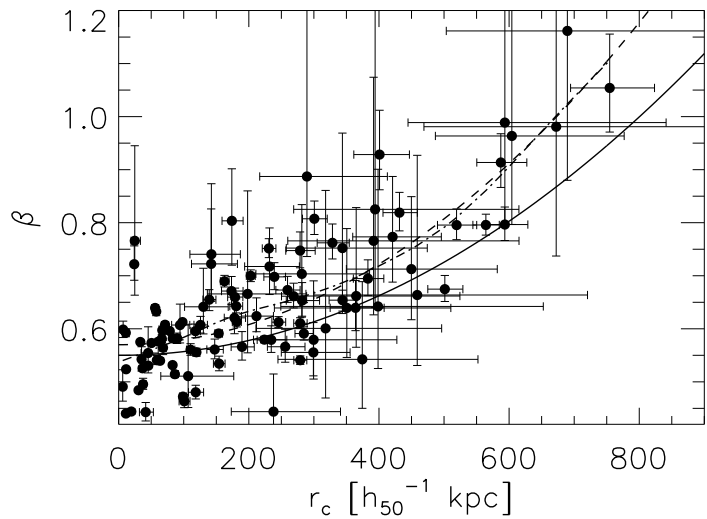,width=11cm,angle=0,clip=}
\end{tabular}
\caption{$\beta$ versus core radius. Solid line: Best fit parabolic model ($\beta_0=0.55$, $r_{\rm
s}=885$\,kpc) of \citet{na99}, dashed line: same model with
$\beta_0=0.57$, $r_{\rm
s}=775$\,kpc. Dot-dashed line: smoothed spline fit.}\label{rc-beta}
\end{figure}
A correlation between the two fit parameters $\beta$ and $\rc$ as induced by
the fitting process is indicated
by the confidence contours (Fig.~\ref{a3532rcb}). However, the typical
uncertainty is much smaller than the range of $\beta$ and $\rc$ values
spanned by the sample. The origin of the correlation present
in Fig.~\ref{rc-beta} may therefore be physical. Using a sample
of 26 galaxy clusters \citet{na99} showed that a physical correlation
exists and 
determined best fit parameters of the function
\begin{equation}
\beta=\beta_0\left[1+(\rc/r_{\rm s})^2\right]
\label{eq:rcb}
\end{equation}
as given in the caption of Fig.~\ref{rc-beta}.
The smoothed spline fit shown in Fig.~\ref{rc-beta} indicates that this
parameterization accounts well for the trend seen. However, for the 
106 clusters used here a different set of parameter values appears to
provide a better description of the data.

\begin{figure}[thp]
\centering\begin{tabular}{l}
\psfig{file=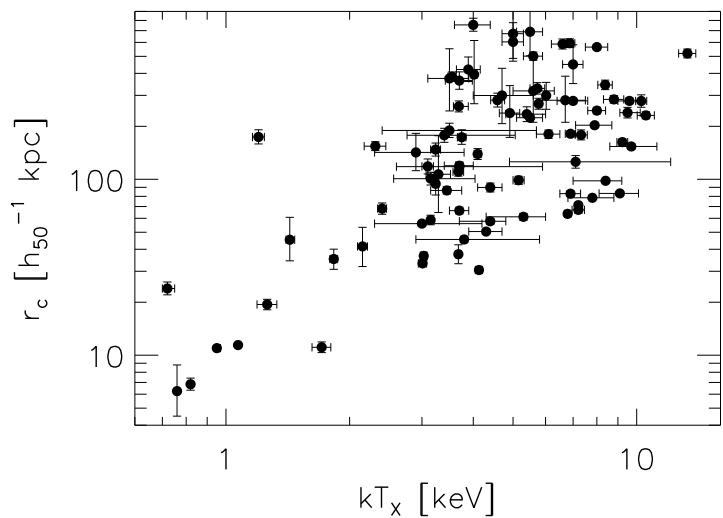,width=11cm,angle=0,clip=}
\end{tabular}
\caption{Core radius versus temperature.
The two parameters are clearly correlated.}\label{rc-tem}
\end{figure}
\begin{figure}[thp]
\centering\begin{tabular}{l}
\psfig{file=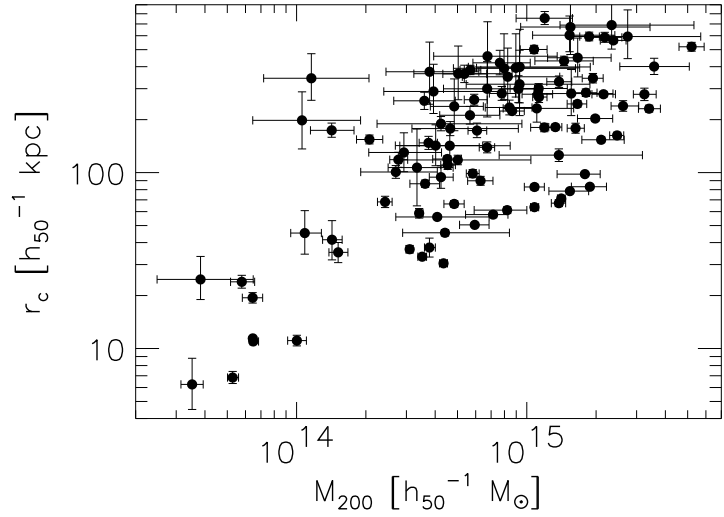,width=11cm,angle=0,clip=}
\end{tabular}
\caption{Core radius versus gravitational mass. The positive
correlation is obvious.}\label{rc-mt}
\end{figure}
Figs.~\ref{rc-tem} and \ref{rc-mt} clearly show that the core radius
is correlated with cluster temperature and mass. The temperature and core
radius are determined completely independent from one another. Also the
gravitational mass estimate is effectively independent of $\rc$
at large radii ($r \gg \rc$; Eq.~\ref{back:ma2}).
A significant portion of the scatter
present in these Figures is probably due to the presence (smaller $\rc$)
or absence (larger $\rc$) of a central excess emission. First results
of a cooling flow study of the \gcs\ clusters indicate a respective trend
(Y. Chen et al., in preparation).

\begin{figure}[thp]
\centering\begin{tabular}{l}
\psfig{file=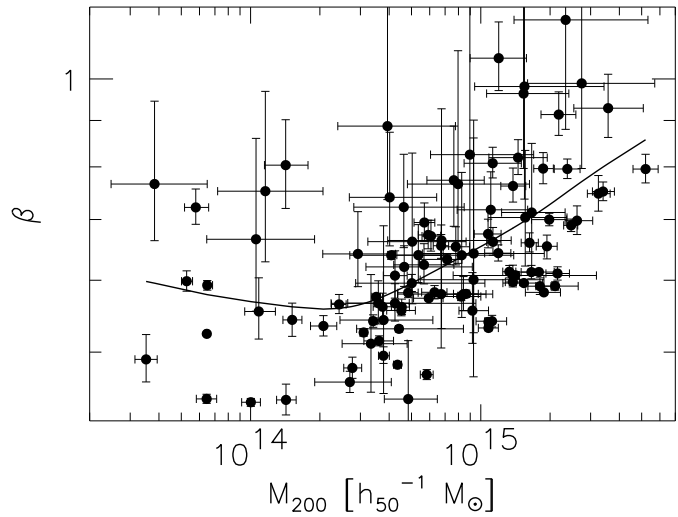,width=11cm,angle=0,clip=}
\end{tabular}
\caption{The fit parameter $\beta$ is plotted versus the gravitational
mass. A smoothed
spline fit is shown as solid line, indicating a possible trend at the 
high mass side.}\label{mb}
\end{figure}
\begin{figure}[thp]
\centering\begin{tabular}{l}
\psfig{file=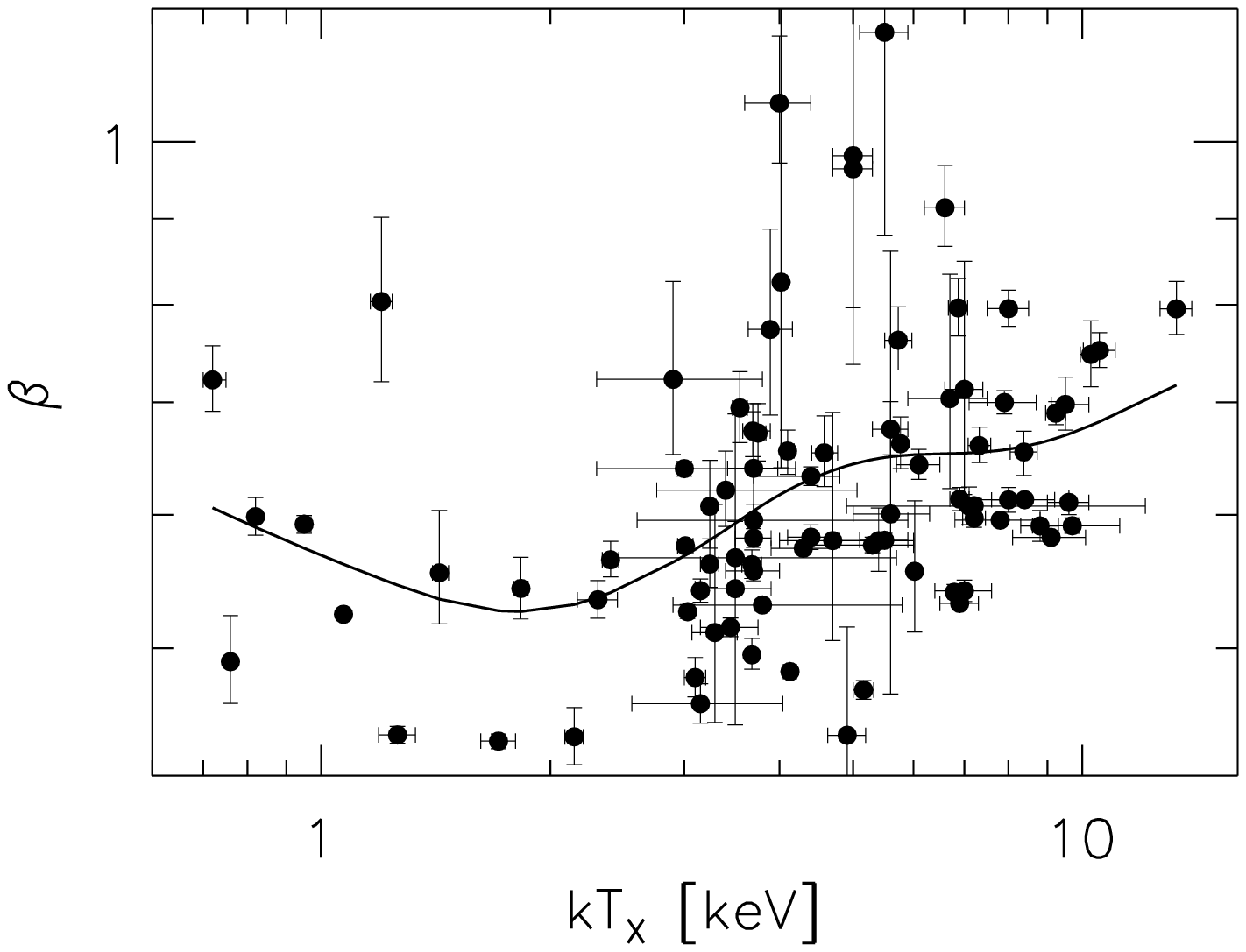,width=11cm,angle=0,clip=}
\end{tabular}
\caption{$\beta$ versus the intracluster gas temperature. A smoothed
spline fit is shown as solid line. No clear trends are indicated, except
a very weak indication for the hottest clusters.}\label{tb}
\end{figure}
In Fig.~\ref{mb} $\beta$ is plotted as a function of $\mtz$.
Also shown is the result of a smoothed spline fit.
For $\mtz\lesssim 3\times 10^{14}\,\msu$ there is no indication of any trend of
$\beta$ with mass. For larger masses $\beta$ appears to increase with
increasing mass.
Note that
$\beta$ is used for the calculation of the gravitational mass
(Eq.~\ref{back:ma2}). The X-ray temperature determination does
not depend on $\beta$ (except possibly through the modeling of the
surface brightness profile often performed to calculate the
effective area for \as\ observations). Shown in Fig.~\ref{tb} is
$\beta$ versus $\tx$, indicating that the two parameters
are almost independent of one another. For the hottest clusters there is,
however, a weak suggestion of a trend in the same sense as in the previous
plot.

\begin{figure}[thp]
\centering\begin{tabular}{l}
\psfig{file=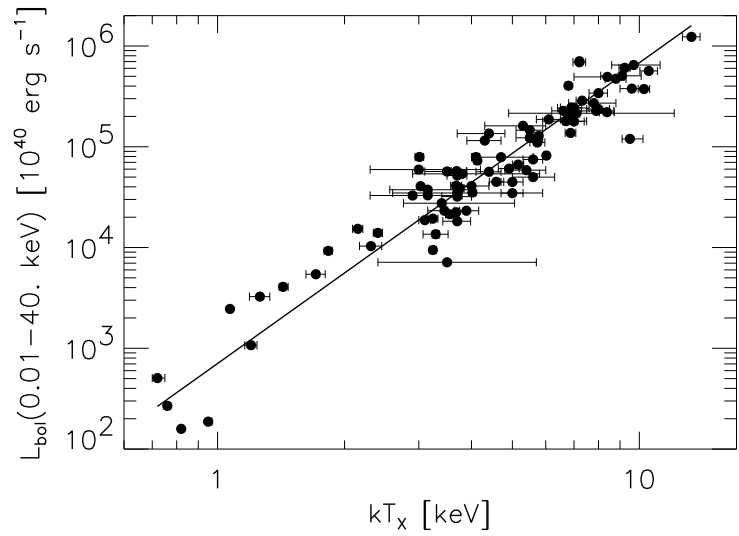,width=11cm,angle=0,clip=}
\end{tabular}
\caption{Bolometric X-ray luminosity versus temperature.
The solid line represents the best
fit result using the bisector method.}\label{lbol-tem}
\end{figure}

The plot $\lx(\eb)$
versus $\tx$ has already been shown in Sect.~\ref{a1835_res}. The
corresponding best fit relations, determined using the linear regression method
described in Sect.~\ref{relat},
are given in Tab.~\ref{tab:lt}.
In Fig.~\ref{lbol-tem} the bolometric X-ray luminosity is shown as a function
of the gas temperature.
The relation seems to be well represented by a power law of the form
$\lbol\propto\tx^3$
(Tab.~\ref{tab:lbolt}) over the entire temperature range.

\begin{figure}[thp]
\centering\begin{tabular}{l}
\psfig{file=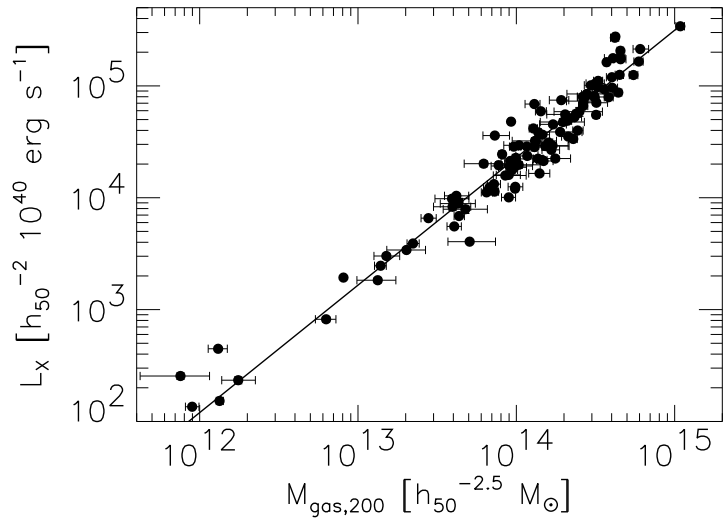,width=11cm,angle=0,clip=}
\end{tabular}
\caption{X-ray luminosity versus gas mass in the energy band ($\eb$).
The solid line represents the best
fit result using the bisector method.}\label{mg-lx}
\end{figure}

\begin{figure}[thp]
\centering\begin{tabular}{l}
\psfig{file=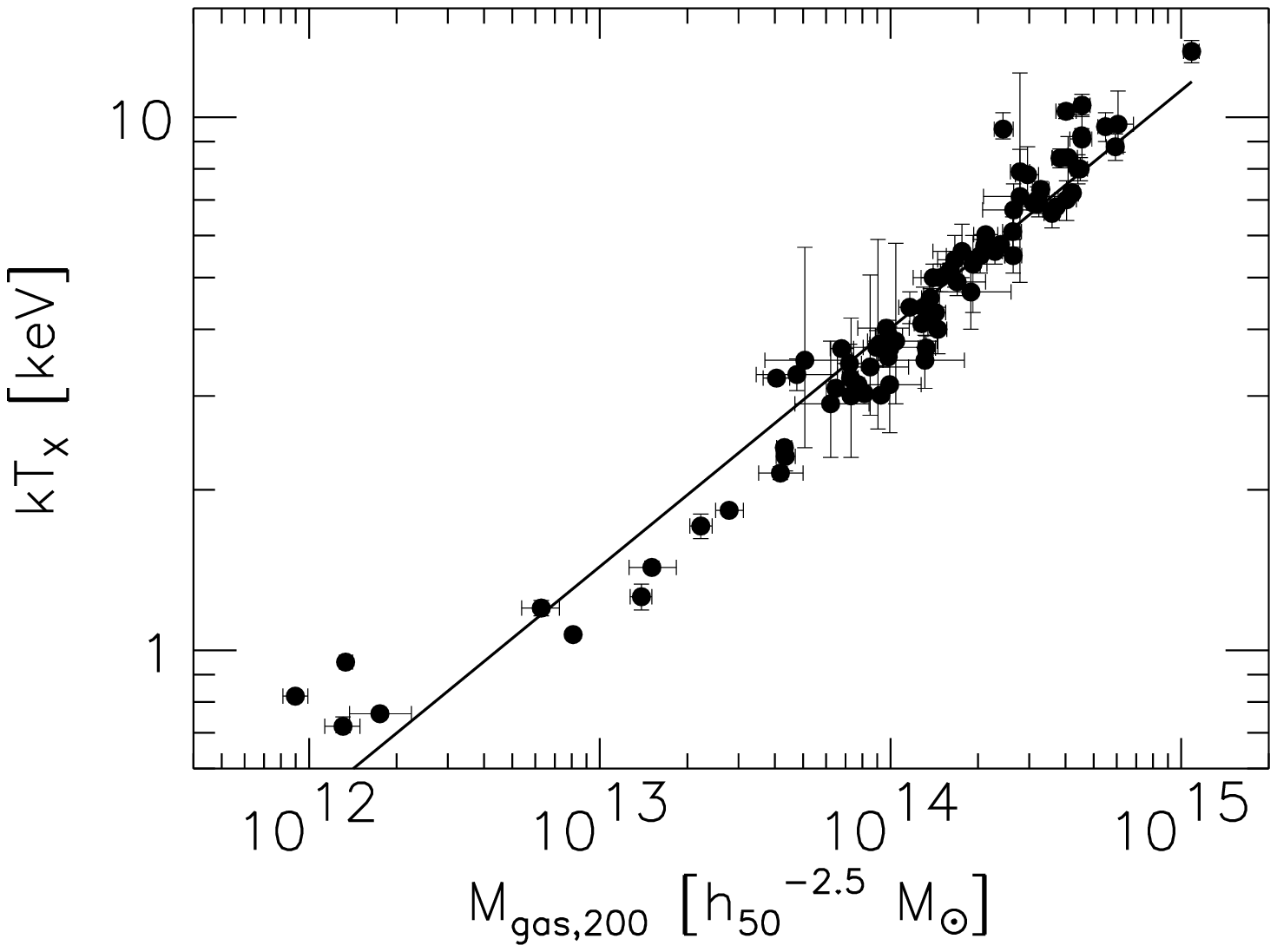,width=11cm,angle=0,clip=}
\end{tabular}
\caption{Gas temperature versus gas mass. The solid line represents the best
fit result using the bisector method.}\label{mg-tem}
\end{figure}

Fig.~\ref{mg-lx} shows that the X-ray luminosity and the gas mass are
tightly correlated. The best fit relations are given in Tab.~\ref{tab:lmg}.
Knowing already that there exist correlations between luminosity and
temperature and between luminosity and gas mass it is not a big surprise that
the gas mass correlates well with the temperature. This relation is
graphically shown in Fig.~\ref{mg-tem} and tabulated in Tab.~\ref{tab:tmg}.
Note that in principal $\mg$ and $\tx$ are determined almost independent
from one another.
However, since $r_{200}$ depends strongly on $\tx$ clearly also the
calculation of $\mgz$ depends on $\tx$.

\begin{figure}[thp]
\centering\begin{tabular}{l}
\psfig{file=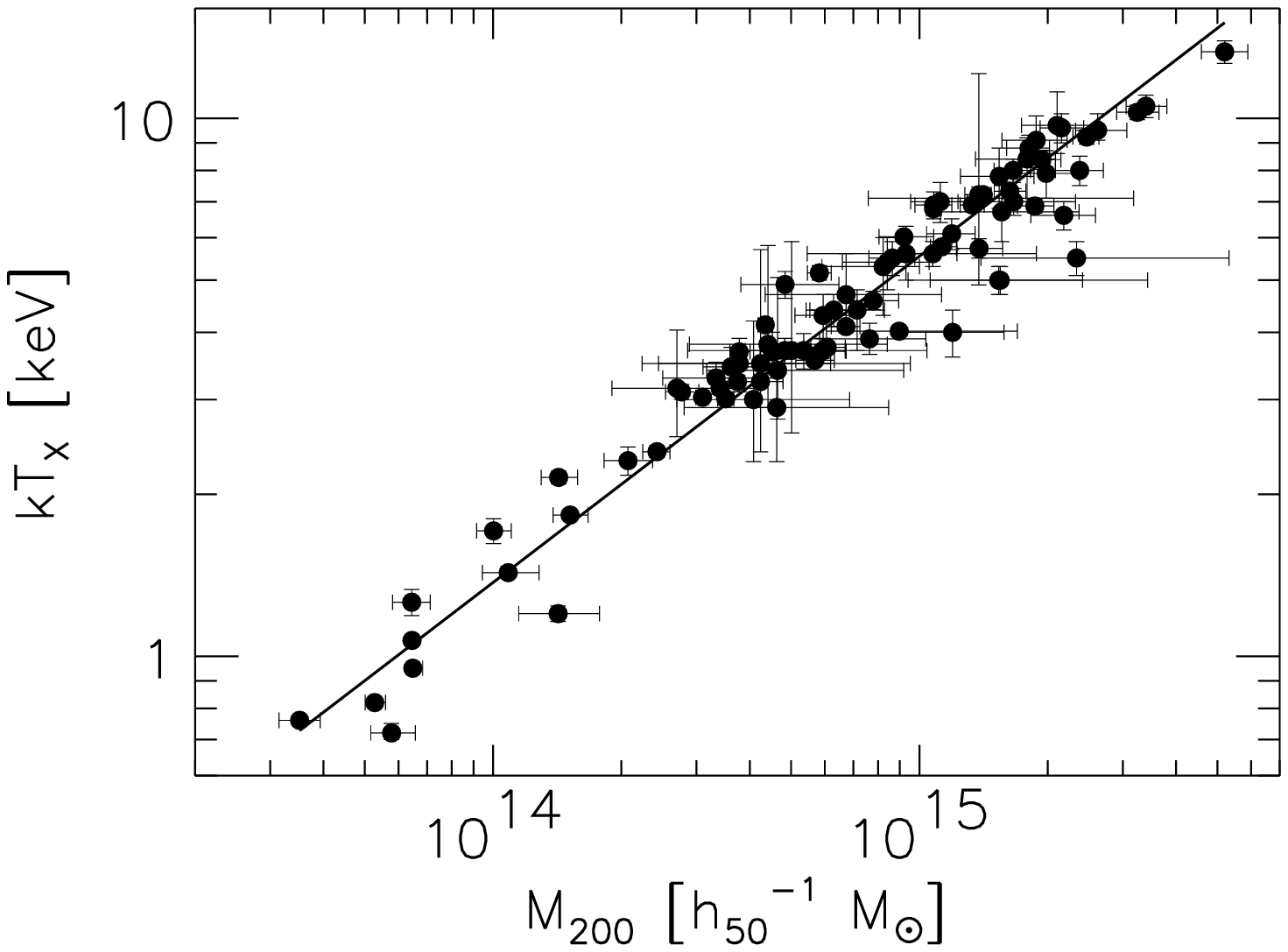,width=11cm,angle=0,clip=}
\end{tabular}
\caption{Gas temperature versus gravitational mass. The solid line represents
the best
fit result using the bisector method.}\label{mt-tem}
\end{figure}

Another relation of specific interest is the $\mt$--$\tx$ relation.
Its relevance derives not only from its importance in the study of
physical processes but also from its role in the context of the interpretation
of temperature functions (analogous to mass functions, see Sect.~\ref{funct}).
If one wants to use temperature functions to determine cosmological
parameters, in some way the connection between temperature and mass has to
be made. This may be done using theoretical predictions, simulations, or
observations. It is therefore important to test the consistency
of the different $\mt$--$\tx$ relations.
In Fig.~\ref{mt-tem} the data points and the
corresponding best fit relation (Tab.~\ref{tab:tmt}) are shown.
\begin{deluxetable}{lcccc}
\tabletypesize{\footnotesize}
\tablecaption{Fit parameter values\label{tab:lt}}
\tablewidth{0pt}
\tablehead{
\colhead{Fit}    &  {$\alpha$} &   \colhead{$\Delta\alpha$}   &
\colhead{$A$} & \colhead{$\Delta A$}
}
\startdata
BCES($L\mid T$) & 2.481 & 0.130 & 2.862 & 0.093 \\ 
bootstrap & 2.367 & 0.132 & 2.934 & 0.095 \\ 
BCES($T\mid L$) & 2.724 & 0.128 & 2.708 & 0.091 \\ 
bootstrap & 2.766 & 0.131 & 2.682 & 0.093 \\ 
BCES-Bisector & 2.598 & 0.125 & 2.788 & 0.089 \\ 
bootstrap & 2.552 & 0.124 & 2.817 & 0.089 \\ 
BCES-Orthogonal & 2.692 & 0.129 & 2.728 & 0.092 \\ 
bootstrap & 2.712 & 0.132 & 2.716 & 0.094 \\ 
\enddata
\tablecomments{Best fit parameter values and standard deviations for
the 88 clusters with measured temperatures for a fit of the form \\
$\log \left(\frac{\lx (\eb )}{\esl
}\right)=A+\alpha\log\left(\frac{\tx}{{\rm keV}}\right)$.\\
The rows denoted `bootstrap' give the results obtained for 10\,000
bootstrap resamplings.}
\end{deluxetable}
\begin{deluxetable}{lcccc}
\tabletypesize{\footnotesize}
\tablecaption{Fit parameter values\label{tab:lbolt}}
\tablewidth{0pt}
\tablehead{
\colhead{Fit}    &  {$\alpha$} &   \colhead{$\Delta\alpha$}   &
\colhead{$A$} & \colhead{$\Delta A$}
}
\startdata
BCES($L\mid T$) & 2.906 & 0.126 & 2.898 & 0.090 \\ 
bootstrap & 2.776 & 0.126 & 2.981 & 0.091 \\ 
BCES($T\mid L$) & 3.067 & 0.122 & 2.797 & 0.088 \\ 
bootstrap & 3.109 & 0.125 & 2.770 & 0.090 \\ 
BCES-Bisector & 2.984 & 0.121 & 2.849 & 0.087 \\ 
bootstrap & 2.933 & 0.120 & 2.881 & 0.087 \\ 
BCES-Orthogonal & 3.050 & 0.123 & 2.808 & 0.088 \\ 
bootstrap & 3.074 & 0.125 & 2.792 & 0.090 \\ 
\enddata
\tablecomments{Best fit parameter values and standard deviations for
the 88 clusters with measured temperatures for a fit of the form \\
$\log \left(\frac{\lbol (\ebol )}{\esl
}\right)=A+\alpha\log\left(\frac{\tx}{{\rm keV}}\right)$.}
\end{deluxetable}
\begin{deluxetable}{lcccc}
\tabletypesize{\footnotesize}
\tablecaption{Fit parameter values\label{tab:lmg}}
\tablewidth{0pt}
\tablehead{
\colhead{Fit}    &  {$\alpha$} &   \colhead{$\Delta\alpha$}   &
\colhead{$A$} & \colhead{$\Delta A$}
}
\startdata
BCES($L\mid M$) & 1.118 & 0.034 & -11.291 & 0.482 \\ 
bootstrap & 1.104 & 0.033 & -11.096 & 0.460 \\ 
BCES($M\mid L$) & 1.162 & 0.038 & -11.907 & 0.533 \\ 
bootstrap & 1.184 & 0.043 & -12.221 & 0.606 \\ 
BCES-Bisector & 1.140 & 0.035 & -11.596 & 0.500 \\ 
bootstrap & 1.143 & 0.036 & -11.647 & 0.515 \\ 
BCES-Orthogonal & 1.143 & 0.036 & -11.637 & 0.513 \\ 
bootstrap & 1.149 & 0.039 & -11.727 & 0.546 \\ 
\enddata
\tablecomments{Best fit parameter values and standard deviations for
the 106 clusters of the extended sample for a fit of the form \\
$\log \left(\frac{\lx (\eb )}{\esl
}\right)=A+\alpha\log\left(\frac{\mgz}{\msug}\right)$.}
\end{deluxetable}
\begin{deluxetable}{lcccc}
\tabletypesize{\footnotesize}
\tablecaption{Fit parameter values\label{tab:tmg}}
\tablewidth{0pt}
\tablehead{
\colhead{Fit}    &  {$\alpha$} &   \colhead{$\Delta\alpha$}   &
\colhead{$A$} & \colhead{$\Delta A$}
}
\startdata
BCES($M\mid T$) & 2.230 & 0.106 & 12.654 & 0.075 \\ 
bootstrap & 2.126 & 0.113 & 12.721 & 0.081 \\ 
BCES($T\mid M$) & 2.246 & 0.115 & 12.644 & 0.081 \\ 
bootstrap & 2.276 & 0.115 & 12.626 & 0.081 \\ 
BCES-Bisector & 2.238 & 0.108 & 12.649 & 0.077 \\ 
bootstrap & 2.199 & 0.112 & 12.675 & 0.080 \\ 
BCES-Orthogonal & 2.243 & 0.112 & 12.645 & 0.080 \\ 
bootstrap & 2.250 & 0.116 & 12.643 & 0.082 \\ 
\enddata
\tablecomments{Best fit parameter values and standard deviations for
the 88 clusters with measured temperatures for a fit of the form \\
$\log\left(\frac{\mgz}{\msug}\right)=A+\alpha\log
\left(\frac{\tx }{\rm keV}\right)$.}
\end{deluxetable}
\begin{deluxetable}{lcccc}
\tabletypesize{\footnotesize}
\tablecaption{Fit parameter values\label{tab:tmt}}
\tablewidth{0pt}
\tablehead{
\colhead{Fit}    &  {$\alpha$} &   \colhead{$\Delta\alpha$}   &
\colhead{$A$} & \colhead{$\Delta A$}
}
\startdata
BCES($M\mid T$) & 1.710 & 0.058 & 13.735 & 0.041 \\ 
bootstrap & 1.641 & 0.052 & 13.778 & 0.037 \\ 
BCES($T\mid M$) & 1.591 & 0.052 & 13.811 & 0.036 \\ 
bootstrap & 1.633 & 0.055 & 13.784 & 0.038 \\ 
BCES-Bisector & 1.649 & 0.046 & 13.774 & 0.033 \\ 
bootstrap & 1.637 & 0.046 & 13.781 & 0.033 \\ 
BCES-Orthogonal & 1.622 & 0.046 & 13.791 & 0.032 \\ 
bootstrap & 1.635 & 0.048 & 13.782 & 0.034 \\ 
\enddata
\tablecomments{Best fit parameter values and standard deviations for
the 88 clusters with measured temperatures for a fit of the form \\
$\log\left(\frac{\mtz}{\msu}\right)=
A+\alpha\log \left(\frac{\tx }{{\rm keV}}\right)$.}
\end{deluxetable}
\section{Mass Function}\label{funct}

In Sect.~\ref{relat} it has been shown that the X-ray luminosity is
closely correlated with cluster mass. Therefore the $\vmax$ estimator
(Sect.~\ref{back:omf}) can
also be applied to estimate the mass function; $\vmax$ then being a
function of mass.
Three different methods have been employed here to correct for the scatter
present in the $\lx$--$\mt$ relation. If $\vmax (\lx)$ is used instead
of $\vmax (\mt)\equiv \vmax (L(\mt))$, where $L(\mt)$ is
the luminosity estimated from the $\lx$--$\mt$ relation using the
determined cluster mass $\mt$, the scatter is automatically taken into
account. This method has been widely used in the construction of X-ray
temperature functions, recently, e.g., by \citet{h00}.
If $\vmax (\mt)$ is used and the utilized $\lx$--$\mt$ relation is
assumed to be the `true' relation then the scatter in this relation
has to be taken into account explicitly.
Therefore following the
method employed for the temperature function by
\citet{m98} and \citet{irb01} the mass function may also be estimated
by determining
$\vmax^\ast (\mt)$, where the measured scatter in $\log\lx$ is included.
Specifically 
\begin{eqnarray}
\vmax^\ast (\mt)\equiv \int^\infty_{-\infty}\vmax(L')\,(2\pi\sll^2)^{-1/2}
\nonumber \\
\times\ \exp\left({-\frac{(\log L'-(A+40)
-\alpha\log\mt)^2}{2\sll^2}}\right)\,d\log L'
\label{eq:vmax}
\end{eqnarray}
has been used, where $A$ and $\alpha$ are the best fit parameter
values taken from the appropriate $\lx$--$\mt$ relation of the form
(\ref{eq:lmform}) and $\sll$ is the
corresponding measured
standard deviation in $\log \lx$ given in Tab.~\ref{tab:scat}.

Now say that for a cluster at redshift $z_1$ to be included in the
flux-limited sample it must have $\lx\ge L_1$. Then
consider the redshift shells
$z_1-dz$ (shell A) and $z_1+dz$ (shell B). The volume enclosed by
shell B is larger than
the volume enclosed by shell A. This means that shell B contains more clusters
with mass $M_1$ which corresponds to $L_1$. However due to measurement and
intrinsic scatter clusters with $M_1$ may be assigned a range of different
luminosities $L_0\le L_1\le L_2$. As long as $L_1-L_0\le L_2-L_1$, which is
satisfied here since the scatter in $L$ is roughly Gaussian in log space,
this implies that more clusters with $M_1$ from shell B will be included in
the sample than clusters with $M_1$ from shell A. This further implies that
the counted clusters with $M_1$ have a mean $L$ slightly larger than $L_1$.
Therefore the measured normalization of an $L$--$M$ relation
constructed with these clusters is expected to be slightly higher
than the measured normalization of an $L$--$M$ relation
constructed with a volume-limited sample.
This also implies that if the former $L$--$M$ relation is used to
calculate $\vmax$ one gets on average a larger $\vmax$. This means
the effect of the scatter in the $L$--$M$ relation
is already taken into account if the
$L$--$M$ relation determined using the flux-limited sample is used for
the calculation of $\vmax(\mt)$.

The drawback of using $\vmax(\lx)$ is that a small number of clusters per
mass bin possibly does not represent the true scatter well. To
minimize this effect here at least ten clusters per mass bin are used. The
drawback of using $\vmax^\ast(\mt)$ or $\vmax(\mt)$, as noted by
\citet{m98}, is the
reliance on the validity of the measured relation over the
entire mass range.
The first method and the method that accounts for the scatter
explicitly (Eq.~\ref{eq:vmax}) have been tested by using Monte Carlo simulations for a
precisely known $\lx$--$\tx$ relation
and scatter  and have been shown to give accurate estimates of
$\phi(T)$ for a large number of clusters
in the study of the \gcs\ temperature function by \citet{irb01}.

\begin{figure}[thp]
\centering\begin{tabular}{l}
\psfig{file=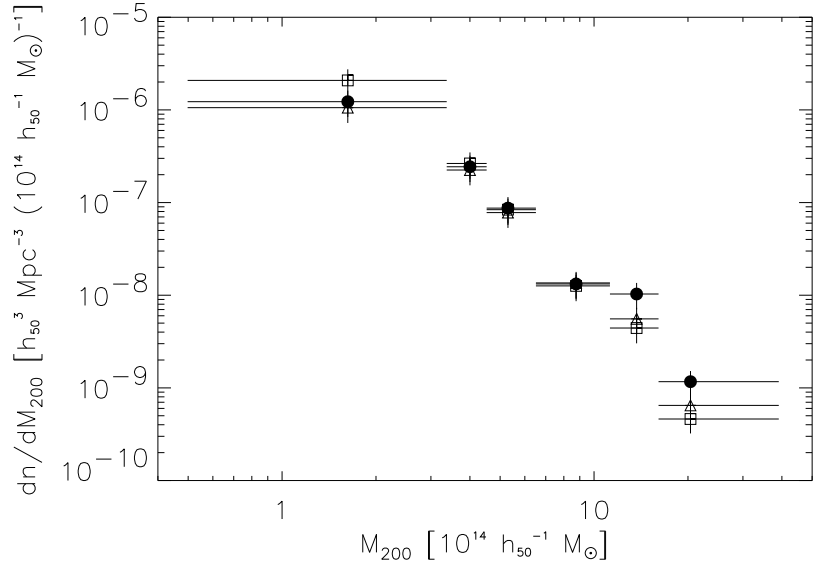,width=11.5cm,angle=0,clip=}
\end{tabular}
\caption{Gravitational mass functions for \gcs .
The mass function plotted with filled circles has been determined
using $\vmax(\lx)$, the ones with open symbols using $\vmax(\mt)$, where 
triangles correspond to the relation ($L\mid M$) and squares to ($M\mid
L$)
both for the flux-limited sample (see text Sect.~\ref{funct}). Vertical error bars
correspond to the formal 1-$\sigma$ Poisson errors, the horizontal
bars indicate the mass intervals covered.}\label{vm_cp_ns}
\end{figure}

In Fig.~\ref{vm_cp_ns} the \gcs\ mass functions determined
by the two approaches using $\vmax(\lx)$ and $\vmax(\mt)$ are shown.
Each bin contains 10
clusters, apart from the highest mass bin, which contains 11
clusters. The highest and lowest mass clusters of the 63 clusters contained
in the sample have been used to
calculate the highest and lowest mass intervals. E.g., the
high mass boundary of the highest mass interval is determined by
adding half the difference between the
largest and second largest mass to the second largest mass.

As expected the method
employing $\vmax(\lx)$ prompts a mass function exhibiting a larger
scatter, because in this case the scatter is accounted for by the
actual scatter of the ten or eleven clusters in each mass bin.
For comparison the two extreme mass functions calculated using $\vmax
(\mt)$ are shown. Extreme is meant in the sense of using the steepest (A)
and shallowest (B) $\lx$--$\mt$ relation for the \gcs\ sample, i.e.\
($M\mid L$) with $\alpha=1.538$ and ($L\mid M$) with $\alpha=1.310$
(Tab.~\ref{tab:lm2}). At the low mass end (A) predicts a lower
luminosity for a given mass than (B) resulting in a smaller $\vmax$ and
therefore a higher $dn/dM$. At the high mass side the effect is
opposite resulting in a lower $dn/dM$ for (A). The differences of these mass
functions to the mass function calculated using $\vmax (\lx)$ can be
understood in a similar way and are caused partly by the indication of a
deviation from a power law shape of the $\lx$--$\mt$ relation.
Using $\vmax^\ast(\mt)$ results in similar mass functions as shown for
the open symbols in Fig.~\ref{vm_cp_ns} but the points lie systematically
lower because the scatter is accounted for twice; first by using an
$\lx$--$\mt$ relation derived from a flux-limited sample and secondly
explicitly by weighting with a Gaussian function.
For the comparison of the
observational mass function to mass functions predicted by certain
cosmological models $\vmax(\lx)$ is used because it is independent of
the precise shape of the $\lx$--$\mt$ relation
and also because $\lx$ has a much smaller measurement uncertainty than
$\mt$. The influence of the choice of the $\vmax$ calculation on the
estimation of cosmological parameters is investigated in
Sect.~\ref{func_pred}. 

Fig.~\ref{vm_cp_ns} shows that the mass function is decreasing fast with
increasing mass, indicating that massive clusters are very rare objects.
This fact shows an advantage of using a flux-limited sample for the
construction of the mass function. The survey volume
is largest for the most luminous -- and therefore most massive -- clusters.
Thus the survey volume is largest for the rarest objects. Construction
of a volume-limited cluster sample that samples the whole mass range
at least equally well would require adding $\gtrsim 10\,000$ clusters
(for instance sampling 10 clusters with $\mtz\approx2\times 10^{15}\,\msu$
implies sampling $\sim 10\,000$ clusters with
$\mtz\approx2\times 10^{13}\,\msu$ in a volume-limited sample).

\chapter{Discussion}
\label{disc}

A precise determination of distribution functions requires a high
sample completeness.
In Sect.~\ref{sample_c} several completeness tests
for \gcs\ are discussed, indicating a high completeness.
In Sect.~\ref{mest_d} the cluster masses determined here are
compared to independent determinations.
The observed
$\lx$--$\mt$ relation is compared to expectations in
Sect.~\ref{relat_d} and possible applications are indicated.
Implications from the measured mean cluster mass to light ratio
are outlined.
Physical properties of the cluster sample are discussed in
Sects.~\ref{fg_d} and \ref{tlm_d}.
The
cluster mass function is compared to previous determinations and to
predictions of cosmological models in
Sect.~\ref{funct_d}.
Cosmological inferences from the mean cluster gas mass fraction are given.
The total gravitational mass contained in galaxy
clusters is compared to the total mass in the universe
in Sect.~\ref{func_dens}.

\section{Sample Completeness}\label{sample_c}

The sample completeness is important for the accuracy of the mass
function. 
The selection criteria detailed in Sect.~\ref{sample} are met by 63
clusters with mean redshift $\langle z \rangle
= 0.05$ and with two  clusters having $z>0.1$. 
The sample is constructed from surveys with much deeper flux limits
and high completenesses. A possible remaining incompleteness in these surveys is
likely to be present at low fluxes close to their flux limits, which
therefore would not effect \gcs .
Nevertheless
four completeness tests have been performed and are described in this
Section; they all
indicate a high completeness of \gcs . The $\log N-\log S$ and $\lx-z$
diagram are compared to 
expectations, the luminosity function is compared to luminosity
functions of deeper surveys, and the $V/\vmax$ test is performed.

\begin{figure}[thbp]
\begin{center}
\psfig{file=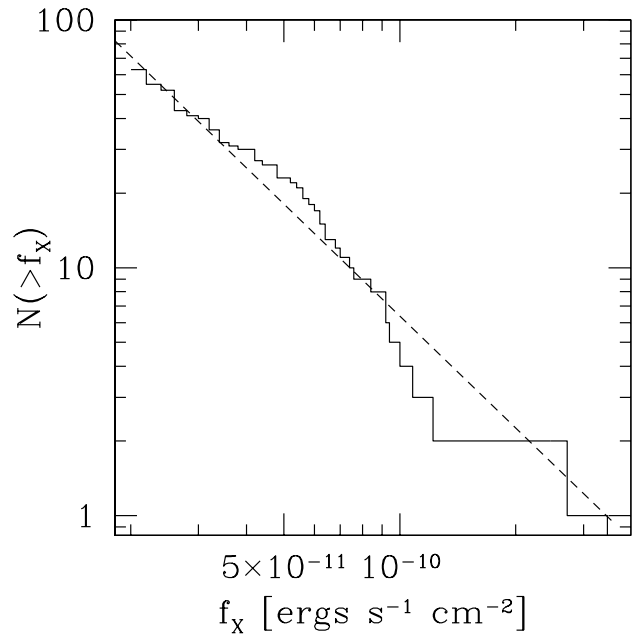,width=9cm,angle=0,clip=}
\end{center}
\caption{$\log N(>\fx)-\log \fx$ diagram. Fluxes are measured in the \ro\
energy band ($\eb$). The dashed line has a slope $-1.5$, expected for a uniform
distribution of clusters in a static Euclidean universe
(`three-halves-law'), the line is normalized to produce the same cluster number
at $\fx = 8\esc$.}\label{lnls}
\end{figure}
Figure~\ref{lnls} shows the integral number counts as a function of
X-ray flux (`$\log N-\log S$'). The slope in the $\log N-\log S$
diagram is very close to the value $-1.5$ expected in a static Euclidean
universe for uniformly distributed clusters.
Due to the small number of clusters (4) the
deviation is not significant for $\fx \gtrsim 1\escc$. 
Since the average redshift is
smallest for the highest fluxes large scale structure is
not completely washed out at the high flux end, therefore the slight bump visible around
$\fx \sim 6\esc$ in Fig.~\ref{lnls} suggests a deviation caused by cosmic
variance. The effect of an
expanding and finite universe on the $\log N-\log S$ -- flattening of the slope
towards low
fluxes -- is small for the redshift range covered by the sample. The
slope consistent with $-1.5$ towards the flux limit therefore
indicates a high
completeness of \gcs .

\begin{figure}[thp]
\begin{center}
\psfig{file=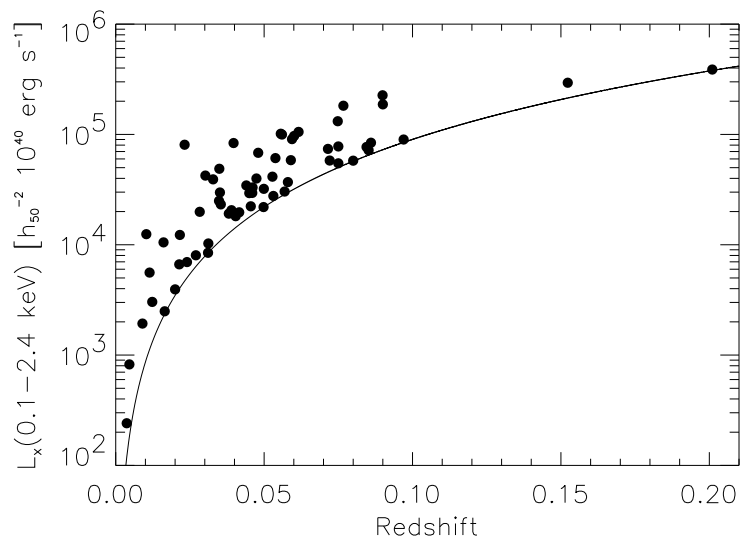,width=11cm,angle=0,clip=}
\end{center}
\caption{X-ray luminosity as a function of redshift. The flux limit is shown as
a solid line.}\label{lz}
\end{figure}

In Fig.~\ref{lz} the X-ray luminosity is plotted as a function of
redshift. The flux limit is shown as a solid line\footnote{The correction
$K(\tg,z)$ for converting observer rest
frame luminosities to source rest frame luminosities
(Eq.~\ref{eq:gr:dlm})
depends on redshift
\emph{and}
source spectrum ($\tg$). For source rest frame luminosities it is therefore not
possible to plot the flux limit as one line in 2 dimensions ($\lx ,z$), but
rather as an
area in 3 dimensions ($\lx ,z,\tg$). For consistency therefore in this
2d plot the observer rest
frame luminosity is given (the correction is less than 6\,\% for 90\,\% of
the clusters anyway).}.
One notes the increase in rare luminous systems with increasing redshift
(volume). Because of the seeming underdensity of clusters in the redshift range
$0.10<z<0.15$ a comparison with the expected number of
clusters as derived from $N$-body simulations has been performed.
An OCDM simulation,
carried out for analysis of the power spectral densities of REFLEX clusters
\citep{sbg00a}, adjusted to the \gcs\ survey volume in the southern
hemisphere (roughly half of the total volume sampled by \gcs ) has
been used. The simulation yields 39 clusters while 33 \gcs\ clusters have been
detected in this region. It is found
that in fact not even one cluster with $z>0.1$ is expected for
this volume based on this
simulation and the \gcs\ subsample also does not contain any cluster with
a redshift larger than 0.1.
This is a further piece of evidence for the high completeness of the sample.

\begin{figure}[thbp]
\begin{center}
\psfig{file=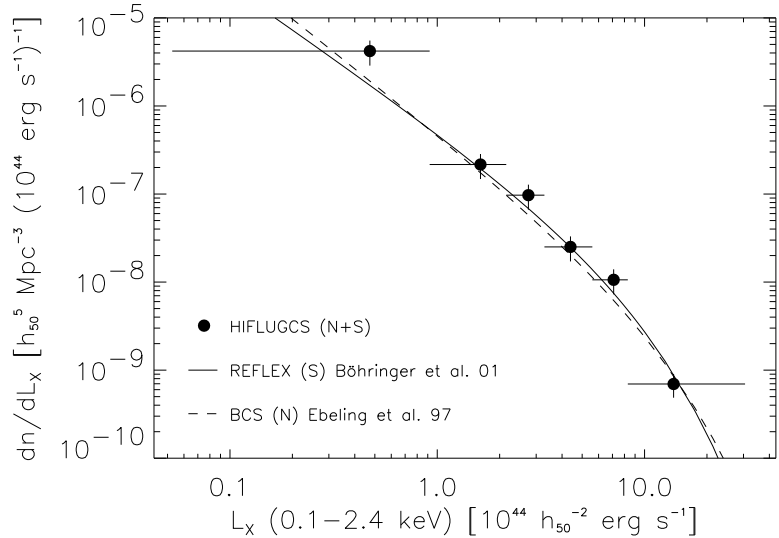,width=11cm,angle=0,clip=}
\end{center}
\caption{X-ray luminosity function for \gcs\ compared to surveys with deeper
flux limits in the northern (N) and southern (S) hemisphere. Vertical error bars
correspond to the formal 1-$\sigma$ Poisson errors (no
cosmic variance), the horizontal bars indicate the
luminosity intervals
covered.}\label{lxfunc}
\end{figure}

In Fig.~\ref{lxfunc} the \gcs\ X-ray luminosity function is compared to 
luminosity functions of larger surveys in the southern
(REFLEX,  \citealt{bcg01}) and 
northern (BCS, \citealt{eef97}) hemisphere. These surveys have much deeper flux
limits (Sect.~\ref{sample}) and contain many more clusters. Very good
agreement is found, which shows the high
completeness and homogeneous selection of \gcs .

The $V/\vmax$ test (e.g., \citealt{r68,s68,av80,p99}, Sect.~14.5) can
be used to asses a possible sample incompleteness.
Assuming a uniform distribution of clusters a value 1/2 is expected on
average.
For \gcs\ $\langle V/\vmax \rangle = 0.47\pm 0.04$, which is
consistent with the 
expectation and this result is interpreted as a clear sign that \gcs\
covers a large enough volume for most of the $\lx$ range to be
representative of the local
Universe with a high sample completeness. The local nature of \gcs\
becomes obvious by noting that the result of the comoving $V/\vmax$ test
is almost  identical to the result of the equivalent test assuming a
Euclidean and 
non expanding space, i.e.\ $\langle (\fx/\fxl)^{-3/2} \rangle = 0.46$.

\section{Comparison of Mass Determinations}\label{mest_d}

Next to a high sample completeness, reliable mass estimates are
obviously important for the construction of the cluster mass
function. In Sect.~\ref{back:dm} it has been outlined that simulations
show that X-ray mass estimates as performed in this work generally
yield unbiased results with a relatively small scatter. In this
Section masses for clusters determined using a similar method by
different authors and using a completely independent method are
directly compared to the results obtained here.
Below a method for a proper comparison is described. In Sects.~\ref{mest_x} and
\ref{mest_o} the results of the comparisons with masses determined using X-ray
and optical data, respectively, are given.

Let us assume having two different measurement methods, $A$ and $B$.
For a sample of  objects simply calculating the mean of the ratios of
the results, $M^A$ and $M^B$, of
these two measurement methods,
i.e.\ $\left\langle\frac{M^A}{M^B}\right \rangle$
(and clearly the ratio of the means, $\frac{\langle
M^A\rangle}{\langle M^B\rangle}$), is not the ideal way
for a comparison. Consider for instance the measurement results for two
objects $(X_1,X_2)$, $M^A=(a,1/a)$ and $M^B=(1,1)$. Both following
means are always equal in
this case and depending on $a$ they can be arbitrarily different from 1, i.e.\
\begin{equation}
\left\langle\frac{M^{A}}{M^{B}}\right \rangle =
\left\langle\frac{M^{B}}{M^{A}}\right \rangle
\left\{\begin{array}{l@{\quad:\quad}l}
= 1 & a=1\,, \\
> 1 & a>1\,, \\
< 1 & a<1\,.\end{array} \right .
\label{eq:means}
\end{equation}
E.g.\ for $a=2$ the former mean would imply that method $A$
yields on average results which are a factor of 1.25 larger than
method $B$. A calculation of the latter mean would imply exactly the
opposite. Using $\log M^i$ instead of $M^i$ to calculate the above
means avoids these apparently qualitatively contradicting results, but
the results still differ quantitatively depending on
which mean is calculated and they cannot be interpreted in a
way that
the values for the different measurement methods differ by a constant
factor. Therefore here results of different measurements are compared
using the
mean and standard deviation of the differences of the
logarithmic values, i.e.\
\begin{equation}
\md (M^A, M^B)\equiv \langle \log M^A - \log M^B \rangle\,,
\label{eq:MD}
\end{equation}
which is straightforward to interpret
($\md (M^A,M^B)=b\Rightarrow M^A = 10^b M^B$ on average)
and gives self consistent ($\md (M^A, M^B)=-\md (M^B, M^A)$)
results.  

\subsection{X-Ray Masses}\label{mest_x}

\begin{figure}[thbp]
\begin{center}
\psfig{file=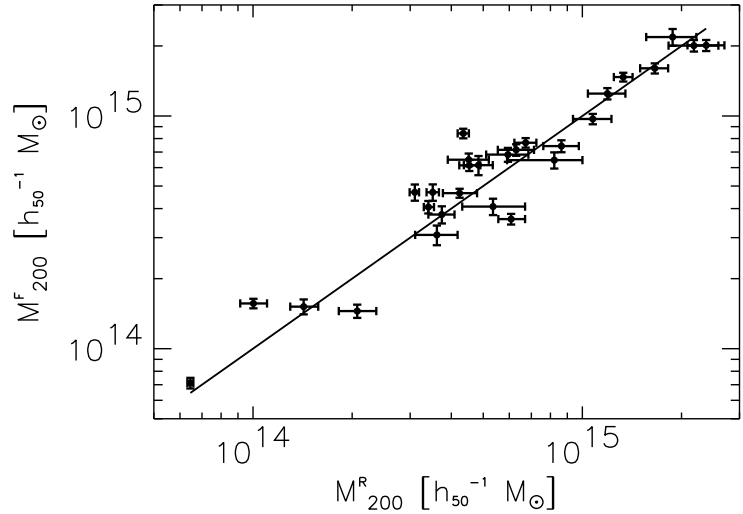,width=11cm,angle=0,clip=}
\end{center}
\caption{Mass determination for 28 groups and clusters. $M^{\rm R}$
denotes masses determined in this work and $M^{\rm F}$ masses
determined by \citet{frb00}.}\label{mm}
\end{figure}

\begin{figure}[thbp]
\begin{center}
\psfig{file=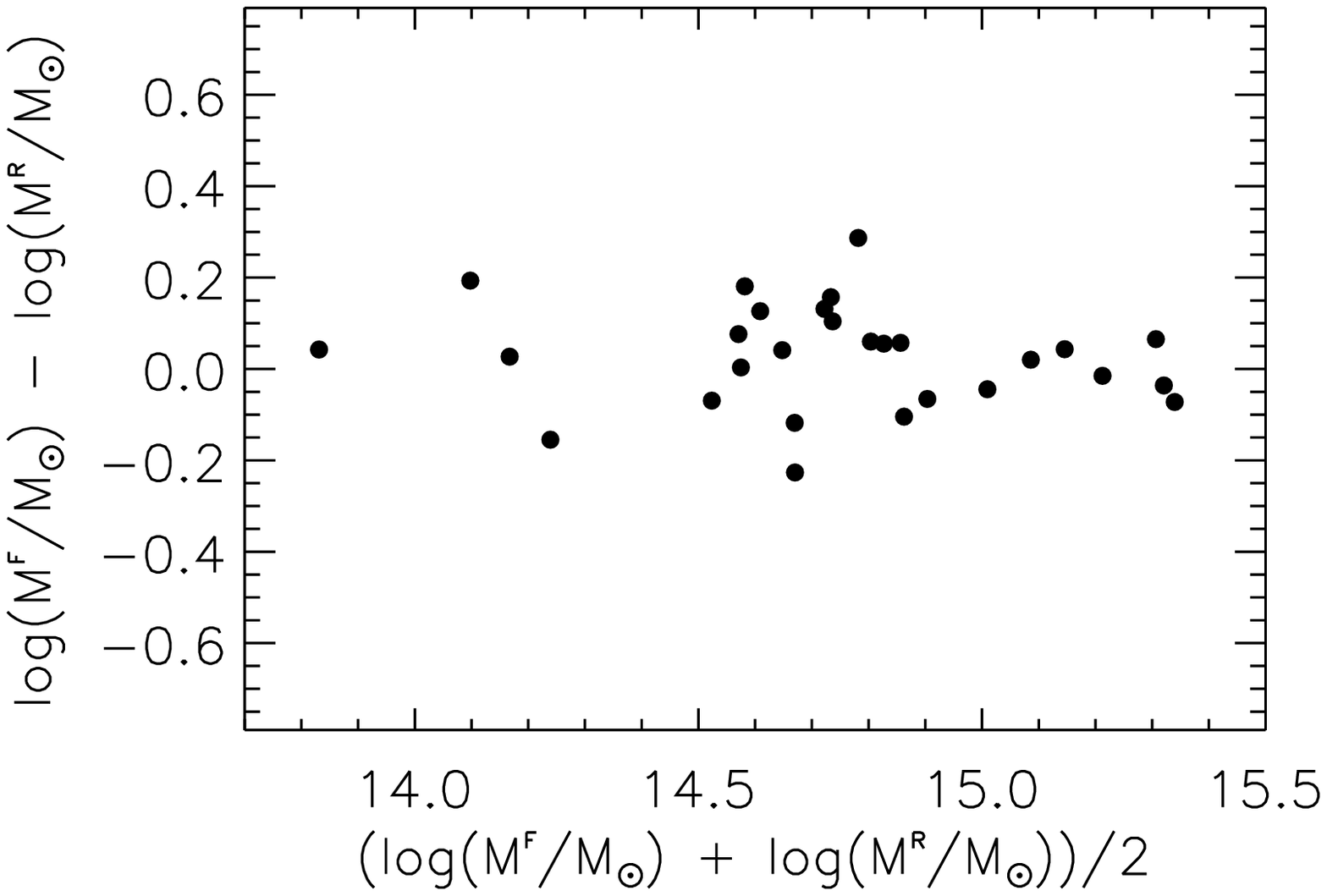,width=11cm,angle=0,clip=}
\end{center}
\caption{Differences of individual mass estimates.}\label{MDmrmf}
\end{figure}

The extended sample has 28 clusters in common with the sample
analyzed independently by \citet{frb00}. Independent is meant in the
sense that different people have reduced and analyzed the data while
the data themselves were obtained from the same satellite
observatories, \ro\ and \as .
Figure~\ref{mm} shows a comparison of
isothermal mass estimates within $r_{200}$ for these clusters.
The differences of the logarithmic values are plotted as a function of
the mean logarithmic mass in Fig.~\ref{MDmrmf}. There is no indication of a
correlation between these two quantities.
Calculation of the mean difference according to (\ref{eq:MD}) yields
$\md (M^{\rm F},M^{\rm R})=0.027\pm 0.113$, indicating that $M^{\rm
F}=1.06^{+0.32}_{-0.24}\,M^{\rm R}$ on average for masses determined
within $r_{200}$.
The standard deviation is almost exactly equal
to the size of the mean statistical mass measurement errors found in
Sect.~\ref{massd}. This result shows that even comparing mass
determinations for individual clusters
not many significant differences are found. On average the results
show very good agreement.

\citet{frb00} determined masses not only by employing the assumption
of isothermality, $M^{\rm F,iso}$, but also using measured cluster gas
temperature profiles, $M^{\rm F,grad}$. A comparison for 38 clusters
included in their sample yields 
$\md (M^{\rm F,grad},M^{\rm F,iso})=-0.097\pm 0.099$, indicating
$M^{\rm F,grad}=0.799^{+0.21}_{-0.16}\,M^{\rm F,iso}$ on average.
The influence of a possible overestimation of masses determined
here on the final determination of cosmological parameters is investigated
in Sect.~\ref{func_pred}.

\subsection{Optical Masses}\label{mest_o}

\begin{figure}[thbp]
\begin{center}
\psfig{file=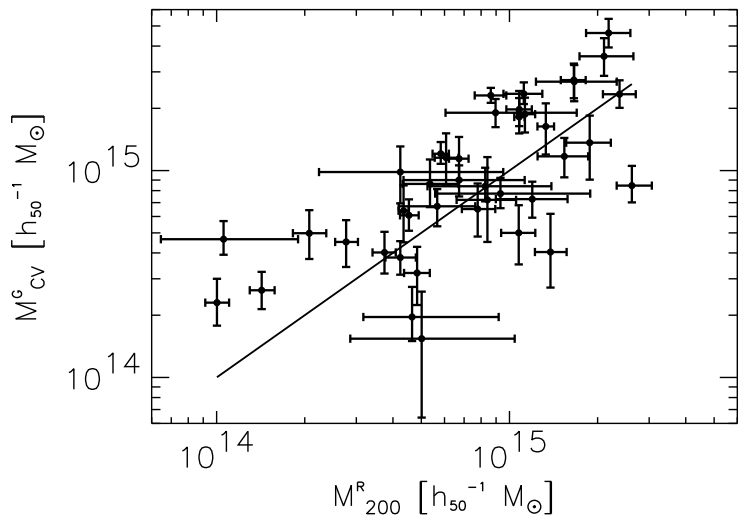,width=11cm,angle=0,clip=}
\end{center}
\caption{Mass determination for 42 clusters. $M^{\rm R}$
denotes masses determined in this work and $M^{\rm G}$ masses
determined by \citet{ggm98}.}\label{mrmg}
\end{figure}

\begin{figure}[thbp]
\begin{center}
\psfig{file=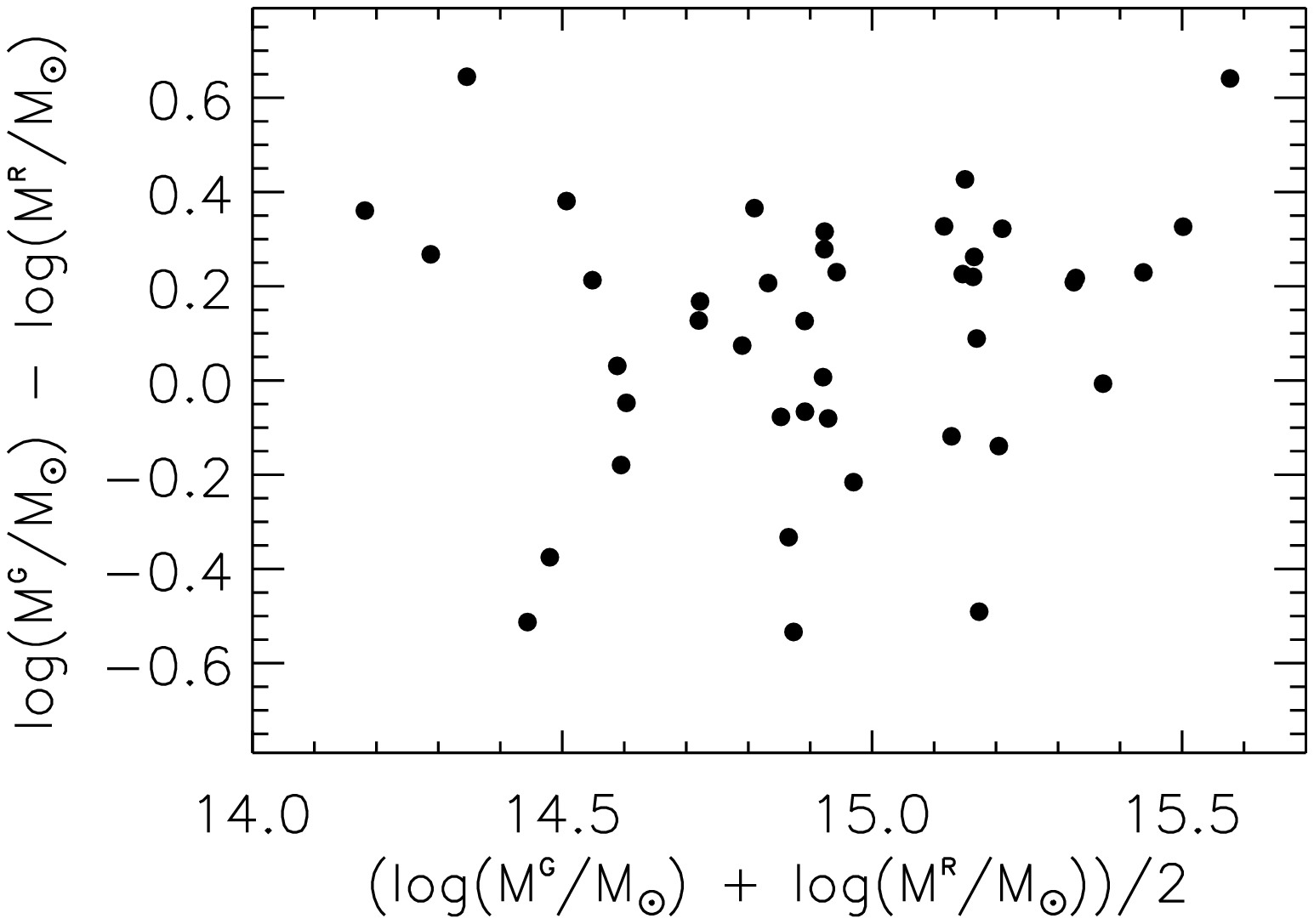,width=11cm,angle=0,clip=}
\end{center}
\caption{Differences of individual mass estimates.}\label{MDmrmg}
\end{figure}

\citet{ggm98} determined virial masses for a sample of nearby galaxy
clusters by compiling optical velocity dispersions of cluster galaxies
from the literature. Figure~\ref{mrmg} shows a comparison of the mass
determinations for 42 clusters common to their sample and the extended
sample presented here. No trend in the difference of the logarithmic
values with mass is seen in Fig.~\ref{MDmrmg}. A calculation of the mean
difference yields
$\md (M^{\rm G},M^{\rm R})=0.098\pm 0.283$, indicating that
$M^{\rm G}=1.25^{+1.15}_{-0.60}\,M^{\rm R}$ on average.
This result shows that comparing mass
determinations for individual clusters
one may find significant differences. The average offset between the
methods is of about the same size as the mean statistical errors found
for the extended sample. The optical masses are about 25\,\%
higher on average. Note, however, that since both measurements aim for
determination of the virial mass, no rescaling of radii has been performed,
i.e.\ masses are not  determined at the same radius for each method, since
this would underestimate the offset and standard deviation of the
measurements as needed here. Moreover \emph{all} clusters in common are
compared, i.e.\ critical cases like A548w, which is part of the larger
complex structure of A548 \citep[e.g.,][]{dbm95}, and 
A2319, known to host a subcluster \citep[e.g.,][]{1995AJ....110...32O},
have not been excluded. These
two cases constitute the two upper extreme clusters in Fig.~\ref{MDmrmg}.
The three lower extreme cases are 
A754, a famous cluster undergoing
a major merger event \citep[e.g.,][]{hb95},
A2589, which shows no obvious indications of irregularity, except a
velocity of the central cD galaxy offset of the cluster core
velocity by about 50\,\% of the
cluster velocity dispersion \citep[e.g.,][]{cmp91}, and 
A3921, which has been found to be in the state of merging
\citep[e.g.,][]{arb96}.
Note also that the simple calculation of $\md$ assigns each cluster the same
weight,
i.e.\ the uncertainty of the mass
measurements is neglected.
It is worth noting that
$\left\langle\frac{M^{\rm R}}{M^{\rm G}}\right \rangle=1.001\pm 0.787$,
misleadingly implying the complete absence of any systematic
differences between the two measurement methods. 
The influence of a possible underestimation of masses as determined
here on the determination of cosmological parameters is investigated
in Sect.~\ref{func_pred}. Note that the isothermal mass measurements
as performed here
lie almost exactly in between the results using gas temperature
gradients and using velocity dispersions on average.

\section{Mass--Luminosity Relation}\label{relat_d}

The close correlation between the X-ray luminosity and the
gravitational mass found in Sect.~\ref{relat} is not
surprising.
If clusters form by collapse when exceeding a common overdensity threshold
(neglecting any redshift dependence for now),
e.g., by spherical collapse (Eqs.~\ref{eq:ks1} and \ref{eq:ks2})
but also any other kind of collapse
from an overdensity independent of mass, then all clusters will share
the same mean density. This
picture implies that the cluster mass is proportional to its volume.
Therefore
\begin{equation}
\mt\propto \rch^3\,,
\label{eq:ss1}
\end{equation}
where $\rch$ is the virial radius,
but for self similar clusters it could be any characteristic radius,
e.g., the core radius.
With the assumption of virial equilibrium one has
\begin{equation}
\tg\propto \mt \rch^{-1}\,.
\label{eq:vir}
\end{equation}
Combining (\ref{eq:ss1}) with bremsstrahlung
emission,
\begin{equation}
\lbol\propto\rog^2 T^{1/2} \rch^3
\quad {\rm and}\quad
\lx(\eb)\propto\rog^2
\rch^3
\label{eq:brems}
\end{equation}
(Sect.~\ref{back:rog}), where $\lbol$ is the bolometric luminosity, one has
\begin{equation}
\lbol\propto\rog^2 T^{1/2} \mt
\quad {\rm and}\quad
\lx(\eb)\propto\rog^2 \mt\,.
\label{eq:lma}
\end{equation}
Using $\rog\propto\mg \rch^{-3}$, (\ref{eq:ss1}), and (\ref{eq:vir})
the relations
\begin{equation}
\lbol\propto\fg^2 \mt^{4/3}\quad
{\rm and}\quad \lx(\eb)\propto\fg^2 \mt
\label{eq:lm}
\end{equation}
follow from (\ref{eq:lma}).
In this simple picture one therefore expects a correlation between
luminosity and mass. For later use note that from (\ref{eq:ss1})
and (\ref{eq:vir}) one finds
\begin{equation}
\mt\propto\tg^{3/2}
\label{eq:tm}
\end{equation}
and, combined with (\ref{eq:lm}),
\begin{equation}
\lbol\propto\fg^2 \tg^2\quad
{\rm and}\quad \lx(\eb)\propto\fg^2 \tg^{3/2}\,.
\label{eq:lt}
\end{equation}

Observationally from the tight correlations between X-ray luminosity
and temperature (e.g., \citealt{m98}), and temperature and mass (e.g.,
\citealt{frb00}) a correlation between luminosity and mass clearly is
expected. Also correlations found between X-ray luminosity and galaxy
velocity dispersion (e.g., \citealt{es91b}) and X-ray luminosity and mean shear
strength from weak lensing studies (e.g., \citealt{sed97}) indicate a
correlation between $\lx$ and $\mt$.

The X-ray luminosity has been compared directly to
gravitational mass estimates by
\citet{rb99b,rb00c}, \citet{s99}, \citet{jf99}, \citet{mlb99}, \citet{ef00},
and \citet{bg01}, where good correlations have been found in all these
studies.

\begin{figure}[thbp]
\begin{center}
\psfig{file=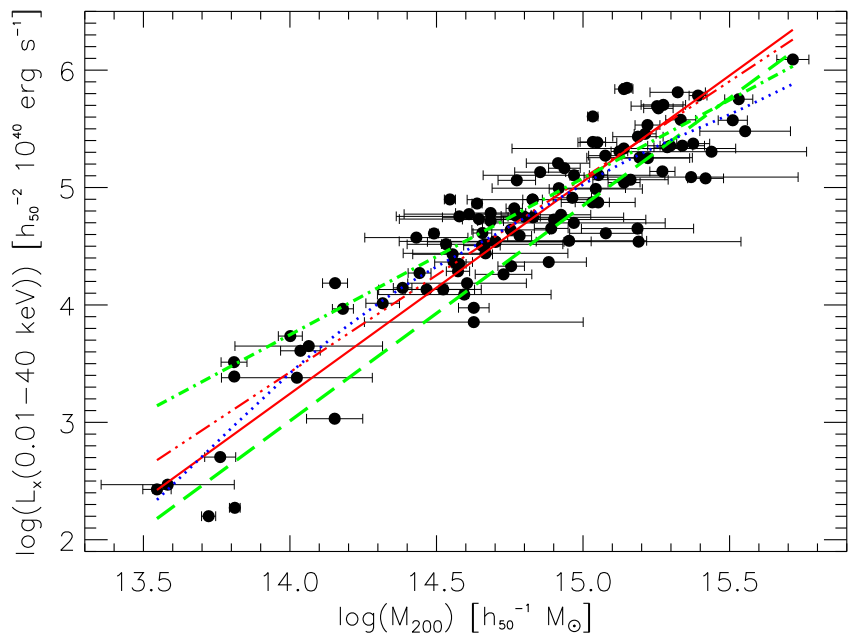,width=13cm,angle=0,clip=}
\end{center}
\caption{Gravitational mass--bolometric X-ray luminosity
relation compared to predicted relations. Shown are:
best fit relation for the extended sample (solid line),
best fit relation determined using \gcs\ (triple-dot-dashed line),
self similar relation normalized by
simulations of \citet{nfw95} (dot-dashed line),
pre-heated relation given by
\citet{eh91}, using a normalization taken from the simulations of
pre-heated clusters by \citet{nfw95} (dashed line). The dotted line gives
the result of a smoothed spline fit.}\label{mtlbol}
\end{figure}
In order to compare the empirical $\lx$--$\mt$ relation with predictions a
quasi bolometric luminosity, $\lbol$, has been calculated in the
source rest frame energy range
$\ebol$ (for the relevant range of cluster gas temperatures
at least 99\,\% of the flux is contained in this energy range).
In Fig.~\ref{mtlbol} this $\lbol$--$M_{200}$ relation
is compared to predicted relations. 
The solid line shows the best fit relation for the 106 clusters in the
extended sample and the 
triple-dot-dashed line shows the best fit relation determined using
\gcs . Here the bisector fit results have been used in order to treat
variables symmetrically, which is the appropriate method for a
comparison to theory \citep[e.g.,][]{ifa90}.
The dot-dashed line
shows the self similar relation ($\lbol\propto \mt^{4/3}$, Eq.~\ref{eq:lm})
normalized by the
simulations of \citet{nfw95} and the dashed line shows the
`pre-heated' relation given by
\citet{eh91} ($\lbol\propto \mt^{11/6}$, see below), using a normalization taken from the simulations of
pre-heated clusters by \citet{nfw95}. 
The idea of pre-heating is that the intracluster gas is not
cold initially, as in the self similar case, but is heated
to a characteristic temperature by some
form of non gravitational heat input, e.g., from SNe, early
during cluster formation. Assuming the central regions of all clusters
to have the same entropy yields the latter relationship.
This can be seen by expressing the bolometric luminosity in terms
of the central gas density, i.e.\
\begin{equation}
\lbol\propto\rog(0)^{2-1/\beta }\, T^{1/2}\,\mt
\label{eq:lgc}
\end{equation}
\citep[e.g.,][]{eh91}.
If the central specific entropy,
\begin{equation}
s\propto\frac{3}{2}\,\kb\,\ln{\frac{\tg}{\rog(0)^{\gamma -1}}}\,,
\label{eq:ent1}
\end{equation}
is set to a constant value and taking $\gamma=5/3$ one finds
\begin{equation}
\rog(0)\propto \tg^{3/2}\,.
\label{eq:ent2}
\end{equation}
Plugging this into (\ref{eq:lgc}) and using
$\beta=2/3$, which is close to the mean value in the extended sample,
yields
\begin{equation}
\lbol\propto \tg^{5/4}\,\mt\,.
\label{eq:ent3}
\end{equation}
With (\ref{eq:tm}) one therefore has
\begin{equation}
\lbol\propto \mt^{11/6}\,.
\label{eq:ent4}
\end{equation}

Fig.~\ref{mtlbol} shows that measured and predicted
relations are in rough agreement,
the difference between the predicted relations being larger than the
difference to the observed relations.
Note, however, that the X-ray luminosity is one of the most uncertain
quantities to be derived from simulations. \citet{fwb99} recently
showed in a comparison of 12 different cosmological hydrodynamics
codes that a factor of 2 uncertainty is a realistic estimate of the
current accuracy. Including gas cooling in simulations worsens the
situation \citep[e.g.,][]{bpb01}.
The slopes of the observed relations are closer to the
pre-heated relation.
Observationally the effect of pre-heating can also result in a decrease of the gas
mass fraction for low temperature systems.
This has actually been observed for
the clusters in the sample (Sect.~\ref{fg_d}). The possibility
that winds from SNe -- originally invoked to explain the
apparent low gas content of elliptical galaxies \citep[e.g.,][]{mb71,l74} -- pre-heat and dilute the central gas and thereby break
the self similarity has been pointed out by various authors (e.g.,
\citealt{k86}). Such a process would work most efficiently on the least massive
clusters (e.g., \citealt{w91,me97,pcn99}). The spline fit (dotted line in
Fig.~\ref{mtlbol}) indicates a weak bending, 
i.e.\ a slight deviation from a power law, over the entire $\mtz$ scale.
There is some indication that the bend is strongest around
$2\times10^{14}\,\msu$, suggesting a possible break.
In the middle part, where most of the clusters are located, the slope
of the dotted line
is close to 4/3, while at the low mass end the slope appears even steeper
than 11/6. 

\begin{figure}[thbp]
\hspace{2.9cm}
\psfig{file=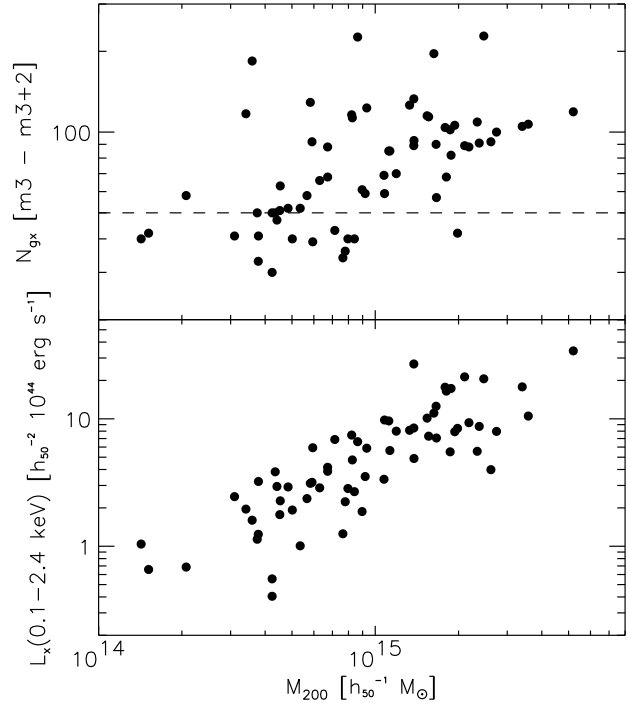,width=14cm,angle=0,clip=}
\caption{Measured number of cluster member galaxies, $\ngx$, as
taken from \cite{aco89}
and X-ray luminosity of the same (66) clusters
as a function of the gravitational mass. Double clusters,
whose components have been treated separately here,
e.g., A3528n/s, are removed. Above the dashed
line all clusters have an Abell richness $R\ge 1$.}\label{nmlxm}
\end{figure}

The Abell catalog \citep{a58,aco89}, as one of the first and largest
systematic cluster catalogs, continues to be a widely used database.
The Abell richness (Sect.~\ref{back:galax})
has been used frequently in the past as a
selection criterium for optical cluster samples
\citep[e.g.,][]{ch87,1996A&A...310....8K}.
The construction of the cluster mass function requires a 
selection closely related to the gravitational cluster mass.
The X-ray luminosity has been shown above to exhibit a tight correlation
with mass.
In Fig.~\ref{nmlxm} the measured number of cluster member galaxies, $\ngx$, as
taken from \citet{aco89} is compared to $\lx$ as gravitational mass tracer.
One notes that there exists a trend for higher mass clusters to have higher
values of $\ngx$. However, there is a huge scatter. One of the main reasons
for this poor correlation may be the much less homogeneous
background compared to the X-ray case.
It is clearly seen that a selection by
X-ray luminosity is much more efficient than a selection by Abell richness in
terms of mass.
Even though only the X-ray surface brightness profile and neither its
normalization nor the X-ray luminosity are directly used in the X-ray mass
determination via the hydrostatic equation, it is
nonetheless reassuring that a similar result is obtained when $\lx$
and richness are compared to masses estimated from optical velocity
dispersions \citep{bg01}.

\begin{figure}[thbp]
\begin{center}
\psfig{file=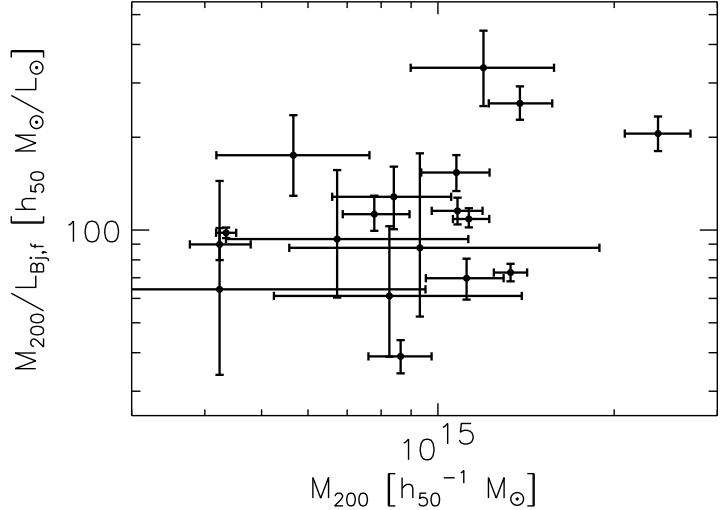,width=11cm,angle=0,clip=}
\end{center}
\caption{Mass to light ratio as a function of mass for 18 clusters in common
with the sample of \citet{gbg00}, where the integrated blue band luminosity
values within the virial radius have been taken from.}\label{mlb}
\end{figure}

In Fig.~\ref{mlb} a plot is shown using a quantity somewhat related to $\ngx$.
\citet{gbg00} provide integrated blue band luminosities, $\lbj$, within the
virial radius. 
Shown is the mass to light ratio, $\mtz/\lbj$, in units of
$\mlsu$ (the mass to light ratio of the sun), as a function of
$\mtz$ for the 18 clusters included
in the extended sample used here and the $C$ sample of \citet{gbg00}.
One notes that this ratio exhibits
a large scatter. Furthermore a very weak trend of a positive correlation with
the mass is indicated in agreement with the findings of \citet{gbg00}.
The median mass to light ratio found here equals $103\,\mlsu$, implying that
the stars in clusters account only for a small fraction of the cluster mass.
Taking into account the slight systematic differences in the mass estimates
(Sect.~\ref{mest_o}) this value agrees with the value quoted by \citet{gbg00}.

One old and simple technique to estimate
the mean density in the universe is to assume this median mass
to light ratio to be representative of the universe
\citep[see, e.g.,][]{bop99}. If $\om=1$ one would expect $\mtz/\lbj\approx
675\,\mlsu$ based on estimates for the total blue band luminosity
density \citep[see refs in][]{gbg00}. The value found here therefore indicates
$\om\approx 0.15$, independent of the Hubble constant.
This value should be considered as a rough estimate
because of the fairly small number of clusters and the scatter present,
because of the difference in the determination of the virial radius for
$\mtz$ and $\lbj$, because of the dependence on the accuracy and
representativeness of the measured galaxy luminosity functions (which lead
to the above mass to light ratio for $\om=1$),
and because of the possibility that the mass to light ratio
may increase or decrease on even larger scales.
Nevertheless the estimate $\om\approx 0.15$ obtained here is in very
good agreement with what will be found
in Sect.~\ref{func_pred} when the mass function is compared to predictions.

A wide range
of possible applications becomes available with the quantification of the $\lx$--$\mt$
relation and its scatter. For large X-ray cluster surveys, where individual mass
determinations are currently not feasible, luminosities can be directly
converted to masses. No combination of observations, simulations, and theory is
then needed, like the frequently used approach of relating X-ray luminosities
to X-ray temperatures by an
observed relation, and converting X-ray temperatures to masses using a relation
where the slope is taken from theoretical arguments and the normalization from
hydrodynamical simulations (e.g., \citealt{mmd00}). The observational
$\lx$--$\mt$ relation
has first been applied directly in this sense in the power spectral analysis of REFLEX
clusters \citep{sbg00a}. An example of another direct application is given in
Sect.~\ref{lmtest}.
The $\lx$--$\mt$ relation may also be applied to
convert theoretical or simulated mass functions to luminosity functions for
comparison with observations, which is currently being performed in
the interpretation of the REFLEX luminosity function.

At this point it is important to note that even for the highest
redshift cluster in the sample used here ($z=0.2$) the dependence of the
observational determination of $\mt$ and $\lx$ on the chosen
cosmological model is very weak. For instance at $z=0.2$ the increase
in the luminosity distance, $D_L$, and the diameter distance,
$D_A$, is less than 5\,\% going from ($\om=1.0,\ol=0$) to
($\om=0.1,\ol=0$).
From (\ref{back:ma2}) one finds that
$\mt(<r)\propto r$ and therefore $\mt(<r)\propto D_A$, implying an
increase of $\mt$ by less than 5\,\% for the two models above.
For $\lx$ one has an increase of less than 10\,\%.
This means that the $\lx$--$\mt$ relation given here can be used unchanged for various
cosmological applications (unless redshift ranges are probed where
evolution becomes important, in this case a model dependent
redshift correction has to be introduced).
A similar calculation for $\vmax$ shows that for the
extreme case $\zm=0.2$ the increase of $\vmax$ is less than
14\,\%, which is
less than the size of the Poissonian error bars in Fig.~\ref{PS0}.

More detailed investigations on the shape of the relation
and the construction of a
volume-limited sample, spanning a reasonably large range in luminosity/mass,
to test how much the $\lx$--$\mt$ relation given
here is affected by being estimated partly from a flux-limited sample
are beyond the scope of this work but are envisaged for the future.

\section{Intracluster Gas Fraction}\label{fg_d}

\begin{figure}[thp]
\centering\begin{tabular}{l}
\psfig{file=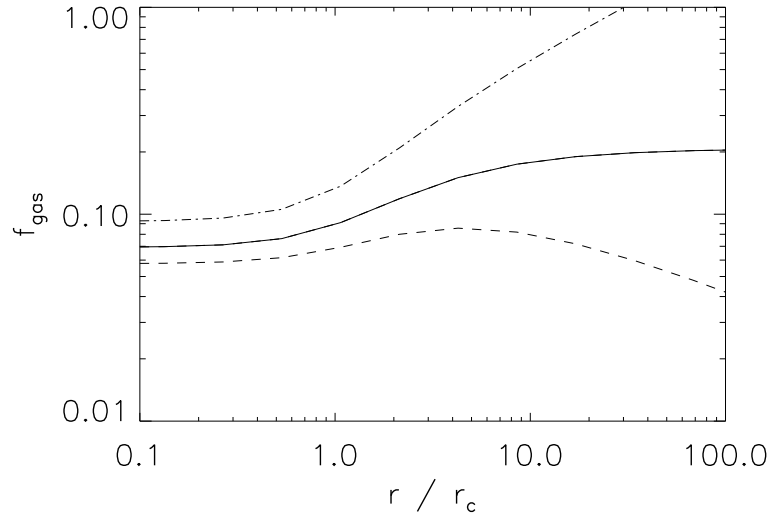,width=11cm,angle=0,clip=}
\end{tabular}
\caption{Radial variation of the gas fraction. Solid line:
$\beta=2/3$, dashed line: $\beta=0.8$, dot-dashed line: $\beta=0.5$.
$\rc=150$\,kpc, $\kb
\tg=5$\,keV, $\nel(0)=0.01\,\rm cm^{-3}$.}\label{fg_b}
\end{figure}

Even though caused by only 5--6 clusters
Figs.~\ref{fg_mt}--\ref{fg_tem} clearly suggest a drop in the gas
fraction at low cluster masses/temperatures ($\sim 2$\,keV).
Note, however, that the
X-ray emission of the low temperature clusters can in general be
traced out to less large (compared to hotter clusters)
radii relative to $r_{200}$.
Thus more extrapolation is performed rendering the (gas and
total) mass estimates more uncertain. On the other hand if the gas
mass fraction is indeed significantly lower one would of course just
expect to observe emission from the gas less far out. Furthermore, as
\citet{vfj99} showed, the gas density profile steepens ($\beta$
increases) only slightly in the
outer cluster parts and they concluded that mass estimates are not
significantly affected. Nevertheless a possibly underestimated $\beta$
value for the low temperature clusters would rather lead to an
\emph{overestimated} value for $\fg$ (Eq.~\ref{back:fg1} and
Fig.~\ref{fg_b}), which indicates that the low gas fractions for the low
$\tx$ systems found here are not an artefact.
In any case deeper
observations of low temperature clusters are needed to resolve this
issue. Despite the problems of a high background level
(Chap.~\ref{a1835}), \xmm\ with its large effective area seems currently to
be best suited for this purpose.

The importance of more and especially deeper observations becomes even
more obvious when the results from other works are
compared.
\citet{djf95} showed for a sample of 11 clusters that $\fg$ increases
continuously between ellipticals, groups, and clusters.
The analysis by \citet{e97}, who combined $\fg$ measurements from
\citet{wf95} and \citet{djf95} ($1\lesssim \tx \lesssim 14$\,keV),
showed that a  homogeneously analyzed
cluster sample is vital if trends in the gas fraction are to be
studied.
First a trend of decreasing $\fg$ with
increasing $\tx$ was indicated, but after
the cluster gravitational masses had been
homogeneously redetermined this trend disappeared.
\citet{af98} found that $\fg$ is decreasing with increasing $\tx$ for
a sample of 30 high luminosity clusters after they had corrected the
temperatures for cooling flows. However, their sample contains rather hot
clusters mainly (5.5--26.4\,keV).
\citet{ef99} found no statistically significant dependence of $\fg$ on
$\mt$ or $\tx$ for a sample of 36 high luminosity clusters once the
sample was corrected for a redshift dependence\footnote{The gas
fraction depends on the diameter distance. Therefore choosing a
`wrong' cosmological model introduces an artificial redshift
dependence for cluster samples spanning a large redshift range.
This effect may
be used to constrain model parameters under the
assumption of a constant gas fraction \citep[e.g.,][]{sa96}.}. When the
redshift dependence was neglected they found results in agreement with
\citet{af98}.
\citet{ae99} found indications for an increase of $\fg$ with increasing
$\mt$ using a sample of 24 clusters in the range $2.2\le \tx \le
14.6$\,keV.
\citet{mme99} found a trend of increasing $\fg$ with
increasing $\tx$ for a sample of 45 galaxy clusters ranging in $\tx$
from 2.4 to 10.1\,keV.
\citet{rsb00} analyzed 33 clusters in the temperature range
1--14\,keV and found no clear trend of $\fg$ with $\tx$.
\citet{gcr01}, who determined the shape of the
gas density profile utilizing interferometric measurements of the
Sunyaev--Zeldovich effect, found no indication for a dependence of
$\fg$ on $\tx$ for a sample of 18 galaxy clusters. Their sample includes
only rather hot (5.7--13.2\,keV) clusters, however, whereas the effect
appears to be strongest for cool clusters.

It is worth noting that while \citet{djf95} found that the gas
fraction comprises only a few percent of the total mass in ellipticals
and small groups,
\citet{d97}, inspired by the early findings that a large fraction of the galactic
mass may be in the form of massive compact halo objects (MACHOs;
\citealt{aaa97}), argued that the total \emph{baryon} fraction may actually
stay constant at about 50\,\% independent of mass scale (from
ellipticals to rich clusters) if MACHOs mainly consisted of
stellar remnants \citep[e.g.,][]{fms97,aaa97}, e.g., white dwarfs,
which would also explain the
abundance of heavy elements in the intergalactic medium.
However, the galactic halo mass fraction contained in MACHOs has been
significantly reduced recently \citep{aaa00}.
\citet{daf90} and \citet{d97} also showed that the ratio gas mass over
stellar mass increases with gas temperature, confirmed by
\citet{arb92} who found the same trend with cluster richness.
\citet{rsb00}, however, did not find strong evidence of an increase of
this ratio with $\tx$. The main part of the discrepancy seems to arise
from the fact that \citet{rsb00} did not find an increase of $\fg$ with
$\tx$. As mentioned later in more detail \citet{b00} compiled evidence
for not only an
increase of $\fg$ with $\tx$ but also for a decrease of the mass fraction
contained in stars with increasing $\tx$.

In summary it is concluded that the large spanned temperature range
available in this work has allowed to find that the gas mass fraction
A) does not vary systematically for clusters with $\tx\gtrsim 2$\,keV and
B) drops abruptly for groups and clusters with lower gas temperatures
in agreement with some of the previous works. The fact that the X-ray
extent relative to the virial radius is usually smaller
for groups than for clusters may be interpreted as a consequence
of the lower gas fraction. Since proposed scenarios to explain a systematic
variation of the gas fraction with temperature or mass, e.g., pre-heating,
should be considered also in the light of relations between other cluster
parameters, they are further discussed in Sect.~\ref{tlm_d}.

\begin{figure}[thp]
\centering\begin{tabular}{l}
\psfig{file=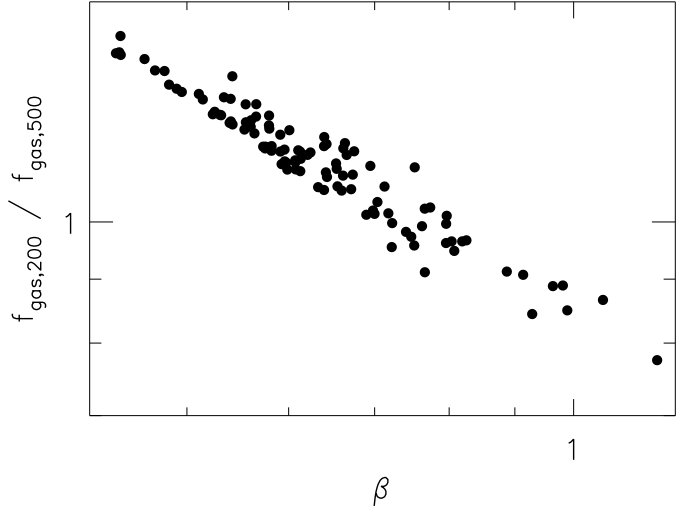,width=11cm,angle=0,clip=}
\end{tabular}
\caption{Variation of the gas fraction as a function of $\beta$. Error
bars have been omitted.}\label{fg2_fg5_b}
\end{figure}

In Fig.~\ref{fg2_fg5_m} it has been shown that for most clusters the gas
mass fraction appears to vary with radius. For the majority of those clusters
$\fg$ seems to increase with radius, i.e.\ between $r_{500}$ and $r_{200}$.
There appears to be no significant trend of this variation with system
mass for the low to medium mass clusters. However, the gas
fraction apparently becomes
more likely to increase less fast or decrease with radius for the massive
clusters. How may this be understood?
Since $\beta$ is an important parameter both for the gas and gravitational
mass determination let us first estimate the behavior of $\fg$ as a
function of $\beta$.
Dividing (\ref{back:gm2}) by (\ref{back:ma2}) one gets an approximate
expression for the gas fraction
\begin{equation}
\fg(<r)\approx f(\tg,\rog(0),\beta,\rc) \left (\frac{r}{\rc}\right )^{-3\beta
+2} \quad : \quad \frac{r}{\rc}\gg 1 \wedge \beta < 1\,,
\label{back:fg1}
\end{equation}
where the intracluster gas has been assumed in hydrostatic equilibrium and
isothermal. Therefore one has by construction an increasing gas
fraction with increasing radius for $\beta\lesssim 2/3$ and a decreasing gas
fraction for $\beta\gtrsim 2/3$. This is illustrated more precisely (keeping
only the latter two assumptions for the intracluster gas) in
Fig.~\ref{fg_b} for some typical parameter values.
It is worth noting that from this Figure it
is also clear that the gas fraction rises with radius out to $\gtrsim
5\,\rc$  as long as $\beta \lesssim 0.8$, i.e.\ for most of the clusters.
Plotting the gas fraction ratios
as determined for the clusters in this work
as a function of the observed $\beta$
values (Fig.~\ref{fg2_fg5_b}) confirms the decrease of $\fg$ with
increasing radius in the outer cluster parts for $\beta\gtrsim 2/3$.
As expected therefore $\beta$ is the relevant parameter here.
The variation of $\beta$ as a function of mass therefore determines
the variation of $\fgz/\fgf$ with mass. A quick look to
Fig.~\ref{mb} confirms this. So also trends of the radial variation of
the gas mass fraction need to be discussed in the context of other relations,
which is done in the next Section.
\section{Relations between Shape Parameters, Temperature, Luminosity,
and Mass}\label{tlm_d}

In this Section the relations found in Sect.~\ref{tlm} are discussed.
Suggested models to explain the observed behavior -- including the variation
of the gas mass fraction -- are described. It is discussed whether the
predictions of a single scenario fit all relations.

It has long been known that a correlation exists between $\lx$ and $\tx$
\citep[e.g.,][]{mic77}.
Most authors find best fit relations similar to $\lbol\propto\tx^3$,
as has been found here, too.
This is significantly steeper than the simple self similar prediction
$\lbol\propto\tx^2$.
Note that inclusion of line emission -- more important
for cooler systems-- would further flatten this relation.

In principle a dependence of the gas fraction on temperature could account
for a steeper $\lbol$--$\tx$ relation (Eq.~\ref{eq:lt}).
And really it has been found in Sect.~\ref{fgas} that
the gas fraction increases with increasing temperature. But this 
is true only for the lowest temperature clusters (Fig.~\ref{fg_tem}).
So clearly this trend
of the gas fraction cannot account for the steepness of the
$\lx$--$\tx$ relation over the entire $\tx$ range.

In addition to this deviation from self similarity over the entire
$\tx$ range, interestingly some authors have found a bend in the
$\lx$--$\tx$ relation in the sense that low $\tx$ clusters give
rise to an even steeper $\lx$--$\tx$ relation \citep[e.g.,][]{hp00}.
Other authors find no evidence for such a feature \citep[e.g.,][]{mz98}.
In this work there is also no clear indication of any bend ranging
from 0.7--13.3\,keV.

The $\mt$--$\tx$ relation has in general been found observationally
to be steeper than the self similar prediction $\mt\propto \tg^{3/2}$
(Eq.~\ref{eq:tm})
and the normalization has been found by previous authors to be lower than
the normalization expected from simulations \citep[e.g.,][and references
therein]{frb00}. Both trends are seen here, too,
indicating that the interpretation of temperature functions in a
cosmological context depends systematically on the used method for the
conversion between $\tx$ and $\mt$, in agreement with the findings
of, e.g., \citet{irb01}. Furthermore
no clear deviation from a power law is found here.
Note that $\tg$ is needed for the calculation of $\mt$ (Eq.~\ref{back:ma2}).
These two quantities are therefore not determined independently
of one another.

A possible dependence of the gas density profile, i.e.\ of the shape
parameter $\beta$, on temperature has been
indicated by observations of previous authors
\citep[e.g.,][]{daf90,w91,moe97,ae99,jf99,lpc00,hp00}. However,
e.g., \citet{mme99}, \citet{vfj99}, and \citet{rsb00} have found no
clear or only a weak trend. Fig.~\ref{tb} shows that for the 88
clusters with measured temperature analyzed in this work there is no
trend found, especially not at the low temperature side. There may be
indications of increasing $\beta$ values with increasing $\tx$ at the
high temperature side, but the evidence is only weak. Stronger
indications indeed are found if $\beta$ is plotted as a function of
gravitational mass (at the high mass side). Interpretations of this
apparent trend, however, must be taken cautiously since $\beta$ enters
the calculation of $\mt$ (Eq.~\ref{back:ma2}).
There are two effects which are likely to enhance the very weak
trend seen in Fig.~\ref{tb} as compared to Fig.~\ref{mb}. First of all
$\mtz\propto\tx^{1.65\pm 0.05}$ (Tab.~\ref{tab:tmt}) and therefore a
possible trend with $\mtz$ will appear weaker in $\tx$. Secondly, the
temperatures of the clusters with higher
$\beta$ values at around 4--6\,keV (causing the slight bumb in the spline fit)
are translated into relatively higher masses (Eq.~\ref{back:ma2}).

The trend of $\beta$ at high masses may be understood by
considering that the core radius clearly correlates with
gravitational mass
(Fig.~\ref{rc-mt}, even though a large mass range is needed
to detect this correlation due to substantial scatter).
For large core radii $\beta$ increases with increasing $\rc$
(Fig.~\ref{rc-beta}).
It is therefore not surprising that for large masses $\beta$ increases
with increasing mass.

A tight correlation exists between the gas mass and the gas temperature
(Fig.~\ref{mg-tem}). The lowest temperature clusters have gas masses
below the prediction by the best fit relation. This is consistent
with the finding that the gas mass fraction drops suddenly for the lowest $\tx$
clusters. A similar behavior, though less obvious, is seen in
Fig.~\ref{mg-lx}:
low luminosity clusters have a smaller gas mass then expected from the
best fit relation.

As outlined above and in the previous Section
measurements by different authors show different
strengths of deviations from expected relations. This is obviously due
to the fact that the measurements are difficult and several effects
may lead to slight distortions of the results.
Some of the main problems are summarized
here. For low temperature clusters the gas is traced out least far
relative to the virial radius, which affects the uncertainty of almost
all measured quantities. Combination of heterogeneous
samples, e.g., poor groups from author X combined with rich clusters
analyzed by
author Y, may introduce artificial trends.
Purely X-ray flux-limited samples may be less (or more)
sensitive to deviations 
in relations compared to volume-limited samples (which are currently
difficult to assemble with the necessary quality and size),
especially when the X-ray luminosity is involved\footnote{Consider
for instance the case of a true $\lx$--$\tg$ relation
which has A) a larger scatter for lower $\tg$ clusters and B)
a steeper slope for lower $\tg$ clusters. Both deviations may appear
weaker then they are if a flux-limited sample is used because the low $\lx$
clusters for a given $\tg$ have a smaller chance to be included in the sample
than the high $\lx$ clusters.}.
The presence of central
excess emission may bias the determination of the fit parameter values
in the sense that a too small core radius and therefore possibly a too
small $\beta$ value is measured \citep[e.g.,][]{mme99}. This effect,
if present, seems not to
bias the determination of the gas mass, however, as has been shown by
comparison of gas masses determined with single and double $\beta$
models \citep[e.g.,][]{r98}. Since the excess emission is often
connected with a drop in temperature towards the very center,
overall temperatures may be biased low if this effect is not taken
into account (e.g., Chap.~\ref{a1835}; for more subtle effects on
temperature estimates due to
accretion of small cool clumps of gas see \citealt{me01}). On the other
hand the presence of substructure
may increase the temperature, the core radius and therefore possibly
also $\beta$ \citep[e.g.,][but see \citealt{na99}]{jf99}. Mass
estimates based on the hydrostatic assumption are also uncertain in
major mergers and additionally the X-ray luminosity may be strongly
affected, at least for a short time \citep[e.g.,][]{rs01}.
A possible dependence of the surface brightness
profile and gas temperature determination on the covered energy
range may introduce
artificial systematic trends, e.g., between $\beta$ and $\tx$.
If the
gas temperature depends on radius in the outer parts, the assumption
of isothermality leads to a biased mass estimate (Sect.~\ref{mest_x},
see also Chap.~\ref{a1835}). 
The contribution
of line emission becomes non negligible for low $\tx$ clusters
\citep[e.g.,][]{bh89}, therefore
deviations from the simplest self similar relations (assuming pure
bremsstrahlung emission) are expected even if clusters were self
similar. As a consequence also possible metallicity variations between groups
and clusters (\citealt{dmm99} for instance find lower metallicities for low
$\tx$ groups)
may affect the group luminosities if typical cluster
metallicities are assumed. Note that the deviations from
the self similar relations are at least partly rather small, e.g., in the case
of the $\mt$--$\tg$ relation, therefore it is necessary to be aware even
of the partially probably small effects that have just been outlined.
However, they are not likely to strongly distort the results obtained from the
mass function and some of the realistic possibilities
have explicitly been shown to cause only deviations smaller than the
statistical uncertainty, e.g., the possible presence of strong gradients in the
gas temperature (Sect.~\ref{func_pred}).

Accordingly a wide variety of models has been suggested to explain the
observations. The currently most popular models concern heating. This
heating may have occured universally prior to the assembly of groups, just
before, during or after collapse
\citep[e.g.,][]{wfn98,vs99,l00,ky00,wfn00,tn01,bm01,wfn01,bbl01}.
Since the required energy input depends on the gas density at the time
of injection and the assumed fraction of the released energy that goes
into gas heating, the cited amounts vary by an order of magnitude from
0.1--0.3 to 1--3\,keV/particle. Various heat sources have been
suggested. Among them SNe, AGN,
and population III stars. 

But also other mechanisms have been proposed. For instance \citet{mtk01}
find in simulations that cooling of the
intracluster gas alone may account for observed
deviations, due to removal of low entropy gas. Based on some
evidence (compiled from
\citealt{mdm96}, \citealt{hmb99}, and \citealt{cnt97}) that while the gas
fraction in clusters
decreases with decreasing temperature the mass fraction in stars increases
(but see, e.g., \citealt{rsb00}, who do not find a similar trend),
\citet{b00} shows that a higher efficiency of galaxy formation in groups
compared to clusters \citep[e.g.,][]{daf90} suffices to explain observations without the need of
additional heating. A comparison of the predictions of this model to 
indications for a metallicity decrease in low temperature groups
\citep[e.g.,][]{dmm99} may be interesting.

\citet{mem99} investigated the effects of clumping and substructure on
measurements of $\mg$. Using an ensemble of 48 simulated clusters they
measured a mean mass weighted clumping factor $C\equiv \langle \rog^2
\rangle / \langle \rog \rangle ^2 \approx$ 1.3--1.4 within
$r_{500}$. They showed that $\langle \mg^{\rm measured}/\mg^{\rm true}
\rangle \approx \sqrt{C}$, indicating that gas masses estimated
assuming a uniform density profile at a given radius overestimate the
true gas mass by about 16\,\%, which is of the same order as the
statistical uncertainties in this work ($\langle \Delta \mgf / \mgf
\rangle = 15\,\%$). Furthermore they found no trend with cluster bulk
properties 
like temperature or mass. The possibility of scale dependent clumping
therefore seems to be excluded as causing deviations from self similar
relations. They showed further that the presence of moderate
substructure does not bias the gas mass measurements significantly.

A synopsis of the main trends indicated from the relations presented in
this work
is now
given. The gas fraction seems to be constant for $\tx \gtrsim
2$\,keV. For lower temperatures a steep decrease of $\fg$ is
indicated. A straightforward conclusion is that smaller systems contain
comparatively less gas. At the same time there are no indications for a
flattening of the gas density profile for these groups. Neither
$\beta$ becomes smaller, that is no flattening
of the surface brightness profile in the outer part is
observed (at least out to $\rx$), nor does $\rc$ become larger,
that is no flattening in the very central regions is observed. 
The observed $\lx$--$\tx$ and $\mt$--$\tx$ relations are steeper
than the self similar
expectation over the entire $\tx$ range, whereas variations in $\fg$ seem
to occur only at low temperatures. It is therefore not necessarily clear
that these deviations (observed at different scales) are caused by one and
the same physical process.

Heating of the \icg\ must have occured at some point as evidenced by
the observed metallicities.
It is possible that the energy provided by SNe is sufficient to account
for the decrease in the gas mass fraction for the lowest temperature systems,
but possibly does not suffice to steepen the $\lx$--$\tx$ and $\mt$--$\tx$
relations over the entire $\tx$ range \citep[e.g.,][]{wfn00}.
A flatter gas density profile for low temperature clusters as one might
expect in this scenario is not observed here.
The steeper slope and lower normalization of the $\mt$--$\tx$ relation may
be caused by an increase in average formation redshift for decreasing
system mass \citep[e.g.,][]{frb00},
i.e.\ by hierarchical growth of structure. Since in this case
low mass systems would have formed in a higher density environment
this implies comparatively higher temperatures (and smaller core radii,
in agreement with the steep relation between $\rc$ and $\tx$ observed
here). However,
neglecting cooling this effect
would also \emph{increase} the luminosity on average for lower
temperature systems
thereby tending to cause a \emph{flattening} of the $\lx$--$\tx$ relation.
Since the emissivity depends on the density squared
cooling may be more important for systems formed in a higher density
environment resulting in a more efficient removal of gas and therefore a larger
\emph{decrease} in luminosity in line with observations.
In this work no attempt has been made to model or simulate these effects
in detail, but the aim has rather been to determine observationally
relevant relations in a homogeneous way for a cluster
sample not dominated by subjective selection criteria for a large
temperature interval and describe
possible scenarios accounting for them.
Even though the overall physical processes determining the main observational
appearance of galaxy clusters are well understood,
at the moment it seems unclear which effect contributes most to the
observed deviations from self similarity. It
appears unlikely that a single suggested effect can account for all of them.

\section{Mass Function}\label{funct_d}

This Section starts out with a comparison of the mass function determined
here with previous determinations. In Sect.~\ref{func_pred} constraints
on cosmological parameters are derived. And in Sect.~\ref{lmtest} 
the $\lx$--$\mt$ relation is applied to convert a luminosity function
into a `mass' function as an example.

\subsection{Comparison to Previous Estimates}\label{func_comp}

\begin{figure}[thbp]
\begin{center}
\psfig{file=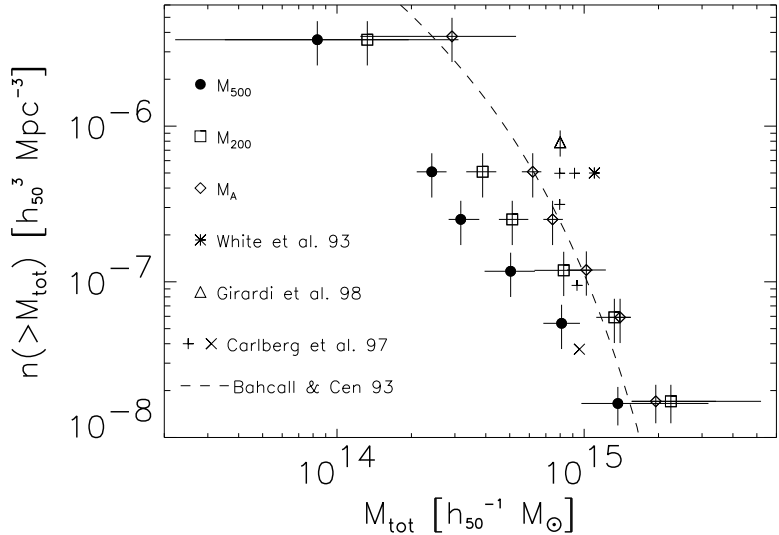,width=11.5cm,angle=0,clip=}
\end{center}
\caption{Cumulative gravitational mass function for \gcs\ using three
different outer radii. Vertical error bars give the
Poissonian errors. Horizontal bars indicate the individual
bin sizes. Each bin contains 10 clusters, apart from the highest mass
bin, which contains 13 clusters. The abundances from previous works
are determined for
$\mab$.}\label{mfunc}
\end{figure}

\citet{bc93} give a mass function constructed a) from optically selected
clusters with masses determined from the galaxy richness and b) from the
cluster X-ray temperature function given by \citet{ha91}. Very good
agreement is found for masses determined within $\rab$
between the \citeauthor{bc93} and the \gcs\ mass function
(Fig.~\ref{mfunc}).

\citet{wef93} constrain the cluster abundance by using published
values for the abundance and median velocity dispersion of richness
class $R\geq 1$ Abell clusters. It is not surprising that their density is
significantly higher than the \gcs\ density since they
intentionally use conservative mass estimates, which are overestimates
of the true cluster mass.

\citet{bgg93} and \citet{gbg98} have determined
the cluster mass function using optically selected cluster samples with masses
determined from published line-of-sight
velocity dispersions of cluster galaxies. 
At the completeness limit given by \citet[triangle in
Fig.~\ref{mfunc}]{gbg98} the cluster density given by \citet{bgg93} is
about a factor of two higher than the density given by
\citet{gbg98}. The latter authors explain this difference in density
by their on average
40\,\% smaller mass estimates due to an improved technique for removing
interlopers and the use of a surface-correction term in the virial theorem
compared to \citet{bgg93}.
The value given by \citeauthor{gbg98} itself lies significantly higher
than the comparable \gcs\ density. The reason could lie in the fact
that their optically estimated masses are on average slightly larger 
than the X-ray masses or that the external normalization for
$R\geq 1$ ($\ngx \ge 50$) clusters which they applied to their mass function is
intrinsically higher than the normalization obtained through \gcs\
directly, or both.
In Sect.~\ref{mest_o} it has been found that for
a common subsample of 42 clusters the virial masses determined
by \citet{gbg98} are on average 25\,\% larger than the masses determined
in this work. This difference might be smaller if masses out to the Abell
radius were compared, since common radii would be used in this case.
Increasing the X-ray masses artificially by 25\,\%, the diamonds
in Fig.~\ref{mfunc} shift towards higher masses, but the shift is too small
to account for the difference to the triangle.
The large scatter in the $\ngx$--$\mt$ diagram (Fig.~\ref{nmlxm}) makes
a reliable estimate of a best fit relation between these two
quantities very
difficult. Nevertheless, in order to get a rough idea of the mass
within $\rab$ that
corresponds to $\ngx=50$, I have performed fits using the minimization
methods outlined earlier and find $5.1\lesssim\mab (\ngx=50)\lesssim8.8\times
10^{14}\,\msu$. Note that this range is in agreement with the ranges
$5$--$8\times 10^{14}\,\msu$ and $5$--$7\times 10^{14}\,\msu$ 
given by \citet{bc93} and \citet{gbg98}, respectively, for $N_{\rm
gx}=50$. This mass range corresponds to a cumulative number 
density obtained through \gcs\ in the range $1.7\lesssim
n(>\mab)\lesssim 8.7\times 10^{-7}\,h_{50}^3\,\rm Mpc^{-3}$. The 
external density estimate applied to normalize the \citeauthor{gbg98}
mass function therefore is a factor 1.2--6.2 higher than the
corresponding estimate obtained here. It is therefore concluded that
both effects (masses and normalization) are important but the latter
factor is responsible for a larger fraction of the
discrepancy. Assuming both normalizations to be determined from
samples that are highly complete and representative of the local
universe this may indicate that either X-ray and optical clusters
are drawn from
different populations or that projection effects (e.g., line of sight galaxy
overdensities, which do not form a bound structure in three
dimensions) possibly bias optically determined normalizations high.

\citet{gg00} recently extended the mass function to loose galaxy
groups, finding
$n(>1.8\times10^{13}\,\msu)=1.6$--$2.4\times 10^{-4}
\,h_{50}^3\,\rm
Mpc^{-3}$, which is outside the mass range that can be tested here.
They find that
the group mass function can be described by a smooth extension of the
cluster mass function by \citet{gbg98}. Consistently this abundance is
higher than the abundance given by \citet{bc93} at that mass scale.

\cite{cmy97} have compiled and partially reestimated abundances for
local cluster samples \citep{ha91,mkd96,wef93,ecf96} for comparison with
higher redshift samples (the `$\times$' shows the density for a sample with
higher mean redshift and therefore it cannot be compared directly). In
general the comparison to the \gcs\ mass
function shows better agreement than the abundances given by
\citet{gbg98} and \citet{wef93}. 

The obvious importance of the definition of the cluster outer radius for the cluster
mass function can be directly appreciated by noting the large
differences between the mass functions determined for \gcs\ for
$\mtf$, $\mtz$, and $\mab$ in Fig.~\ref{mfunc}, especially towards
lower masses. Since for self similar clusters the mass scales with the
third power of the characteristic radius (Sect.~\ref{relat_d}),
determination of the mass within a characteristic overdensity is the
natural choice. The formally determined $\mab$ mass function is mainly
given for the comparison
with previous mass functions and it is recalled again that
especially for the low mass systems the assumption of hydrostatic
equilibrium
may not be justified out to $\rab$, and therefore the mass
estimates for $\mtf$ and $\mtz$ should be considered much
more precise than the estimates for $\mab$.
Note that it also seems likely that especially small clusters and
groups are not in virial equilibrium out to $\rab$.

\subsection{Comparison to Predicted Mass Functions}\label{func_pred}

In this Section one of the main results of this work is derived and
discussed. The mean density of the universe is determined by a comparison
of the observed and predicted mass function.
The formalism outlined in Sect.~\ref{back:tmf} is used for the
calculation of mass functions for given cosmological models.
As mentioned earlier due to the
low redshift range spanned by \gcs , the effect of a redshift
correction is very small and therefore $z=0$ is set for all
calculations, unless noted otherwise.

Similarly to the work of \citet{irb01} the statistical
uncertainty in the mass determination is incorporated
in the model mass function as
\begin{eqnarray}
\frac{d\tilde n(M)}{dM}\equiv \frac{1}{\vmax(M) }\,\int^\infty_{-\infty}\frac{dn(M')}{dM'}\,\vmax(M')\, \nonumber\\
\times\,(2\pi\,\bar\sigma_{\mt ,\log}^2)^{-1/2}\,\exp\left(\frac{-(\log M' -
\log M)^2}{2\,\bar\sigma_{\mt ,\log}^2}\right) d\log M'\,,
\label{eq:nschl}
\end{eqnarray}
where $\bar\sigma_{\mt ,\log} =
0.12$ represents the logarithmic mean
mass measurement uncertainty. Note that since \gcs\ is flux-limited and not
volume-limited the weighting has to be performed on the
mass distribution, $N(M)/dM$, rather than on the mass function
itself. The effect of this weighting on the model mass function is a
slight amplitude increase at the high mass end.

\begin{figure}[thbp]
\begin{center}
\psfig{file=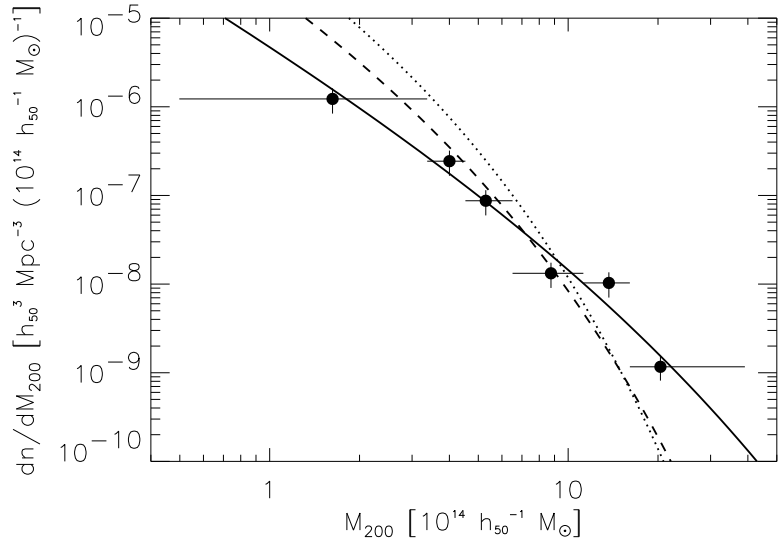,width=11cm,angle=0,clip=}
\end{center}
\caption{\gcs\ mass function compared to the best fit model mass function
with $\om=0.12$ and $\sigma_8=0.96$ (solid line). Also shown are the best fit model
mass functions for fixed $\om=0.5$ ($\Rightarrow \sigma_8=0.56$,
dashed line) and $\om=1.0$ ($\Rightarrow \sigma_8=0.43$, dotted
line).}\label{PS0}
\end{figure}

\begin{figure}[thbp]
\begin{center}
\psfig{file=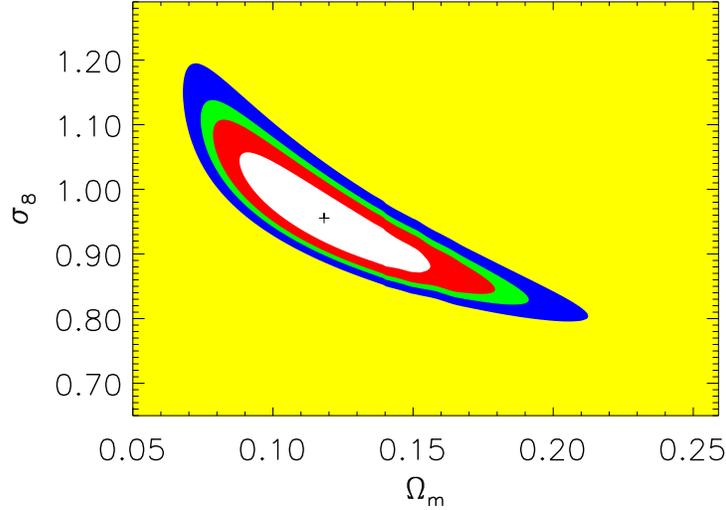,width=11cm,angle=0,clip=}
\end{center}
\caption{Statistical confidence contours for the $\chi^2$
procedure. The cross indicates the minimum. Ellipses indicate 68\,\%, 90\,\%,
95\,\%, and 99\,\% confidence levels for two interesting
parameters.}\label{banana}
\end{figure}

For the modeling to be independent of the precise knowledge of the
$\lx$--$\mt$ relation the quantitative comparison has been performed
using a standard $\chi^2$ procedure on the
differential binned mass function given in Fig.~\ref{PS0} (rather than
using a maximum likelihood approach on the mass distribution).
The $\chi^2$ values have been calculated in a fine grid of $\om$ and
$\sigma_8$ values assuming a flat cosmic geometry. The cosmological
constant enters the calculation only through $\dc$ and therefore
has a negligible influence here. The minimum and statistical error ellipses for some
standard confidence levels (c.l.)\ are given in Fig.~\ref{banana}. The tight
constraints obtained show that with \gcs\ one can go beyond determining an
$\om$--$\sigma_8$ relation and put limits on $\om$ and $\sigma_8$
individually.  It is found that
\begin{equation}
\om=0.12^{+0.06}_{-0.04}\quad {\rm and}\quad  \sigma_8=0.96^{+0.15}_{-0.12}
\label{eq:omres}
\end{equation}
(90\,\% c.l.\ statistical errors for two interesting parameters),
indicating a  relatively low value for the density parameter.
The large covered mass range and the specific region in
$\om$/$\sigma_8$ parameter space allow to derive these tight
constraints from a local cluster sample.
For comparison for a given $\sigma_8$ value the
$\om$ value which minimizes $\chi^2$ is calculated.
These pairs can then
roughly be described by a straight line in log space given by
(using the usual notation)
\begin{equation}
\sigma_8=0.43\,\om^{-0.38}\,.
\label{eq:omsig}
\end{equation}
In
Fig.~\ref{PS0} also the best fit model mass functions for
given $\om=0.5$ and $\om=1.0$ are plotted and one notes immediately
that these value pairs give a poorer description of  the shape of
the mass function. Previous estimates have
generally yielded a combination of slightly higher $\om$/$\sigma_8$
values (e.g., \citealt{p99}, Sect.~17.2). It has to be noted,
however, that for instance in the important work of \citet{wef93}, who
find $\sigma_8\approx 0.57\,\om^{-0.56}$, the authors explicitly
state that their estimates of
$\sigma_8$ are probably biased high due to their conservative mass
estimates.

Due to the large given ranges of several orders of magnitude in mass and
especially density the $\chi^2$
values have been determined from comparison model/data naturally in
log space. However, it has been verified that the same calculation in
linear space yields best fit values lying within the 68\,\% error ellipse.

In Sect.~\ref{funct} arguments have been given why
$\vmax(\lx)$ has been used for the determination of the mass function.
Nevertheless if $\vmax(\mt)$ is used instead (see Fig.~\ref{vm_cp_ns}),
consistent results are obtained. Since in this case one wants to
estimate $L$ from $M$ the relation ($L\mid M$) is the appropriate one
\cite[e.g.,][]{ifa90}. Performing a fit to a mass function constructed
with $\vmax(\mt)$ results in best fit values $\om=0.14$ and
$\sigma_8=0.86$, which is consistent with the error range given in
(\ref{eq:omres}).

As shown in Sect.~\ref{mest_x} it is possible that the assumption
of isothermality leads to an overestimate of the cluster masses on
average. Therefore the robustness of the
results has been tested by multiplying the isothermal cluster
masses by 0.799 and
recalculating the minimum. As expected values for both $\om$ and
$\sigma_8$ are found to be lower. But the new minimum is contained
well within the error ranges given in (\ref{eq:omres}). On the other
hand in Sect.~\ref{mest_o} it has been shown by comparison with optical mass
measurements that masses could possibly be underestimated. After
increasing all cluster masses by 25\,\% a fit shows that both 
$\om$ and $\sigma_8$ become slightly larger, but again they lie well
within the range (\ref{eq:omres}).
Therefore these tests indicate that the constraints obtained here are fairly
insensitive against systematic uncertainties in the mass measurements.

As mentioned in Sect.~\ref{massd} it has been shown that the small
range of low redshifts covered here ensures that no redshift
corrections need to be applied. Nevertheless it has been tested whether or
not the best fit parameter values change if $\mtz$ is calculated using 
$\roc =\roc (z)$ for the extreme (strong evolution) case
($\om=1,\ol=0$), i.e.\ $\roc =
4.6975\times10^{-30}\,(z+1)^3\,\rm g\,cm^{-3}$ for each cluster
redshift (Eqs.~\ref{roc} and \ref{h0e}). The model mass function is then calculated for the mean
redshift of \gcs , $\langle z\rangle=0.05$, using the formulae
outlined in Sect.~\ref{back:tmf}. It has been found that within
the grid used here the best fit values do not change at all and also the error
ellipses are almost not affected, thereby confirming that the
application of redshift corrections does not affect the results.

The value $H_0=71\,\rm km\,s^{-1}\,Mpc^{-1}$ has been adopted for the
calculation of model mass functions based on
the recent combination of constraints obtained using the Hubble Space
Telescope \citep{mhf00}. Setting the Hubble constant to their lower
limit, $H_0=65\,\rm km\,s^{-1}\,Mpc^{-1}$, does not affect the best fit
parameter values significantly. Using even $H_0=60\,\rm km\,s^{-1}\,Mpc^{-1}$
changes the results for $\om$ and $\sigma_8$ only well within the
68\,\% error ellipse. Therefore the constraints obtained here on
these cosmological
parameters do not depend significantly on the specific choice of $H_0$.

The data point in Fig.~\ref{PS0} that may be affected most by cosmic
variance is the one at the lowest mass, since the maximum search
volume is smallest for the clusters in this bin. Therefore
the sensitivity of the best fit results on this last point has been
tested by ignoring
it. The decrease in the covered mass range of course increases the
resulting error ellipse, but the best fit values vary only within
the (smaller) 68\,\% error ellipse. It may be worth noting that
leaving out the highest mass bin or leaving out the highest \emph{and}
lowest mass bin changes the best fit values only within the 90\,\%
error ellipse.

Since it has been found that for the estimate of the statistical errors one
needs to explore ranges $\om < 0.1$ for $\ob\sim 0.04$ I regarded it
necessary to check
if the approximation to the transfer function as given in
Sect.~\ref{back:tmf} is still applicable. 
Recently \citet{eh98} derived a fitting function, that includes, e.g.,
also the oscillations induced by the baryons, which gives a better
description of transfer functions computed with CMBFAST \citep{sz96} than
fitting functions for zero or small baryon contribution to the
total matter density derived previously. Therefore
this improved version of an analytic transfer function has been
incorporated in the $\chi^2$
procedure. It has been found that within the used grid the minimum does not
change at all, the choice of the \citet{bbk86} fitting function
combined with the shape parameter given by \citet{s95} therefore seems
to be accurate enough for the purposes here. However,
the confidence contours towards low
$\om$ are getting compressed when the \citet{eh98} fitting function is
used, thereby slightly decreasing the area of the error ellipse for a given
confidence level. Since the latter statistical error
ellipse can be regarded as more realistic, this one is shown
in Fig.~\ref{banana}.

It has also been tested whether or not the recently found deviations of the PS
formalism compared to large $N$-body simulations
\citep[e.g.,][]{gbq99,jfw01} have a significant
influence on the results obtained here. The best fit
PS model ($\om=0.12$, $\sigma_8=0.96$) has been compared to the model
obtained using the
`universal' mass function (fit to $N$-body simulations,
\citealt{jfw01}) for the same parameter values. These two models agree
well for $M\lesssim 10^{15}\,\msu$. The differences become larger than
the size of the Poissonian error bars (Fig.~\ref{PS0}) for $M\gtrsim
2\times 10^{15}\,\msu$, in the sense that the \citeauthor{jfw01} mass function
predicts higher cluster abundances than PS. For larger values of $\om$
the differences become comparable to the size of the error bars at
lower masses, e.g., for $\om =\sigma_8=0.5$ around $M\sim
5\times 10^{14}\,\msu$. To estimate the influence of these differences
on the best fit values derived using the PS mass function,
the parameter values of the \citeauthor{jfw01} model have been adjusted
to reproduce the PS mass
function, where $\om =0.15$ and $\sigma_8=0.86$ have been found.
The value for $\om$
becomes slightly larger but the combination of both values is still
contained within the 90\,\% error ellipse. It is therefore
concluded that the differences between the model mass functions do not
significantly affect the interpretation of the \gcs\ mass
function. Moreover this test can be regarded as confirmation of the validity
of the PS mass function for the accuracy needed here.

Also in this work the normalized baryon density may be estimated, using
the mean gas mass fraction determined in Sect.~\ref{fgas}. One may
set the gas mass fraction in clusters, being the largest collapsed objects,
equal to the baryon fraction in the universe, i.e.\ $\fg=\ob/\om$.
It has been tested if consistent results are obtained if the value
determined for $\ob$ here
is used
for the calculation of model mass functions
instead the one given in Sect.~\ref{back:tmf}.
Using $\ob=0.19\pm 0.08\,h_{50}^{-3/2}\,\om$
results in model mass functions very similar to the ones calculated using the
baryon fraction given by \citet{bt98}. It is therefore not surprising
that the best fit values for $\om$ and $\sigma_8$ vary only well within the
68\,\% error ellipse if the former $\ob$ determination is used.

It is worth noting that combining these two measurements of
$\ob$, i.e.\ $0.19\pm 0.08\,h_{50}^{-3/2}\,\om=0.0772\,h_{50}^{-2}$,
gives further evidence for a low value for $\om$ by yielding an
estimate $\om=0.34^{+0.22}_{-0.10}$ using $H_0=71\,\rm
km\,s^{-1}\,Mpc^{-1}$ \citep{mhf00}, where the error has been determined
from the standard deviation of $\ob$ given above. This value for $\om$
is an upper limit since baryons contained
in, e.g., the cluster galaxies have been neglected. A low value for
$\om$ has been indicated by
this method for smaller cluster samples by several works previously (e.g.,
\citealt{wf91,bhb92,b93,wne93,wf95,djf95,e97,ef99}; but see \citealt{sb01}).
One has
to keep in mind, however,
that this estimate extrapolates the gas fraction from cluster scale to
cosmic scales. For the clusters in the sample used here
it has been found that the gas fraction is not constant but varies with
radius and cluster mass (Sect.~\ref{fgas}), therefore further observational
tests of this assumption may be useful. 

In summary, even though a conservative estimate has been made
by neglecting the possible presence of gas temperature gradients,
previous estimates obtained from cluster abundances mostly yielded
higher values for $\om$ and $\sigma_8$ \citep[e.g.,][]{wef93,gbg98}.
For these works one could expect this already from
Fig.~\ref{mfunc} and possible reasons
for this have been discussed in Sect.~\ref{func_comp}.
The results are, however, in good agreement with the
results from the power spectral analysis of the REFLEX
clusters. \citet{sbg00a} find for a given $\Lambda$CDM model
($\om=0.3$) that $\sigma_8=0.7$ represents the data well, which is
very close to $\sigma_8 =0.68$ expected using relation (\ref{eq:omsig})
found here. Moreover the (1-$\sigma$) range $0.17\leq \om \leq 0.37$ (using
$h=0.71$ in their Eq.\ 18) quoted for $\om$ directly is also
consistent with the 90\,\% range determined here. Furthermore
\citet{irb01}, who analyzed the \gcs\ temperature function using
temperatures from homogeneously reanalyzed \as\ data, find 
$\om=0.19_{-0.06}^{+0.08}$ and $\sigma_8=0.96_{-0.10}^{+0.11}$
(90\,\% c.l.\ statistical uncertainty; assuming an open cosmology) by
comparison with Press--Schechter 
models, which is in good agreement with the results presented here.

\subsection{Mass Function estimated using $\lx$--$\mt$
Relation}\label{lmtest}

\begin{figure}[thbp]
\begin{center}
\psfig{file=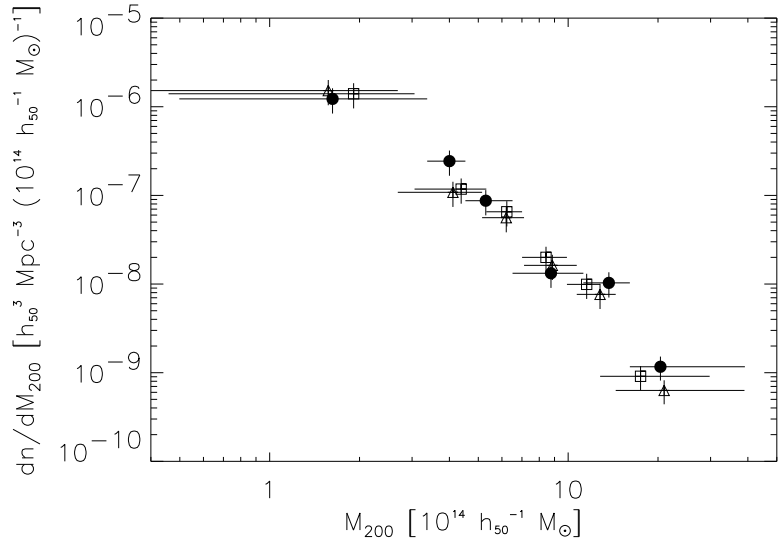,width=11cm,angle=0,clip=}
\end{center}
\caption{\gcs\ mass function (filled circles) compared to `mass'
functions estimated using measured luminosities and luminosity--mass
relations (open symbols). Squares have been calculated using the
($M\mid L$) relation and triangles using the ($L\mid M$) relation  for
\gcs\ clusters.}\label{mf_lm}
\end{figure}

To show consistency and the power of the $\lx$--$\mt$ relation,
fits to `mass' functions, where masses have been
estimated from the measured X-ray luminosity have also been performed.
Relations for the flux-limited sample have been used to get the best
mass estimate for the cluster luminosities included in \gcs . Mass
functions for the two extreme relations 
($M\mid L$) with $\alpha=1.538$ and ($L\mid M$) with $\alpha=1.310$
are shown in Fig.~\ref{mf_lm}. First of all one notes the fairly good
agreement between the three mass functions. In detail the differences
between the two mass functions estimated from different
luminosity--mass relations
can be understood by considering that
at the low luminosity end the steeper relation predicts a higher
mass for a given luminosity than the shallower relation, resulting in a
shift towards higher masses of the mass function. At the high mass side the effect is
opposite, resulting in a shift towards lower masses for the steeper
relation. On average the points for the steeper relation lie higher
which is caused by the fact that a steeper relation results in a
smaller $dM$ on average, which gives rise to an increased $dn/dM$.
The differences to the mass function calculated using the measured
masses are again understood by a similar comparison and are partially
caused by a
possible deviation of the shape of the $\lx$--$\mt$ relation from 
a pure power
law.

Despite these small differences performing an actual
fit\footnote{For the fit the corresponding scatter in
$\log M$ for the two relations ($L\mid M$) and ($M\mid L$),
$\slm=0.22$ and $0.21$,
respectively (Tab.~\ref{tab:scat}), has replaced the mass
measurement error, $\bar\sigma_{\mt ,\log}=0.12$, in (\ref{eq:nschl}).}
results in the ($\om$, $\sigma_8$) values (0.14, 0.85) for ($L\mid M$)
and (0.22, 0.74) for ($M\mid L$). The first case is consistent with the error
range given in (\ref{eq:omres}) and in the second case the 90\,\%
statistical error ellipse overlaps with the 90\,\% ellipse in Fig.~\ref{banana}.
From the above and Fig.~\ref{PS0} it is clear that using steeper
luminosity--mass relations results in
higher values for $\om$ and lower values for $\sigma_8$.
Here one wants to estimate $M$ from $L$ and therefore ($M\mid L$) is the
appropriate relation to use.
This test demonstrates that with in general easy to
obtain X-ray luminosities of a statistical cluster sample and with
the knowledge of the $\lx$--$\mt$ relation (even if
approximated as simple power law) and its scatter realistic constraints on
cosmological parameters can be set.
\section{Total Gas and Gravitating Mass in Clusters}\label{func_dens}

\begin{figure}[thbp]
\begin{center}
\psfig{file=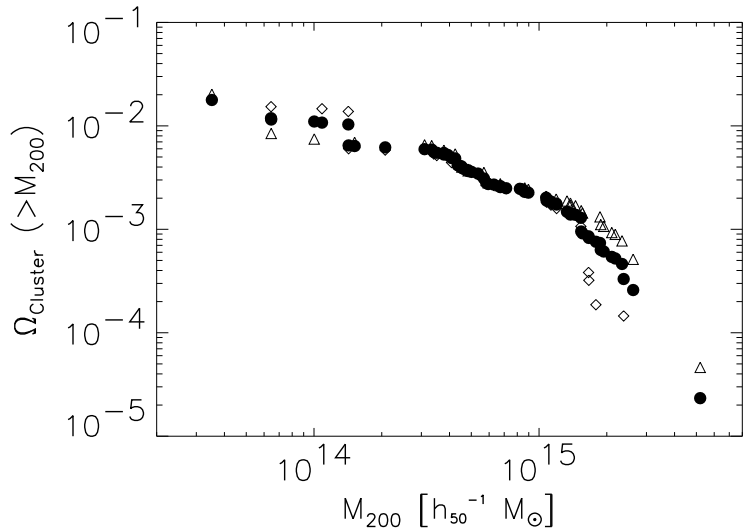,width=11cm,angle=0,clip=}
\end{center}
\caption{Gravitational mass density contained in galaxy clusters as a function of
minimum mass. Filled circles indicate the complete \gcs , open
triangles indicate the 34 clusters north of the galactic plane, and
open diamonds the
29 clusters at southern galactic latitudes included in \gcs .}\label{mdens}
\end{figure}

To estimate the fraction of the gravitational mass density relative to the
critical density contained in galaxy clusters,
\begin{equation}
\oc(>\mti) =\frac{1}{\roc} \int_{\mti}^\infty \mt\,\phi(\mt)\,d\mt\,,
\label{eq:oclmint}
\end{equation}
the individual cluster masses divided by the
corresponding maximum search volumes have been summed up for \gcs , i.e.\ 
\begin{equation}
\oc =\frac{1}{\roc} \sum_i \frac{\mtzi}{\vmaxi}\,.
\label{eq:oclm}
\end{equation}
Note that the determination of $\oc$ is
independent of the Hubble constant. The cumulative diagram for $\oc(>\mtz)$
is shown in Fig.~\ref{mdens}. In order to perform a conservative error estimate,
\gcs\ has been split into two parts with $\bii\ge +20\,\rm deg$ and $\bii\le
-20\,\rm deg$, and the results for these subsamples are also shown in the
Figure. This estimate is
conservative because \gcs\ is about twice as large as each
subsample. Taking the second and third lowest mass clusters together with
the maximum mass range given by their individual uncertainties, one obtains
\begin{equation}
\oc = 0.012^{+0.003}_{-0.004}
\label{eq:ocl}
\end{equation}
for masses larger than $6.4^{+0.7}_{-0.6}\times
10^{13}\,\msu$, i.e.\ the total gravitating mass
contained within the virial radius of clusters amounts only to 
$1.2^{+0.3}_{-0.4}$ percent of the total mass in a critical density
universe. Combined with the best estimate from Sect.~\ref{func_pred},
$\om=0.12$, this implies that
about 90\,\% of the total mass in the universe resides outside
virialized cluster regions above the given minimum mass. If galaxies
trace mass it also follows that by far most of the galaxies do not sit
in clusters. This result is consistent with the general presumption
that clusters are rare objects, rare peaks in the density distribution
field.

Comparing the diagram to the 
mass fraction $\oc = 0.028^{+0.009}_{-0.008}$ in clusters with masses
larger than $2\times 10^{14}\,\msu$ given by \citet{fhp98},
based on the mass function determined by
\citet{bc93}, one finds that their estimate is a factor 4--5
higher. However, the \citeauthor{bc93} mass function is given for
$\mab$ and one gets a consistent result if $\oc$ is calculated
using the cluster masses formally determined within $\rab$ here. It needs
to be pointed out that at $\mab\sim 2\times 10^{14}\,\msu$
the typical virial radius is $\sim 1\,\mpc$ and a mass
determination at $3\,\mpc$ based on the assumption of virial
equilibrium may therefore be rather uncertain and possibly leading to
overestimates of $\oc$. This becomes more crucial if mass functions
for $\mab$ are extrapolated even down to galaxy masses.
For instance for the lowest mass group contained in \gcs\ one finds that
$\mtz=3.5\times 10^{13}\,\msu$ while $\mab=1.2\times 10^{14}\,\msu$, i.e.\
$\mab$ is a factor 3.4 larger.
This way
\citet{fhp98} find $\oc = 0.12\pm 0.02$ within the mass range $2\times
10^{12}$--$2\times 10^{14}\,\msu$, which, compared to the results
from Sect.~\ref{func_pred}, would account already for almost all mass
in the universe.

\begin{figure}[thbp]
\begin{center}
\psfig{file=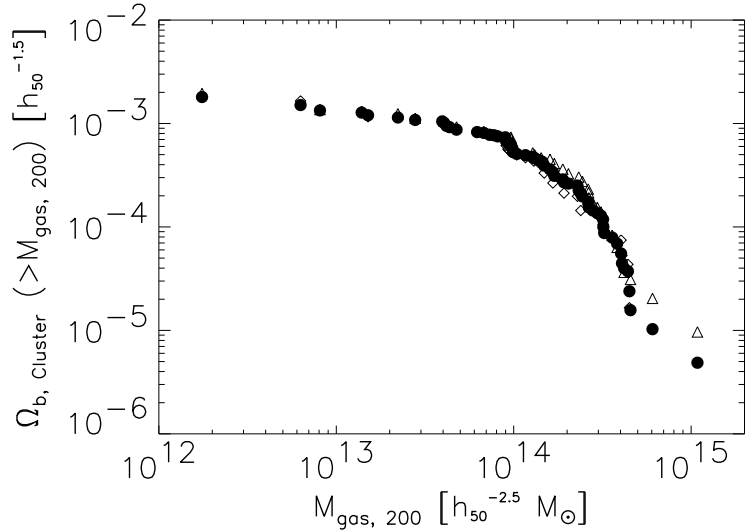,width=11cm,angle=0,clip=}
\end{center}
\caption{Gas mass density contained in galaxy clusters as a function of
minimum gas mass. Symbols have the same meaning as in
Fig.~\ref{mdens}.}\label{mgdens} 
\end{figure}

Replacing $\mtz$ with $\mgz$ in (\ref{eq:oclm}) and performing an
analogous calculation yields the fraction
of the total gravitating mass in the universe contained in the intracluster
gas of galaxy clusters,
$\obc$. The result is shown in Fig.~\ref{mgdens}. One notes
that the cumulative curve flattens out with decreasing gas mass more
strongly than the analogous plot in Fig.~\ref{mdens}, which is caused by
the decreasing gas mass fraction with decreasing mass (Fig.~\ref{fg_mt}).
The two highest gas mass triangles lie a factor
of two higher than the two highest gas mass solid circles by construction,
since the same cluster masses (originating in one hemisphere)
are divided by volumes which differ by about a factor of two.
Combining the results for the second and third
lowest gas mass clusters in the same way
as for (\ref{eq:ocl}) one obtains
\begin{equation}
\obc = 0.0015^{+0.0002}_{-0.0001}\,h_{50}^{-1.5}
\label{eq:obcl}
\end{equation}
for gas masses larger than $6.9^{+1.4}_{-1.5}\times
10^{12}\,\msug$. 
\citet{fhp98} give a gas mass fraction 
$\obc = 0.0044^{+0.0028}_{-0.0021}\,h_{50}^{-1.5}$
in clusters with gravitational
masses larger than $2\times 10^{14}\,\msu$. Using the value they
quote for $\fg$ this corresponds to gas masses larger than $3.2\times
10^{13}\,\msug$ and is again a factor of 4--5 higher than the value
determined at this minimum gas mass from Fig.~\ref{mgdens}. Their
result is still higher but consistent if gas masses determined within
$\rab$ are compared. If $\obc$ is taken as an estimate of the baryon
content in clusters it has to be regarded as a lower limit since
especially for the low mass clusters the cluster galaxies may
contribute a non negligible amount of baryons \citep[e.g.,][]{daf90}.

\chapter{Conclusions}
\label{conc}

In this work tight constraints on the mean mass density in the universe --
one of the fundamental cosmological parameters -- have been obtained.
Use has
been made of high quality X-ray observations of a well defined sample of
galaxy clusters.
The strongest constraints obtained from the cluster mass function
$\om=0.12^{+0.06}_{-0.04}$
are well consistent with the contraint from the cluster gas mass fraction
$\om\lesssim0.34$ and the estimate from cluster mass to light ratios
$\om\approx 0.15$.
Previous measurements of cluster abundances
indicated somewhat higher values for $\om$ than found here.
Possible reasons for this have been
discussed in Sects.~\ref{func_comp} and \ref{func_pred}. However, the results
obtained here are consistent with the results from the
power spectral analysis based on
the 462 galaxy clusters of the REFLEX survey.

\begin{figure}[thbp]
\vspace{0.0cm}
\psfig{file=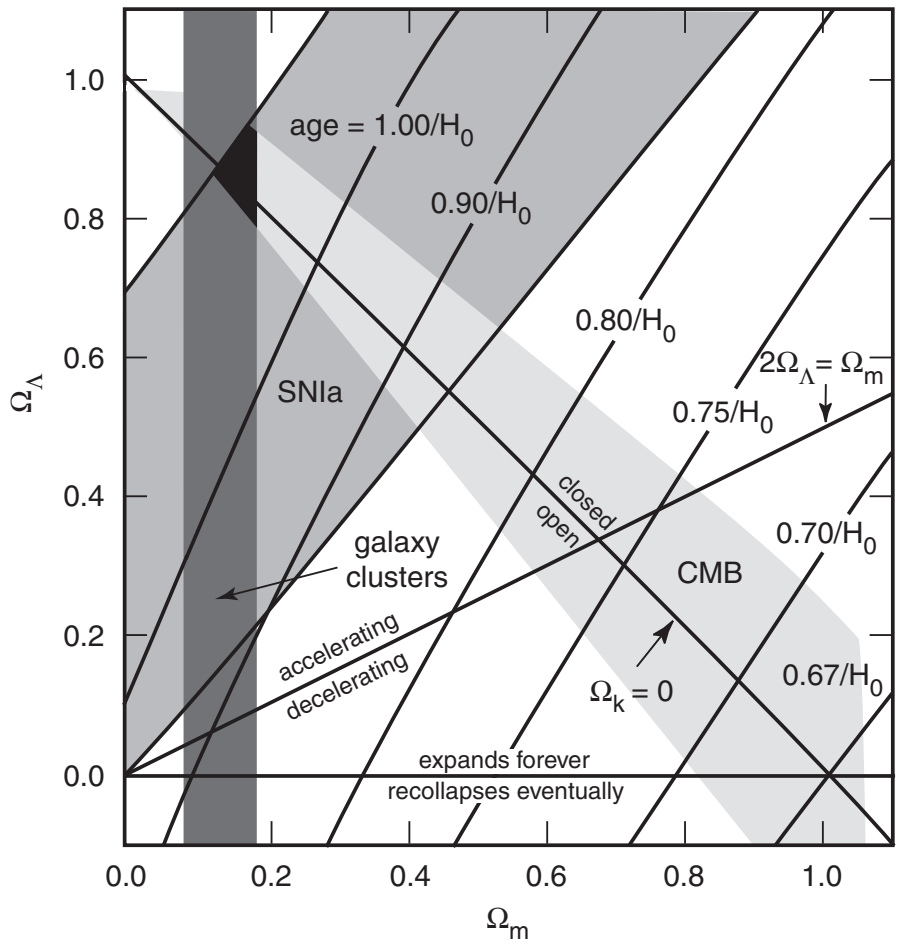,width=14.5cm,angle=0,clip=}
\caption{Observational constraints on two fundamental cosmological
parameters. The area labeled galaxy clusters shows the constraints
obtained in this work.}\label{const} 
\end{figure}

How do the results of this work fit into the general
picture? How do they compare to constraints derived from completely
independent measurements und what conclusions can be drawn from a
comparison? Among the various methods
that have been applied \citep[e.g.,][]{pe99},
currently two more methods seem especially
encouraging for constraining the relevant parameters
\citep[e.g.,][]{bop99}.
As mentioned in the Introduction one approach uses measurements
of temperature fluctuations of the CMB,
another distant SNe as standard candles.
The constraints (95\,\% c.l.)\ achieved recently by these two methods
\citep[e.g.,][]{dab00,pag99}
are compared in an $\om$--$\ol$ diagram to the results obtained
here in Fig.~\ref{const}.
The first noteworthy aspect is that these independent approaches
nicely complement each other in the way
they constrain different regions in parameter space.
Secondly it is found that all three error
ranges overlap; there is an area for which all measurements are consistent. 
This consistency is very encouraging.

Note that Fig.~\ref{const} shows the statistical (90\,\%) error range of the
measurements performed here.
Possible systematic uncertainties arise from the cluster mass measurements,
from the sample construction,
and from the formalism used
for the comparison to predicted mass functions.
However, several sources of possible systematic errors have been checked
and the introduced uncertainties have been found to be smaller than
the statistical ones. This gives strong confidence to the results. 

What are the implications of the fact that the concordance
is confined to a specific
region? In Fig.~\ref{const} some additional lines are drawn that separate
special regions. The region of consistency is very close to the
line $\om+\ol=1$, i.e.\
$\ok=0$ (Eq.~\ref{oms}). This indicates that the universe may have a flat
geometry. Furthermore the region is located clearly within a
parameter region where
the universe will never collapse again, but expand infinitely
($\om\le1\wedge\ol\ge0$).
Moreover the region lies well above the line that separates
the states in which the expansion of the universe is
decelerating or accelerating ($2\,\ol=\om$).
This implies that not only will the universe expand forever but it will do so
ever faster.

Further it is assuring to note that the concordance region indicates
that the universe is older than $H_0^{-1}=13.8$\,Gyr
(for $H_0=71\,\rm km\,s^{-1}\,Mpc^{-1}$), since the age inferred
for the oldest globular clusters ranges between 11.5--15.8\,Gyr
\citep[e.g.,][]{bh95,sdw97,cgc00}.

The results on the mass to light ratios and gas mass fractions indicate
that the nature of most of the matter in the universe is as yet unknown.
Moreover most recent measurements of the CMB anisotropies indicate
$\om + \ol \approx 1$, i.e.\ a geometry close to being flat
\citep[e.g.,][]{dab00,bab00,jab01,phl01,wtz01,dab01}.
Combining these results with
the constraint on $\om$ obtained in this work one finds
$\ol \approx 0.88$.
This implies that the observed matter accounts only for a
tiny fraction of the essence that determines the geometry of
the universe.

Observational progress continues in all wavelengths.
Improvements on the accuracy of the derived parameters from
the X-ray side are achievable in the near future, e.g., by a comparison
of the
local mass function presented here with mass functions of well defined
higher redshift samples of massive clusters.
A large spanned redshift interval allows to further weaken the degeneracy
between $\om$ and $\sigma_8$, enabling a more independent measurement of
$\om$ \citep[e.g.,][]{pe80}. The reason lies in the fact that the evolution of
clusters depends on $\om$.
For this purpose 
observations of the most luminous REFLEX clusters by \xmm\ are
already scheduled.
\chapter{Summary}
\label{summ}

An X-ray selected and X-ray flux-limited sample
comprising the 63 X-ray brightest galaxy clusters in the
sky (excluding the galactic band, called \gcs ) has been constructed based on the
\ro\ All-Sky Survey.
The flux limit
has been set at $2\,\esc$ in the energy band $\eb$. It has been shown
that a high completeness is indicated by several tests. Due to the
high flux limit this sample can be used for a variety of
applications requiring a statistical cluster sample without any
corrections to the effective survey volume.

Mainly high quality pointed observations have been used to determine
fluxes and physical cluster parameters. It has been shown that a tight correlation
exists between the X-ray luminosity and the gravitational mass using
\gcs\ and an extended sample of 106 galaxy clusters. The
relation and its scatter have been quantified using different fitting
methods. A comparison to theoretical and numerical predictions shows
an overall agreement. This relation may be directly applied in large X-ray
cluster surveys or dark matter simulations for conversions between
X-ray luminosity and gravitating mass.

Data from the performance verification phase of the recently launched
X-ray satellite observatory \xmm\ on the galaxy cluster Abell 1835 has been 
analyzed, in order to test the assumption of isothermality of the cluster
gas in the outer parts applied throughout the work.
It has been found that the measured outer temperature profile is consistent
with being isothermal. 
In the inner regions a clear drop of the temperature by a factor
of two has been found.

Physical properties of the cluster sample have been studied
by analyzing relations between different cluster parameters.
The overall properties are well understood but in detail deviations
from simple expectations have been found.
It has been found that the gas mass fraction does not vary as
a function of intracluster gas temperature. For galaxy groups
($\kb\tx\lesssim 2$\,keV), however, a steep drop of $\fg$ has been observed.
No clear trend of a variation of the shape of the surface brightness
profile, i.e.\ $\beta$, has been observed as a function of temperature.
The $\lx$--$\tx$ relation has been found to be steeper than expected
from simple self similar models, as has been found by previous authors.
But no clear deviations from a power law shape down to $\kb\tx=0.7$\,keV have
been found.
The $\mt$--$\tx$ relation found here is steeper than expected
from self similar models and its 
normalization is lower compared to hydrodynamic simulations, in
agreement with previous findings.
Suggested scenarios to account for these deviations, including heating
and cooling processes, and observational difficulties have been
described.
It appears that a blend of 
different effects, possibly including a variation of mean formation
redshift with system mass, is needed to account for the observations
presented here.

Using \gcs\ the gravitational mass function has been determined for
the mass interval $3.5\times 10^{13}< \mtz < 5.2\times
10^{15}\,\msu$. Comparison with Press--Schechter mass functions
has yielded tight constraints on
the mean matter density in the universe and the amplitude
of density fluctuations. The
large covered mass range has allowed to put constraints on the parameters
individually. Specifically
it has been found that
$\om=0.12^{+0.06}_{-0.04}$ and $\sigma_8=0.96^{+0.15}_{-0.12}$ (90\,\%
c.l.\ statistical uncertainty). 
This result is consistent with two more estimates of $\om$ obtained in this
work using different methods. The mean intracluster gas fraction of the 
106 clusters in the extended sample combined with predictions from 
the theory of nucleosynthesis indicates $\om\lesssim 0.34$. The cluster
mass to light ratio multiplied by the mean luminosity density implies
$\om\approx 0.15$.
Various tests for
systematic uncertainties have been performed, including comparison of
the Press--Schechter mass function with the most recent results from
large $N$-body simulations, yielding deviations smaller than the
statistical uncertainties. For comparison
the best fit $\om$ values for fixed $\sigma_8$ values
have been determined yielding the relation $\sigma_8=0.43\,\om^{-0.38}$.

The mass function has been integrated to obtain the
fraction of the total gravitating
mass in the universe contained in galaxy clusters. Normalized to the critical
density it has been found that
$\oc = 0.012^{+0.003}_{-0.004}$ for cluster masses larger than $6.4^{+0.7}_{-0.6}\times
10^{13}\,\msu$. With the value for $\om$ determined here
this implies that about 90\,\% of the mass
in the universe resides outside virialized cluster regions.
Similarly it has been found that the fraction of the
total gravitating mass which is contained
in the intracluster gas,
$\obc = 0.0015^{+0.0002}_{-0.0001}\,h_{50}^{-1.5}$
for gas masses larger than $6.9^{+1.4}_{-1.5}\times
10^{12}\,\msug$, is very small.

\pagenumbering{Roman}
\bibliographystyle{apalike}
\addcontentsline{toc}{chapter}{Bibliography}
\bibliography{$tx/inputes/aamnem99,$tx/inputes/bibo_engl}
\clearpage
\addcontentsline{toc}{chapter}{List of Figures}
\listoffigures
\clearpage
\appendix
\chapter{Acknowledgements}
\thispagestyle{empty}

During the preparation of this work I have benefitted from
many people. Here I would like to express my gratitude to a few of
them\footnote{Titles are omitted for brevity.}.

Hans B\"ohringer is an excellent scientist and I have gained a 
lot from discussions with him. He has supported this work
in various ways by
giving advice, encouraging independent initiatives, and providing
the chance to participate in exciting related projects.
Gregor Morfill has facilitated the start of this work and has been
supporting it continuously.
Joachim Tr\"umper has made possible a valuable research visit to Tokyo
Metropolitan University.
My officemate Peter Schuecker has introduced me to a variety
of interesting subjects by sharing
his knowledge during innumerable discussions and his kind
attitude has ensured a friendly relation in the tiny `container'
as well as in the spacious new building.
With the members of the galaxy cluster group at
MPE, especially Yasushi Ikebe, Paul Lynam, Alexis Finoguenov,
and J\"org Retzlaff, I have enjoyed exciting scientific and
worldly discussions. My wife Angelika has given me
continuous encouragement especially during difficult stages.

\as\ temperatures for two clusters in this work have been determined
while enjoying the outstanding hospitality of the cluster group
at Tokyo Metropolitan University,
especially Takaya Ohashi, Ken'ichi Kikuchi, Akihiro Kushino, Tae Furusho,
and Yoshitaka Ishisaki.
Useful discussions about the
\xmm\ background determination have been held with
S\'ebastien Majerowicz, Doris Neumann, and
Monique Arnaud at CEA-Saclay.
The person who initially drew my attention to 
galaxy clusters was 
Simon White by advising a seminar talk at the University of Munich.

This work has benefitted a lot from the use
of unpublished catalogs from the REFLEX and NORAS teams.
Yasushi Ikebe has provided three cluster temperatures prior to
publication. Volker Mueller, J\"org Retzlaff and Peter
Schuecker have provided results from OCDM simulations. 
Marisa Girardi provided electronic tables of the mass determinations
based on galaxy velocity dispersions. Alexis Finoguenov provided
electronic tables of partly unpublished X-ray mass measurements.
Robert Schmidt has made available electronic tables
of the Chandra results on A1835.
Some of the basic programs used in this work are extensions
of software developped by former cluster group members, especially
Raimund Schwarz.
The code for calculation of the universal mass
function has been provided by
Adrian Jenkins. The code for calculation of the \citeauthor{eh98} transfer
function has been provided by these authors.
Software provided by Manami Sasaki has been used to correct individual photons,
detected by \xmm , for vignetting, and arf files have been obtained from 
Kyoko Matsushita.
The BCES regression software has been provided by Michael Akritas, 
Tina Bird, and Matthew Bershady.
A style file for the
deluxetable environment based on the AASTEX document class
has been obtained from Alexey Vikhlinin.
Figure~\ref{const} has been prepared based on Figures presented in
\citeauthor{dab00} and \citeauthor{bop99} with the help of
Barbara Mory.

The author acknowledges the dedicated work of and the
benefit from discussions with the \ro\ and \xmm\
hardware, software, and calibration teams at MPE.
Extensive use has been made of the \ro\ Data Archive at MPE,
of the NASA/IPAC Extragalactic Database (NED) 
which is operated by the Jet Propulsion Laboratory, Caltech, under contract
with the National Aeronautics and Space Administration, and
of observations obtained with \xmm\, an ESA
science mission with instruments and contributions directly funded by
ESA Member States and the USA (NASA).

\end{document}